\documentclass[openany,graybox,envcountsame,envcountsect]{svmono}

\usepackage{hyperref}
\usepackage{mathptmx}       
\usepackage{helvet}         
\usepackage{courier}        
\usepackage{type1cm}        
\usepackage{makeidx}         
\usepackage{graphicx}        
\usepackage{multicol}        
\usepackage[bottom]{footmisc}

\usepackage[small,nohug]{diagrams}
\diagramstyle[labelstyle=\scriptstyle]

\makeindex

\usepackage{amscd,amssymb,amsmath,latexsym,bm}
\usepackage{enumerate}
\usepackage[mathcal,mathscr]{euscript}
\usepackage{epsfig}
\usepackage{fancybox}
\usepackage{verbatim}
\usepackage{mathtools}
\usepackage{marginnote}
\usepackage{mathrsfs}
\usepackage{phonetic}
\usepackage{ulem}

\usepackage{paralist}

\newcommand{\CM}{{\mathbb C}}
\newcommand{\NM}{{\mathbb N}}
\newcommand{\RM}{{\mathbb R}}
\newcommand{\SM}{{\mathbb S}}
\newcommand{\TM}{{\mathbb T}}
\newcommand{\ZM}{{\mathbb Z}}
\newcommand{\PM}{{\mathbb P}}

\newcommand{\Aa}{{\cal A}}
\newcommand{\Ee}{{\cal E}}
\newcommand{\Pp}{{\cal P}}

\newcommand{\BB}{{\bf B}}

\newcommand{\Bb}{{\cal B}}

\newcommand{\Ff}{{\cal F}}

\newcommand{\Ww}{{\cal W}}
\newcommand{\Uu}{{\cal U}}

\newcommand{\Ss}{{\cal S}}
\newcommand{\Oo}{{\cal O}}
\newcommand{\Tr}{\mbox{\rm Tr}}
\newcommand{\Tt}{{\cal T}}
\newcommand{\Rr}{{\cal R}}

\newcommand{\Cc}{{\cal C}}

\newcommand{\Ll}{{\cal L}}

\newcommand{\Kk}{{\cal K}}
\newcommand{\Hh}{{\cal H}}

\newcommand{\one}{{\bf 1}}

\newcommand{\TR}{{\rm Tr}} 
\newcommand{\tr}{{\rm tr}} 
\newcommand{\ev}{{\mbox{\rm ev}}}
 
\newcommand{\Ch}{{\rm Ch}} 
\newcommand{\Ind}{{\rm Ind}} 
\newcommand{\Ker}{{\rm Ker}} 
\newcommand{\Ran}{{\rm Ran}} 
\newcommand{\sgn}{{\rm sgn}} 
\newcommand{\sign}{{\rm sign}} 
\newcommand{\Sig}{{\rm Sig}} 
\newcommand{\Exp}{{\rm Exp}} 
\newcommand{\PI}{{\rm \Pi}} 

\newcommand{\diag}{{\mbox{\rm diag}}}

\newcommand{\ess}{{\mbox{\rm\tiny ess}}}
\newcommand{\sym}{{\mbox{\rm\tiny sym}}}
\newcommand{\Wind}{{\mbox{\rm Wind}}}
\newcommand{\GFunc}{f_{\mbox{\rm\tiny Ind}}}
\newcommand{\FFunc}{f_{\mbox{\rm\tiny Exp}}}
\newcommand{\tFFunc}{\tilde{f}_{\mbox{\rm\tiny Exp}}}

\newarrow {Congruent} 33333

\begin{document}

\title{Bulk and Boundary Invariants for Complex Topological Insulators}

\subtitle{From $K$-Theory to Physics}

\author{Emil Prodan and Hermann Schulz-Baldes}


\maketitle

\frontmatter

\begin{dedication}
To our families.
\end{dedication}

\preface

Topological insulators are crystalline solids with supposedly very special properties. If stumbling upon such a crystal, which is possible because topological insulators are known to occur naturally on earth \cite{GBW}, a curious investigator will discover that the electrons deep inside the material are locked and they do not flow under electric field excitations. The immediate conclusion will be that the crystal is an insulator. However, when examining the surface of the crystal, our fictitious character will discover that the surface electrons are free to move like in a metal. Perhaps the first reaction will be to assign this odd behavior to surface contaminants and other factors like that, and the natural course of action will be to cleave a new surface and see what happens. To ones surprise, no matter how careful the new surface is cleaved, the metallic character is still present. There are many untold details to the story, but, broadly speaking, this is what a topological insulator ought to be. As the story suggests, the special properties must be determined by the bulk characteristics of the material, but there must be a bulk-boundary correspondence principle which tells how these bulk characteristics determine the metallic character of the surface. We should specify here, at the beginning, that although the properties of the topological insulators are ultimately determined by the number, type and arrangements of the atoms in the repeating cell of the crystal, the topology guaranteeing the metallic surface states is actually routed in the abstract space of electron ground states and has, for instance, nothing to do with the appearance and shape of the sample.

\vspace{.2cm}

One may be reminded of the Integer Quantum Hall Effect (IQHE) \cite{KDP}, where robust conducting channels occur along the edges of a specially prepared sample immersed in a relatively large magnetic field. In contradistinction, no magnetic fields were mentioned in the above story. The special properties of the topological insulators are intrinsic to the materials, which presumably will enable a broader range of applications.  It was Haldane \cite{Hal} who realized in 1988 that all the characteristics of the IQHE can occur naturally in materials with special unit cells and hopping matrices. The next milestone of the field occurred much later, in 2005, when Kane and Mele revealed that these special hopping matrices can be induced by the spin-orbit interaction \cite{KM1,KM2}. At the same time, they discovered a new class of topological materials, the quantum spin-Hall insulators in two space dimensions which have topologically non-trivial time-reversal symmetric ground states. These developments gathered momentum with the theoretical prediction \cite{BHZ} and then the experimental confirmation \cite{KWB} of the first quantum spin-Hall insulator, and then further with the theoretical prediction of new topological insulators in three space dimensions \cite{MB, FK1,FK2,Roy} and their experimental realization \cite{HQW}. The field of topological insulators is now fairly mature and there are several very good surveys \cite{QZ1,QZ2,HK,HM} and excellent monographs \cite{She,Ber,FMo,ORV}, where the reader can also find extensive literature on the subject. We want to mention in particular the short survey by Ando \cite{And}, which includes a table of 34 topological materials that were synthesized and characterized in laboratories, together with a summary of the findings for each compound. Ando's analysis reveals that, while the patterns seen in the surface electronic band structure agree quite well with the theoretical predictions, the transport experiments indicate a weak bulk metallic character ({\it i.e.} large, but nevertheless finite resistivity at low temperatures) for all these materials (excepting the 2-dimensional ones). Because of it, the transport characteristics of the surfaces were impossible to measure and the main conjecture about their metallic character is yet to be confirmed. The lack of insulating bulk character is usually attributed to the disorder in the samples, which is difficult to control for materials with such large and complex unit cells. A great deal of experimental effort was invested in overcoming this last hurdle, and one success has been recently reported for thin films \cite{BKS}. On the theoretical front, these issues prompted the need for theoretical methods which can handle more realistic models of topological insulators, in particular, to incorporate the effects of disorder. On the fundamental level, a rigorous proof of the conjectures on topological insulators in such real world conditions is highly desirable. 

\vspace{.2cm}

\noindent {\bf What are the main aims?} The present monograph is a mathematically rigorous contribution to the theory of so-called complex classes of topological insulators, namely those classes which are not specified by symmetries invoking a real structure, such as time-reversal or particle-hole symmetries (see Chapter~\ref{Chap-Physics} for a concise description). The main objectives are:

\begin{enumerate}[{Aim} 1:]
\item Construct the observable algebras within an effective one-particle framework.
\item Encode the non-trivial topology in bulk and boundary invariants which are robust against disorder and magnetic fields.  
\item Establish the equality between the bulk and the boundary invariants.
\item Determine the range of the invariants using generalized Streda formulas which connect different invariants.
\item Establish local index theorems for the so-called strong bulk and boundary invariants.
\item Prove the defining property of topological insulators, {\it i.e.} the immunity of the boundary states against Anderson localization.
\item Connect the invariants to response coefficients and other physical observables.
\end{enumerate}

\noindent {\bf Which mathematical tools are used?} The C$^*$-algebras describing the bulk systems are those introduced by Bellissard for the description of the quantum Hall effect and quasicrystals \cite{Bel}. Algebras describing half-space models and their boundaries are the extensions of the bulk algebras introduced in \cite{SKR,KRS}. These algebras form a short exact sequence of C$^*$-algebras, which is central to the bulk-edge correspondence. In the mathematical literature, these algebras are respectively well-known  as (twisted) crossed product algebras \cite{Ped, Wil} and their Toeplitz extensions, as given by Pimsner and Voiculescu \cite{PV}. In a first step, the topological invariants are encoded in the $K$-theory of these algebras. Based on the Pimsner-Voiculescu $6$-term exact sequence \cite{PV} and on \cite{Ell,Rie1,RiS}, these $K$-groups and their generators can be determined completely. In a second step, the $K$-theoretic content is extracted via pairings with the cyclic cohomology of the observables algebras, the latter being a key element of Connes' non-commutative geometry \cite{Con}. At this step, numerical invariants are generated and, for bulk systems, these invariants extend those known in the physics literature \cite{SRFL,QHZ,RSFL}. It is then possible to prove duality results for the connecting maps of $K$-theory, such as the suspension map, Bott map, index map, exponential map, and their counterparts in cyclic cohomology \cite{Pim,Nes,ENN,KRS}. This allows to connect various invariants. In particular, the bulk invariants are equal to the boundary invariants as well as the Volovik-Essin-Gurarie invariants calculated in terms of the Green functions \cite{Vol,EG1}. Another technique used here is that of Fredholm modules for index calculations, as introduced by Atiyah \cite{Ati} and further developed by Kasparov \cite{Kas} and Connes \cite{Con0}. This technique leads by rather elementary means to index theorems for the so-called strong invariants of topological insulators. Alternative mathematical approaches to the duality results behind the bulk-boundary correspondence were given in \cite{Had} and  \cite{BCR,Bou}. To achieve Aim 4, we use another technical tool, namely the Ito derivative w.r.t. the magnetic field, as introduced by Rammal and Bellissard \cite{RB} and further elaborated in \cite{ST}. Resuming, this monograph shows how a variety of abstract mathematical tools, ranging from C$^\ast$-algebras and their $K$-theories to non-commutative geometry, can be put to work on very concrete problems coming from solid state physics, and help resolve issues which are presently addressed in the physics community.

\vspace{.2cm}

\noindent {\bf What is new and what was known before?} The real space versions of the bulk invariants in arbitrary dimensions already appeared in our prior works \cite{PLB,MSHP,PS}, where also the index theorems for these invariants were proved. These works paralleled the much earlier work of Bellissard on two-dimensional quantum Hall systems \cite{Bel,Bel2,BES}. This approach to topological invariants allows to go beyond their definition based on Bloch theory, as it is usually done in the physics literature \cite{SRFL,QHZ}. The use of $K$-theory to connect with the invariants of Volovik \cite{Vol} and Essin-Gurarie \cite{EG1} is new. Also, the definitions of the boundary invariants for arbitrary dimensions is new, as are the index theorems for them. In the context of condensed matter physics, the connecting maps of $K$-theory were first put to work for integer quantum Hall systems, where they provided a structural framework for the proof of the equality between the bulk and edge Hall conductances, under quite general assumptions \cite{SKR,KRS,KS}. This series of works was heavily inspired by Hatsugai's work \cite{Hat1} on edge states for the Harper model. Actually, these works only used the exponential connecting map which is also applied to higher even dimensions here. A key new element of the present work is the use of the index map for chiral systems (see Sections~\ref{sec-Kenter} and \ref{sec-indBBC}), and actually for the much wider class of approximately chiral systems. The index map is the key to a sound definition of the boundary invariants and is also instrumental for the proof of the bulk-edge correspondence for chiral systems, from which the delocalized character of the boundary states follows (see Aim~6). 

\vspace{.2cm}

Another important new result is a generalized Streda formula and its corollaries on the ranges of the pairings of $K$-theory with cyclic cohomology. The classic Streda formula refers to the equality between the variation w.r.t. magnetic field of $0$-cocycle pairings (particularly, the density of states) and $2$-cocycle pairings (particularly, the Hall conductance) \cite{Str,RB,ST}. This equality will be generalized to cocycles of arbitrary dimensions and this will enable us to attach physical content to the abstractly defined topological invariants. As we shall see in Chapter~\ref{Chap-Conclusions}, the generalized Streda formula has numerous physical applications and unifies other results obtained in the literature \cite{QHZ,SF}. Further new results in Chapter~\ref{Chap-Conclusions} concern the stroboscopic interpretation of the orbital polarization, the connection of orbital polarization to spectral flow of boundary states and the prediction of a quantum Hall effect in approximately chiral systems in dimension $d=3$. Interestingly, the Hall conductance of these surface states is dictated by the bulk invariant.

\vspace{.2cm}

\noindent {\bf  What is left out?} There is no attempt here to deal with systems having time-reversal symmetry, particle-hole symmetry or reflection symmetries. There is an exhaustive physical literature on such systems starting with \cite{RSFL,QHZ,HPB}, and a few more mathematical oriented works \cite{HL,Pro0,ASV,GP,SB,FM,KZ,Thi1,GS} which  already proposed topological invariants for such systems. However, the bulk-boundary correspondence for these systems has only been established for very special situations \cite{ASV,GP,MT1,MT2}. Based on \cite{SB,Thi1,GS}, we expect that these symmetries can be accommodated in the framework developed here and that the bulk-boundary correspondence will follow for these systems, too, but this definitely requires further investigations. Even for the complex classes extensively treated here, $K$-theoretic techniques can supply further interesting results not included in the monograph. For example, in \cite{DS} it is shown that the Laughlin argument (piercing of a flux through a quantum Hall system and inducing an associated spectral flow) can be described by an exact sequence of C$^*$-algebras. This exact sequence is a mapping cone and is hence different from the exact sequence of the bulk-boundary correspondence. Nevertheless, the $K$-theory associated to that sequence links Hall conductance ({\it i.e.} Chern numbers) to a spectral flow and hence captures again the essence of Laughlin's argument. Implementing symmetries in this sequence allows to derive criteria for the existence of zero modes attached to flux tubes in dirty superconductors or Kramers bound states at defects in quantum spin Hall systems \cite{DS}. Another example are boundary forces \cite{Kel,KZo,Kel2,Pro7}. It is actually the firm belief of the authors that other defects, {\it e.g.} as described in \cite{TK,ITT,RGF}, can also be described by adequate sequences of C$^*$-algebras and the associated $K$-theoretic sequences can be used to uncover new interesting topological effects. From this perspective, the bulk-boundary correspondence can be seen as one particular situation where these ideas can be implemented, albeit probably the most important one. To further support this belief, we included here the stroboscopic interpretation of the orbital polarization as a further example. Behind it is a natural exact sequence associated to the suspension construction in $K$-theory. Let us mention that the use of exact sequences to connect topological invariants in physics is not restricted to solid state systems, but has also been successfully implemented in scattering theory to prove Levinson's theorem \cite{KR1,KR2,BS}.

\vspace{.2cm}

Concerning the index theory, let us first point out that it has very recently been shown \cite{Bou} how to obtain index theorems as stated in Chapter~\ref{Chap-IndexTheorems} by evaluating the general Connes-Moscovici local index formula \cite{CM} in a form proved in \cite{CGX} under much broader assumptions. The argument is close to \cite{Andr} and avoids using the intricate geometric identities discovered in \cite{Con,PLB,PS} and presented in Section~\ref{Sec-KeyId}, but the price are other technicalities. We decided to stay with the more direct arguments which rely on the Calderon-Fedosov formula \cite{Cal,Fed} for the Fredholm index and the above-mentioned geometric identities. On another front, we did not attempt any (generalized) index theorems for the so-called weak topological invariants. Such results are possible \cite{Pro4}, but one has to leave the realm of finitely summable Fredholm modules and work with semifinite spectral triples \cite{CPRS1,CPRS2,CPRS3}. Actually, the latter framework was shown to be fruitful in much broader contexts, in some cases even for correlated quantum systems \cite{PR,PRS1,PRS2,CPR1,CPR2}. This brings the hope that the electron-electron interaction can be treated by these techniques. This is one of the big open issues in the field and is not dealt with in the present work. We also decided not to include any numerical evaluation of the invariants. This will be presented elsewhere. While completing the manuscript, we came across the following works \cite{BCR,MT1,MT2} which open new directions and partially overlap with our presentation.

\vspace{.2cm}

\noindent {\bf How is the monograph organized?} Chapter~\ref{Chap-Illustration} illustrates the key concepts on perhaps the simplest of all topological systems, a lattice model with chiral symmetry in space dimension $d=1$. In this case, the bulk invariant is provided by the winding number of the so-called Fermi unitary operator and the edge effect consists in the emergence of zero-energy quantum states localized near the edge, called zero edge modes. The space of zero edge modes is invariant under the chiral symmetry, hence the zero modes have a specific chirality assigned to them. The bulk-boundary principle then asserts that the bulk invariant is equal to the number of zero edge modes with positive chirality minus the number of zero edge modes with negative chirality. As a result, if the bulk invariant is not zero, there will always be zero edge modes and their number is necessarily larger or equal to the value of the bulk invariant. This statement, which is proved here using a $K$-theoretic approach, holds in the presence of disorder and regardless of how the lattice is terminated at the edge, provided the chiral symmetry is always present. Along the way, many of the concepts used later on in the monograph are introduced. Actually the key ideas on how to use the index map for the bulk-boundary correspondence in chiral systems is already exposed in Chapter~\ref{Chap-Illustration}.

\vspace{.2cm}

Chapter~\ref{Chap-Physics} gives a brief overview of the classification table of topological insulators and superconductors \cite{SRFL,RSFL,Kit}, which is now accepted by the majority of the condensed matter physics community. The present work only deals with the first two rows of this table, the so-called unitary symmetry class A and the chiral unitary symmetry class AIII. They are also called the complex classes since they are classified by the complex $K$-theory while the remaining 8 classes are classified by real $K$-theory. The physics and the conjectures for the complex classes are presented in detail in Sections~\ref{Sec-UnitaryClass} and \ref{Sec-ChiralUnitaryClass}. These sections also provide simple models in arbitrary dimensions where the bulk-boundary principle can be witnessed first hand. The last section of Chapter~\ref{Chap-Physics} introduces the physical models which are studied in the remainder of the manuscript, together with technical conditions on these models.

\vspace{.2cm}

Chapter~\ref{Chap-Observables} introduces the operator algebras for bulk, half-space and boundary observables. Section~\ref{Sec-BulkAlgebra} describes the disordered non-commutative torus which plays the role of bulk algebra. This $C^\ast$-algebra can be presented as a $d$-fold iterated crossed product ($d $ is the dimension of the physical space) and it has a canonical representation on $\ell^2(\mathbb Z^d)$ which generates the bulk models discussed in Chapter~\ref{Chap-Physics}. Section~\ref{Sec-HalfSpaceAlgebra} then introduces the disordered non-commutative torus with a boundary. Here one of the unitary generators becomes a partial isometry which can be seen as introducing a defect. This algebra plays the role of the half-space algebra and it has a canonical representation on $\ell^2(\mathbb Z^{d-1}\times \mathbb N)$ which generates the physical models on a half-space. The algebra of boundary observables is a prime ideal of the half-space algebra. The elements of this algebra generate the boundary conditions. The exact sequence between the bulk, half-space and boundary algebras is also discussed in this chapter. The last sections of the chapter present the non-commutative analysis tools for the observables algebras and the smooth sub-algebras where this calculus actually takes place.

\vspace{.2cm}

Chapter~\ref{Chap-KTheory} presents the $K$-theory of the observables algebras. It begins with a concise description of the basic principles of $K$-theory. The exact sequence of Chapter~\ref{Chap-Observables} is shown to be isomorphic to the Pimsner-Voiculescu exact sequence \cite{PV} and the latter is then used to compute the $K$-groups. In particular, the $K$-groups of the bulk algebra and of the non-commutative torus coincide. For the latter, the generators of the $K$-groups have been computed explicitly by Elliott  \cite{Ell} and Rieffel \cite{Rie1} and we reproduce them in Section~\ref{Sec-GeneratorsKGroups}. Section~\ref{Sec-ConnectingMaps} computes various connecting maps between the $K$-groups of observables algebras. This section is central for the whole book.

\vspace{.2cm}

Chapter~\ref{Chap-TopologicalInvariants} invokes the cyclic cohomology and its pairing with the $K$-theory to define the bulk and the boundary topological invariants in terms of the Chern characters paired with the appropriate elements of the $K$-groups. It is shown how to suspend these invariants and that this suspension does not alter the values of the invariants. The equality between the bulk and boundary invariants is established using the duality between the pairings for bulk and boundary algebras. The range of these pairings is calculated using a generalized Streda formula. Detailed proofs are provided for all of these central results.

\vspace{.2cm}

Chapter~\ref{Chap-IndexTheorems} constructs finitely summable Fredholm modules canonically associated with the observables algebras. The pairings of the associated Connes-Chern characters with the $K$-groups are expressed as Fredholm indices. Section~\ref{Sec-Equality} establishes the equality between the Chern and Connes-Chern characters based on two remarkable geometric identities, which in turn provide the index formulas for the bulk and boundary invariants. The metallic character of the boundary states is established as a direct consequence of these index formulas.  

\vspace{.2cm}

Chapter~\ref{Chap-Conclusions} presents a series of corollaries which describe our physical predictions based on the mathematical statements from the previous chapters. The chapter starts with a brief introduction to the bulk and boundary transport coefficients of homogeneous disordered systems. These linear and non-linear coefficients are then connected to the bulk and boundary topological invariants for systems of class A. Predictions about the quantized values and the robustness of these physically measurable properties are provided. Similar results are presented for the spontaneous electric polarization and the magneto-electric response coefficients. For chiral symmetric solid state systems, the physically relevant quantities are the spontaneous chiral electric polarization and its variations w.r.t. magnetic fields, which are shown to be of topological nature and connected to the bulk and boundary invariants constructed for systems from class AIII. Again several of these measurable quantities have quantized vales. The chapter also includes a prediction and discussion of an IQHE at the surface of chiral or at least approximately chiral symmetric systems. The generalized Streda formula developed in Chapter~\ref{Chap-TopologicalInvariants} is an essential tool for the analysis in Chapter~\ref{Chap-Conclusions}.

\vspace{\baselineskip}
\begin{flushright}\noindent
New York, August 2015,\hfill {\it Emil Prodan}\\
Erlangen\hfill {\it Hermann Schulz-Baldes}\\
\end{flushright}

\extrachap{Acknowledgements}

The seed of this project was implanted when both authors attended the program ``Topological Phases of Quantum Matter" organized at Erwin Schršdinger International Institute for Mathematical Physics (ESI), Vienna, by Martin Zirnbauer in the summer of 2014. Their hospitality is greatly acknowledged. At that time, we could already sketch how a number of already available technical tools could be put to work to complete the bulk-boundary correspondence program for the complex classes of topological insulators in any space dimension. From discussions with the participants, we learned that there was a broad interest in the applications of $K$-theory in the field of topological insulators. This eventually convinced us to consider the book format, which enabled us to expand the presentation and to give it a more pedagogical style. Over the years and during the preparation of the manuscript, we benefited from great conversations with Jean Bellissard, Johannes Kellendonk,  Taylor Hughes, Giuseppe De Nittis, Martin Zirnbauer, Alan Carey, Terry Loring, Gian-Michele Graf, Stefan Teufel, Motoko Kotani, Carlos Villegas-Blas, and many others. We thank them all at this occasion.

\vspace{.3cm} 

Let us add that we are grateful for having received financial support. The work of E.~P. was supported by U.S. NSF grant DMR-1056168, that of H.S.-B. by the DFG through various grants.

\tableofcontents

\extrachap{Acronyms and Notations}

\begin{description}[CABRTTTT]

\item[BGH]{bulk gap hypothesis }

\item [MBGH]{mobility bulk gap hypothesis}

\item [CH]{chiral symmetry hypothesis}

\item [ACH]{approximate chiral symmetry hypothesis}

\item [CCR]{conventions on the Clifford representations}

\item [IQHE]{integer quantum Hall effect}

\item [${\rm sgn}$]{sign function}

\item [$\chi$]{characteristic function of a set}

\item [$\I$]{imaginary unit $\sqrt{-1}$}

\item [$d$]{dimension of physical space}

\item [$N$ and $2N$]{dimension of the fibers}

\item [$M_N(\CM)$]{$N\times N$ matrices with complex entries}

\item [{\rm diag}(A,B)]{block diagonal matrix built from matrices $A$ and $B$}

\item [$\mbox{\rm tr}$]{trace over finite dimensional Hilbert spaces}

\item [$\TR$]{trace over an infinite dimensional space}

\item [$| \cdot |$]{absolute value and matrix norm}

\item[$\| \cdot \|_{(s)}$]{$s$-Schatten norm}

\item [$\langle\,,\,\rangle$]{Euclidean scalar product and pairing of $K$-group with $K$-cycle}

\item [$\one$]{identity operator}

\item [$x,y$]{points on the lattice $\ZM^d$}

\item [$\Ff$]{discrete Fourier transform}

\item [$k$]{quasimomentum}

\item [$S_j,\widehat{S}_j$]{bilateral and unilateral shift in the $j$th space direction}

\item [$\one_N$]{identity in $M_N(\CM)$ or $M_N(\CM) \otimes \Aa$}

\item [$(\Omega,\ZM^d,\PM)$]{disorder config. space with $\ZM^d$-action and probability measure}

\item [$\omega\in\Omega$]{disorder configuration}

\item [$\Aa_d$, $\widehat{\Aa}_d$, $\Ee_d$]{C$^*$-algebras of bulk, half-space and boundary observables}

\item [$\partial_j$, $\widehat \partial_j$, $\widetilde \partial_j$]{derivation over $\Aa_d$, $\widehat \Aa_d$ and $\Ee_d$ algebras}

\item[$\Tt$, $\widetilde \Tt$]{canonical traces over $\Aa_d$ and $\Ee_d$}

\item[$\mathscr A_d$, $\mathscr E_d$]{smooth bulk and boundary sub-algebras}

\item[$L^s(\Aa_d,\Tt)$]{non-commutative $L^s$-spaces for bulk observables}

\item[$L^s(\Ee_d,\widetilde \Tt)$]{non-commutative $L^s$-spaces for boundary observables}

\item[$\mathcal W_{s,k}(\Aa_d,\Tt)$]{Sobolev spaces of first kind for bulk observables}

\item[$\mathcal W'_{s,k}(\Aa_d,\PM)$]{Sobolev spaces of second kind for bulk observables}

\item[$\mathcal W_{s,k}(\Ee_d,\widetilde \Tt)$]{Sobolev spaces of first kind for boundary observables}

\item[$\mathcal W'_{s,k}(\Ee_d,\PM)$]{Sobolev spaces of second kind for boundary observables}

\item[$\| \cdot \|_s$]{norm over non-commutative $L^s$-spaces}

\item[$\| \cdot \|_{s,k}$]{norm over Sobolev spaces of first kind}

\item[$\| \cdot \|'_{s,k}$]{norm over Sobolev spaces of second kind}

\item [$C^k(\Aa_d)$]{domain of the $k$-fold derivations}

\item [$\Aa^+$]{unitization of algebra $\Aa$}

\item [$\pi_\omega$, $\widehat \pi_\omega$, $\widetilde \pi_\omega$]{representations of  $\Aa_d$, $\widehat{\Aa}_d$ and $\Ee_d$ on the physical Hilbert space}

\item [$h$, $\hat h$, $\tilde h$] {bulk, half-space and boundary Hamiltonians}

\item[$H_\omega$, $\widehat H_\omega$, $\widetilde H_\omega$]{physical representations $\pi_\omega (h)$, $\widehat \pi_\omega (\hat h)$ and $\widetilde \pi_\omega (\tilde h)$ of Hamiltonians}

\item [$p_F$]{Fermi projection $\chi(h\leq \mu)$}

\item [$P_\omega$]{physical representation $\pi_\omega(p_F)$ of Fermi projection}

\item [$u_F$]{Fermi unitary operator}

\item [$U_\omega$]{physical representation $\pi_\omega(u_F)$ of the Fermi unitary operator}

\item [$\tilde p_\Delta$]{chiral boundary projection provided by the index map}

\item [$\GFunc$]{function used for lift in the index map}

\item [$\tilde u_\Delta$]{boundary unitary operator provided by the exponential map}

\item [$\FFunc$]{function used for lift in the exponential map}

\item [$J$]{chirality operator}

\item [$\BB$]{the magnetic field}

\item [$V^x$, $U^x$]{magnetic and dual magnetic translations by $x \in \ZM^d$}

\item [$\sigma_j$]{Pauli matrices and representation of odd complex Clifford algebra}

\item [$\gamma_j$]{representation of even complex Clifford algebra}

\item [$\Ss_n$]{symmetric group over $n$ points}

\item [$D$]{Dirac operator defining $K$-cycle}

\item [$E$]{Hardy projection $\chi(D>0)$}

\item [$F$]{phase of Dirac operator $\sgn(D)$}

\item [$G$]{unitary encoding chiral Dirac phase of even $K$-cycle}

\end{description}

\mainmatter

\chapter{Illustration of key concepts in dimension $d=1$}
\label{Chap-Illustration}

\abstract{This introductory chapter presents and illustrates many of the key concepts developed in this work on a simple example, namely the Su-Schriefer-Heeger model \cite{SSH} of a conducting polymer. This model has a chiral symmetry and non-trivial topology, given by a non-commutative winding number which is remarkably stable against perturbations like a random potential  \cite{MSHP}. Hence this is a relatively simple example of a topological insulator. Here the focus is on the bulk-boundary correspondence in this model, which connects the winding number to the number of edge states weighted by their chirality. This connection will be explained in a $K$-theoretic manner. These arguments constitute a rather mathematical introduction to the bulk-edge correspondence and the physical motivations and insights will be given in the following chapters. }

\section{Periodic Hamiltonian and its topological invariant}
\label{sec-period1d}

As a general rule, the topology in topological insulators is always inherited from periodic models and this topology can be shown in many instances to be stable under perturbations which also break the periodicity. It is therefore instructive to start out with a detailed analysis of the periodic models and to identify their topological invariants. The one-dimensional periodic Hamiltonian $H$ considered here acts on the Hilbert space $\CM^2\otimes \CM^N\otimes\ell^2(\mathbb Z)$ and is given by
\begin{equation}
\label{Model1d}
H
\; =\;
\tfrac{1}{2}(\sigma_1+\I \sigma_2)\otimes\one_N\otimes S + \tfrac{1}{2}(\sigma_1-\I \sigma_2)\otimes\one_N\otimes S^\ast +m \,\sigma_2\otimes\one_N\otimes\one
\;,
\end{equation}
where $\one_N$ and $\one$ are the identity operators on $\CM^N$ and $\ell^2(\ZM)$ and the $2\times 2$ Pauli matrices are
$$
\sigma_1\;=\;\begin{pmatrix} 0\; & 1 \\ 1 & 0 \end{pmatrix}
\;,
\qquad
\sigma_2\;=\;\begin{pmatrix} 0 & -\I \\ \I & \;0 \end{pmatrix}
\;,
\qquad
\sigma_3\;=\;\begin{pmatrix} 1 & \;0 \\ 0 & -1 \end{pmatrix}
\;,
$$
and $S$ is the right shift on $\ell^2(\mathbb Z)$ while $m\in\RM$ is the mass term. The component $\CM^2\otimes \CM^N$ of the Hilbert space will be referred to as the fiber. This Hamiltonian goes back to Su, Schrieffer and Heeger \cite{SSH} and its physical origin will be discussed in Section~\ref{SubSec-Exp2}. It has a chiral symmetry w.r.t. the real unitary $J=\sigma_3\otimes\one_N\otimes\one$ squaring to the identity
\begin{equation}
\label{eq-1dchiral}
J^*\,H\,J\;=\;-\,H
\;.
\end{equation}
The Fermi level $\mu$ is always assumed positioned at $0$ for chiral symmetric systems, see Chapter~\ref{Chap-Physics}. Note that a model with chiral symmetry can display a spectral gap at $\mu=0$ only if the fiber has even dimension, which is obviously the case here.

\vspace{.2cm}

The discrete Fourier transform $\Ff:\ell^2(\mathbb Z)\to L^2(\SM^1)$ defined by
$$
(\Ff\phi)(k)
\;=\;
(2\pi)^{-\frac{1}{2}}
\;\sum_{x\in\ZM}\phi_{x}\;e^{-\I\langle x| k\rangle}
\;,
$$
partially diagonalizes the Hamiltonian to $\Ff H\Ff^*=\int_{\SM^1}^\oplus dk\;H_k$ with
$$
H_k
 =\;
\tfrac{1}{2}(\sigma_1+\I \sigma_2)\otimes \one_N \,e^{-\I k}\; +\; \tfrac{1}{2}(\sigma_1-\I \sigma_2)\otimes \one_N \,e^{\I k} \;+\;m\, \sigma_2 \otimes \one_N
$$
or
$$
H_k \;=\;
\begin{pmatrix}
0 & e^{-\I k}-\I m \\ e^{\I k} +\I m & 0
\end{pmatrix}
\otimes \one_N
\;.
$$
Also the chiral symmetry operator diagonalizes $\Ff J\Ff^*=\int_{\SM^1}^\oplus dk\,J_k$, even with constant fibers $J_k=\sigma_3\otimes\one_N$. The two eigenvalues of $H_k$ are 
$$
E_\pm(k)\;=\;
\pm \sqrt{m^2+1-2m\sin(k)}
\;,
$$
and both are $N$-fold degenerate. Their symmetry around $0$ reflects the chiral symmetry $J_k H_kJ_k=-H_k$ which, as for any Hamiltonian with chiral symmetry, implies $\sigma(H_k)=-\sigma(H_k)$. The central gap around $0$ is $\Delta=[-E_g,E_g]$ with $E_g=\big ||m|-1 \big |$. Hence it  is open as long as $m\not\in\{-1,1\}$. Let us also note that for $m=0$, one has $E_\pm(k)=\pm 1$ for all $k$, namely the two bands are flat. In fact, one readily checks that the eigenfunctions of $H$ are supported on two neighboring sites each.

\vspace{.2cm}

In the mean-field approximation, which will be assumed throughout, the electron ground state is encoded in the Fermi projection $P_F=\chi(H \leq \mu)$ and we recall that in the chiral symmetric models one fixes  $\mu = 0$ to ensure the charge neutrality of the system. Since we are in dimension one, this projection cannot be used to define a topological invariant (other then the electron density), and we should rather look for a unitary operator.  Note that $J P_FJ=\one-P_F$ and therefore the so-called flat band Hamiltonian  
$$
Q\;=\;\one-2P_F\;=\;\mbox{\rm sgn}(H)
$$
satisfies again $J^* QJ=-Q$. It also satisfies $Q^2=\one$, hence its spectrum consists of only two eigenvalues, $1$ and $-1$, which are both infinitely degenerate. The chiral symmetry combined with $Q^2=\one$ implies the existence of a unitary $U_F$ on $ \CM^N\otimes \ell^2(\mathbb Z)$ such that
\begin{equation}
\label{eq-QUrep}
Q
\;=\;
\begin{pmatrix}
0 & U_F^\ast \\ U_F & 0
\end{pmatrix}
\;.
\end{equation}
In analogy with the Fermi projection, this unitary operator $U_F$ will be called the Fermi unitary operator. The existence of the Fermi unitary operator is a generic characteristic of chiral symmetric gapped Hamiltonians. Note that $U_F$ can be constructed entirely from the electron ground state and, reciprocally, the electron ground state can be reconstructed entirely from $U_F$. Also, note that in the physics literature and in our previous work \cite{PS} $U_F$ and $U_F^\ast$ are interchanged. The choice in \eqref{eq-QUrep} will prove more convenient here, especially when computing the index map, see below.

\vspace{.2cm}

For the Hamiltonian \eqref{Model1}, one readily calculates $\Ff Q\Ff^*=\int_{\SM^1}^\oplus dk\,Q_k$, with
$$
Q_k\;=\;
\begin{pmatrix}
0 & \frac{e^{-\I k} +\I m }{|e^{-\I k} +\I m|} 
\\ 
\frac{e^{\I k} +\I m }{|e^{\I k} +\I m|}  & 0
\end{pmatrix}
\otimes \one_N
\;.
$$
In general, every flat band Hamiltonian of a periodic chiral Hamiltonian with open central gap is fibered as
$$
Q_k
\;=\;
\begin{pmatrix}
0 & U_k^\ast \\ U_k & 0
\end{pmatrix}
\;,
$$
with some unitary matrix $U_k\in M_N(\CM)$ acting on $\CM^N$ which is supposed to be differentiable in $k$. It is now natural to consider the winding number associated to the Fermi unitary operator, which for reasons explained further below will be called the first odd Chern number:
\begin{equation}
\label{eq-windnumber}
\Ch_{1}(U_F)
\;=\;
\I\;\int_{\SM^1} \frac{dk}{2\pi}\;\tr\bigr(U_k^*\partial_kU_k\bigr)
\;.
\end{equation}
For the Hamiltonian \eqref{Model1} one finds
$$
\Ch_{1}(U_F)
\;=\;
\left\{
\begin{array}{cc}
 -\,N \;,\;\;\;\;\;\;\; & m\in(-1,1)\;,
\\
\;\;0\;,\;\;\;\;\;\;\; & m\not\in [-1,1]\;.
\end{array}
\right.
$$
This integer $\Ch_{1}(U_F)$ is the bulk invariant associated to the ground state of Hamiltonian \eqref{Model1d}. The term {\sl invariant} reflects the fact that $\Ch_{1}(U_F)$ does not change for sufficiently small perturbations of the Hamiltonian, even though $U_F$ itself does change. In particular, the following perturbations are of interest:

\begin{enumerate}[\rm (i)]

\item Next nearest hopping terms.

\item A random potential or random hopping elements.

\item Terms breaking the chiral symmetry  \eqref{eq-1dchiral}.

\end{enumerate}

\noindent The perturbations (i) and (iii) can be dealt with in the framework of periodic operators where a Bloch Floquet transform is applicable. If the chiral symmetry is broken, then the flat-band Hamiltonian is not described as in \eqref{eq-QUrep} by a unitary anymore, but it  may still have invertible off-diagonal entries of which a winding number is well-defined as well. For the random perturbations in (ii) one is forced out of the realm of Bloch theory. One of the main points to be developed further down is to show how this can be accomplished. Of course, another question addressed is to find the adequate replacement for $\Ch_{1}(U_F)$ for higher dimensions.

\section{Edge states and bulk-boundary correspondence}
\label{sec-edge1d}

In this section, an edge or boundary for the one-dimensional periodic Hamiltonian \eqref{Model1d} is introduced. This can be achieved by simply restricting \eqref{Model1d} to the half-space Hilbert space $\CM^2\otimes\CM^N\otimes\ell^2(\NM)$, {\it e.g.} by imposing the Dirichlet boundary condition
$$
\widehat H
\; =\;
\tfrac{1}{2}(\sigma_1+\I \sigma_2)\otimes\one_N\otimes \widehat S 
\;+\; 
\tfrac{1}{2}(\sigma_1-\I \sigma_2)\otimes\one_N\otimes \widehat S^\ast 
\;+\;
m \,\sigma_2\otimes\one_N\otimes\one
\;.
$$
All half-space operators will carry a hat from now on. For example, $\widehat S$ above is the unilateral right shift on $\ell^2(\NM)$ and there is the half-space chirality operator $\widehat{J}=\sigma_3\otimes\one_N\otimes\one$. The half-space Hamiltonian still has the chiral symmetry $\widehat{J}\;\widehat{H}\;\widehat{J}=-\widehat{H}$.  Again the chiral symmetry implies that the spectrum satisfies $\sigma(\widehat{H})=-\sigma(\widehat{H})$. Furthermore, the direct sum of two copies of $\widehat{H}$ is a finite dimensional perturbation of $H$. Hence the essential spectra coincide $\sigma_\ess(H)=\sigma_\ess(\widehat{H})$, but $\widehat{H}$ may have additional point spectrum, corresponding to the edge states which are also called bound or boundary states.  

\begin{example}
Let us consider the Hamiltonian $\widehat H$ for $m=0$. It takes the form
$$
\widehat H
\;=\;
\begin{pmatrix}
0 & \one_N\otimes \widehat S \\
\one_N\otimes \widehat S^* & 0
\end{pmatrix}
\;.
$$
The spectrum is now $\sigma(\widehat{H})=\{-1,0,1\}$ with infinitely degenerate eigenvalues $\pm 1$ having compactly supported eigenstates on two neighboring sites, and a kernel of multiplicity $N$ containg vectors supported in the upper entry over the boundary site $0$. They result from the fact that $|0\rangle\in\ell^2(\NM)$ lies in the kernel of the unilateral left shift $\widehat S^*$. For $N=1$, this zero mode is simple and perturbations of the Hamiltonian $\widehat H$ within the class of half-sided chiral Hamiltonians cannot remove it since the symmetry of the spectrum has to be conserved and a simple eigenvalue cannot split into two by perturbation theory. The same stability actually holds for $N>1$ because the signature of $\widehat{J}$ on the kernel is $N$ and also this signature is conserved during a homotopy of chiral Hamiltonians. Note also that the signature is equal to $N=-\Ch_1(U_F)$. Due to the stability of both quantitities, the equality $\Ch_1(U_F)=-\mbox{Sig}(\widehat{J}|_{\Ker(\widehat{H})})$ holds also in a neighborhood of the Hamiltonian $\widehat H$ with $m=0$.
\hfill $\diamond$
\end{example}

Now let us go on with a more strucutral analysis of the edge states which is not as tightly linked to the special model under considertion. Suppose that $\psi\in\CM^2\otimes\CM^N\otimes\ell^2(\NM)$ is such a normalized bound state with energy $E$, namely $\widehat{H}\psi=E\psi$. Then $\widehat{H}\;\widehat{J}\,\psi=-E\,\widehat{J}\,\psi$, which implies that the span $\Ee$ of all eigenvectors with eigenvalues in $[-\delta,\delta]\subset \Delta$ is invariant under $J$. Therefore $\widehat{J}$ can be diagonalized on $\Ee$  leading to a splitting $\Ee=\Ee_+\otimes\Ee_-$ such that $\widehat{J}$ is $\pm 1$ on $\Ee_\pm$.  Accordingly, the spectral projection $\widetilde{P}(\delta)=\chi(|\widehat{H}|\leq \delta)$ can be decomposed into an orthogonal sum $\widetilde{P}(\delta)=\widetilde{P}_+(\delta)+\widetilde{P}_-(\delta)$ and $\widehat J \; \widetilde P(\delta) = \widetilde{P}_+(\delta)-\widetilde{P}_-(\delta)$. The difference of the dimensions of $\Ee_\pm$ spaces is the boundary invariant of the system 
$$
{\rm Tr}(\widehat J \; \widetilde{P}(\delta))
\;=\;N_+-N_-
\;,
\qquad
N_\pm\;=\;\dim(\Ee_\pm)
\;.
$$ 
This invariant is also equal to the signature of $\widehat{J}|_\Ee$ and such signatures are again well-known to be homotopy invariants, as already pointed out in the example above. The invariant is independent of the choice of $\delta>0$ as long as $\delta$ lies in the gap of $H$, hence its value must be determined entirely by the spectral subspace of the zero eigenvalue, known also as the space of the zero modes. Zero modes in $\Ee_+$ and $\Ee_-$ are said to have positive and negative chirality, respectively. The following result now connects the bulk invariant $\Ch_1(U_F)$ to the boundary invariant ${\rm Tr}(\widehat J \; \widetilde{P}(\delta))$.

\begin{theorem}
\label{theo-1d}
Consider the Hamiltonian $H$ on $\CM^2\otimes\CM^{N}\otimes\ell^2(\ZM)$ given by \eqref{Model1d} and let $\widehat{H}$ be its half-space restriction. If $U_F$ is the Fermi unitary operator defined via {\rm \eqref{eq-QUrep}} and if its winding number is defined by {\rm \eqref{eq-windnumber}}, then the bulk-edge correspondence in the following form holds
\begin{equation}
\label{eq-bulkedge1d}
\Ch_1(U_F)
\;=\;
-\,{\rm Tr}(\widehat J \; \widetilde{P}(\delta))
\;.
\end{equation}
\end{theorem}

This result can be proved by various means (see the above example and \cite{EG1,EG2}, but likely there are other references). However, in the following, a detailed $K$-theoretic proof will be provided.  Such a structural argument stresses the robust nature of the above equality. In particular, stability under the perturbations listed at the end of Section~\ref{sec-period1d} will be covered. Furthermore, it will be possible to extend the structural argument to higher dimensional systems.

\section{Why use $K$-theory?}
\label{sec-Kenter}

There have been numerous works that use $K$-theory for topological condensed matter systems. Pioneering were the papers by Bellissard on the integer quantum Hall effect \cite{Bel,Bel2}, which were reviewed and extended to the regime of dynamical Anderson localization in \cite{BES}. $K$-theory can be used to obtain gap labelling \cite{Bel}. Starting with the Kitaev's paper \cite{Kit}, $K$-theory and $KR$-theory (which is $K$-theory in presence of symmetries) were more recently used as a tool to classify topological insulators \cite{SCR,FM,DG,Thi1,KZ,MFM} or define topological invariants in the absence of periodicity \cite{HL,LH}. Here the main objective is a different one:

\begin{enumerate}[$\bullet$]

\item Use the connecting maps of $K$-theory to relate different invariants.

\end{enumerate}

This was first achieved in \cite{SKR,KRS,KS} for integer quantum Hall systems, where the equality of bulk and edge Hall conductivity was proved using the exponential map of $K$-theory. There are other connecting maps in $K$-theory though, in particular the index map, the suspension map and the Bott periodicity map. In this work it will be shown how they can be put to work as well and produce interesting identities. In this introductory section on the one-dimensional Su-Schrieffer-Heeger model, the $K$-theoretic index map of the so-called Toeplitz extension will be used to prove Theorem~\ref{theo-1d}. Along the lines, quite a few things about $K$-theory will said and used without proof. These are all standard facts that are well-known in the mathematics community and can be found in the introductory books on $K$-theory \cite{WO,RLL} or the more advanced textbook \cite{Bla0}, but for the convenience of the reader they will be briefly reviewed in Section~\ref{sec-Ktheory} of Chapter~\ref{Chap-KTheory}.

\vspace{.2cm}

The Toeplitz extension is at the very heart of $K$-theory. The reader familiar with all this can jump directly to Proposition~\ref{prop-1d}. The Toeplitz extension is the following short exact sequence of C$^*$-algebras:
\begin{equation}
\label{ToeplitzExtension}
\begin{diagram}
0 & \rTo & \Kk & \rTo{i}  & T\big(C(\SM^1) \big ) \,\cong\, C^\ast(\widehat S \, ) & \rTo{ \mathrm{ev} } & C(\SM^1) \,\cong \,C^\ast(S) & \rTo & 0
\end{diagram}
\end{equation}
Here $\Kk$ denotes the algebra of compact operators on $\ell^2(\NM)$, $C(\SM^1)$ is the algebra of continuous functions over the unit circle which, by the discrete Fourier transform, is isomorphic with the algebra generated by the shift operator $S$ on $\ell^2(\ZM)$, and $T(C(\SM^1))$ is the algebra of Toeplitz operators. The latter can be presented as the C$^\ast$-algebra of operators on $\ell^2(\NM)$ which can be approximated in operator norm by polynomials in $\widehat{S}$ and $\widehat{S}^\ast$, that is, by finite sums
$$
\sum_{n,m\geq 0} a_{n,m} (\widehat S)^n (\widehat S^\ast)^{m} \;.
$$
Since 
\begin{equation}
\label{HatS}
\widehat S^\ast \, \widehat S \;=\; \one \quad {\rm and} \quad \widehat S \;\widehat S^\ast \;=\;\one - \widetilde{P}
\;,
\end{equation} 
where $\widetilde{P}=|0\rangle \langle 0|$ is the one-dimensional projection on the state $|0\rangle\in\ell^2(\NM)$ at the boundary,
the operators from $T(C(\SM^1))$ can be uniquely expressed as:
\begin{equation}
\label{HatS2}
\sum_{n\geq 0} a_n (\widehat S)^n 
\;+\; 
\sum_{n < 0} a_n (\widehat S^\ast)^{-n} 
\;+\; 
\sum_{n,m \geq 0} c_{n,m} (\widehat S)^n \widetilde{P} (\widehat S^\ast)^m 
\;.
\end{equation} 
One can now see explicitly the connection between the Toeplitz operators and the half-line observables. Indeed, the first two terms in \eqref{HatS2} represent the restriction of the bulk operator $\sum_{n\in \mathbb Z} a_n  S^n$ to the half-line via the Dirichlet boundary condition, while the third term redefines the boundary condition. The latter is just a compact operator on $\ell^2(\mathbb N)$, hence $\Kk$ is a sub-algebra of $T(C(\SM^1))$ and $i$ in \eqref{ToeplitzExtension} denotes the associated inclusion map. The second morphism in \eqref{ToeplitzExtension} is defined by ${\rm ev}(\widehat S) = e^{-\I k}$ and ${\rm ev}({\widehat S}^*) = e^{\I k}$, or equivalently ${\rm ev}(\widehat S) = S$ and ${\rm ev}({\widehat S}^*) = S^\ast$. Since $\widetilde{P}= \widehat S^\ast \widehat S-\widehat S \widehat S^\ast$, one has ${\rm ev}(\widetilde{P}) = 0$ which means that the compact operators are sent to zero by the second morphism. As a consequence, the sequence \eqref{ToeplitzExtension} is exact, namely the image of each of the three maps is equal to the kernel of the following map.

\vspace{.2cm}

All the operators appearing above lie in matrix algebras over one of the algebras in the Toeplitz extension \eqref{ToeplitzExtension}. Indeed, the Hamiltonian $H$ as well as $P_F$ and $Q$ belong to the algebra $M_{2N}(C(\SM^1))\cong \CM^{2N\times 2N} \otimes C(\SM^1)$ of $2N\times 2N$ matrices with coefficients in $C(\SM^1)$, and the half-line Hamiltonian $\widehat{H}$ is an element of  $M_{2N}(T(C(\SM^1)))$. Actually, $\widehat{H}$ is a so-called lift of $H$, namely, one has ${\rm ev}(\widehat{H})=H$. The Fourier transform of the Fermi unitary operator $U_F$ lies in $M_N(C(\SM^1))$. Finally, the finite dimensional projections $\widetilde{P}(\delta)$ and $\widetilde{P}_{\pm}(\delta)$ are projections in $M_{2N}(\Kk)$.

\vspace{.2cm}

\noindent {\bf Warning:} {\it From here on, $K$-theoretic concepts will be used and only explained on an intuitive level. Details are found in Chapter~\ref{Chap-KTheory}.}

\vspace{.2cm}

The proof of Theorem~\ref{theo-1d} will show how the equality \eqref{eq-bulkedge1d} results from a $K$-theoretic index theorem associated to the Toeplitz extension. The definitions of $K$-groups and of the index map are recalled in Section~\ref{sec-Ktheory}. Roughly stated, for each C$^*$-algebra $\Aa$ there exist two groups $K_0(\Aa)$ and $K_1(\Aa)$ given by homotopy classes of projections and unitaries, respectively, in the matrix algebras over $\Aa$. The group operation in $K_0(\Aa)$ is given by the direct sum of projections, while in $K_1(\Aa)$ by the multiplication of unitaries. The $K$-groups of all algebras in the Toeplitz extension \eqref{ToeplitzExtension} are well-known: $K_0(\Kk)\cong \ZM$ generated by the rank one projection $\widetilde{P}=|0\rangle\langle 0|$, $K_0(T(C(\SM^1)))\cong \ZM$ and $K_0(C(\SM^1))\cong \ZM$ both generated by the identity, $K_1(\Kk)=0$ and $K_1(T(C(\SM^1)))=0$, and finally $K_1(C(\SM^1))\cong \ZM$ generated by $e^{-ik}$ (or $S$) which is a function with unit winding number. The elements $K_1(C(\SM^1))$ can be uniquely labeled by their winding number, namely the first odd Chern number. It is also worth pointing out that the class $[\widetilde{P}]_0$ in $K_0(T(C(\SM^1)))$ is trivial  because the isometry $\widehat{S}$ satisfies $\widehat{S}^*\,\widehat{S}=\one$ and $\widehat{S}\,\widehat{S}^*=\one-\widetilde{P}$. Hence $\one$ and $\one-\widetilde{P}$ are Murray-von Neumann equivalent and are therefore in the same $K_0$-class, to that $[\widetilde{P}]_0=[\one]_0-[\one-\widetilde{P}]_0=0$. On the other hand, in $K_0(\Kk)$ the projection $\widetilde{P}$ defines a non-trivial class which is actually the generator of $K_0(\Kk)$.

\vspace{.2cm}

The central result of $K$-theory used for the bulk-boundary correspondence is that, for every exact sequence of C$^*$-algebras, there is a $6$-term exact sequence of the $6$ associated $K$-groups. For the Toeplitz extension, this sequence is
\begin{equation}\label{SixTermDiagram}
\begin{diagram}
& K_0(\Kk)=\ZM & \rTo{ i_\ast } & K_0(T(C(\SM^1)))=\ZM & \rTo{\ \ {\rm ev}_\ast \ \ } & K_0(C(\SM^1))=\ZM &\\
& \uTo{\rm Ind} & \  &  \ & \ & \dTo{\rm Exp} & \\
& K_1(C(\SM^1))=\ZM  & \lTo{{\rm ev}_\ast} & K_1(T(C(\SM^1)))=0 & \lTo{i_\ast} & K_1(\Kk)=0 &
\end{diagram}
\end{equation}
Here the maps $i_\ast$ and ${\rm ev}_\ast$ are push-forward maps naturally induced by the maps in \eqref{ToeplitzExtension}. Interesting are the so-called boundary maps $\Exp$ and $\Ind$. The exponential map $\Exp$ has to be trivial for the Toeplitz extension as $K_1(\Kk)=0$. Focus will therefore be on the index map ${\rm Ind}$, which has to be an isomorphism. First of all, note that it maps classes of unitaries from the bulk algebra $C(\SM^1)$ to projections in the boundary algebra $\Kk$. Hence it establishes a link between the topology of the bulk and the boundary, which is precisely what we are looking for. Let us first recall the general definition of the index map (as already pointed out, more details and a more stringent formulation using utilizations are given in Section~\ref{sec-Ktheory}) and then evaluate it explicitly. Given a class $[U]_1\in K_1(C(\SM^1))$ associated to a unitary $U\in M_N(C(\SM^1))$, one first constructs a unitary lift
$$
\widehat{W}
\;=\;
\mbox{\rm Lift} \,\begin{pmatrix} U\; & 0 \\ 0\; & U^* \end{pmatrix}
\;\in\;
M_{2N}(T(C(\SM^1)))
\;,
$$
which is by definition a unitary satisfying ${\rm ev}(\widehat{W})=\mbox{\rm diag}(U,U^*)$, and then defines
\begin{equation}
\label{eq-IndIntro}
\Ind([U]_1)
\;=\;
\left[
\widehat{W}\begin{pmatrix} \one_N & 0 \\ 0 & 0 \end{pmatrix}\widehat{W}^*
\right]_0
\;-\;
\left[
\begin{pmatrix} \one_N & 0 \\ 0 & 0 \end{pmatrix}
\right]_0
\;.
\end{equation}
In general, it can be shown that the lift exists and that the r.h.s. of \eqref{eq-IndIntro} really specifies an element in $K_0(\Kk)$ and not in $K_0(T(C(\SM^1)))$, as one may think at first sight.

\vspace{.2cm}

Let us first calculate $\Ind([S^n]_1)$ for the bilateral left shift $S^n$ by $n$ sites. These unitaries generate $K_1(C(\SM^1))=\{[S^n]_1\,|\,n\in\ZM\}$. A unitary lift for $n\geq 0$ is
$$
\mbox{\rm Lift} \,\begin{pmatrix} S^n & 0 \\ 0 & (S^n)^* \end{pmatrix}
\;=\;
\begin{pmatrix} \widehat{S}^n & \ \widetilde{P}_n \\ 0 & (\widehat{S}^n)^* \end{pmatrix} 
\;,
$$
where as above $\widehat{S}$ is the unilateral right shift and $\widetilde{P}_n=\sum_{k=1}^{n}|k\rangle\langle k|$ is the projection on the $n$ states localized at the boundary of $\ell^2(\NM)$. Hence $\widehat{S}^n(\widehat{S}^*)^n=\one-\widetilde{P}_n$ and $\widetilde{P}_n\widehat{S}^n=0$.  Evaluating \eqref{eq-IndIntro} now shows
\begin{equation}
\label{eq-IndShift}
\Ind([S^n]_1)
\;=\;
\left[
\,
\begin{pmatrix}
\widehat{S}^n\,(\widehat{S}^*)^n & 0  \\ 0 & 0
\end{pmatrix}
\,
\right]_0
\;-\;
\left[
\,
\begin{pmatrix}
\one & 0  \\ 0 & 0
\end{pmatrix}
\,
\right]_0
\;=\;
-\,[\widetilde{P}_n]_0
\;,
\end{equation}
which is the explicit form of the isomorphism between $K_1(C(\SM^1))$ and $K_0(\Kk)$. This concludes our description of the $K$-theory associated to the Toeplitz extension \eqref{ToeplitzExtension}.

\vspace{.2cm}

Now let us come to the application to the model \eqref{Model1}. First of all, the Fermi unitary $U_F$ in \eqref{eq-QUrep} defines a class in $K_1(C(\SM^1))$, and the finite dimensional projections $\widetilde{P}(\delta)$ and $\widetilde{P}_{\pm}(\delta)$ specify classes in $K_0(\Kk)$. Hence they lie in the l.h.s. of the six term exact sequence \eqref{SixTermDiagram} for the Toeplitz extension \eqref{ToeplitzExtension} and they are connected via the index map. In fact, the following holds.

\begin{proposition}
\label{prop-1d}
Let $U_F\in M_N(C(\SM^1))$ be given by {\rm \eqref{eq-QUrep}}. Further let us choose an odd and non-decreasing smooth function $\GFunc:\RM\to [-1,1]$ such that $\GFunc(E)=-1$ for $E\leq -E_g$ and $\GFunc(E)=1$ for $E\geq E_g$. Then
\begin{equation}
\label{eq-bulkedge1dInd0}
\Ind([U_F]_1)
\;=\;
\left[
e^{-\I\frac{\pi}{2}\GFunc(\widehat{H})}
\,
\begin{pmatrix}
\one_N & 0  \\ 0 & 0 
\end{pmatrix}
\,e^{\I\frac{\pi}{2}\GFunc(\widehat{H})}
\right]_0
\;-\;
\left[
\begin{pmatrix}
\one_N & 0  \\ 0 & 0
\end{pmatrix}
\right]_0
\;.
\end{equation}
\end{proposition}

\noindent {\bf Proof.} For the evaluation of the index map \eqref{eq-IndIntro} one needs the lift
$$
\widehat{W}
\;=\;
\mbox{\rm Lift} \begin{pmatrix} U_F & 0 \\ 0 & U_F^* \end{pmatrix} 
\;=\;
\mbox{\rm Lift}\left( \,\begin{pmatrix} 0 & \one_N \\ \one_N & 0 \end{pmatrix} \begin{pmatrix} 0 & U_F^\ast \\ U_F & 0 \end{pmatrix} \,\right)
\;=\;
\begin{pmatrix} 0 & \one_N \\ \one_N & 0 \end{pmatrix} \; \mbox{\rm Lift}(Q)
\;.
$$
Now recall that $Q=\mbox{\rm sgn}(H)$ is a self-adjoint unitary that will now be expressed as a smooth function of $H$ with values on the unit circle. Actually, with the function $\GFunc$ defined in the proposition, one has $Q=\I e^{-\I\frac{\pi}{2}\GFunc(H)}$. Hence a lift is given by
$$
\mbox{\rm Lift}(Q)
\;=\;
\I\, e^{-\I\frac{\pi}{2}\GFunc(\widehat{H})}
\;.
$$
As it is obtained by smooth functional calculus from $\widehat{H}$, it follows that $\mbox{\rm Lift}(Q)\in M_{2N}(T(C(\SM^1)))$ as required. We arrived at
$$
\widehat{W}
\;=\;
\I \, \begin{pmatrix} 0 & \one_N \\ \one_N & 0 \end{pmatrix} \; e^{-\I\frac{\pi}{2}\GFunc(\widehat{H})}
\;.
$$
Plugging into the definition \eqref{eq-IndIntro} of the index map
\begin{align*}
\Ind([U_F]_1)
& =\!
\left[
\begin{pmatrix} 0 & \one_N \\ \one_N & 0 \end{pmatrix}
e^{-\I\frac{\pi}{2}\GFunc(\widehat{H})}
\begin{pmatrix} \one_N & 0  \\ 0 & 0 \end{pmatrix} 
e^{\I\frac{\pi}{2}\GFunc(\widehat{H})}
\begin{pmatrix} 0 & \one_N \\ \one_N & 0 \end{pmatrix}
\right]_0
\!-
\left[
\begin{pmatrix}
\one_N & 0  \\ 0 & 0
\end{pmatrix}
\right]_0   ,
\end{align*}
and the projection appearing in the first term is homotopic to the projection appearing in the statement.
\hfill $\Box$

\vspace{.2cm}

The previous argument did not require the presence of any spectral gaps in the spectrum of $\widehat{H}$ and will therefore also apply to higher dimensional models, see Proposition~\ref{IndMap} below. In presence of  spectral gaps, however, one can further refine the argument.

\begin{proposition}
\label{prop-1dbis}
Let $U_F\in M_N(C(\SM^1))$ be given by {\rm \eqref{eq-QUrep}}.  Then for $0<\delta<E_g$
\begin{equation}
\label{eq-bulkedge1dInd}
\Ind([U_F]_0)
\;=\;
[\widetilde{P}_{+}(\delta)]_0\;-\;[\widetilde{P}_{-}(\delta)]_0
\;.
\end{equation}
\end{proposition}

\noindent {\bf Proof.} Let $\GFunc$ be as in Proposition~\ref{prop-1d} and, moreover, let it be such that $\GFunc(E)\in\{-1,0,1\}$ for any $E\in\sigma(\widehat{H})$. For sake of concreteness, suppose $\GFunc(E)=0$ only for $E=0$ and no other $E\in\sigma(\widehat{H})$. Recall that, in dimension $d=1$, the spectrum of $\widehat H$ is discrete inside $[-E_g,E_g]$. Now,
$$
e^{-\I\frac{\pi}{2}\GFunc(\widehat{H})}
\,
\diag(\one_N,0_N)
\,e^{\I\frac{\pi}{2}\GFunc(\widehat{H})}
\;=\;
e^{-\I\frac{\pi}{2}\GFunc(\widehat{H})}
\,
\tfrac{1}{2}\,(\widehat J+\one_{2N} ) 
\,e^{\I\frac{\pi}{2}\GFunc(\widehat{H})}
\;.
$$
The chiral symmetry of $\widehat{H}$ combined with $\GFunc(-E)=-\GFunc(E)$, for $E \in \sigma(\widehat H)$, implies
$$
e^{-\I\frac{\pi}{2}\GFunc(\widehat{H})} \,\widehat J\;=\; \widehat J\,e^{\I\frac{\pi}{2}\GFunc(\widehat{H})} 
\;,
$$
so that
$$
e^{-\I\frac{\pi}{2}\GFunc(\widehat{H})}
\,
\tfrac{1}{2}\,(\widehat J + \one_{2N}) 
\,e^{\I\frac{\pi}{2}\GFunc(\widehat{H})} 
\;=\;  
\tfrac{1}{2}\;\widehat J(e^{\I \pi \GFunc(\widehat{H})}+\one_{2N}) + \diag(0_N,\one_N) .
$$
With the choice made for $\GFunc$ one has $e^{\I \pi \GFunc(\widehat{H})}+\one_{2N}=2\widetilde P(\delta)$, so that
$$
\tfrac{1}{2}\;\widehat J \; (e^{\I \pi \GFunc(\widehat{H})}\,+\,\one_{2N})
\;=\; 
\widehat J \; \widetilde P(\delta) 
\;=\; 
\widetilde P_+(\delta) \,-\, \widetilde P_-(\delta).
$$
Then, by noticing that $\widetilde P_+(\delta)$ and $\diag(0_N,\one_N)- \widetilde P_-(\delta)$ are orthogonal projections and that $\diag(\one_N,0_N)$ and $\diag(0_N,\one_N)$ are homotopic,
\begin{align*}
\Ind([U_F]_1] & \;=\; [\widetilde P_+(\delta) \,+\, \diag(0_N,\one_N)\,-\, \widetilde P_-(\delta)]_0\;-\;
[\diag(\one_N,0_N)]_0 \\
 & \;=\; [\widetilde P_+(\delta)] \;+\;[\diag(0_N,\one_N) \,-\, \widetilde P_-(\delta)]_0 \;-\; [\diag(0_N,\one_N)]_0 \; .
 \end{align*}
The statement now follows from the rule 3. of the standard characterization of the $K_0$ group, listed in Section~\ref{sec-K0props}.
\hfill $\Box$

\section{Why use non-commutative geometry?}

Theorem~\ref{theo-1d} results by extracting a numerical identity from the $K$-theoretic idendity \eqref{eq-bulkedge1dInd}. This is done via a pairing of the $K$-groups with adequate cohomology theory, which is the cyclic cohomology developed by Connes since the early 1980's \cite{Con0,Con}. This was at the heart of the early developments of non-commutative geometry. Actually, it could also be referred to as non-commutative differential topology as topological invariants are calculated by tools of non-commutative differential and integral calculus. In the simple framework of periodic models, the relevant pairings of $K$-theory with cyclic cohomology are established by the two maps
\begin{align}
& \widetilde{\Ch}_0\;:\;K_0(\Kk)\to\ZM
\;,
\qquad
& &
\widetilde{\Ch}_0([\widetilde{P}]_0-[\widetilde{P}']_0)\;=\;
\Tr(\widetilde{P})-\Tr(\widetilde{P}')\;,
\label{eq-Ch0per}
\\
& \Ch_1\;:\;K_1(C(\SM^1))\to\ZM
\;,
\qquad
& &
\Ch_1([U]_1)\;=\;
\I\,\int_{\SM^1} \frac{dk}{2\pi}\;\tr\bigr(U(k)^*\partial_kU(k)\bigr)
\;,
\label{eq-Ch1per}
\end{align}
where in the second line it is supposed that $k\mapsto U(k)$ is differentiable. Any continuos path $k\mapsto U(k)$ can be approximated by a differentiable one, which means that any $K$-theory class in $K_1(C(\SM^1))$ has differentiable representatives simply because the smooth functions C$^\infty(\SM^1)$ are dense in $C(\SM^1)$. Such arguments are always needed in differential topology, and also in non-commutative differential topology, where it is necessary to work with dense  subalgebras (of smooth elements) of C$^*$-algebras. This issue will be discussed in detail in Section~\ref{Sec-Sobolev}. The term pairing expresses the fact that $\widetilde{\Ch}_0([\widetilde P]_0)$ and  $\Ch_1([U]_1)$ do not depend on the choice of the representative of the two classes. The following result now connects the two pairings.

\begin{proposition}
\label{prop-1dpairings} The maps $\widetilde \Ch_0$ and $\Ch_1$ are well-defined group homomorphisms into the additive group $\ZM$, and
\begin{equation}
\label{eq-1dpairing}
\Ch_1([U]_1)
\;=\;
-\;\widetilde \Ch_0(\Ind([U]_1))
\;.
\end{equation}
\end{proposition}

\noindent {\bf Proof.} Neither of the pairings depends on the representatives, namely, norm continuous paths of projections and unitaries, respectively, have constant pairings. Furthermore, $\widetilde \Ch_0([\widetilde{P}]_0+[\widetilde{P}']_0)=\widetilde \Ch_0([\widetilde{P}]_0)+\widetilde\Ch_0([\widetilde{P}']_0)$ holds by definition and  elementary properties of the winding number imply $\Ch_1([UU']_1)=\Ch_1([U]_1)+\Ch_1([U']_1)$. Finally the equality \eqref{eq-1dpairing} follows once it is verified for every class. But 
$$
\Ch_1([S^n]_1)
\;=\;
n
\;=\;
\Tr(\widetilde{P}_n)
\;=\;\widetilde\Ch_0([\widetilde{P}_n])
\;=\;
-\,\widetilde\Ch_0(\Ind([S^n]_1))
\;,
$$
where in the last equality \eqref{eq-IndShift} was used. Actually, it would have been sufficient to check the above equality for the (sole) generator $n=1$.
\hfill $\Box$

\vspace{.2cm}

\noindent {\bf Proof} of Theorem~\ref{theo-1d}. This follows by combining Propositions~\ref{prop-1d} and \ref{prop-1dpairings}.
\hfill $\Box$

\section{Disordered Hamiltonian}

The next step is to add a random perturbation to the Hamiltonian \eqref{Model1}, just as in \cite{MSHP}. Let  $\omega'_x,\omega''_x\in[-\frac{1}{2},\frac{1}{2}]$ be independent and uniformly distributed random variables and define a disorder configuration in the Tychonov space $\Omega=([-\frac{1}{2},\frac{1}{2}]\times[-\frac{1}{2},\frac{1}{2}])^{\ZM}$ by $\omega=(\omega'_x,\omega''_x)_{x\in\ZM}$. The probability measure on $\Omega$ is just the product measure. The associated Hamiltonian $H_\omega$ for two coupling constants $\lambda',\lambda''\geq 0$ is still acting on $\ell^2(\mathbb Z,\CM^2\otimes \CM^N)$ and is given by
\begin{align}
\label{eq-Hamdis}
H_\omega
\;=\; &
\sum_{x\in\ZM}
\tfrac{1}{2}(1+\lambda'\omega'_x)
\big (
(\sigma_1+\I \sigma_2)\,|x\rangle\langle x+1| 
\;+\; (\sigma_1-\I \sigma_2)\,|x+1\rangle\langle x|
\big)
\nonumber
\\
&\;+\;m(1+\lambda''\omega''_x) \,\sigma_2\;|x\rangle\langle x|
\;.
\end{align}
For $\omega=0$ or $\lambda' = \lambda''=0$, the Hamiltionian $H_\omega$ is exactly the same as \eqref{Model1}. From now on, the letter $H$ will be used for the full family $H=\{H_\omega\}_{\omega\in\Omega}$ of random Hamiltonians. The spectra $\sigma(H_\omega)$ of these operators are known to be almost surely and given by $\sigma(H_\omega)=\sigma(H_{\omega=0})+[-\lambda',\lambda']+[-\lambda'',\lambda'']$.

\vspace{.2cm}

As we have already seen, the periodic model exhibits a non-trivial topological phase and, according to \cite{ABF,MSHP,SP,PS,ABK}, this phase is stable against disorder. This means that the trivial and topological phases continue, in the presence of disorder, to be separated by a sharp phase boundary where a localization-delocalization transition must occur. This phase boundary is characterized by a divergence of the Anderson localization length and it can be mapped using transport experiments. The existence of such sharp phase boundary can be established by an analytical calculation, which we reproduce below from \cite{MSHP}. To simplify notations, let us use $t_x=(1+\lambda'\omega'_x)$ and $m_x=m(1+\lambda''\omega''_x)$, in which case the Schr\"{o}dinger equation at the Fermi level $E =0$ for \eqref{eq-Hamdis} reads
$$
\begin{pmatrix} 0 & t_x \\ 0 & 0 \end{pmatrix} \begin{pmatrix} \psi_{x+1,+1} \\ \psi_{x+1,-1} \end{pmatrix} 
\;+\; 
\begin{pmatrix} 0 & 0 \\ t_x & 0 \end{pmatrix} \begin{pmatrix} \psi_{x-1,+1} \\ \psi_{x-1,-1} \end{pmatrix} 
\;+\; \I \begin{pmatrix} 0 &  - m_x \\ m_x & 0 \end{pmatrix} \begin{pmatrix} \psi_{x,+1} \\ \psi_{x,-1} \end{pmatrix} 
\;=\; 0
\;.
$$ 
On the components, $t_x\psi_{x-\alpha,\alpha}+\I\alpha m_x \psi_{x,\alpha} =0$, $\alpha = \pm1$, hence the solution is
$$
\psi_{\xi_\alpha+x,\alpha}
\;=\;
\I^x \prod_{j=1}^{x} \left( \frac{t_j}{m_j}\right ) ^\alpha \psi_{\xi_\alpha,\alpha}
\;,
$$
where $\xi_\alpha = 0,1$ for $\alpha=\pm 1$, respectively. The inverse of Anderson localization length is given by
$$
\Lambda^{-1} 
\;=\;
\max_{\alpha=\pm 1}\big [- \lim\limits_{x\rightarrow \infty} \frac{1}{x}\log|\psi_{\xi_\alpha+x,\alpha}| \big ] 
\;=\;
\left|\lim\limits_{x\rightarrow \infty} \frac{1}{x}\sum_{j=1}^{x}(\ln |t_j|-\ln|m_j|)\right|
\;.
$$
Using Birkhoff's ergodic theorem \cite{Bir} on the last expression,  
$$
\Lambda^{-1} 
\;=\;
\left | \ \int_{-1/2}^{1/2}d\omega' \int_{-1/2}^{1/2}d\omega'' \ (\ln|1+\lambda' \omega'| - \ln|m+\lambda'' \omega''|)\right |
\;.
$$
The integrations can be performed explicitly and, in the regime of large $\lambda$'s where the arguments of the logarithms (inside the absolute values) take negative to positive values as $\omega$'s are varied, the result is
\begin{equation}\label{LocLength1}
\Lambda^{-1}
\;=\;
\left | \ln \left [\frac{|2+\lambda'|^{\frac{1}{\lambda'}+\frac{1}{2}}}{|2-\lambda'|^{\frac{1}{\lambda'}-\frac{1}{2}}} \  \frac{| 2m-\lambda''|^{\frac{m}{\lambda''}-\frac{1}{2}}}{|2m+\lambda''|^{\frac{m}{\lambda''}+\frac{1}{2}}}  \right ] \right |
\;.
\end{equation}
One can now check that, indeed, the Anderson localization length diverges for certain values of $\lambda'$ and $\lambda''$. A plot of the manifold where this occurs can be found in \cite{MSHP} and there one can see that the topological phase is indeed fully enclosed by this manifold. In other words, the only way to cross from the topological to the trivial phase is to go through a localization-delocalization quantum transition. As we shall see, it is exactly this divergence of the localization length which triggers an abrupt change in the quantized values of the bulk topological invariant.

\vspace{.2cm}

While the bulk analysis, just by itself, can be carried in the regime of strong disorder, the bulk-boundary correspondence will be established under the following assumption:

\vspace{0.2cm}

\noindent {\bf Bulk Gap Hypothesis} {\sl $E_g=\inf \sigma(H_\omega)\cap\RM_{\geq 0}$ is positive, namely $0\not\in\sigma(H_\omega)$.}

\vspace{.2cm}

\noindent Each $H_\omega$ still has the chiral symmetry \eqref{eq-1dchiral}, that is $J H_\omega J=-\,H_\omega$, and therefore also the flat band Hamiltonian  $Q_\omega=\one-2P_\omega=\mbox{\rm sgn}(H_\omega)$ satisfies $J Q_\omega J=-Q_\omega$ and $Q_\omega^2=\one$. This implies as in \eqref{eq-QUrep}
\begin{equation}
\label{eq-Udisdef}
Q_\omega
\;=\;
\begin{pmatrix}
0 & U_\omega^\ast \\ U_\omega & 0
\end{pmatrix}
\; ,
\end{equation}
with a unitary operator $U_\omega$ on $\ell^2(\mathbb Z,\CM^N)$. The aim in the following is to show that Theorem~\ref{theo-1d} remains valid provided that the disorder does not close the gap and the invariant $\Ch_1(U)$ is adequately defined.

\vspace{.2cm}

Neither of the operators $H_\omega$, $U_\omega$ and $Q_\omega$ is periodic anymore, but this lack is replaced by the so-called covariance relation, explained next. First of all, on $\Omega$ one has an $\ZM$-action $\tau:\ZM\times \Omega\to\Omega$ given by 
$$
\omega\;=\;(\omega'_x,\omega''_x)_{x\in\ZM} 
\;\;\mapsto \;\;
\tau \omega\;=\;(\omega'_{x-1},\omega''_{x-1})_{x\in\ZM}\; ,
$$ 
and with this action one has
\begin{equation}
\label{eq-cova}
S\,H_\omega\,S^\ast
\;=\;
H_{\tau\omega}
\;.
\end{equation}
Similar covariance relation applies to any function of the Hamiltonian (such as $Q_\omega$) or to operators extracted from such functions (such as $U_\omega$). 

\section{Why use operator algebras?}
\label{sec-Algenter}

A fruitful point of view \cite{Bel} is to consider the whole C$^*$-algebra $\Aa_1$ of one-dimensional covariant operator families on $\ell^2(\ZM)$, which is constructed as follows. One starts with the set $\Aa_{1,0}$ of families $a=\{A_\omega\}_{\omega\in\Omega}$ of operators on $\ell^2(\ZM)$ satisfying the covariance relation $SA_\omega S^*=A_{\tau \omega}$ as well as the finite range condition $\langle x|A_\omega|y\rangle = 0$ for all $|x-y|>C$ for some $C<\infty$. Then  $\Aa_{1,0}$ is a $*$-algebra because the product and adjoint of finite range covariant operator families is again such a family.  A C$^*$-norm on $\Aa_{1,0}$ is defined by
$$
\|a \|
\;=\;
\sup_{\omega\in\Omega}\;\|A_\omega\|
\;,
$$
where on the right we have the standard operator norm. Then $\Aa_{1}$ is the C$^\ast$-algebra given by the closure of $\Aa_{1,0}$ under this norm. Elements in $\Aa_1$ are covariant families of bounded operators having decaying off-diagonal matrix elements and will still be denoted by $a=\{A_\omega\}_{\omega\in\Omega}$. Note the lower case notation, which will be use throughout for elements of the algebras, while the upper case letters will be reserved for operators on the physical Hilbert space. While $\Aa_1$ was defined as algebra of covariant operator families with certain decay conditions, it is isomorphic to the C$^*$-algebraic  (reduced) crossed product algebra $C(\Omega)\rtimes_\alpha \ZM$ of $C(\Omega)$  w.r.t. the $\ZM$-action $\alpha(f)(\omega)=f(\tau^{-1}\omega)$ on $C(\Omega)$.  The isomorphism is 
$$
\{A_\omega\}_{\omega \in \Omega}\; \mapsto\; a\in C(\Omega\times\ZM)
\;, 
\qquad a(\omega,x)\;=\;\langle 0|A_\omega|x\rangle
\;,
$$
which associates a continuous function over $\Omega\times\ZM$ to every covariant operator family. This identification of $\Aa_1$ with the crossed product algebra will tacitly be used below, and further stressed and explored in the higher dimensional cases. The Hamiltonian $h=\{H_\omega\}_{\omega\in\Omega}$, the flat band Hamiltonian $q=\{Q_\omega\}_{\omega\in\Omega}$ and the Fermi unitary $u_F=\{U_\omega\}_{\omega\in\Omega}$ are all elements of matrix algebras over $\Aa_1$. One crucial fact is that the $1$-periodic (or translation invariant) operators are also covariant, and actually identified with those covariant operator families which do not depend on $\omega$. Hence the algebra of periodic operators $C(\SM^1)$ (in its Fourier transformed representation) is a (closed) subalgebra of $\Aa_1$. This implies that the generators of the $K$-groups of $C(\SM^1)$ also specify elements of the $K$-groups of $\Aa_1$. In fact, even more holds, namely the $K$-groups coincide.

\begin{proposition}
\label{prop-1dhomotop} The $K$-groups of $\Aa_1$ are
$$
K_0(\Aa_1)\;=\;\ZM\;,
\qquad
K_1(\Aa_1)\;=\;\ZM\;,
$$
and the generators are the same as those of $C(\SM^1)$, namely $\one$ and $S$ respectively.
\end{proposition}

\noindent {\bf Proof.} We will check that $C(\SM^1)$ is a deformation retract of $\Aa_1 =C(\Omega)\rtimes \ZM$ and this implies that $K_j(\Aa_1)=K_j(C(\SM^1))$ \cite[Section~6.4]{WO}. The key for this is the contractibility of $\Omega$ to one point which we choose to be $0=(0,0)_{x\in\ZM}$. Indeed, $\gamma_\lambda: \Aa_1\to \Aa_1$ defined by 
\begin{equation}
\label{eq-deform}
(\gamma_\lambda a)(\omega,x)
\;=\;
a(\lambda\omega,x)\;,
\qquad
\lambda\in[0,1]
\;,
\end{equation}
is a continuous family (in $\lambda$) of continuous morphisms which connects $\gamma_1=\mbox{\rm id}_{\Aa_1}$ to a right inverse $\gamma_0:\Aa_1\to C(\SM^1)$ of the inclusion map $i:C(\SM^1)\to\Aa_1$ by a continuous path.
\hfill $\Box$

\vspace{.2cm}

The algebra $\Aa_1$ (and matrix algebras over it) contains covariant operator families on $\ell^2(\ZM)$. The edge algebra is now $\Ee_1=C(\Omega)\otimes \Kk$ and the half-space algebra is $\widehat{\Aa}_1=\Aa_1\oplus \Ee_1$ as a direct sum of vector spaces, but not algebras. Operators in $\widehat{\Aa}_1$ are concretely given by the sum of a half-space restriction of a covariant operator in $\Aa_1$ and a compact operator in $\Ee_1$, namely 
$$
\hat a 
\;=\; (a,\tilde k)
\;=\;
\{\PI A_\omega \PI^*+K_\omega\}_{\omega\in\Omega}
\;,
$$ 
if $a=\{A_\omega\}_{\omega\in\Omega}\in\Aa_1$ and $\tilde k=\{K_\omega\}_{\omega\in\Omega}\in\Ee_1$, and where $\PI:\ell^2(\ZM)\to\ell^2(\NM)$ denotes the partial isometry with $\PI\PI^*=\one_{\ell^2(\NM)}$ and projection $\PI^*\PI$ in $\ell^2(\ZM)$ onto $\ell^2(\NM)\subset\ell^2(\ZM)$. The product and adjoint in $\widehat{\Aa}_1$  and $\Ee_1$ are naturally inherited from the operator product on $\ell^2(\NM)$. Exactly as in \eqref{ToeplitzExtension}, one has an exact sequence of C$^*$-algebras
\begin{equation}
\label{eq-toepextdisoredered}
\begin{diagram}
0 & \rTo & \Ee_1 & \rTo{i}  & \widehat{\Aa}_1 & \rTo{ \mathrm{ev} } & \Aa_1 & \rTo & 0
\end{diagram}
\end{equation}
The detailed construction of these algebras will be given in Chapter~\ref{Chap-Observables}.  Again, various operators constructed from the disordered Hamiltonian $h=\{H_\omega\}_{\omega\in\Omega}\in\Aa_1$ are in this sequence. The half-space restriction $\widehat{h}=\{\widehat{H}_\omega\}_{\omega\in\Omega}$ is an element of a matrix algebra over the Toeplitz extension $\widehat{\Aa}_1$ as is the lift of $q=\{Q_\omega\}_{\omega\in\Omega}\in\Aa_1$. Furthermore, the projections $\tilde{p}_{\pm}(\delta)=\{\widetilde{P}_{\pm,\omega}(\delta)\}_{\omega\in\Omega}$ on bound states, constructed for every $\omega$ just as in Section~\ref{sec-period1d} by splitting $\widetilde{P}_{\omega}(\delta)=\chi(H_\omega\in[-\delta,\delta])$ with $\delta<E_g$  into $\pm 1$ eigenspaces of $\widehat J$, lie in $\Ee_1=C(\Omega)\otimes \Kk$, and they define a class in the $K_0$-group of this C$^*$-algebra. It is worth pointing out that both projections $\widetilde{P}_{\pm,\omega}(\delta)$ are indeed continuous and, in particular, do not change dimension. On the other hand, the covariant family of Fermi unitaries $u_F=\{U_{\omega}\}_{\omega\in\Omega}$ defined in \eqref{eq-Udisdef} specify a class in $K_1(\Aa_1)$. Now the index map of the $K$-theoretic exact sequence associated with \eqref{eq-toepextdisoredered} connects these two classes, namely by exactly the same proof as given for \eqref{eq-bulkedge1dInd}, one shows the following.

\begin{proposition}
\label{prop-1ddisorered}
Let $u_F=\{U_{\omega}\}_{\omega\in\Omega}\in M_N(\Aa_1)$ be given by {\rm \eqref{eq-Udisdef}} and $\tilde{p}_{\pm}(\delta)=\{\widetilde{P}_{\pm,\omega}(\delta)\}_{\omega\in\Omega}$ the projections on the zero energy bound states of positive and negative chirality, respectively. Then, with the $K$-theoretic index map associated to the exact sequence {\rm \eqref{eq-toepextdisoredered}},
\begin{equation}
\label{eq-bulkedge1dIndDisoredered}
\Ind([u_F]_1)
\;=\;
[\tilde{p}_{+}(\delta)]_0-[\tilde{p}_{-}(\delta)]_0
\;.
\end{equation}
\end{proposition}

\section{Why use non-commutative analysis tools?}
\label{sec-NCenter}

The equivalent of Theorem~\ref{theo-1d}, namely Theorem~\ref{theo-1ddisordered} below, will again follow by extracting numbers from the $K$-theoretic identity \eqref{eq-bulkedge1dIndDisoredered}. For this purpose, one has to extend the definitions \eqref{eq-Ch0per} and \eqref{eq-Ch1per} of the cyclic cocycles $\widetilde \Ch_0$ and $\Ch_1$ to the operator algebra $\Aa_1$ describing disordered systems. The generalization of $\Ch_0$ is 
\begin{equation}
\label{eq-Ch0dis}
\widetilde \Ch_0([\tilde{p}]_0-[\tilde{p}']_0)\;=\;
\int \PM(d\omega)\;
\big(\Tr(\widetilde{P}_\omega)\;-\;\Tr(\widetilde{P}'_\omega)\big)\;.
\end{equation}
Actually, by continuity, the map $\omega\mapsto\Tr(\widetilde{P}_\omega)\in\ZM$ is constant and therefore the average $\PM$ over the disorder is not necessary. As to $\Ch_1$, the definition \eqref{eq-Ch1per} involves differentiation in Fourier space and this now has to be replaced by non-commutative differentiation. For any finite range operator $a=\{A_\omega\}_{\omega\in\Omega}\in\Aa_{1,0}$, one defines its derivative $\partial a\in\Aa_{1,0}$ by
$$
\partial a (\omega,x)
\;=\;
-\,\I\,x\,a (\omega,x)
\;.
$$
This definition can be extended to so-called differentiable operators $a\in\Aa_1$ as long as the r.h.s. defines an operator in $\Aa_1$. The set of differentiable operators is denoted by C$^1(\Aa_1)$. By iteration one defines C$^n(\Aa_1)$, and then C$^\infty(\Aa_1)=\bigcap_{n\geq 1}C^n(\Aa_1)$. The latter is a Fr\'{e}chet algebra, clearly dense in $\Aa_1$, that is invariant under holomorphic functional calculus. It follows \cite{GVF} that the algebraic $K$-groups $K_j(C^\infty(\Aa_1))$ are equal to the topological $K$-groups $K_j(\Aa_1)$ for $j=0,1$. Operators in this sub-algebra are sufficiently regular for differential topology. Apart from differentiation,  a non-commutative integration tool is needed. A state $\Tt$ on $\Aa_1$ is defined by
$$
\Tt(a)
\;=\;
\int \PM(d\omega)
\;\langle 0|A_\omega|0\rangle
\;=\;
\int \PM(d\omega)
\;a(\omega,0)
\;,
\qquad
a=\{A_\omega\}_{\omega\in\Omega}
\;.
$$
In fact, it is a trace that is invariant under $\partial$, as shows the following lemma.

\vspace{.5cm}

\begin{lemma}
\label{lem-trace} The following holds.

\begin{enumerate}[\rm (i)]

\item For $a,b\in\Aa_1$, one has $\Tt(ab)=\Tt(ba)$.

\item For $a\in C^1(\Aa_1)$, one has $\Tt(\partial a)=0$.

\item For $a,b\in C^1(\Aa_1)$, one has $\Tt(\partial a\,b)=-\Tt(a\,\partial b)$.

\item For a translation invariant $a\in \Aa_1$ with  Fourier transform $k\in\SM^1\mapsto a(k)$,  one has $\Tt(a)=\int_{\SM^1}\frac{dk}{2\pi}\,a(k)$.

\item For a translation invariant $a\in C^1(\Aa_1)$, one has $(\partial a)(k)=\partial_k a(k)$ where  $k\in\SM^1\mapsto a(k)$ 
and $k\in\SM^1\mapsto (\partial a)(k)$ are the Fourier transforms.

\end{enumerate}

\end{lemma}

The straightfoward proof is left to the reader. Finally, one can introduce
\begin{equation}
\label{eq-ch1}
\Ch_1(u)\;=\;
\I\;
\Tt(u^{-1}\,\partial u)
\;,
\qquad
u\in C^1(\Aa)
\;.
\end{equation}
Let us point out that, for translation invariant $u$, this reduces precisely to \eqref{eq-windnumber}.

\begin{proposition}
\label{prop-Ch1ddisorered}
$\Ch_1$ is a homotopy invariant, namely for any continuous path $\lambda\in [0,1]\mapsto u(\lambda)\in C^1(\Aa)$ the number $\Ch_1(u(\lambda))$ is constant.
\end{proposition}

\noindent {\bf Proof.} First of all, $u\mapsto \Ch_1(u)$ is continuous and therefore the path $\lambda\in [0,1]\mapsto u(\lambda)$ can be approximated by a differentiable one. For such a differentiable path,
\begin{align*}
-\I\;\partial_\lambda\;
\Ch_1(u(\lambda))
& = \;
\Tt(\partial_\lambda u^{-1}\,\partial u)
\;+\;\Tt(u^{-1}\,\partial \partial_\lambda u)
\\
& =\;
-\,\Tt(u^{-1}\partial_\lambda u\,u^{-1}\,\partial u)
\;-\;\Tt(\partial u^{-1}\, \partial_\lambda u)
\;,
\end{align*}
where in the second equality Lemma~\ref{lem-trace}(iii) was used. As $\partial u^{-1}=-u^{-1}\partial u\,u^{-1}$ one concludes that $\partial_\lambda\,\Ch_1(u(\lambda))=0$ and this completes the proof.
\hfill $\Box$

\vspace{0.2cm}

The physical model is defined over $\CM^{2N} \otimes \ell^2(\ZM)$ rather than just $\ell^2(\ZM)$ and $U_\omega$ is actually defined over $\CM^{N} \otimes \ell^2(\ZM)$. As one can see, most of the time we will be dealing with the matrix algebras over $\Aa_1$. The non-commutative calculus can be trivially extended to cover these cases, by replacing $\Tt$ by $\Tt \otimes {\rm tr}$, where $\rm tr$ is the trace over the fiber. We now can finally define the bulk invariant for the disordered chiral system, as ${\rm Ch}_1(u_F)$. Based on the above result, we can state at once that, if $h(\lambda)$ is a smooth deformation of $h$ such that its central spectral gap remains open, then $u_F(\lambda)$ varies smoothly in $M_N(\CM) \otimes C^1(\Aa_1)$ and, consequently, ${\rm Ch}_1(u_F)$ remains unchanged.

\section{Why prove an index theorem?}

Proposition~\ref{prop-Ch1ddisorered} implies that $\Ch_1$ only depends on the $K_1$-class of its argument so that one may write $\Ch_1(u)=\Ch_1([u]_1)$. The homotopy invariance can, in particular, be applied to the homotopy $u_\lambda=\gamma_\lambda(u)$ with $\gamma_\lambda$ defined in  \eqref{eq-deform}. This implies $\Ch_1(u)=\Ch_1(u_0)$ for $u\in C^1(\Aa_1)$. Now $u_0\in C^1(\Aa_1)$ is translation invariant and therefore $\Ch_1(u_0)$ can be calculated by \eqref{eq-windnumber} as a winding number. In particular, this shows that $\Ch_1(u)\in\ZM$. An alternative way to verify the integrality of $\Ch_1(u)$ is to prove an index theorem. This has the advantage that one can also prove that the pairing is well-defined and integral in the regime of a mobility bulk gap, namely, when the Fermi level lies in a region of the essential spectrum which is dynamically Anderson localized. This type of extension is crucial for the understanding of the quantum Hall effect \cite{BES} and will be discussed further in Chapter~\ref{Chap-IndexTheorems}, which also applies to the present one-dimensional example.

\begin{theorem}
\label{theo-1dindex}
Let $\PI:\ell^2(\ZM)\to\ell^2(\NM)$ be the surjective partial isometry as above. For a unitary $u=\{U_\omega\}_{\omega\in\Omega}\in C^1(\Aa_1)$, the operators $\PI U_\omega\PI^*$ are Fredholm operators with an almost sure index given by
$$
\Ch_1(u)
\;=\;
-\;\Ind(\PI U_\omega\PI^*)
\;.
$$
\end{theorem}
 
This is an extension of the Noether-Gohberg-Krein index theorem to covariant operators and its proof can be found in \cite{KRS} as well as \cite{PS}. It assures us that the bulk invariant ${\rm Ch}_1(u_F)$ remains stable and quantized in the regime where the spectral gap of $h$ is replaced by a mobility gap. After all these preparations, the disordered version of Theorem~\ref{theo-1d} can finally be stated and proved.

\begin{theorem}
\label{theo-1ddisordered}
Consider the element $h=\{H_\omega\}_{\omega \in \Omega}\in\Aa_1$ associated to the Hamiltonian {\rm \eqref{eq-Hamdis}} and let $\hat{h} = \{\widehat H_\omega\}_{\omega \in \Omega}\in\widehat \Aa_1$ be a restriction to the half-space given by an arbitrary chiral symmetric boundary condition. Assume $h$ to have a central spectral gap and let $u_F$ be the Fermi unitary element as well as  $N_{\omega,\pm}=\Tr (\widehat{P}_{\pm,\omega}(\delta))$. Then, for all $\omega$,  
\begin{equation}
\label{eq-bulkedge1ddisordered}
\Ch_1(u_F)
\;=\;
-\;N_{\omega,+}\;+\;N_{\omega,-}
\;.
\end{equation}
\end{theorem}

\noindent {\bf Proof.} Set $h(\lambda)=\gamma_\lambda(h)$ with the homotopy $\gamma_\lambda$ given in \eqref{eq-deform}, which induces a smooth deformation $u_F(\lambda)$. By homotopy invariance, $\Ch_1(u_F(\lambda))$ is constant, in particular, $\Ch_1(u_F)=\Ch_1(u_F(0))$.  Furthermore, the projections supplied by the index map define a homotopy of projections and, since the pairing $\widetilde \Ch_0([p]_0)=\int \PM(d\omega)\,\Tr(P_\omega)=\Tr(P_\omega)$ is homotopy invariant, it and can be can be computed at $\lambda=0$. Consequently, the equality \eqref{eq-bulkedge1ddisordered} follows from the equality at $\lambda=0$, which was already proved in Theorem~\ref{theo-1d}. 
\hfill $\Box$

\vspace{.2cm}

\noindent {\bf Second proof} of Theorem~\ref{theo-1ddisordered}, based merely on Theorem~\ref{theo-1dindex}. First of all, the chiral symmetry $J H_\omega J=-H_\omega$ implies that there exists an invertible operator $A_\omega$ such that, in the grading of $J$,
\begin{equation}
\label{eq-HamOffDiag}
H_\omega
\;=\;
\begin{pmatrix}
0 & A_\omega^\ast \\ A_\omega & 0
\end{pmatrix}
\;.
\end{equation}
By homotopy invariance of the index, 
$$
\Ind(\PI U_\omega\PI^*)
\;=\;
\Ind(\PI A_\omega\PI^*)
\;=\;
\dim(\Ker( \PI A_\omega\PI^* ))
\,-\,
\dim(\Ker( \PI A^*_\omega\PI^* ))
\;.
$$
But $\Ker(\PI H_\omega\PI^*)= \bigl(\Ker( \PI A^*_\omega\PI^* )\oplus 0\bigr) \oplus\bigl(0\oplus\Ker( \PI A_\omega\PI^* )\bigr)$, and $\widehat{J}$ is positive definite on the first and negative definite on the second summand. Therefore
$$
\Ind(\PI U_\omega\PI^*)
\;=\;
\Sig\bigl(\widehat{J}|_{\Ker(\PI H_\omega\PI^*)}\bigr)
\;,
$$
where the signature is calculated of the (finite dimensional non-degenerate) quadratic form obtained by restriction of $\widehat{J}$ to $\Ker(\PI H_\omega\Pi^*)$. But this signature is up to a sign precisely the r.h.s. of \eqref{eq-bulkedge1ddisordered}.
\hfill $\Box$

\vspace{.2cm}

Another thing that becomes apparent in the above proof is how to address the stability of the invariants under terms which break chiral symmetry, see Section~\ref{sec-period1d}. Indeed, such terms lead to non-vanishing diagonal entries in the Hamiltonian in the form \eqref{eq-HamOffDiag}. If, however, the off-diagonal entry $A_\omega$ remains invertible, then one can still define its winding number via the pairing with $\Ch_1$. Such systems are called approximately chiral and are further described in Section~\ref{sec-BulkHam}

\section{Can the invariants be measured?}

Of course, it is interesting to link the invariants to quantities that can potentially be measured. The best know example is the quantum Hall effect in which an invariant is linked to the Hall conductance. For the present one-dimensional chiral models the so-called chiral polarization is connected to the bulk invariant $\Ch_1(u_F)$ as is discussed in Section~\ref{sec-ChiralPol}. One of the things that is always true is that the bulk invariant determines the boundary invariant, which is here the chirality of the bound states. This boundary invariant can in principle be measured.

\chapter{Topological solid state systems: conjectures, experiments and models}
\label{Chap-Physics}

\abstract{This chapter reviews the ten classes of topological insulators and superconductors and presents their classifying table. The two complex classes of the table, which are the focus of our work, are then discussed in depth. The emphasis is on the physical properties, experimental achievements and the conjectures put forward by the physics community. The bulk-boundary correspondence principle is exemplified using exactly solvable models in arbitrary dimensions. The chapter also introduces the generic classes of physical models which incorporate the effect of an external magnetic field and disorder. It elaborates the main assumptions and summarizes the behavior of various physical quantities of interest. The reader will find here several technical results from functional analysis used in our work.}

\section{The classification table}

Hereafter, a crystal will be said to be insulating in the bulk if the direct bulk resistivity diverges as the temperature is taken to zero. In what concerns the electron-electron interaction, all insulators mentioned in this work are well described by mean-field approximations, hence the analysis is always carried out in the independent electron picture. Then, a strong topological insulator is a crystal which is insulating in the bulk, but becomes metallic when an edge or a surface (called boundary hereafter) is cut to the crystal. This definition automatically implies that boundary spectrum emerges at the Fermi level and, since disorder is unavoidable in real samples, it also implies that this spectrum is immune to Anderson localization, at least in the regime of weak disorder. For superconductors, the fermionic quasiparticle excitations are assumed to be well described within the Bogoliubov-de Gennes approximation. Then a strong topological superconductor has gapped fermionic quasiparticle excitations in the bulk, but supports gapless excitations modes along any boundary cut to the system. There are other effects appearing in topological insulators, {\it e.g.} the existence of zero modes attached to defects, but this is not in the focus of the present work (except in Chapter~\ref{Chap-Illustration}).

\vspace{.2cm}

\begin{table}\label{Table1}
\begin{center}

\begin{tabular}{|c|c|c|c||c||c|c|c|c|c|c|c|c|}
\hline
$j$ & TRS & PHS & CHS & CAZ & $d=0,8$ & $d=1$ & $d=2$ & $d=3$ & $d=4$ & $d=5$ & $d=6$ & $d=7$
\\\hline\hline
$0$ & $0$ &$0$&$0$& A  & $\ZM$ &  & $\ZM$ &  & $\ZM$ &  & $\ZM$ &  
\\
$1$& $0$&$0$&$ 1$ & AIII & & $\ZM$ &  & $\ZM$  &  & $\ZM$ &  & $\ZM$
\\
\hline\hline
$0$ & $+1$&$0$&$0$ & AI &  $\ZM$ & &  & & $2 \, \ZM$ & & $\ZM_2$ & ${\ZM_2}$
\\
$1$ & $+1$&$+1$&$1$  & BDI & $\ZM_2$ &$\ZM$  & &  &  & $2 \, \ZM$ & & $\ZM_2$
\\
$2$ & $0$ &$+1$&$0$ & D & $\ZM_2$ & ${\ZM_2}$ & $\ZM$ &  & & & $2\,\ZM$ &
\\
$3$ & $-1$&$+1$&$1$  & DIII &  & $\ZM_2$  &  $\ZM_2$ &  $\ZM$ &  & & & $2\,\ZM$
\\
$4$ & $-1$&$0$&$0$ & AII & $2 \, \ZM$  & &  $\ZM_2$ & ${\ZM_2}$ & $\ZM$ & & &
\\
$5$ & $-1$&$-1$&$1$  & CII & & $2 \, \ZM$ &  & $\ZM_2$  & $\ZM_2$ & $\ZM$ & &
\\
$6$ & $0$ &$-1$&$0$ & C&  &  & $2\,\ZM$ &  & $\ZM_2$ & ${\ZM_2}$ & $\ZM$ &
\\
$7$ & $+1$&$-1$&$1$  &  CI &  & &   & $2 \, \ZM$ &  & $\ZM_2$ & $\ZM_2$ & $\ZM$
\\
[0.1cm]
\hline
\end{tabular}
\end{center}
\caption{Classification table of strong topological insulator and superconductors. Each row represents a universal symmetry class, defined by the presence ($1$ or $\pm 1$) or absence ($0$) of the three symmetries: time-reversal (TRS), particle-hole (PHS) and chiral (CHS), and by how TRS and PHS transformations square to either $+\one$ or $-\one$. Each universality class is identified by a Cartan-Altland-Zirnbauer (CAZ) label. The strong topological phases are organized by their corresponding symmetry class and space dimension $d=0,\ldots,8$. These phases are in one-to-one relation with the elements of the empty, $\mathbb Z_2$, $\mathbb Z$ or $2\, \mathbb Z$ groups. The table is further divided into the complex classes A and AIII (top two rows), which are the object of the present study, and the real classes AI, \ldots, CI (the remaining 8 rows).}
\label{tab-class}
\end{table}

One of the first efforts to classify the strong topological insulators and superconductors was undertaken by  Schnyder, Ryu, Furusaki, and Ludwig in \cite{SRFL}. The first accomplishment of their work was to realize that the classification should be performed inside the universality classes. Focussing mainly on random matrices, Altland and Zirnbauer \cite{Zir,AZ} argued that there are ten classes which cover both Fermionic systems of electrons with conserved particle number and systems of the Bogoliubov-de Gennes type. These classes are listed in Table~\ref{Table1}. Each class is characterized by the transformations of their elements, {\it i.e.} the quantum systems themselves, under three generic symmetries, namely, the time-reversal (TRS), particle-hole (PHS) and chiral (CHS) symmetries. The TRS and PHS can square to plus or minus the identity, leading to a total of precisely ten distinct choices. Note that the combination of a TRS and a PHS results in a transformation of CHS type, and this aspect needs to be taken into account when counting the universality classes. As explained in Ref.~\cite{Zir}, these classes are closely connected to Cartan's symmetric spaces, which explains the Cartan labels assigned to them ({\it e.g.} A, AIII, etc.). The separation in universality classes applies to random matrices and disordered metals and insulators alike. Ref.~\cite{SRFL} then went systematically over these ten classes for bulk insulators in dimension $d\leq 3$, by performing an analysis of the localized/delocalized character of the boundary states in the presence of disorder. This analysis was based on the classification of the one- and two-dimensional disordered Dirac Hamiltonians by Bernard and LeClair \cite{BL} and on a complementary field-theoretic argument based on the replica trick, both of which rely on effective theories involving saddle-point approximations (the non-linear sigma models).   The final conjecture of Ref.~\cite{SRFL} was that all topological phases  for $d \leq 3$ (those with a non-vanishing entry in Table~\ref{Table1}) display delocalized boundary spectrum which fills the bulk gap entirely. For the unitary chiral AIII class it is now known that the conjecture is not entirely true, as disorder can localize the entire boundary spectrum except at the Fermi level \cite{EG2}, which for AIII class in pinned at $E=0$, and magnetic fields can even open spectral gaps in the boundary spectrum. Ref.~\cite{SRFL} also introduced a higher winding number for chiral systems in dimension $d=3$ allowing to distinguish so-called strong topological insulators. The possible values of this invariant and its analogues in other dimensions and universality classes appear in Table~\ref{Table1}. For example, the $\ZM$ for class A systems in $d=2$ is the well-known Chern number of quantum Hall systems.

\vspace{.2cm}

The structure of the classifying table reported in Ref.~\cite{SRFL} differed from the one seen in Table~\ref{Table1}. The latter displays an obvious flow-pattern and periodicity with the space dimension and, because of these characteristics, the table is also called the periodic table of topological insulators and superconductors. These features were pointed out by Kitaev \cite{Kit}, who noted that the systems with (without) TRS and PHS are classified by the real (complex) $K$-theories.  Then Bott periodicity alone can explain the patterns seen in Table~\ref{Table1}, as it is nicely explained in Ref.~\cite{SCR,KZ,Thi1,MFM} for the real classes. See also \cite{GS} for an index-theoretic approach which holds in the regime of strong disorder. In the complex case, there are only two available $K$-groups, the $K_0$ and $K_1$ groups, and they classify the two complex classes A and AIII, respectively. One can move between the two groups using the suspensions maps, the $\theta$-map and the Bott map (see Section~\ref{Sec-SuspensionBottPer}), which effectively increase the space dimension by one. As such, to any strong topological insulator from class A one can associate a strong topological insulator from class AIII using Bott map and by doubling the dimension of the fiber to accommodate for the chiral symmetry; and to each topological system from AIII class one can associate a strong topological insulator from class A using the $\theta$-map. Repeating this procedure, starting from $d=0$ where $K_0 \simeq \ZM$, one can get an understanding of the flow-pattern, the periodicity and the counting of the strong complex topological phases listed in Table~\ref{Table1}. Let us mention that Table~\ref{Table1} is adopted from Ref.~\cite{RSFL}, which relied on the same classifying criterium and methods as Ref.~\cite{SRFL}. Further let us point out that the $2\ZM$ entries in Table~\ref{Table1} express that the invariants for the corresponding systems are always even \cite{RSFL,GS}.

\vspace{.2cm}

The complex $K$-groups of the algebras of bulk observables, in the presence of disorder and magnetic fields, are listed in Section~\ref{sec-KObsAlg} and, as one can immediately see from Table~\ref{Table1}, the strong topological insulators account only for a fraction of these groups. As discussed in Section~\ref{Sec-GeneratorsKGroups}, the strong topological systems are generated by the top generators of the $K$-groups, while the rest of the generators generate the so-called weak topological insulators. The same is true for the real classes. An example of weak topological insulator is the quantum Hall effect in three space dimensions \cite{KHW}. As we shall see, the bulk-boundary principle applies to the weak topological insulators too, but with two important modifications: 1) The principle does not work for all boundaries. In other words, boundaries cut along specific crystallographic planes do not carry topological boundary states (see \cite{YMTI} for explicit examples). 2) Their bulk and boundary invariants (see Sections~\ref{Sec-BoundaryTopInvariants} and \ref{Sec-BulkTopInvariants}) do not satsify index formulas and for this reason the bulk invariants cannot be formulated in the regime of strong disorder and the delocalization of the topological boundary spectrum cannot be established by the present methods. The latter remains an important open issue because, in certain circumstances, the weak topological insulators were shown to display metallic boundary states in the presence of disorder \cite{BF,MBM,RKS,KOI} and robust conducting channels along line-defects \cite{ITT}. We want to mention that new mathematical tools, targeting precisely this issue, were put forward in Ref.~\cite{Pro4}.

\vspace{.2cm}

Lastly, let us point out that there are additional classes of topological insulators which received substantial attention from both theoretical and experimental physics communities. These are the crystalline topological insulators \cite{Fu,AF}, which are stabilized by the TRS and a space point-symmetry of the crystal, and furthermore the spin-orbit and TRS free topological insulators \cite{AFGB}, which are stabilized just by a space point-symmetry. By stabilized we mean that interesting topological classifications of phases emerges when these constraints are enforced, at least in the periodic case. 

\section{The unitary class}
\label{Sec-UnitaryClass}

The systems in the unitary class have no symmetry constraints except for the requirement that the time evolution is unitary. As a consequence, the generators of the time evolution, which are the Hamiltonians if the discussion is about the quantum systems, are self-adjoint operators. This means, for example, that open or dissipative quantum systems are excluded from the unitary class or, putted differently, the topological characteristics associated with the unitary class may brake down when unitarity is lost. As such, the self-adjoint property of the Hamiltonians can be regarded as a ``symmetry" which, like all the other symmetries in the classification table, stabilizes the topological properties of the systems from class A. In this section we introduce the models and their physical characteristics, both for bulk and half-space. We formulate the bulk-boundary principle for periodic systems and demonstrate this principle using an exactly solvable model in arbitrary dimensions. The existing experimental results are briefly surveyed.

\subsection{General characterization}
\label{Sec-UClassGeneral}

The most general translation invariant ({\it i.e.} $1$-periodic) lattice model from the unitary class in $d$ space dimensions takes the form:
\begin{align}
\label{UGenericModel}
H: \mathbb C^N \otimes \ell^2(\mathbb Z^d)   \rightarrow \mathbb C^N \otimes \ell^2(\mathbb Z^d)
\;,
\qquad
H \, =\,\sum_{y\in \mathbb Z^d} W_y \otimes S^y,
\end{align}
where $S^y$ is the shift operator by $y$ on $\ell^2(\mathbb Z^d)$ given by $S^y|x\rangle = |x+y \rangle$, and the $N \times N$ matrices $W_y$, called tunneling or hopping matrices, satisfy only the constraint
$$
W_y^* \;=\; W_{-y}\;,
$$
ensuring that $H$ is self-adjoint. Throughout, we denote the space of $N \times N$ matrices with complex entries by $M_N(\mathbb C)$.  Also, ${\rm tr}$ will denote the trace of matrices, such as those from $M_N(\mathbb C)$, or more general the trace over finite dimensional Hilbert spaces. The trace over infinite Hilbert spaces, such as $\ell^2(\mathbb Z^d)$ or $\mathbb C^N \otimes \ell^2(\mathbb Z^d)$, will be denoted as usual by ${\rm Tr}$. 

\vspace{.2cm}

The dimension $N$ of the fiber is determined by the number of molecular orbitals per unit cell of the material included in the model, and the larger this number the more precise the model is. Let us make it clear from the beginning that \eqref{UGenericModel} are not toy models, but rather the {\it models of choice} in materials science. Given a concrete material, such lattice Hamiltonians can be generated empirically by fitting available experimental data or using first-principle calculations \cite{LQZDFZ,ZYZDF,WDF,ORV}. The main tool for generating lattice models from first principles continuous model calculations is the maximally localized Wannier basis set. The reader can find in \cite{MMY} impressive demonstrations of how effective and accurate this tool can be. Even when working empirically, the lattice models can be finely tuned to accurately reproduce a broad range of experiments and, once such fine tuning is achieved, the models can be used for predictions. An example of this sort is the discovery of the first topological insulator \cite{BHZ}. The {\it quantitative} predictions based on a lattice model made in \cite{BHZ} were later shown to be extremely accurate by the experiment \cite{KWB}.

\vspace{.2cm}

Typically, the hopping matrices $W_y$ in \eqref{UGenericModel} decay rapidly with $y$ and in practice the summation over $y$ is restricted to a finite number $\Rr\subset\ZM^d$ of terms, and this will be done from now on. We refer to such Hamiltonians as having finite hopping range. If adequate conditions are imposed on the fall-off of $W_y$ in $y$, the case $\Rr=\ZM^d$ can be also managed with some further technical effort, but it will not be pursued here. For the periodic models, one can use the Bloch-Floquet decomposition
\begin{equation}
\label{eq-BlochFloquet}
\Ff H\Ff^*\;=\;
\int^\oplus_{\mathbb T^d} dk \; H_k
\end{equation}
over the Brillouin torus $\mathbb T^d$, to reduce the analysis to that of a smooth family of  $N \times N$ matrices
$$
H_k : \mathbb C^N \rightarrow \mathbb C^N\;, 
\qquad 
H_k \,=\, \sum_{y\in \Rr} e^{\I \langle y|k\rangle} W_y
\;.
$$
Throughout, $\langle\,,\,\rangle$ will denote the Euclidean scalar product. Examining the classification table, we see that the topological phases in the unitary class are conjectured to occur only in even space dimensions, and for each such dimension there is an infinite sequence of topological phases. It is also conjectured that these phases can be distinguished from one another by tagging them with just one integer number. In the bulk, this number is given by the top even Chern number, which is a measurable physical coefficient (see Chapter~\ref{Chap-Conclusions}) and  takes the form \cite{ASSS}
\begin{equation}\label{EvenChernK}
\mathrm{Ch}_{d}(P_F)
\;=\;
\frac{(2 \pi \I)^\frac{d}{2}}{ (\frac{d}{2})!} \sum_{\rho \in \Ss_d} (-1)^\rho\int_{\mathbb{T}^d} \frac{dk}{(2\pi)^d} \ \mathrm{tr}\Big ( P_F(k) \prod_{j=1}^d \frac{\partial P_F (k)}{\partial k_{\rho_j}} \Big )
\;,
\end{equation}
for the periodic crystals. Throughout, $\Ss_d$ will denote the group of permutations and $\I = \sqrt{-1}$. In \eqref{EvenChernK},
$$
P_F (k)\; =\; \chi(H_k \leq \mu)
$$
is the spectral projection onto the energy bands below the Fermi level $\mu$. The standard terminology for it is the Fermi projection. Because we are dealing with insulators, the Fermi level is assumed to be located in a spectral gap of $H$. Throughout, $\chi(A)$ will denote the characteristic function of a set $A$. We will present an explicit topological model shortly, but let us mention at this point that the periodic models with ${\rm Ch}_d(P_F) \neq 0$ are ubiquitous. For example, if one generates the hopping matrices $W_y$ randomly, assuming $\Rr$ and $N$ large, then the chances of obtaining a topological system are far greater than the chances of obtaining a trivial one.

\vspace{.2cm}

Our analysis, while limited to lattice models, will include uniform magnetic fields and disorder. The presence of a uniform magnetic field is incorporated in the lattice models using the Peierls substitution \cite{Pei}, which amounts to replacing the ordinary shift operators with the dual magnetic translations
\begin{equation}
\label{DualMagTransSymmetric}
\one \otimes S^y  \;\;\mapsto \;\; U_\sym^y \;=\; \one \otimes e^{\frac{\I}{2}\langle y|\bm B| X \rangle}S^y \;=\; \one \otimes S^y e^{\frac{\I}{2}\langle y|\bm B| X \rangle}
\;.
\end{equation}
Here, $\bm B$ is a real anti-symmetric $d\times d$ matrix representing the magnetic field and $X$ is the position operator on $\ell^2(\mathbb Z^d)$. The label ``sym" indicates that the so-called symmetric gauge has been used above. After the substitution, the lattice Hamiltonians take the form
\begin{equation}
\label{UGenericModelBSymmetric}
H_\sym
\;=\;
\sum_{y\in \Rr}  W_y \otimes U_\sym^y 
\;=\;
\sum_{y\in \Rr} \; \sum_{x \in \mathbb Z^d} e^{\frac{\I}{2}\langle y|\bm B| x \rangle} \, W_y \otimes |x\rangle \langle x -y| 
\;.
\end{equation}
The Hamiltonian \eqref{UGenericModelBSymmetric} is no longer invariant to the ordinary lattice translations. Nevertheless, $H_\sym$ is invariant relative to the magnetic translations
\begin{equation}
\label{MagTransSymmetric}
V_\sym^x \,H_\sym \,(V_\sym^x)^\ast\, = \,H_\sym\,, 
\qquad  V_\sym^x \,= \,\bm 1 \otimes e^{-\frac{\I}{2}\langle x|\bm B| X \rangle}S^x \,= \,\bm 1 \otimes  S^x e^{-\frac{\I}{2}\langle x|\bm B| X \rangle}
\,,
\end{equation}
written here also in the symmetric gauge. 

\vspace{.2cm}

A Landau gauge can be defined so that no Peierls phase is generated when the lattice is shifted in the $d$-th direction. While the symmetric gauge is more convenient for the bulk analysis, the Landau gauge is obviously more convenient for systems with a boundary in the $d$th direction. The dual and the direct magnetic translations in the Landau gauge can be obtained from the symmetric ones via the transformations
\begin{equation}
\label{DualMagTransLandau}
U^y 
\;=\; 
e^{-\frac{\I}{2}\langle y |\BB_+|y \rangle } \, e^{\frac{\I}{2} \langle X|\BB_+|X \rangle } \,U_\sym^y \,e^{ -\frac{\I}{2} \langle X|\BB_+|X \rangle }
\;=\; 
S^y e^{\I \langle y|\BB_+ | X\rangle}
\end{equation}
and
\begin{equation}
\label{MagTransLandau}
V^x 
\;=\; 
e^{\frac{\I}{2} \langle x|\BB_+|x\rangle } \, e^{\frac{\I}{2} \langle X|\BB_+|X\rangle}\,V^x_\sym\,e^{-\frac{\I}{2} \langle X|\BB_+|X\rangle}
\;=\;
e^{\I \langle X | \BB_+|x\rangle} S^x
\;,
\end{equation}
where $\BB_+$ is the lower triangular part of $\BB$. Note that, indeed, if $x$ and $y$ are strictly along the $d$-th direction, both $U^y$ and $V^x$ reduce to ordinary shifts. By the conjugation of \eqref{UGenericModelBSymmetric} with the local unitary operator $e^{\frac{\I}{2} \langle X|\BB_+|X\rangle}$ (namely, by a gauge transformation), the Hamiltonian becomes
\begin{equation}
\label{UGenericModelBLandau}
H
\;=\;
e^{\frac{\I}{2} \langle X|\BB_+|X\rangle}
\,H_\sym\,
e^{-\frac{\I}{2} \langle X|\BB_+|X\rangle}
\;=\;
\sum_{y\in \Rr}  e^{\frac{\I}{2}\langle y |\BB_+|y \rangle } \, W_y \otimes U^y
\;.
\end{equation}
It is unitarily equivalent to \eqref{UGenericModelBSymmetric} and satisfies $V^x H (V^x)^\ast = H$. As a consequence, $H$ in \eqref{UGenericModelBLandau} is periodic in the $d$-th direction. We will refer to \eqref{UGenericModelBLandau} as the representation of the Hamiltonian in the Landau gauge. 

\vspace{.2cm}

A homogeneous disorder will be described by a dynamical system $(\Omega,\tau,\ZM^d,\PM)$. Here, $\Omega$ is a compact metrizable topological space representing the disorder configuration space and $\tau$ is a homeomorphic action of $\mathbb Z^d$ on $\Omega$, describing the behavior of the disorder configurations under the lattice translations. Furthermore $\PM$ is an invariant and ergodic probability measure on $\Omega$ w.r.t. $\tau$, which defines the disorder averaging procedure. A more detailed description of the space of disorder configurations is given in Section~\ref{sec-disorder} below. If disorder is present, all the coefficients in the Hamiltonian develop a random component and its generic form becomes
\begin{align}
\label{UGenericModelBDSymmetric}
H_{\sym,\omega}
& \;=\;
\sum_{y\in \Rr} \;\sum_{x \in \mathbb Z^d} \,W_y(\tau_x \omega) \otimes |x \rangle \langle x|U_\sym^y \\
& \;=\;
\sum_{y\in \Rr}\; \sum_{x \in \mathbb Z^d} \,e^{\frac{\I}{2}\langle y|\bm B| x \rangle} \;W_y(\tau_x \omega) \otimes |x \rangle \langle x -y |
\;, \nonumber 
\end{align}
in the symmetric gauge. The hopping matrices $W_y$ are now continuous functions over $\Omega$ with values in $M_N(\mathbb C)$. The models with disorder are no longer invariant to the magnetic translations, but this property is replaced by the following covariance relation:
\begin{equation}
\label{CovRelSymmetric}
V_\sym^x \,H_{\sym,\omega}\, (V_\sym^x)^\ast 
\;= \;
H_{\sym,\tau_x \omega}
\;,
\qquad
x\in\ZM^d
\;.
\end{equation}
The Landau representation is obtained by conjugating \eqref{UGenericModelBDSymmetric} by $e^{\frac{\I}{2} \langle X|\BB_+|X\rangle}$, which gives
\begin{equation}
\label{UGenericModelBDLandau}
\boxed{
\;H_{\omega}
 \;=\;
\sum_{y\in \Rr} \sum_{x \in \mathbb Z^d} e^{\frac{\I}{2}\langle y|\bm B_+| y \rangle} W_y(\tau_x \omega) \otimes |x \rangle \langle x|U^y
\;,\;
}
\end{equation}
and the covariance relation becomes
\begin{equation}
\label{CovRelLandau}
\boxed{
\;V^x \,H_\omega\, (V^x)^\ast 
\;= \;
H_{\tau_x \omega}
\;,
\qquad
x\in\ZM^d
\;.\;
}
\end{equation}
The bulk-boundary analysis will be carried in the Landau gauge, hence we will primarily work with the Hamiltonian \eqref{UGenericModelBDLandau}.

\vspace{.2cm}

Since the origin of the lattice is completely arbitrary and will change every time a crystal is put down and picked up again in the lab, the model of the disordered crystal must include the whole family of covariant Hamiltonians $H=\{H_\omega\}_{\omega \in \Omega}$. The notation is appropriate because in the absence of disorder, the entire family consists of just one element, the $H$ itself. Systems with the covariance property \eqref{CovRelLandau} are called homogeneous and have remarkable properties. As we shall see later, there exists  a Fourier calculus for them which can be used to define a (non-commutative) differential calculus. Also, any such covariant family $F=\{F_\omega\}_{\omega\in\Omega}$ posses the self-averaging property that for $\PM$-almost every configuration $\omega$ one has
\begin{equation}
\label{eq-BrikhoffTV}
\lim_{V \rightarrow \infty} \tfrac{1}{|V|} \;{\rm Tr}\big (\PI_V \, F_\omega \, \PI_V^* \big ) 
\;=\; 
\int_\Omega  \mathbb P(d\omega') \ {\rm tr}(\PI_0 \, F_{\omega'} \, \PI_0^* )
\;,
\end{equation}
where $\PI_V:\mathbb C^N \otimes \ell^2(\mathbb Z^d)\to \mathbb C^N \otimes \ell^2(V \cap \mathbb Z^d)$ is a partial isometry onto the quantum states $|\alpha\rangle \otimes |x \rangle$ with $x$ located inside $V$. In particular, $\PI_0$ is the partial isometry onto the quantum states $|\alpha\rangle \otimes |0 \rangle$. Identity \eqref{eq-BrikhoffTV} follows directly from Birkhoff's ergodic theorem \cite{Bir}. The quantity on the l.h.s. of \eqref{eq-BrikhoffTV} is called the trace per volume of the covariant observable $F$. It is hence, with probability one, independent of the disorder configuration and is equal to the disorder average of the trace of its matrix elements computed at the origin (or any other point of the lattice). In the following we will use the notation $\Tt(F)$ for the trace per volume of a family covariant observables. The top even Chern number can be formulated for the generic models \eqref{UGenericModelBDSymmetric} or \eqref{UGenericModelBDLandau} using a real-space representation and the trace per volume \cite{PLB}
\begin{equation}
\label{EvenChernR}
\boxed{
\;
\mathrm{Ch}_{d}(P_F)
\;=\;
\frac{(2 \pi \I)^\frac{d}{2}}{\frac{d}{2}!} 
\;
\sum_{\rho \in \Ss_d}(-1)^\rho\; \Tt \Big ( P_{\omega} \prod_{i=1}^{d} \big (\I[P_{\omega},X_{\rho_i}]\big ) \Big )
\;.
\;
}
\end{equation}
Here, 
$$
P_F 
\;=\; 
\{ P_{\omega}\}_{\omega \in \Omega}
\;=\;
\{ \chi(H_\omega \leq \mu)\}_{\omega \in \Omega}
$$
is the covariant family of spectral projections onto the energy spectrum below $\mu$, that is, the family of Fermi projections. The top even Chern number, as defined in \eqref{EvenChernR}, is known to remain quantized, non-fluctuating from one disorder configuration to another, and be homotopically stable as long as the Fermi level resides in a region of Anderson-localized spectrum \cite{BES,PLB}. These statements will be re-examined in Chapter~\ref{Chap-IndexTheorems}.
 
\vspace{.2cm}

To model a boundary, the physical space and the models are restricted to the half-space $\mathbb Z^{d-1} \times \mathbb N$. The half-space Hamiltonian $\widehat{H}$ then acts on the Hilbert space $\mathbb C^N \otimes \ell^2(\mathbb Z^{d-1} \times \mathbb N)$. For the moment being, it can just be thought of as the restriction of $H$ which corresponds to Dirichlet boundary conditions. In Section~\ref{Sec-HSHamilt} other allowed boundary conditions will be described. When the bulk Chern number $\mathrm{Ch}_{d}$ does not vanish, the energy spectrum of the half-space Hamiltonian $\widehat{H}$ extends inside the bulk insulating gap, covering it completely \cite{QHZ,EG}. The electron states corresponding to the spectrum inside the bulk insulating gap are exponentially localized near the boundary, hence the terminology boundary states and boundary spectrum (see Section~\ref{Sec-HSHamilt} below for an explicit example). For periodic crystals with a planar boundary, say $x_d \geq 0$, the spectrum can be represented as energy bands rendered as functions of the momentum $k \in \mathbb T^{d-1}$ parallel to the boundary. The hallmark feature of the topological phases from the unitary class is the existence of boundary energy bands that connect the bulk valence and conduction bands. For $d>2$, the boundary bands display one or more singularities called Weyl points. Around a Weyl point, denoted by $k^W\in \mathbb T^{d-1}$ in the following, the spectrum and the states are well described by a Weyl operator
\begin{equation}
\label{eq-Weyl}
\sum_{j=1}^{d-1} {\rm v}_j  (k_j-k_j^W) \sigma_j,
\end{equation}
where $\sigma=(\sigma_1,\ldots,\sigma_{d-1})$ are the generators of an irreducible representation of the odd complex Clifford algebra $Cl_{d-1}$ and ${\rm v}=({\rm v}_1,\ldots,{\rm v}_{d-1})$ are the non-vanishing slopes of the bands in different directions parallel to the boundary, which can be positive or negative. 

\begin{remark} 
In the literature, the singular points \eqref{eq-Weyl} are sometimes also called Dirac points, which is not appropriate for the following reasons. In 4 dimensions, for example, the zero mass Dirac operator takes the form $\langle k,\gamma \rangle $ and has a chiral symmetry w.r.t. the product $\gamma_1\cdots\gamma_d$. This splits it into two chiral sectors and, in each of those chiral sectors, one gets the classical Weyl operator $\langle k,\sigma \rangle$ when the ''time'' direction is separated out. Here, $\gamma$ and $\sigma$ denote the Dirac and Pauli matrices. This pattern can be recognized in any dimension, and in general, the Weyl operator involves an odd number of Clifford generators and does not have a chiral symmetry, but rather a chirality that will be introduced below. Throughout, we will be consistent and use the notation $\sigma$ ($\gamma$) for the generators of the odd (even) complex Clifford algebras, and refer to the operators $\langle k,\sigma \rangle$ \big ($\langle k, \gamma \rangle $\big) as Weyl (Dirac) operators when the dimension of $k$ is odd (even), respectively.
\hfill $\diamond$ 
\end{remark}

Now, suppose that all the Weyl singularities have been identified from the boundary band spectrum and that the asymptote \eqref{eq-Weyl} of the Hamiltonian has been extracted for each singularity (dimension $d=2$ is special in this respect, see below). Then the chirality of a Weyl point 
\begin{equation}\label{eq-chirality}
\nu_W
\;=\;
\prod^{d-1}_{j=1}\mbox{\rm sgn}({\rm v}_j)
\;\in\; \{-1,1\},
\end{equation}
is a well defined topological invariant, provided the Weyl point remains separated from the rest. The central conjectures for the unitary class is the following bulk-boundary principle \cite{QHZ}
\begin{equation}
\label{BB1}
\mathrm{Ch}_d(P_F) 
\;=\; 
\chi \sum \nu_W
\;,
\end{equation}
where the sum on the left goes over all Weyl points. In other words, under deformations of the model, the Weyl singularities will move and possibly collide and annihilate, yet the sum of their chiralities remains the same and equal to the bulk invariant. Above, $\chi$ is a sign factor which depends on the normalization (or the sign convention) of the bulk invariant and on the specific representation $\sigma$ of the odd $d-1$ dimensional Clifford algebra (recall that there are two inequivalent representations).

\vspace{.2cm}

In dimension $d=2$, the chiralities are given by the signs of the slopes of the boundary bands traversing the bulk insulating gap. The slopes are computed at a fixed (but arbitrarily chosen) energy level. If a slope of a band happens to be zero, then this band is excluded. The bulk-boundary principle \eqref{BB1} was first demonstrated by Hatsugai \cite{Hat1} for the special case of the Harper operator with rational magnetic field. In higher dimension, the bulk-boundary principle will be exemplified on an exactly solvable model in Section~\ref{Sec-UExactModel}. A proof of \eqref{BB1} will be given in Section~\ref{sec-PairingsDuality}, combined with the evaluation of the boundary invariants for periodic systems in Section~\ref{Sec-BoundaryTopInvariants}.

\vspace{.2cm}

One of the main goals of the present work is to formulate $\sum \nu_W$ as a boundary topological invariant which makes sense in the presence of disorder and magnetic fields, and to derive an index theorem for it.  In dimension $d=2$, this was achieved in \cite{KRS} and will be reviewed and expanded in Chapter~\ref{Chap-Conclusions}. The boundary invariant and the bulk-boundary equation takes the form
\begin{equation}\label{BB0}
2 \pi \,\widetilde{\Tt} \big(J_\shortparallel \ \rho(\widehat H)\big)
\;=\;
\mathrm{Ch}_2(P_F)
\;,
\end{equation}
where $\widetilde{\Tt}$ is the trace per length, taken in the direction parallel with the boundary, $J_\shortparallel$ is the current operator along the boundary and $\rho$ is a distribution which integrates to one and has support inside the bulk insulating gap, but is otherwise arbitrary. As above, the Hamiltonian $\widehat H$ describes the system with a boundary. Physically, the invariant on the l.h.s. of \eqref{BB0} gives the charge current spontaneously carried by the boundary states when they are populated with the distribution $\rho$. If the bulk invariant $\mathrm{Ch}_2(P_F)$ is nonzero, \eqref{BB0} automatically ensures that the boundary spectrum cannot display gaps or be localized by disorder. This statement will be generalized to higher dimensions in this work.

\subsection{Experimental achievements} 

The prototypical example of a topological condensed matter system from the unitary class is the two-dimensional electron gas subjected to a perpendicular uniform magnetic field for which the integer quantum Hall effect (IQHE) is observed \cite{KDP}. In this case, the Chern number $\Ch(P_F)$ equals the Hall conductance of the system and all the characteristics described above have been mapped experimentally with amazing precision. We have been careful not to use the word ``material" because this topological state of matter is stabilized by a magnetic field which needs to be externally maintained. It was Haldane \cite{Hal} who first realized that two-dimensional materials can display characteristics similar to IQHE without the need of an external magnetic field. The minimal yet not sufficient requirements for this to happen is a unit cell containing two (chemically active) molecular orbitals and complex tunneling matrices between these molecular orbitals. The decisive step towards the experimental realization of a topological material from the unitary class were taken in 2013 in the series of works \cite{Cha1,Cha2} where a thin film of (Bi,Sb)$_2$Te$_3$, which in the pristine bulk phase is a time-reversal symmetric topological insulator, was doped with chromium magnetic atoms to induce a gapped ferromagnetic ground state. In the short period since then, there have been quite a number of experimental refinements \cite{Kou1,BWM,KRR,Col,Kou2,JJ}, notably the achievement of the quantum critical regime at the transition between the presumed topological and trivial phases \cite{CYT}. The scaling analysis with the temperature revealed the existence of the critical point and confirmed beyond any doubt that a new topological state of matter was indeed achieved (see also \cite{XP} for numerical simulations and discussion). Other materials  \cite{LHH} and experimental paths have been explored. For example, a topologically non-trivial state was realized in a system of one-dimensional array of optical guides which implemented literally the one-dimensional Aubry-Andre model \cite{KLR,VZK}. The condensed matter system proposed by Haldane \cite{Hal}, in its exact form, was finally realized experimentally with ultra-cold fermions in a periodically modulated optical honeycomb lattice \cite{JMD}. Here the complex tunneling matrices were tuned using time-modulated  pulses. Strong two-dimensional topological insulators were also theoretically predicted \cite{HR,RH} and then realized in photonic crystals \cite{WYJ}. Furthermore, they were also theoretically predicted \cite{PP,WLZ,Yan,WLB} and then realized in phonon or acoustic crystals \cite{Nas}. Lastly, we should mention that driving a condensed matter systems with time-periodic potentials \cite{KBR,LRG,Rec} or by considering incommensurate potentials \cite{KRZ,Pro6} opens the possibility of experimental realizations of topological states which mimic topological insulators in space dimensions higher than three. Such a system will be discussed in details in Section~\ref{Sec-VirtualTI}.

\subsection{Conventions on Clifford representations}
\label{SSec-Clifford}

To give a firm meaning to the invariants and also to the index theorems presented in Chapter~\ref{Chap-IndexTheorems}, the following conventions will be adopted throughout.

\vspace{0.2cm}

\noindent {\bf Conventions on the Clifford representations (CCR).} Since only the complex classes of topological insulators are investigated, we will only be dealing with the complex Clifford algebras $Cl_n$. They are defined by $n$ generators obeying the commutation relations 
\begin{equation}
\label{eq-CliffordCR}
\nu_i \nu_j \;+\; \nu_j \nu_i \;= \;2\, \delta_{i,j}\, \one\;, 
\qquad \nu_i^\ast \;=\; \nu_i\;, 
\qquad i,j=1,\ldots,n\;.
\end{equation}
As previously mentioned, when the parity of $n$ is important, we will use for the generators the symbols $\sigma$ ($n$ odd) and $\gamma$ ($n$ even) but, if the parity is not important and the discussion can be carried in parallel for the two cases, then we will use the symbol $\nu$. The commutation relations \eqref{eq-CliffordCR} are invariant to the operations
\begin{equation}
\label{eq-transf}
\nu'_i \;=\; u \nu_i u^{-1}\;,  
\qquad 
\nu'_i \;=\; \sum_j A_{i,j} \nu_j
\;,
\end{equation}
and their combinations, where $u$ is any unitary element from the Clifford algebra and $A$ is an orthogonal matrix form $M_{n\times n}(\RM)$, that is $AA^T = A^T A = \one$. Below, we list our conventions.  

\begin{enumerate}[\rm (i)]

\item The orientation of the physical space is fixed once and for all. In other words, one is allowed to redefine the space directions using only proper orthogonal transformations. For example, the reflections are excluded.

\item The orientation of the generators $\nu_i$ is also fixed once and for all. This means that all systems of generators can be connected to a reference one using the transformations in \eqref{eq-transf} with $A$ an orthogonal matrix.

\item Once the previous convention is adopted, we can unumbiguously define a chiral element (up to a harmless unitary conjugation), for which we adopt the following normalization
$$
\nu_0 
\;=\; 
(-\I)^{[\frac{n}{2}]} \,\nu_1 \nu_2 \cdots \nu_n\;, 
\qquad 
\nu_0^\ast \;=\; \nu_0
\;, 
\qquad \nu_0^2 \;=\; \one
\;.
$$

\item For $n=2k+1$, the commutation relations accept two inequivalent irreducible representations on $\CM^{2^k}$. In this odd case, the chiral element commutes with the entire $Cl_{2k+1}$, hence in an irreducible representation it will be sent to a matrix proportional to unity. Our convention is that $\nu_0$ is sent exactly into the identity. In other words, our odd representations are uniquely defined (up to proper isomorphisms) by the previous conventions and by
\begin{equation}
\label{eq-OddClifford}
\sigma_1 \sigma_2 \cdots \sigma_{2k+1} \;=\; \I^k\, \one
\;.
\end{equation}
For example, the Pauli matrices obey this convention.

\item For $n=2k$, the commutation relations accept a unique irreducible representations on $\CM^{2^k}$. In this case, the chiral element anti-commutes with the generators, hence it provides a grading, which we spell again below
\begin{equation}
\label{eq-grading}
\gamma_0 
\;=\; 
(- \I)^k \gamma_1 \gamma_2 \cdots \gamma_{2k}\;, 
\qquad 
\gamma_0^\ast \;=\; \gamma_0\;, 
\qquad \gamma_0^2 \;=\; \one
\;.
\end{equation}

\end{enumerate}

\begin{example}
\label{NiceClifford}  
A well-known particular sequence of irreducible representations can be constructed inductively, starting from the one dimensional representation of $Cl_1$ given by $\sigma_1 =1$. Then, for $Cl_2$, 
$$
\gamma_1 \;=\; \begin{pmatrix} 0 & \;1 \\ 1 & \;0 \end{pmatrix}
\;, 
\qquad 
\gamma_2 \;=\; \begin{pmatrix} 0 & -\I \\ \I & \;\;0 \end{pmatrix}
\;, 
\qquad
\gamma_0 \;=\; \begin{pmatrix} 1 & \;\;0 \\ 0 & -1 \end{pmatrix}
\;,
$$
and then one can continue iteratively by building the representation of $Cl_{2k+1}$ from the one of $Cl_{2k}$ via
$$
\sigma_{i}\;=\;\gamma_i  \quad \mbox{\rm for }i\leq 2k\;,
\qquad
\sigma_{2k+1}\;=\;\gamma_0 \;,
$$
and the representation of $Cl_{2k+2}$ from the one of $Cl_{2k+1}$ by
$$
\gamma_{i} \;=\; \begin{pmatrix} 0 & \sigma_i \\ \sigma_i & 0 \end{pmatrix} 
\quad \mbox{\rm for }i\leq 2k+1\;,
\qquad 
\gamma_{2k+2} \;=\; 
\I \begin{pmatrix} 0 & -\one \\ \one & \;\;0 \end{pmatrix}
\;, 
\qquad 
\gamma_{0} \;=\; \begin{pmatrix} \one & \;\;0 \\ 0 & -\one \end{pmatrix}
\;.
$$
These representations satisfy the normalizations \eqref{eq-OddClifford} and \eqref{eq-grading}. 
\hfill $\diamond$
\end{example}

\subsection{Bulk-boundary correspondence in a periodic unitary model}
\label{Sec-UExactModel} 

We present here a simple model from the unitary class in even dimension $d$ which displays a rich phase diagram and yet can be explicitly solved in the bulk and with a boundary.  Consider the irreducible representation of $Cl_d$ from example \ref{NiceClifford} and let $e_j$ be the generators of $\mathbb Z^d$ and $S_j$ the associated shifts on $\ell^2(\mathbb Z^d)$. The Hilbert space of the model is $\mathbb C^{2^\frac{d}{2}} \otimes \ell^2(\mathbb Z^d)$ and the bulk Hamiltonian is translation invariant and takes the form
\begin{eqnarray}\label{Model1}
H \;=\; 
\tfrac{1}{2\I} \sum_{j=1}^d \gamma_j \otimes (S_j-S_j^\ast) 
\;+\; 
\gamma_0 \otimes \Big(m+\tfrac{1}{2}\sum_{j=1}^d  (S_j+S_j^\ast)\Big)
\;.
\end{eqnarray}
The Fermi level is assumed at $\mu = 0$. The Bloch-Floquet decomposition gives
$$
H_k
\;=\;
\sum_{j=1}^d \gamma_j \, \sin (k_j) \;+\; \gamma_0 \, \Big (m+\sum_{j=1}^d \cos( k_j) \Big)
\;.
$$
As $(H_k)^2$ is proportional to the identity, there are just two eigenvalues of $H_k$
\begin{eqnarray}
\label{BulkEvenE}
E_k^\pm 
\;=\; \pm \,\sqrt{\sum_{j=1}^d \sin^2 (k_j) 
\;+\;
\Big(m+\sum_{j=1}^d \cos (k_j) \Big)^2}
\;,
\end{eqnarray}
hence the model displays two $\frac{d}{2}$-fold degenerate energy bands, arranged symmetrically relative to $E=0$. There is a spectral gap at the Fermi level, except when $m=0$, $\pm 2$, $\ldots$, $\pm d$. These are precisely the points where the topological transitions take place. Due to the simplicity of the spectrum, the Fermi projector can be computed explicitly
$$
P_k\;=\;
(E_k^+ - E_k^-)^{-1}(E_k^+ - H_k),
$$
and the top even Chern number can be evaluated using Eq.~\eqref{EvenChernK}. An analytical calculation is feasible by counting its jumps at the critical values of $m$ where the bulk gap closes. This analysis has been carried out in \cite{GJK}, see also \cite{QHZ}, and is sketched in the following. At the critical values, the band spectrum displays a set of Dirac singularities.

\begin{remark} 
Since the discussion is now about the bulk Hamiltonian, therefore in even $d$ space dimensions, near the singular points the Hamiltonian takes the form of a Dirac operator rather than a Weyl operator. Hence, the appropriate terminology here is Dirac points rather than Weyl points. \hfill $\diamond$ 
\end{remark} 

Both the critical $m$ values and the location of the Dirac points can be derived from \eqref{BulkEvenE} by imposing the gap closing condition
$$
\sum_{j=1}^d \sin^2( k_j) \;=\; 0 
\qquad {\rm and} \qquad   m+\sum_{j=1}^d \cos (k_j) \;=\;0
\;.
$$
These equations have the following solutions:
$$
 \begin{array}{lll}
&  m^0_c = -d\,,   & \;\;\;\;k^D = (0,0,\ldots,0)\;,
\\
&  m^1_c = -d+2\,, &  \;\;\;\;k^D = (\pi,0,0,\ldots,0) \;\;\; \mbox{plus $\left(^d_1\right)$ permutations}\;,\\
&  m^2_c = -d+4\,,  & \;\;\;\;k^D = (\pi,\pi,0,\ldots,0) \;\;\; \mbox{plus $\left(^d_2\right)$ permutations}\;,\\
& \;\;\;\;\;\;\;\; \vdots & \\
&  m^{d-1}_c = d-2\,,  & \;\;\;\;k^D = (\pi,\ldots,\pi,0) \;\;\; \mbox{plus $\left( ^{\ \ d}_{d-1}\right)$ permutations}\;,\\
& m^d_c = d\,,  & \;\;\;\;k^D = (\pi,\ldots,\pi)\;.
\end{array}
$$
The jumps of the Chern number at the gap closings can be explicitly evaluated \cite{GJK,QHZ}, allowing us to ultimately compute the actual Chern numbers. Indeed, when the gap is closed, there will be a number of Dirac singularities in the band spectrum, and the jumps of the bulk invariant result entirely from these Dirac points. When the bulk gap is nearly closed, {\it i.e.} $m=m_c+\epsilon$, $|\epsilon| << 1$, and near such Dirac singularity, the Bloch Hamiltonian takes an asymptotic form,
$$
H_k \;=\; \sum_{j=1}^d \alpha^D_j (k-k^D)_j \gamma_j \;+\;\epsilon \, \gamma_0 + \Oo(k^2)
\;,
$$
where $\alpha^D_j = \pm 1$ if $k^D_j =0,\pi$, respectively.  It will convenient to make the change of variables $\alpha^D_j (k-k^D)_j \rightarrow \xi_j$, in which case
$$
H_k \;=\; \langle \xi, \gamma \rangle \;+\;\epsilon \, \gamma_0
\;.
$$
The contribution to the Fermi projector coming from the band spectrum near the Dirac singularity is
$$
P(\xi)\; =\; \frac{1}{2}\;-\;\frac{1}{2}\frac{\langle \xi, \gamma \rangle +\epsilon \, \gamma_0}{\sqrt{\xi^2 + \epsilon ^2} }\;.
$$
To compute the contribution $\mathcal I$ of the band spectrum near $k^D$ to the total Chern number, we plug $P(\xi)$ into \eqref{EvenChernK}
\begin{align*}
\mathcal I 
\;=\; 
\frac{(2 \pi \I)^\frac{d}{2}}{ (\frac{d}{2})!} \frac{-1}{2^{d+1}} \prod_{i=1}^d 
\;\alpha^D_i\sum_{\rho \in \Ss_d} (-1)^\rho\int \frac{dk}{(2\pi)^d} 
\mathrm{tr}\left ( \frac{\epsilon \, \gamma_0}{\sqrt{\xi^2 + \epsilon ^2}} \prod_{j=1}^d \frac{\gamma_{\rho_j}}{\sqrt{\xi^2 + \epsilon ^2}}\right ) \; ,
\end{align*}
where the simplifications are solely due to the properties of the $\gamma$ matrices. The factor $\prod_{i=1}^d \alpha^D_i$ represents the Jacobian produced by the change of the variable made above. Up to a factor, the integrand converges to the Dirac-delta distribution, hence the domain of integration can be extended to $\RM^d$, in which case the integral can be explicitly evaluated and, with our conventions on $\gamma$'s, the result is
$$
\mathcal I \;=\; \frac{\chi}{2}\;\frac{\epsilon}{|\epsilon|}\;\prod_{i=1}^d \;\alpha^D_i\;, \qquad \chi = (-1)^{\frac{d}{2}+1}\;.
$$
When $\epsilon$ is varied from negative to positive values, $\mathcal I$ will jump by twice this quantity, leading to a total jump of $\chi \sum_D \prod_{i=1}^d \alpha^D_i$ for the bulk invariant, at the gap closing. Here it is assumed that $m$ increases and the sum is over all Dirac singularities present in the boundary band spectrum. Using the information provided above about the number and locations of the Dirac points, we see that the change of the Chern number at a critical value $m_c^n$ is
$$
\Delta_n {\rm Ch}_d(P_F) \;=\; \chi(-1)^n \binom{d}{n}\;.
$$
Finally, one can check that ${\rm Ch}_d(P_F)=0$ for $m<m_c^0$ by sending $m$ to $-\infty$. Hence for $m\in (-d+2n,-d+2n+2)$ with $n=0,\ldots,d-1$,
\begin{equation}\label{ExplicitEvenChern}
{\rm Ch}_d(P_F)\; =\; \chi\sum_{j=0}^n (-1)^j \binom{d}{j}
\;=\;
\chi (-1)^n \binom{d-1}{n}\;,
\end{equation}
and ${\rm Ch}_d(P_F) = 0$ for $m\not\in[-d,d]$.

\vspace{0.2cm}

Let us now consider the case with a boundary. Specifically the Hamiltonian is restricted to the Hilbert space $\mathbb C^{2^\frac{d}{2}} \otimes \ell^2(\mathbb Z^{d-1} \times \mathbb N)$ with Dirichlet boundary condition at $x_d=0$. As before, this restriction is denoted $\widehat{H}$. The Hamiltonian $\widehat{H}$ remains translationally invariant in the first $d-1$ direction, hence one can perform a partial Bloch-Floquet decomposition:
$$
\Ff\widehat{H}\Ff^*
\;=\;
\int^\oplus_{\mathbb T^{d-1}} dk \; \widehat{H}_k
\;,
\qquad
\widehat{H}_k: \mathbb C^{2^\frac{d}{2}} \otimes \ell^2(\mathbb N) \rightarrow \mathbb C^{2^\frac{d}{2}} \otimes \ell^2(\mathbb N)
\;,
$$
with
$$
\widehat{H}_k 
\;=\; 
\sum_{j=1}^{d-1}\sin(k_j) \, \gamma_j\otimes 1 \,+\, \tfrac{1}{2\I} \gamma_d \otimes (\widehat S - \widehat S^*) \,+\, \gamma_0 \otimes \Big  (m+\sum_{j=1}^{d-1}\cos (k_j) + \tfrac{1}{2}(\widehat S+\widehat S^*)\Big)
\,.
$$
Here, $\widehat S$ is the unilateral shift operator on $\ell^2(\mathbb N)$. For $\widehat E_k$ inside the bulk insulating gap, the solutions to the Schr\"odinger equation $H_k \psi_k = \widehat E_k \psi_k$ must be sought in the form
$$
\psi_k(x) 
\;=\; 
\xi_k \otimes (\lambda_k)^x
\;, 
\qquad 
|\lambda_k|<1
\;,
\;\;
\xi_k \in \mathbb C^{2^\frac{d}{2}}
\;.
$$
Writing the Schr\"odinger equation for generic $x_d>0$ and at $x_d=0$ with the Dirichlet boundary condition, leads to two independent constraints:
$$
\Big[\sum_{j=1}^{d-1}\sin(k_j)\gamma_j \;+\; \frac{\lambda_k -\lambda_k^{-1}}{2 \I}\gamma_d \;+\;\Big (m+\sum_{j=1}^{d-1}\cos (k_j )+ \frac{\lambda_k+\lambda_k^{-1}}{2}\Big)\gamma_0 \Big] \xi_k 
\;=\; 
\widehat E_k \xi_k\;,
$$
and
$$
\Big[\sum_{j=1}^{d-1} \sin(k_j)\gamma_j \;+\; \frac{\lambda_k}{2 \I}\gamma_d \;+\;\Big (m+\sum_{j=1}^{d-1}\cos( k_j) + \frac{\lambda_k}{2}\Big)\gamma_0 \Big] \xi_k 
\;=\; 
\widehat E_k \xi_k\;.
$$
Taking the difference of these equation, we obtain the simpler constraints 
$$
(\I \gamma_d + \gamma_0) \xi_k =0 
$$ 
and 
$$
\Big[\sum_{j=1}^{d-1} \sin(k_j)\gamma_j \;+\; \Big (m+\sum_{j=1}^{d-1}\cos( k_j) + \lambda_k\Big)\gamma_0 \Big] \xi_k 
\;=\; 
\widehat E_k \xi_k
\;.
$$
It is not difficult to see that these two constraints can be simultaneously satisfied only if the coefficient of $\gamma_0$ in the last constraint is identically zero. The conclusion is that $\xi_k \otimes (\lambda_k)^x$ solves the Schr\"odinger equation with the Dirichlet boundary condition at $x_d=0$ if and only if
$$
(\I \gamma_d + \gamma_0) \xi_k 
\;=\;0 
\quad \mbox{and} \quad 
\lambda_k 
\;=\; 
-\Big(m+\sum_{j=1}^{d-1}\cos (k_j) \Big)
\;.
$$
This implies that $\xi_k$ is a common eigenvector for two commuting matrices:
$$
\Big[\sum_{j=1}^{d-1} \sin(k_j)\gamma_j\Big] \xi_k 
\;=\; \widehat E_k \xi_k
\;,
$$
and
\begin{equation}\label{C2}
-\;\I\,\gamma_0\gamma_d \xi_k 
\;=\; 
-\;(-\I)^{\frac{d}{2}+1}\, \gamma_1 \cdots \gamma_{d-1} \xi_k 
\;=\; \xi_k\;.
\end{equation}

For $d=2$, the condition \eqref{C2} is equivalent to $\gamma_1 \xi_k =\xi_k$, hence $\xi_k$ is the unique eigenvector corresponding to the positive eigenvalue of $\gamma_1$, denoted by $\xi_+$ in the following (no dependence on $k=k_1$). The Schr\"odinger equation $\widehat{H}_k \psi_k = \widehat E_k \psi_k$ then admits a unique solution inside the insulating gap:
$$
\widehat E_k\;=\;\sin (k)\;, 
\qquad 
\psi_k(x)\; =\; \xi_+ \otimes \frac{(\lambda_k)^x}{\sqrt{2\big (1-(\lambda_k)^2 \big)} }\;, 
\qquad \lambda_k 
\;=\; -(m+\cos (k))
\;,
$$
which leads to an edge state provided the constraint $|\lambda_k|<1$ is satisfied. As one can see, there are no singular points in the boundary band spectrum and $\widehat{E}_k \approx \pm k $ near $E=0$. The sign depends on where the band crosses the $E=0$ mark, which can be at $k=0$ or $\pi$. The chirality $\nu^W$ of the edge band is determined by the constraint $|\lambda_k|<1$ which is equivalent to 
\begin{equation}\label{Co}
\cos( k) \in [-m-1,-m+1]\cap [-1,1]
\;.
\end{equation}
If $|m|>2$, the constraint \eqref{Co} cannot be fulfilled and consequently there are no edge bands. If $m\in(-2,0)$, then $k=0$ does satisfy \eqref{Co}, but $k=\pi$ does not. Hence, the slope of the edge band is positive when it crosses the $E=0$ level, hence the chirality $\nu^W$ is positive. If $m \in (0,2)$, then $k=\pi$ does satisfy \eqref{Co}, but $k=0$ does not. Hence, the slope of the edge band is negative when it crosses the $E=0$ level, hence the chirality is negative. These and the values of the Chern number given in \eqref{ExplicitEvenChern} confirm the bulk-boundary correspondence \eqref{BB1} in two space dimensions.

\vspace{.2cm}

For $d>2$, note that the matrix on the l.h.s. of \eqref{C2} is Hermitean and commutes with all $\gamma_1,\ldots,\gamma_{d-1}$. Hence, the constraint \eqref{C2} reduces the algebra of $\gamma_1,\ldots,\gamma_{d-1}$ to an irreducible representation of the complex odd Clifford algebra $Cl_{d-1}$. Indeed, the dimension of the linear subspace $\mathcal L \subset \mathbb C^{\frac{d}{2}}$ spanned by the $\xi$'s satisfying \eqref{C2} is $2^{\frac{d-2}{2}}$, and this subspace is invariant for the matrices $\gamma_1, \ldots, \gamma_{d-1}$. Hence we can define the linear operators:
$$
\hat{\sigma}_j: \mathcal L \rightarrow \mathcal L\,, 
\qquad 
\hat{\sigma}_j = \gamma_j \downharpoonright_{\mathcal L}\,, 
\qquad j=1,\ldots,d-1\,,
$$
which satisfy the Clifford relations $\hat{\sigma}_i \hat{\sigma}_j + \hat{\sigma}_j \hat{\sigma}_i = 2 \delta_{i,j}$ for $i,j=1,\ldots,d-1$, and the CCR convention $\hat \sigma_1 \cdots \hat \sigma_{d-1}= \I^\frac{d-2}{2}\one_{\mathcal L}$.

\vspace{.2cm}

We can now draw the conclusions for $d>2$: 

\begin{enumerate}[\rm (i)]

\item $\xi_k$'s are eigenvectors of a reduced Hamiltonian which is a Weyl-type operator
$$
\Big[\sum_{j=1}^{d-1} \sin(k_j)\hat{\sigma}_j\Big] \xi_k \;=\; \widehat E_k \xi_k\;.
$$

\item The band spectrum inside the insulating gap is given by
\begin{equation}\label{BoundaryEvenE}
\widehat E_k^\pm \;=\;
\pm \sqrt{ \sum_{j=1}^{d-1} \sin^2 (k_j) }
\;.
\end{equation}
The $\pm$ branches are connected at a singular point which occurs at $E=0$. This singularity is the Weyl point mentioned earlier. The bands are $2^{\frac{d-4}{2}}$-fold degenerate. This degeneracy can be lifted by a small perturbation except at $E=0$ where the bands will remain connected via a singularity. It is, however, possible to move the singularity both in $k$-space and in energy.

\item The $2^{\frac{d-4}{2}}$ eigenstates corresponding to $\widehat E_k^\pm$, respectively, are all of the form
$$
\psi_k(x) 
\;=\; 
\xi_k^\pm \otimes \frac{(\lambda_k)^x}{\sqrt{2(1-(\lambda_k)^2)} }\,, 
\qquad 
\lambda_k \;=\; -\Big(m+\sum_{j=1}^{d-1}\cos (k_j )\Big)
\;.
$$

\item Generically, the boundary bands are not defined over the entire Brillouin zone, but only over the domain determined by the implicit condition $|\lambda_k|<1$. By examining \eqref{BulkEvenE} and \eqref{BoundaryEvenE}, one can see that if $k$ is at the edges of this domain, then $\widehat E_k^+$ is aligned  with $\min\limits_{k_d} (E_{k,k_d}^+)$, and $\widehat E_k^-$ is aligned  with $\max\limits_{k_d} (E_{k,k_d}^-)$, where $E^\pm_{k,k_d}$ are the bulk eigenvalues \eqref{BulkEvenE}. These identities are not generic though as it may happen that edge spectrum overlaps bulk spectrum.

\item From \eqref{BoundaryEvenE}, one sees that the coordinates of the Weyl points are restricted to 
$$
k_j^W \in \{0,\pi\}, \qquad j=1,\ldots,d-1
\;.
$$
For $k$ in a neighborhood of a Weyl point, the reduced Hamiltonian can be approximated by an exact Weyl operator
\begin{equation}\label{W1}
\sum_{j=1}^{d-1}\alpha_j (k_j - k_j^W)\hat{\sigma}_j
\,,
\end{equation}
where the sign factors $\alpha_j=\pm 1$ are determined by the exact location of the Weyl point in the Brillouin zone. For example, if $k_j^W=0$ then $\sin (k_j)\approx k_j-k_j^W$, while for $k_j^W = \pi$ rather $\sin( k_j) = -(k_j -k_j^W)$. We recall that the signs of a pair $(\alpha_i,\alpha_j)$ can always be flipped by a continuous rotation in the $(k_i,k_j)$ plane. As such, if \eqref{W1} contains an even number of negative $\alpha_j$'s, then \eqref{W1} is homotopic with $+\langle(k - k^W)|\hat{\sigma}\rangle$ and will have a positive chirality. If \eqref{W1} contains an odd number of negative $\alpha_j$'s, then \eqref{W1} is homotopic with $-\langle (k - k^W)|\hat{\sigma}\rangle$ and will have a negative chirality.

\item There can be more than one Weyl point. The condition which determines how many Weyl points are there and where are they exactly located is
$$
|\lambda_{k^W} |<1 
\quad\Longleftrightarrow \quad
\sum_{j=1}^{d-1}\cos( k_j^W) \;\in\; [-1 -m,1-m]\cap [-d+1,d-1]
\;.
$$

\end{enumerate} 

Now we can demonstrate the bulk-boundary principle \eqref{BB1} for this particular model. Indeed, let $m \in (-d+2n, -d +2n+2)$. Then there is only one combination (modulo permutations) of $d-1$ numbers equal to $+1$ or $-1$, representing the $\cos(k_j^W)$ appearing in the last equation, such that their sum belongs to the interval $(-1 -m,1-m).$ Indeed, since
$$
(-1 -m,1-m)\; \subset\; (d -2n-1,d-2n+1)\;,
$$ 
$n$ of these numbers have to be $-1$ and $(d-1-n)$ of them have to be $+1$.
There are $\binom{d-1}{n}$ permutations of these signs, corresponding to as many distinct locations of the Weyl points. Furthermore, precisely $n$ of the coordinates $k^W_j$ are equal to $\pi$ while the remaining are zero, hence the chirality of all Weyl points is the same and equal to $(-1)^n$. The conclusion is that the boundary invariant is
$$
\sum \nu_W \;=\; (-1)^n  \binom{d-1}{n}\;,
\qquad
m\in(-d +2n,-d-2n+2)
\;,
$$
and hence, when multiplied by the sign factor $\chi$, it equals the bulk even Chern number given in \eqref{ExplicitEvenChern}.

\section{The chiral unitary class}
\label{Sec-ChiralUnitaryClass}

The solid state systems from the chiral unitary class have a unitary time evolution semi-group and a sub-lattice symmetry to be described in great length below. Following the same format as for the previous section, we introduce the models and their physical characteristics, both for bulk and for half-space. We formulate the bulk-boundary principle for periodic systems and demonstrate this principle using an exactly solvable model in arbitrary odd dimension. The existing experimental results are briefly surveyed. 

\subsection{General characterization}
\label{Sec-CUClassGeneral}

The lattice models for insulators from the chiral unitary class are defined over the Hilbert space $\mathbb C^{2N} \otimes \ell^2(\mathbb Z^d)$, as the dimension of the fiber is necessarily an even integer. A Hamiltonian $H$ displays chiral (or sublattice) symmetry if there exists a symmetry $J$ on $\mathbb C^{2N}$ satisfying $J^* = J$ and $J^2=\one_{2N}$ and having eigenspaces of equal dimension, such that 
\begin{equation}
\label{eq-chiralcont}
(J\otimes \one)\, H\, (J\otimes \one)
\;=\;-\,H 
\;.
\end{equation}
Throughout, we work with a basis of $\mathbb C^{2N}$ such that $J$ takes a diagonal form 
\begin{equation}
\label{eq-Jgrad}
J \;=\; \begin{pmatrix} \one_N & 0 \\ 0 & -\one_N \end{pmatrix},
\end{equation}
We will also write $J$ instead of $J \otimes \one$. The Fermi level is pinned at $0$ for the chiral unitary symmetry class which is a point of reflection symmetry of the spectrum of $H$ by \eqref{eq-chiralcont}. Since we deal with insulators, the Fermi level will also be assumed to be in a spectral gap of the bulk Hamiltonian. In this situation the Fermi projection $P_F=\tfrac{1}{2}(\one-\sgn (H) )$ is given in terms of a unitary $U_F$ on $\CM^N\otimes\ell^2(\ZM^d)$ because \eqref{eq-chiralcont} implies
\begin{equation}
\label{eq-signU}
\sgn(H) 
\;=\; \begin{pmatrix}
0\; & U_F^* \\
U_F\; & 0
\end{pmatrix}
\;.
\end{equation}
We will refer to $U_F$ as the Fermi unitary operator, in analogy with the Fermi projection for the unitary class. It encodes the Fermi projection $P_F=\chi(H\leq 0)$ of a chiral Hamiltonian via
\begin{equation}
\label{eq-PF_UF}
P_F
\;=\; 
\frac{1}{2}
\begin{pmatrix}
\one & -U_F^* \\
-U_F & \one
\end{pmatrix}
\;.
\end{equation}

\vspace{.2cm}

Let us begin by looking at periodic models with vanishing magnetic field. The Hamiltonian $H : \mathbb C^{2N} \otimes \ell^2(\mathbb Z^d)  \rightarrow \mathbb C^{2N} \otimes \ell^2(\mathbb Z^d)$ is given by \eqref{UGenericModel} together with the chirality constraint, which implies $W_y=\begin{pmatrix} 0 & w_y \\ w_{-y}^* & 0 \end{pmatrix}$ with $N\times N$ matrices $w_y$ so that
\begin{align}
\label{CUGenericModel}
H \;=\;\sum_{y\in \mathbb Z^d}
\begin{pmatrix}
0 & w_y \\
w_{-y}^* & 0
\end{pmatrix} \otimes S^y
\;.
\end{align}
Its Bloch-Floquet decomposition \eqref{eq-BlochFloquet} has fiber Hamiltonians
$$
H_k \;=\; \sum\limits_{y \in \mathbb Z^d} 
\begin{pmatrix}
0 &  e^{\I \langle y|k\rangle} w_y \\
e^{-\I \langle y|k\rangle} w_y^* & 0
\end{pmatrix}.
$$
By examining the classification table, we see that the topologically non-trivial phases are conjectured to occur only in odd space dimensions. Furthermore, for each such dimension, there is an infinite sequence of topological phases and the phases can be distinguished from one another by tagging them with just one integer number. In the bulk, this number is given by the top odd Chern number \cite{SRL,RSFL}:
\begin{equation}
\label{OddChernK}
{\mathrm{Ch}}_d  (U_F)
\;=\; 
\frac{\I(\I \pi)^\frac{d-1}{2}}{ d!!} \sum_{\rho \in \Ss_d} (-1)^{\rho}\int_{\mathbb{T}^d}\frac{dk}{(2 \pi)^d} \  
\mathrm{Tr} \Big ( \prod_{j=1}^d U^\ast_F(k) \frac{\partial U_F(k)}{\partial k_{\rho_j}}  \Big )
\;,
\end{equation}
where $U_F(k)$ is the $N\times N$ matrix appearing in the Bloch-Floquet decomposition $\Ff U_F\Ff^*=\int^\oplus_{\mathbb T^d} dk \, U_F(k)$ of the Fermi unitary operator. As we shall see in Chapter~\ref{Chap-Conclusions}, the bulk topological invariant for chiral symmetric solid state systems is a physically measurable coefficient.

\begin{remark} 
We will use the same notation for the bulk invariants, but it will be always understood that ${\rm Ch}_d$ refers to \eqref{EvenChernK} (and its extensions) when $d$ is even, and to \eqref{OddChernK} (and its extensions) when $d$ is odd.\hfill $\diamond$ 
\end{remark}

Next, let us write out the generic chiral models with a magnetic field and disorder. In the symmetric gauge, the systems are again described by covariant families of Hamiltonians of the form \eqref{UGenericModelBDSymmetric}, but with the chirality constraint \eqref{eq-chiralcont}:
\begin{align}
\label{UCGenericModelBDSymmetric}
H_{\sym,\omega} 
& \;=\;
\sum_{y \in \Rr}\sum_{x \in \mathbb Z^d}  \left(
\begin{array}{cc}
0 & w_y(\tau_x \omega) \\
w_{-y}(\tau_x \omega)^* & 0
\end{array}
\right) \otimes | x \rangle \langle x|\,U_\sym^y \\
& \;=\;
\sum_{y \in \Rr}\sum_{x \in \mathbb Z^d} e^{\frac{\I}{2}\langle y|\bm B| x \rangle} \left(
\begin{array}{cc}
0 & w_y(\tau_x \omega) \\
w_{-y}(\tau_x \omega)^* & 0
\end{array}
\right) \otimes | x \rangle \langle x-y|
\;. 
\nonumber
\end{align}
The representation in the Landau gauge, which will be primarily used in the following, is similarly obtained from \eqref{UGenericModelBDLandau}:
\begin{equation}
\label{UCGenericModelBDLandau}
\boxed{
\;
H_\omega 
 \;=\;
\sum_{y \in \Rr}\sum_{x \in \mathbb Z^d}  e^{\frac{\I}{2}\langle y|\bm B_+| y \rangle}\left(
\begin{array}{cc}
0 & w_y(\tau_x \omega) \\
w_{-y}(\tau_x \omega)^* & 0
\end{array}
\right) \otimes | x \rangle \langle x|\,U^y
\;.\;
}
\end{equation}
Here $w_y$ are continuous functions on the space of disorder configurations $\Omega$. The top odd Chern number has a real-space representation \cite{MSHP,PS}, which can be applied to models like \eqref{UCGenericModelBDLandau}. With the notation introduced in the previous section,
\begin{equation}
\label{OddChernR}
\boxed{
\;
{\mathrm{Ch}}_{d}(U_F) 
\;=\; 
\frac{\I(\I \pi)^\frac{d-1}{2}}{ d!!} \sum_\rho (-1)^\rho \;\mathcal{T}\Big ( \prod_{i=1}^{d}   U^\ast_\omega \, \I [U_\omega,X_{\rho_i}] \Big )
\;,\;
}
\end{equation}
where $U_F=\{U_\omega\}_{\omega\in\Omega}$ is the covariant family of Fermi unitary operators. The invariant ${\mathrm{Ch}}_{d}(U_F)$ is known to remained quantized, non-fluctuating from one disorder configuration to another, and be homotopically stable as long as the Fermi level resides in a region of dynamically localized spectrum, see \cite{PS} and Chapter~\ref{Chap-IndexTheorems}.

\vspace{.2cm}

When a chiral symmetry preserving boundary is present and ${\mathrm{Ch}}_{d}(U_F)\neq 0$, the energy spectrum extends inside the bulk insulating gap. The boundary spectrum does not necessarily cover the entire insulating gap. A situation when this doesn't happen is when a magnetic field perpendicular to the surface of a three-dimensional crystal breaks the boundary spectrum into a Hofstadter pattern. The case $d=1$ is special and, since it was already discussed in Chapter~\ref{Chap-Illustration}, it will be excluded from the following discussion. For periodic crystals with a planar boundary, say $x_d \geq 0$, and in the absence of magnetic fields, the boundary states can be determined as a function of momentum $k$ parallel to the boundary. The hallmark feature is the existence of boundary energy bands  displaying Dirac singularities at $E=0$ \cite{SRFL,HRV}. Around a Dirac point $k^D$, the spectrum and the states are well described by a Dirac operator 
\begin{equation}\label{Dirac}
\sum_{j=1}^{d-1} v_j  (k_j-k_j^D) \gamma_j
\;,
\end{equation}
where $\gamma$ are the generators of the irreducible representation of the complex even Clifford algebra $Cl_{d-1}$ (fixed by our conventions) and $v_j$ are the slopes of the bands at $E=0$.  
Now a chirality $\nu_D=\prod_{j=1}^{d-1} \sgn{(v_j)}$ can be defined for each Dirac point, just as for the Weyl points in Section~\ref{Sec-UnitaryClass}. The central conjecture for the chiral unitary class is the following bulk-boundary principle \cite{RSFL}:
\begin{equation}
\label{BB2}
{\mathrm{Ch}}_d(U_F) 
\;=\; 
\chi \sum \nu_D
\;,
\end{equation}
where the sum carries over all Dirac singularities located at $E=0$ of the boundary band spectrum, and $\chi$ is again a sign which depends on the representations of the Clifford algebras and normalization of the bulk invariant. One conclusion that can be drawn from this principle is that, as long as ${\mathrm{Ch}}_d(U_F) \neq 0$, there will always be boundary bands at $E=0$. Hence, unavoidably, the insulator becomes metallic when a boundary is present. Similarly as for the unitary class, it is one of the main goals of the present work to formulate $\sum \nu_D$ as a boundary topological invariant which makes sense in the presence of  magnetic fields and disorder, to derive an index theorem for it and to establish \eqref{BB2}. Among other things, this will enable us to demonstrate that the boundary energy spectrum at $E=0$ remains extended in the presence of disorder whenever ${\mathrm {Ch}}_d(U_F) \neq 0$.

\begin{figure*}
\begin{center}
\includegraphics[width=10cm]{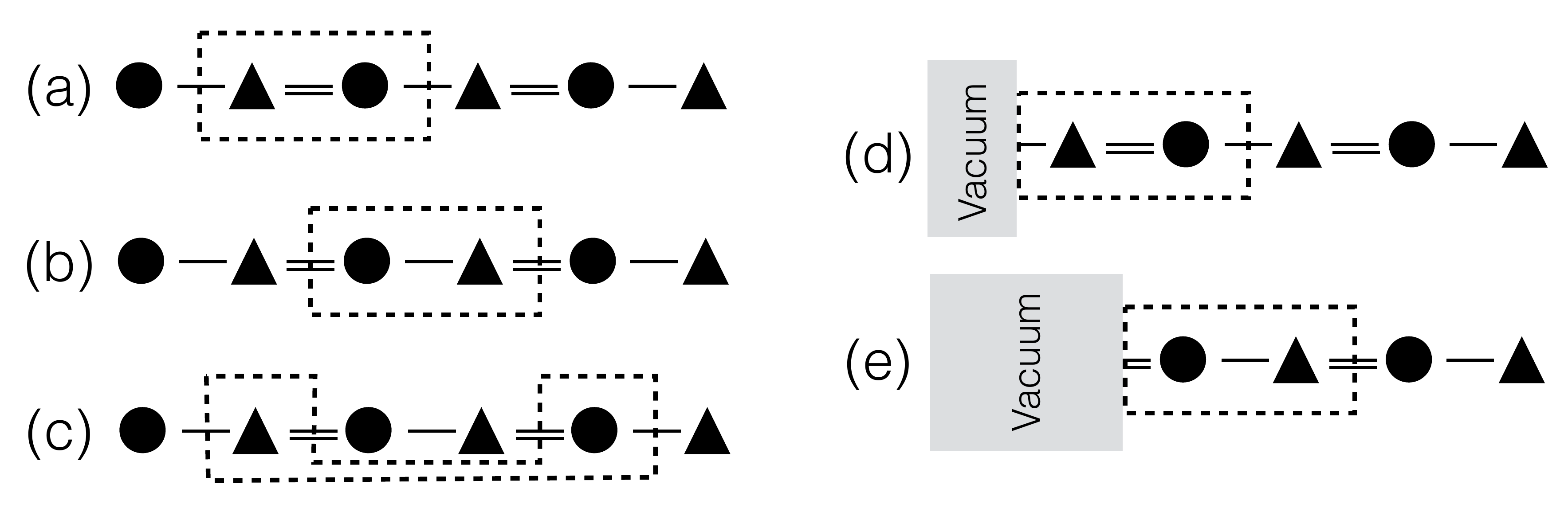}
\caption{Graphical representation of the model \eqref{Model1d} as a molecular chain containing two species of atoms with alternating hopping amplitudes. Panels (a)-(c) show various possibilities to choose the unit cell. Panels (d)-(e) show the unique unit cells compatible with the given boundaries.}
\label{fig-chiral}
\end{center}
\end{figure*}

\vspace{.2cm}

We now come to the extremely important point of choosing the unit cell of the crystal. This determines which states are regrouped in the fibers $\CM^{2N}$ and which are the hopping matrices in the Hamiltonian \eqref{UCGenericModelBDLandau}. It is well-known in the physics community that the value of the bulk invariant for the chiral class depends on this process. We will carry out the discussion on a model in dimension $d=1$ which describes a chain with two different atoms  (as in \cite{HPB}, p.~22, for example). Figure~\ref{fig-chiral} shows two alternating molecular states (or two  alternating atoms) and two alternating hopping matrices (the horizontal links).  Each of the choices (a), (b) and (c) of the unit cell lead to a different chiral unitary operator $U_F^{\mbox{\rm\tiny (a)}}$,  $U_F^{\mbox{\rm\tiny (b)}}$ and  $U_F^{\mbox{\rm\tiny (c)}}$, respectively. For adequate fixed values of the parameters, one finds  $\Ch_d(U_F^{\mbox{\rm\tiny (a)}})=0$, $\Ch_d(U_F^{\mbox{\rm\tiny (b)}})=1$ and $\Ch_d(U_F^{\mbox{\rm\tiny (c)}})=2$, respectively. Furthermore, \cite{Thi2} showed that, using certain isomorphisms defined in momentum space, one can change ${\rm Ch}_d$ by any even number. In the real space representation, one such isomorphism corresponds to redefining the unit cell in panel (a) into the unit cell in panel (c). This arbitrariness is very puzzling at first sight for, given a concrete problem, how are we going to predict the physical surface properties from the bulk invariant? The issue has a very simple resolution: The boundary itself dictates which {\it unique} unit cell is to be used in the computation of the bulk invariant. Thus the rule is that the boundary never cuts through a unit cell, which mathematically means that the fiber subspaces are either erased or kept, entirely, but never split. For example, if the boundary is as in panel (d), then only the unit cell shown in panel (a) obeys this rule, and if the boundary is as in panel (e), then only the unit cell shown in panel (b) obeys the rule. The unit cells of the type shown in panel (c) will always be cut through by a boundary, hence they can be dropped from the beginning. These conclusions apply also in higher space dimensions where one needs $d$ boundaries to uniquely determine the bulk unit cell and hence the Fermi unitary as well as its Chern number.

 \subsection{Experimental achievements} 
 \label{SubSec-Exp2} 
 
We should make clear from the start that the chiral symmetry is never exact in solid state systems. After all, the non-relativistic Schrodinger operators are bounded from below and the spectrum extends all the way to $+ \infty$. The chiral symmetry should be sought in the electron spectrum near the Fermi level, which determines most of the electronic properties of materials. Moreover, it will be shown below that for approximate chiral systems, namely those obtained by a sufficiently small perturbation of an exact chiral system, one can still define a Fermi unitary and its Chern number. Hence non-trivial odd Chern numbers do not require exact chiral symmetry, but in such conditions the delocalized character of the boundary states is lost, in general. There are, however, several materials where chiral symmetry can be assumed virtually exact. The prototypical examples of strong topological materials from the chiral unitary class are the one-dimensional conducting polymers, with poly-acetylene as the prominent representative \cite{BCM}. The Su-Schrieffer-Heeger model \cite{SSH} used in our introductory Chapter~\ref{Chap-Illustration}  was developed precisely for the description of poly-acetylene. The conducting polymers are $\pi$-conjugated organic molecular chains which in the absence of lattice distortions would have extended $\pi$-molecular orbitals and would display a metallic character. The systems, however, are unstable to Peierls lattice distortions which double the original repeating cells \cite{BCM}. These distortions open small gaps at the Fermi level and drive these systems into an insulating chiral topological phase. There is a tremendous interest in these materials, not because of their topological properties, but because these polymers can become again metallic when doped with strong oxidizing or reducing agents \cite{CFP}, thus paving the way for conducting plastics \cite{CFP}. 

\vspace{.2cm}

Graphene \cite{Nov1,Nov2} is a two-dimensional crystal which also displays the chiral symmetry. The band spectrum of graphene is gapped everywhere except at two special points of the Brillouin zone, hence graphene can be considered as a special case of weak topological material from the chiral class. Its honeycomb lattice can be cleaved along the zigzag, the bearded or the arm-chair edges, all of which preserve the chiral symmetry.  Using a partial Bloch-Floquet transformation in the momentum $k$ parallel with the boundary, one obtains families of $k$-dependent one-dimensional chiral symmetric Hamiltonians. Excepting two $k$ values, these Hamiltonians are gapped and hence one can compute the bulk invariant  \cite{Hat2}.  Whenever the invariant takes a non-trivial value, boundary states occur at $E=0$ which ultimately lead to dispersionless boundary bands. It is known that such dispersionless edge states exists along the zigzag edge \cite{FWN}. The bearded edge is unstable for graphene, but it was engineered in photonic crystals and the dispersionless  edge states were confirmed \cite{Plo}.  There are no edge states along the armchair edge. The different behaviors are due to the fact that the unit cell changes from one boundary to another (cf. discussion above). Alternatively, these characteristics of graphene can be explained directly using the boundary invariant  \cite{Hat2}. 

\vspace{.2cm}

In a recent development, Kane and Lubensky \cite{KL} have discovered that, within the harmonic approximation, any isostatic mechanical lattice has a built-in chiral symmetry. They also demonstrated, theoretically, the mechanical analog of the one dimensional Su-Schrieffer-Heeger model and constructed weak chiral symmetric topological mechanical materials in two and three dimensions. These theoretical predictions have recently  been confirmed in the lab \cite{PBV}. 

\vspace{.2cm}

So far, we have only mentioned the weak topological insulators in higher dimensions. The search for the strong topological materials with exact (or weakly broken) chiral symmetry is vigorously underway. For example, there are several feasible proposals to realize such systems with cold atoms trapped in optical lattices \cite{WDD,WDM}. Our Section~\ref{sec-SurfacQHE} should be a helpful theoretical contribution to this search.
 
\subsection{Bulk-boundary correspondence in a periodic chiral model}
\label{CUExactModel} 

Here we present a simple model from the chiral unitary class which displays a rich phase diagram and yet can be explicitly solved in the bulk and with a boundary.  Let $\gamma$ be the generators of the irreducible representation of $Cl_{d+1}$ from example \ref{NiceClifford}. Using the same notations as in Section~\ref{Sec-UExactModel}, the bulk Hamiltonian acting now on $\mathbb C^{2^\frac{d+1}{2}} \otimes \ell^2(\mathbb Z^d)$ is 
$$
H 
\;=\; 
\tfrac{1}{2\I} \sum_{j=1}^d \gamma_j \otimes \big (S_j-S_j^\ast \big) 
\;+\; 
\gamma_{d+1} \otimes \Big(m+\tfrac{1}{2}\sum_{j=1}^d(S_j+S_j^\ast)\Big)
\;.
$$
It has the required chiral symmetry $\gamma_0 H \gamma_0 =-H$. Its Bloch-Floquet fibers
$$
H_k\;=\;
\sum_{j=1}^d   \sin (k_j)\gamma_j\;+\; \Big (m+\sum_{j=1}^d \cos (k_j) \Big)\gamma_{d+1}
$$
have just two eigenvalues
\begin{eqnarray}
\label{CUBulkE}
E_k^\pm 
\;=\; 
\pm \,
\sqrt{\sum_{j=1}^d \sin^2  (k_j)\; +\;\Big(m+\sum_{j=1}^d \cos (k_j) \Big)^2}
\;.
\end{eqnarray}
Hence the model displays two $\frac{d+1}{2}$-fold degenerate energy bands arranged, symmetrically relative to $E=0$. There is a spectral gap at $E=0$, except when $m$ is equal to $\pm 1$,$\pm 3$, $\ldots$, $\pm d$. These are precisely the points where the topological transitions take place. Due to the simplicity of the spectrum, the Fermi unitary matrix $U_k$ can be computed explicitly to be
$$
U_k\; =\; (E_k^+ )^{-1} \left [ \sum_{j=1}^d   \sin (k_j)\,\sigma_j\;+\; \I \Big (m+\sum_{j=1}^d \cos (k_j) \Big) \, \one \right ]
\;,
$$
where $\sigma_j$'s are the irreducible representation of the odd complex Clifford algebra $Cl_d$ on $\mathbb C^{2^{\frac{d-1}{2}}}$ (with our CCR conventions). The top odd Chern number can again be computed by counting the change at the critical values of $m$ where the bulk gap closes, as done for the unitary case. Formally, the gap closing conditions are exactly the same as in the unitary case, and the analysis can be adapted. Near a gap closing, the contribution $\mathcal I$ to the bulk invariant becomes
\begin{align*}
\mathcal I 
\;=\; 
\frac{\I(\I \pi)^\frac{d-1}{2}}{ d!!} \prod_{i=1}^d 
\;\alpha^D_i\sum_{\rho \in \Ss_d} (-1)^\rho\int \frac{dk}{(2\pi)^d} 
\mathrm{tr}\left ( \frac{-\I\epsilon }{\sqrt{\xi^2 + \epsilon ^2}} \prod_{j=1}^d \frac{\sigma_{\rho_j}}{\sqrt{\xi^2 + \epsilon ^2}}\right ) \; ,
\end{align*}
which can be computed explicitly 
$$
\mathcal I \;=\; \frac{\chi}{2}\;\frac{\epsilon}{|\epsilon|}\;\prod_{i=1}^d \;\alpha^W_i\;, \qquad \chi = (-1)^{\frac{d-1}{2}}\;,
$$
where one should note that now we have Weyl singularities at the gap closing. By literally repeating the counting done for the unitary case, we conclude 
\begin{equation}\label{ExplicitOddChern}
{\rm Ch}_d (U_F)
\;=\; 
\chi (-1)^n \binom{d-1}{n}\;,
\end{equation}
for $m\in (-d+2n,-d+2n+2)$ with $n=0,\ldots,d-1$, and ${\rm Ch}_d(U_F) = 0$ otherwise.

\vspace{0.2cm}

We now impose the Dirichlet boundary condition at $x_d=0$. As before, a partial Bloch-Floquet decomposition has fibers
$$
\widehat H_k 
\;=\; 
\sum_{j=1}^{d-1}\sin (k_j)  \, \gamma_j \otimes 1 + \tfrac{1}{2\I}\gamma_d \otimes (\widehat S - \widehat S^*) +\gamma_{d+1} \otimes \Big (m+\sum_{j=1}^{d-1}\cos (k_j) + \tfrac{1}{2}(\widehat S+\widehat S^*)\Big)
\;,
$$
and the solutions to the Schr\"odinger equation $\widehat H_k \psi_k = \widehat E_k \psi_k$ are sought in the form
$$
\psi_k(x)\; = \;\xi_k \otimes (\lambda_k)^x\;, 
\qquad 
|\lambda_k|<1\;,\;\; \xi_k \in \mathbb C^{2^\frac{d+1}{2}},
$$
Due to the Dirichlet boundary condition at $x_d=0$ this leads to the two independent constraints
$$
\Big [\sum_{j=1}^{d-1}\sin(k_j)   \gamma_j \;+\; \frac{\lambda_k -\lambda_k^{-1}}{2 \I}\gamma_d  \;+\;  \Big (m+\sum_{j=1}^{d-1}\cos (k_j) + \frac{\lambda_k+\lambda_k^{-1}}{2}\Big)\gamma_{d+1} \Big ] \xi_k
\; =\; \widehat E_k \xi_k
\;,
$$
and
$$
\Big[\sum_{j=1}^{d-1} \sin(k_j)\gamma_j \;+\; \frac{\lambda_k}{2 \I}\gamma_d \;+\;\Big (m+\sum_{j=1}^{d-1}\cos (k_j) \,+\, \frac{\lambda_k}{2}\Big)\gamma_{d+1} \Big ] \xi_k
\; =\; \widehat E_k \xi_k
\;.
$$
They can be simultaneously satisfied if only if
$$
(\I \gamma_d + \gamma_{d+1}) \xi_k \;=\;0 \quad \mbox{and} \quad \lambda_k \;=\; -\Big(m+\sum_{j=1}^{d-1}\cos( k_j)\Big)
\;.
$$
This implies that $\xi_k$ is a common eigenvector for two commuting matrices:
\begin{equation}\label{C3}
\Big [\sum_{j=1}^{d-1} \sin(k_j)\gamma_j\Big ] \xi_k \;=\; \widehat E_k \xi_k\;,
\end{equation}
and
\begin{equation}\label{C4}
-\I \gamma_{d+1} \gamma_d  \xi_k \;=\; \xi_k\;.
\end{equation}

Let $\mathcal L \subset \mathbb C^{2^\frac{d+1}{2}}$ be the linear space spanned by the $\xi$'s satisfying \eqref{C4} whose dimension is $2^\frac{d-1}{2}$. This linear space is invariant for the matrices $\gamma_1,\ldots,\gamma_{d-1}$ so that one can define
$$ 
\hat \gamma_j \;=\; \gamma_j \downharpoonright_{\mathcal L}\;, \qquad  j=1,\ldots,d-1\;,
$$
as well as $ \hat \gamma_0 = \gamma_0 \downharpoonright_{\mathcal L} =(-\I)^\frac{d-1}{2}\,\hat \gamma_1 \cdots \hat \gamma_{d-1}$. This provides an irreducible representation of the even complex Clifford algebra $Cl_{d-1}$ on $\mathcal L$, satisfying our conventions. We are now ready to draw our conclusions for $d>1$:

\begin{enumerate}[\rm (i)]

\item $\xi_k$'s are eigenvectors of a reduced Hamiltonian which is of Dirac-type
$$
\Big [\sum_{j=1}^{d-1} \sin(k_j)\hat \gamma_j \Big] \xi_k \;=\; \widehat E_k \xi_k
\;.
$$

\item The band spectrum inside the insulating gap is given by
\begin{equation}\label{CUBoundaryE}
\widehat E_k^\pm \;=\;\pm \sqrt{ \sum_{j=1}^{d-1} \sin^2  (k_j) }
\;.
\end{equation}
The $\pm$ branches are connected at a singular point which occurs at $E=0$. This singularity is the Dirac point mentioned earlier. The bands are $2^{\frac{d-3}{2}}$-fold degenerate. This degeneracy can be lifted by a small periodic perturbation except at the Weyl point  where the bands will remain connected via a singularity.

\item The $2^{\frac{d-3}{2}}$ eigenstates corresponding to $\widehat E_k^\pm$ are all of the form
$$
\psi_k(x) \;=\; \xi_k^\pm \otimes \frac{(\lambda_k)^x}{\sqrt{2(1-(\lambda_k)^2)} }\;, 
\qquad 
\lambda_k \;=\; -\Big(m+\sum_{j=1}^{d-1}\cos (k_j) \Big)
\;.
$$

\item Generically, the boundary bands are not defined over the entire Brillouin zone, but only over the domain determined by the implicit condition $|\lambda_k|<1$. 

\item From \eqref{CUBoundaryE}, the $d-1$ coordinates $k_j^D$ of the Dirac points can only be equal to $0$ or $\pi$. For $k$ in a neighborhood of such a Dirac point, the reduced Hamiltonian can be approximated by an exact Dirac operator
\begin{equation}\label{W2}
H_k \;\approx\; \sum_{j=1}^{d-1}\alpha_j (k_j - k_j^D)\, \hat \gamma_j\;,
\end{equation}
where the sign factors $\alpha_j=\pm 1$ are determined by the exact location of the Dirac point in the Brillouin zone. We can always flip the signs of a pair $(\alpha_i,\alpha_j)$ by a continuous rotation in the $(k_i,k_j)$ plane. As such, the Hamiltonians \eqref{W2} fall into two homotopy classes, one of positive chirality for which \eqref{W2} contains an even number of negative $\alpha_j$'s, and one of negative chirality for \eqref{W2} which contains an odd number of negative $\alpha_j$'s.
\item There could be more than one Dirac point. The condition which determines how many Dirac points are there and where are they exactly located is:
$$
|\lambda_k^D |<1 \;\;\;\Longleftrightarrow\;\;\; \sum_{j=1}^{d-1}\cos (k_j^D) \;\in\; [-1 -m,1-m]\cap [-(d-1),d-1]\;.
$$
\end{enumerate} 
The bulk-boundary correspondence can now be established following line by line the arguments provided for the unitary case.

\section{Main hypotheses on the Hamiltonians}
\label{Sec-MainH}

This section translates the settings and the assumptions in a mathematically precise language and presents the behavior of various quantities of interest under such circumstances. Most of the statements are well-known or can be found in the literature, hence some are presented without a proof. Having all these statements listed in one place will be useful because they are referenced often throughout the book. 

\subsection{The probability space of disorder configurations} 
\label{sec-disorder}

Here an explicit mathematical definition of the dynamical system  $(\Omega,\tau,\ZM^d,\PM)$ describing the disorder configurations of the models is given. Throughout it will be assumed that this particular set-up is given. Recall that the allowed hopping range $\Rr\subset\ZM^d$ is supposed to be finite.

\begin{definition}\label{DisorderConfigSpace} 
Suppose that the randomness in the individual hopping process by $y\in \mathbb Z^d$ can be described by a compact and convex (hence contractible) space $\Omega_0^y$ equipped with the probability measure $\mathbb P_0^y$. Then the dynamical system $(\Omega,\tau,\ZM^d,\PM)$ is defined by:
\begin{enumerate}[\rm (i)]
\item The compact and metrizable Tychonov space
\begin{equation}
\label{eq-Omega} 
\Omega\;=\;\big(\prod_{y\in \mathcal R}\Omega^y_0\big)^{\ZM^d} 
\;.
\end{equation}
\item The family of homeomorphisms
$$
(\tau_z \omega)_x^y\;=\;\omega_{x - z}^y
\;,
\qquad
\omega\;=\; \Big (\omega_x^y\Big )_{x\in\ZM^d}^{y\in \Rr}\in\Omega
\;, \quad z\in \mathbb Z^d.
$$
In particular, the homeomorphisms corresponding to the generators $e_j$ of $\mathbb Z^d$ will be dented by $\tau_j$, so that $\tau_z = \tau_1^{\circ z_1} \ldots \tau_d^{\circ z_d}$.
\item The product probability measure
\begin{equation}
\label{eq-OmegaProb} 
\PM(d\omega) 
\;=\; 
\prod_{y \in \Rr}\; \prod_{x\in \mathbb Z^d} \,\PM^y_0(d \omega_x^y)
\;,
\end{equation}
which is invariant and ergodic w.r.t. the $\ZM^d$ action $\tau$.
\end{enumerate}
\end{definition}

For sake of concreteness, let us give a very concrete and simple example of $\Omega$ and also the matrix functions $W_y$ entering into the Hamiltonian \eqref{UGenericModelBDLandau}. One may choose $\Omega^y_0=[-\tfrac{1}{2},\tfrac{1}{2}]$ with $\PM_0(d \omega_x^y)=d \omega_x^y$, and 
$$
W_y(\omega)
\;=\; 
(1+\lambda_y \omega_0^y) W_y
$$ 
with real coefficients $\lambda_y$ which can be seen as a measure of disorder strength. 

\vspace{0.2cm}

One last but important observation spurs from the fact that the space $\Omega$ is contractible. In this case, all the maps are homotopic with the constant map. As a consequence, the map $\tau$ and the identity map are homotopically equivalent. This will have an important consequence for the K-theory of the observables algebras.

\subsection{The bulk Hamiltonians}
\label{sec-BulkHam}

The analysis carried out in this book applies to the families of Hamiltonians $H=\{H_\omega\}_{\omega \in \Omega}$ defined in \eqref{UGenericModelBDLandau} and \eqref{UCGenericModelBDLandau}, and indexed by the disorder probability space $(\Omega,\tau,\ZM^d,\PM)$ described in Definition~\ref{DisorderConfigSpace}. These families of Hamiltonians satisfy the covariance relation \eqref{CovRelLandau}. The bulk analysis can be carried out as well in the symmetric gauge but, to avoid confusion, we consider only the Landau gauge from now on. Almost surely, the spectra of $H_\omega$ are identical non-random sets (see {\it e.g.} \cite{CFKS}). This non-random set can be regarded as the spectrum of $H$, the family of covariant Hamiltonians. 

\vspace{.2cm}

\noindent {\bf Bulk Gap Hypothesis (BGH):} {\it The Fermi level $\mu\in\RM$ lies in a gap $\Delta\subset\RM$ of the spectrum of $H$.}

\vspace{.2cm}

The gap mentioned above will be referred as the bulk or insulating gap. By a well-known Combes-Thomas estimate ({\it e.g.} \cite{DDS}) one deduces the following estimate on the Fermi projection.

\begin{proposition}\label{FermiProjection1} If BGH holds, then the Fermi projection has exponential decay
\begin{equation}
\sup_{\omega \in \Omega} \big |\langle x |\chi(H_\omega \leq \mu) | y \rangle \big | 
\;\leq\; \gamma \,e^{-\beta |x-y|}\;,
\end{equation}
for some strictly positive and finite constants $\gamma$ and $\beta$.
\end{proposition}

A periodic insulator has, by definition, always a bulk gap. Turning on a disordered perturbation will ultimately close the bulk gap. Nevertheless, it is possible that the Fermi level lies in a region of dynamically Anderson localized spectrum. In this regime, the Fermi level is located in the essential spectrum, but the spectrum is dense pure point and the eigenvectors decay exponentially at infinity. This regime can nicely be characterized by requiring the means square replacement to be bounded \cite{BES}, however, for sake of simplicity and because it holds in many random models anyway (in particular, those considered here, see \cite{DDS}), we choose to characterize this regime by the stronger Aizenmann-Molchanov bound \cite{AM}.

\vspace{.3cm}

\noindent {\bf Mobility Bulk Gap Hypothesis (MBGH):} {\it The Fermi level $\mu\in\RM$ lies in an interval $\Delta\subset\RM$ of the spectrum of $H$ which is Anderson localized, in the sense that the Aizenmann-Molchanov bound on the resolvent
\begin{equation}\label{AizenmannnMolchanov}
\int_\Omega \PM(d\omega) \; \big| \langle x | (E + \I \epsilon - H_\omega )^{-1}|y\rangle \big |^s 
\;\leq \;
\gamma_s \,e^{-\beta_s |x-y|}
\end{equation}
holds uniformly as $\epsilon \rightarrow 0$, for all $E \in \Delta$ and any $s \in (0,1)$. Above, $\gamma_s$ and $\beta_s$ are strictly positive and finite parameters which depend only on $s$.
}

\begin{definition} 
We say that the energy spectrum is delocalized at energy $E$ if the  uniform Aizenmann-Molchannov bound \eqref{AizenmannnMolchanov} cannot be established. 
\end{definition}

The physical regime where BGH is replaced by MBGH is often referred to as the strong localization regime. The existence of a mobility gap also induces a special behavior on the Fermi projection.

\begin{proposition}[\cite{AG,PLB,DDS}]
\label{FermiProjection11} If MBGH holds, then, on average, the Fermi projection is exponentially localized
\begin{equation}
\int_\Omega \PM(d\omega)  \big|\langle x |\chi(H_\omega \leq \mu) | y \rangle \big | 
\;\leq \;
\gamma \;e^{-\beta |x-y|}
\end{equation}
for some strictly positive and finite constants $\gamma$ and $\beta$.
\end{proposition}

Next we describe the behavior of the Fermi projections under homotopies. To describe the deformations of the covariant Hamiltonians properly, recall that the hopping matrices are continuous functions over $\Omega$ with values in $M_N(\mathbb C)$. As such, it is natural to view $W_y$ as elements of the  C$^\ast$-algebra $M_N(\mathbb C) \otimes C(\Omega)$, where $C(\Omega)$ is equipped with the supremum norm 
$$
\|\phi\|_{C(\Omega)} 
\;=\; 
\sup_{\omega \in \Omega} |\phi(\omega)|
\;.
$$

\begin{definition}\label{ContDeformation} 
We call $t\in [0,1]\mapsto H(t)$ a continuous deformation of a family of a covariant Hamiltonians $H$ if $H(t)$ are covariant families of Hamiltonians obtained by continuous variations of $W_y$ in $M_N(\mathbb C) \otimes C(\Omega)$, for every $y \in \mathcal R$. \end{definition}  

Here it is understood that $\mathcal R$ is sufficient large (but finite) to account for all the non-zero hopping matrices during the variation of $t\in[0,1]$. Note that the alignment of the Fermi level with respect to the spectrum can  be changed by adding a constant to $H$, and this can be done by modifying $W_0$. In other words, the above definition of deformations includes also the continuous variations of the Fermi level relative to the spectrum.

\begin{proposition}\label{FermiProjection2} The following holds:

\begin{enumerate}[\rm (i)]

\item Let $t\in [0,1]\mapsto H(t)$ be a continuous deformation such that BGH holds for all $t$. Then 
$$
\sup_{\omega \in \Omega} \big | \big \langle x \big | \chi \big (H_\omega (t') \leq \mu ) - \chi \big (H_\omega (t) \leq \mu \big ) \big | y \big \rangle \big | 
\; \leq\;
 C(t,t')\; e^{-\beta|x-y|}\;,
$$
where $\beta$ is a strictly positive constant (hence independent of $t$ or $t'$) and $C(t,t')$ is a continuous function of the arguments, such that $C(t,t)=0$ for all $t \in [0,1]$.
\item If BGH is replaced by MBGH above, then \cite{RS}
$$
\int_\Omega \PM(d\omega)  \big | \big \langle x \big | \chi \big (H_\omega (t') \leq \mu ) - \chi \big (H_\omega (t) \leq \mu \big ) \big | y \big \rangle \big | 
\; \leq\;
 C(t,t') \;e^{-\beta|x-y|}\;.
$$

\end{enumerate}
\end{proposition}

The above statements apply to both the unitary and chiral unitary Hamiltonians. The latter class posses a chirality operator, which is a selfadjoint operator $J: \mathbb C^{2N} \otimes \ell^2(\mathbb Z^d) \rightarrow \mathbb C^{2N} \otimes \ell^2(\mathbb Z^d)$, with $J=J^*$ and squaring to $J^2=\one$ and commuting with the position operator, {\it i.e.} $J$ is local.  

\vspace{.2cm}

\noindent {\bf Chirality Hypothesis (CH):} {\it  The family $H$ of covariant Hamiltonians has an (exact) chiral symmetry if and only if $J H_\omega J=-H_\omega$ for all $\omega \in \Omega$.}

\vspace{.2cm}

Throughout, we will chose a basis for $\mathbb C^{2N}$ such that the chirality operator is in the diagonal form given in \eqref{eq-Jgrad}. Then all chiral symmetric Hamiltonians take the form shown in Eq.~\eqref{UCGenericModelBDLandau}. We recall that the Fermi level is fixed at $\mu=0$ for the chiral unitary class. 

\begin{proposition}
\label{FermiUnitary1} 
Suppose BGH and CH hold. Then:

\begin{enumerate}[\rm (i)]
\item The family $\sgn(H)$ is chiral symmetric and is of the form
$$
\sgn(H_\omega) \;=\; \begin{pmatrix} 0 & U_\omega^* \\ U_\omega & 0 \end{pmatrix}\;.
$$

\item The family $U_F=\{U_\omega\}_{\omega \in \Omega}$ is covariant and unitary on $\mathbb C^N \otimes \ell^2(\mathbb Z^d)$. In analogy with the Fermi projection, $U_F$ will be called the Fermi unitary operator.

\item The matrix elements of $U_\omega$ are exponentially localized
$$
\sup_{\omega \in \Omega} \big |\langle x |U_\omega | y \rangle \big | 
\;\leq \;
\gamma \;e^{-\beta |x-y|}
\;,
$$
for some strictly positive and finite constants $\gamma$ and $\beta$.

\item If BGH is replaced by MBGH, then {\rm (i)-(iii)} hold with the modification
$$
\int_\Omega \PM(d \omega) \big |\langle x |U_\omega | y \rangle \big |
\; \leq \;\gamma \;e^{-\beta |x-y|}
\;.
$$
\end{enumerate}

\end{proposition}

\noindent {\bf Proof.}
(i) We have $\sgn(H) = \one_{2N}-2 P_F$. Since $J P_F J = \one_{2N} - P_F$, the first part of the statement follows. The second part is a consequence of the chirality. (ii) Because $U_F$ is obtained by functional calculus form a covariant family of operators, it is itself covariant. Since $ \sgn(H) ^2 =\one$, one has $U_\omega U_\omega^\ast = U_\omega^\ast U_\omega =\one_N$. The statements (iii) and (iv) follow from Propositions~\ref{FermiProjection1} and \ref{FermiProjection11} and the formula in (i).
\hfill $\Box$

\vspace{0.2cm}

When discussing the continuous deformations for models from the chiral unitary class, we use Definition \ref{ContDeformation} with the added assumption that, at all times, $H(t)$ remains chiral symmetric relative to the same $J$.

\begin{proposition}
\label{FermiUnitary2} 
The following holds:

\begin{enumerate}[\rm (i)]

\item Let $t\in [0,1]\mapsto H(t)$ be a continuous deformation of $H$ and assume that BGH  and CH hold for all $t$. Then 
$$
\sup_{\omega \in \Omega} \big |\langle x | U_\omega (t') - U_\omega (t) | y \rangle \big |  
\;\leq\; C(t,t')\; e^{-\beta|x-y|}
\;,
$$
where $\beta$ is a strictly positive constant (hence independent of $t$ or $t'$) and $C(t,t')$ is a continuous function of the arguments, such that $C(t,t)=0$ for all $t \in [0,1]$. 
\item If BGH is replaced by MBGH above, then
$$
\int_\Omega \PM(d\omega)\; \big |\langle x | U_\omega (t') - U_\omega (t) | y \rangle \big |  \;\leq\; C(t,t')\; e^{-\beta|x-y|}
\;.
$$
\end{enumerate}
\end{proposition}

\noindent {\bf Proof.}  Both statements follow from Propositions~\ref{FermiProjection2} and \ref{FermiUnitary1}.\hfill $\Box$

\vspace{.2cm}

As already pointed out, for the physical materials, the chiral symmetry does not hold exactly but only approximately. In the following we introduce a notion of approximate chirality, which will ultimately allow us to define topological invariants for such systems. Let us write a general covariant Hamiltonian $H_\omega$ on $\CM^{2N}\otimes\ell^2(\ZM^d)$ in the grading of $J$ given in \eqref{eq-Jgrad}
\begin{equation}
\label{eq-hgrad}
H_\omega
\;=\;
\begin{pmatrix} B_\omega\; & A_\omega^* \\ A_\omega\; & C_\omega \end{pmatrix}
\;.
\end{equation}
Then the CH is equivalent to saying that the self-adjoint covariant operators $B_\omega$ and $C_\omega$ vanish. Given the CH, the BGH is then equivalent to the invertibility of $A_\omega$.  The invertibility of $A_\omega$ will turn out to be sufficient to define invariants, so let us state it as a generalization (of a combination of  BGH and CA):

\vspace{.2cm}

\noindent {\bf Approximate Chirality Hypothesis (ACH):} {\it The off-diagonal entry $A_\omega$  in {\rm \eqref{eq-hgrad}} is invertible and, moreover, $\|B_\omega A_\omega^{-1}\|<1$ and $\|C_\omega (A_\omega^*)^{-1}\|<1$ uniformly in $\omega$. The Fermi unitary operator of a Hamiltonian $H_\omega$ satisfying the ACH is given by $U_\omega=A_\omega|A_\omega|^{-1}$.}

\vspace{.2cm}

\noindent Under the ACH, there exists a continuous deformation of Hamiltonians with ACH
\begin{equation}
\label{eq-homotopyACA}
\lambda\in [0,1]\;\;\mapsto \;\;
H_\omega(\lambda)
\;=\;
\begin{pmatrix} \lambda\,B_\omega & A_\omega^* \\ A_\omega & \lambda\,C_\omega \end{pmatrix}
\;,
\end{equation}
connecting the Hamiltonian $H_\omega=H_\omega(1)$ to an exact chiral Hamiltonian $H_\omega(0)$. Furthermore one has:

\begin{proposition}
\label{prop-gapACA}
Let $H_\omega$ satisfy the  ACH. Then each operator $H_\omega(\lambda)$ on the path \eqref{eq-homotopyACA} also satisfies the BGH.
\end{proposition}

\noindent {\bf Proof.} The invertibility of $H_\omega(\lambda)$ is equivalent to the invertibility of 
$$
H_\omega(\lambda)\,
\begin{pmatrix} 0 & A_\omega^{-1} \\  (A_\omega^*)^{-1} & 0 \end{pmatrix}
\;=\;
\begin{pmatrix} \one & \lambda\, B_\omega A_\omega^{-1}  \\ \lambda\,C_\omega (A_\omega^*)^{-1} & \one\end{pmatrix}
\;,
$$
This is guaranteed because the Schur complement $\one- \lambda^2\, B_\omega A_\omega^{-1} C_\omega (A_\omega^*)^{-1}$ is invertible. 
\hfill $\Box$

\subsection{The half-space and boundary Hamiltonians}
\label{Sec-HSHamilt}

The half-space lattice Hamiltonians are restrictions of the bulk Hamiltonians to the half-space, hence to the Hilbert space $\mathbb C^N \otimes \ell^2(\mathbb Z^{d-1} \times \mathbb N)$. The surjective partial isometry $\PI_d$ from $\ell^2(\mathbb Z^d)$ onto $\ell^2(\mathbb Z^{d-1} \times \mathbb N)$ will become useful in the following. We want the half-space Hamiltonians to be realistic models of disordered crystals with a {\it homogeneous} boundary. The latter means that the covariance property w.r.t. magnetic translations along the first $(d-1)$-directions is preserved. For the unitary class, we claim that this can be achieved within the following generic class of half-space Hamiltonians
\begin{equation}
\label{HalfSpaceGenericH}
\widehat H_\omega \;=\;  \PI_d H_\omega \PI_d^*  \;+ \;
\widetilde H_\omega
\;,
\end{equation}
where the first term represents the restriction of the generic bulk Hamiltonians \eqref{UGenericModelBDSymmetric} to half-space via a simple Dirichlet boundary condition and the second term will be referred to as the boundary Hamiltonian. Supposing again a finite range condition, its most general covariant expression in the symmetric gauge is
\begin{align*}
\label{HalfSpaceGenericHSymmetric}
\widetilde H_{\sym,\omega} & \;=\;   \sum_{n,m=0}^R \sum_{y \in \mathcal R'}  \sum_{x \in \mathbb Z^{d-1}}
 \widetilde W^{y}_{n,m}(\tau_{x,n} \omega) \otimes |x,n\rangle \langle x,n| U_\sym^{y,n-m} \\
& \;=\;   \sum_{n,m=0}^R \sum_{y \in \mathcal R'}  \sum_{x \in \mathbb Z^{d-1}} e^{\frac{\I}{2} \langle y,n-m|\BB|x,n\rangle}
 \widetilde W^{y}_{n,m}(\tau_{x,n} \omega) \otimes |x,n\rangle \langle x-y,m|\;, \nonumber
\end{align*}
where $\Rr'$ is a finite subset of $\ZM^{d-1}$, $R$ a finite number and $\widetilde W^{y}_{n,m} \in M_N(\mathbb C) \otimes C(\Omega)$. The representation in the Landau gauge is obtained by conjugating $\widetilde H_{\sym,\omega}$ with $e^{\frac{\I}{2} \langle X|\BB_+|X\rangle}$, which gives
\begin{equation}
\label{UHalfSpaceGenericHLandau}
\boxed{
\;
\widetilde H_\omega \;=\;   \sum_{n,m=0}^R \sum_{y \in \mathcal R'}  \sum_{x \in \mathbb Z^{d-1}}
e^{\frac{\I}{2}\langle y,n-m |\BB_+|y,n-m \rangle } \, \widetilde W^{y}_{n,m}(\tau_{x,n} \omega) \otimes |x,n\rangle \langle x,n| U^{y,n-m} 
\;.\; }
\end{equation}
The Landau gauge representation will be primarily used in the following.

\vspace{.2cm}

Let us further discuss the terms above. The first term $\PI_d H_\omega \PI_d^*$ models the idealized situation where a boundary was physically created and the remaining hopping matrices are not effected at all by the process of cutting the boundary. Of course, this is not what happens in reality and this is why the boundary Hamiltonian is needed. Note that its hopping matrices depend on $m$ and $n$ instead of $n-m$, which enables us to model practically any homogeneous distortion occurring near the boundary. These distortions will eventually become experimentally undetectable far away from the boundary, hence we imposed the cut-off at $m,n\leq R$, where $R$ can be arbitrarily large but nevertheless finite. 

\vspace{.2cm}

We now turn our attention to the chiral unitary class. The chiral symmetry on $\mathbb C^{2N} \otimes \ell^2(\mathbb Z^{d-1} \times \mathbb N)$ is given by 
\begin{equation}\label{HalfSpaceJ}
\widehat J \;= \;\PI_d\; J\; \PI_d^*\;.
\end{equation}
Since $J$ is local, $\widehat J$ inherits the basic properties $\widehat J^\ast = \widehat J$ and $\widehat{J}\,^2 = \one$. The bulk-boundary principle derived in our work applies strictly to pairs $(H,\widehat H)$ of bulk and half-space Hamiltonians which are chiral symmetric with respect to $(J,\widehat J)$. The generic half-space Hamiltonians which remains chiral symmetric, $\widehat J \;\widehat H\, \widehat J = - \widehat{H}$, takes the form \eqref{HalfSpaceGenericH} with  \eqref{UCGenericModelBDLandau} and \eqref{UHalfSpaceGenericHLandau}, and the boundary hopping matrices assume the chiral form
\begin{equation}
\label{ChiralW}
\widetilde W^{y}_{n,m}(\tau_{x,n} \omega)
\;=\;\begin{pmatrix} 0 & \widetilde w^{y}_{n,m}(\tau_{x,n} \omega) \\ \widetilde w^{y}_{n,m}(\tau_{x,n} \omega)^\ast & 0 \end{pmatrix}
\;.
\end{equation}
For chiral systems, one should also keep in mind the discussion at the and of Section~\ref{Sec-CUClassGeneral} where we have seen that the bulk unit cell needs to be adapted to a given boundary. 

\vspace{0.2cm}

We now present the behavior of various quantities of interest. Recall the decomposition Eq.~\eqref{HalfSpaceGenericH}, which justifies the notation $\widehat H = (H,\widetilde H)$ for the covariant families of half-space Hamiltonians. Below, the components $H$ and $\widetilde H$ are assumed of the generic forms \eqref{UGenericModelBDLandau} and \eqref{UHalfSpaceGenericHLandau}, respectively. When we say that BGH holds for $\widehat H$ we are referring specifically to the bulk component $H$.  The following estimates are by now standard with proofs based on the functional calculus introduced by Dynkin \cite{Dyn}, often also referred to as the Helffer-Sjorstrand formula \cite{HS}.

\begin{proposition}[\cite{EG,SB,Pro5}] \label{HFSmoothFunc1} 
 Assume that the BGH holds for the half-space Hamiltonian $\widehat H$. Then, for any smooth function $\phi$ with support in the bulk insulating gap, 
$$
\sup_{\omega \in \Omega} \big | \langle x,n | \phi(\widehat H_\omega) | y,m \rangle \big | 
\;\leq\; 
\frac{A_M}{1+ |x-y|^M}\; e^{-\beta (n+m)}
\;,
\qquad
n,m\in\NM\;,\;\;x,y\in\ZM^{d-1}\;.
$$
where $M$ is any integer and $A_M$ and $\beta$ are strictly positive constants.
\end{proposition}

Definition~\ref{ContDeformation} of continuous deformations extends literally to the half-space Hamiltonians. By similar proofs, one obtains the following:

\begin{proposition}[\cite{EG,SB,Pro5}]  \label{HFSmoothFunc2} Let $\widehat H(t)$ be a continuous deformation of a family of covariant half-space Hamiltonians. Then, for any smooth function $\phi$ with support in a common insulating gap, 
$$
\sup_{\omega \in \Omega} \big | \langle x,n | \phi \big (\widehat H_\omega(t)\big) - \phi\big(\widehat H_\omega(t')\big ) | y,m \rangle \big | 
\;\leq\; 
\frac{C_M(t,t')}{1+ |x-y|^M}\; e^{-\beta (n+m)}
\;,
$$
where $M$ is any integer, $\beta$ is a strictly positive constant, and $C_M(t,t')$ is a continuous function of the arguments such that $C_M(t,t)=0$ for all $t\in [0,1]$.
\end{proposition}


\chapter{Observables algebras for solid state systems}
\label{Chap-Observables}

\abstract{This chapter introduces the C$^\ast$-algebras of bulk, half-space and boundary observables, together with their canonical representations which generate the physical models for topological insulators presented in Chapter~\ref{Chap-Physics}. Then the exact sequence connecting these algebras is discussed. In particular, it is shown to be isomorphic to the Pimsner-Voiculescu exact sequence. This chapter also introduces the non-commutative analysis tools and the smooth sub-algebras to be used in the remainder of the book. }

\section{The algebra of bulk observables}
\label{Sec-BulkAlgebra}

In this section, the operator algebra in arbitrary dimension $d$ are first studied as mathematical objects, without any mentioning of the connection to Hamiltonians. The canonical covariant representations are given in Sections~\ref{SubSec-URep} and \ref{sec-SymmetricGauge}, and the connection to  concrete physical bulk models introduced in the previous chapter is then established in Section~\ref{sec-examplesbulkalg}.

\subsection{The disordered non-commutative torus}
\label{sec-rotalg}

Let $\BB=(B_{i,j})_{1\leq i,j\leq d}$ be the anti-symmetric real matrix of a constant magnetic field, with entries from $[0,2\pi)$. We will need the decomposition of $\BB$ into its lower and upper triangular parts, hence we introduce the notation $\BB_+$ for the lower triangular part and $\BB_- = \BB_+^T$, such that $\BB=\BB_+ - \BB_-$. Recall the  $d$-dimensional non-commutative torus, defined as the universal C$^*$-algebra generated by $u_1,\ldots,u_d$ satisfying the commutation relations 
\begin{equation}
\label{BComm1}
u_i u_j\;=\;e^{\I B_{i,j}}u_ju_i\;, 
\qquad  
u_j^\ast u_j \;=\; u_j^\ast u_j \;=\; \one\;.
\end{equation}
The non-commutative torus algebra is sufficient to describe periodic tight-binding models of solid state systems submitted to an external magnetic field, in which case the representation of the $u_j$ will be that of the (dual) magnetic translations. Information about the non-commutative torus and its $K$-theory can be found in \cite{Ell,Rie1,Wal} and in the remainder of the book. The Morita equivalence of higher dimensional non-commutative tori was solved in \cite{RS, EL}. Notes on the non-commutative geometry of these spaces can be found in \cite{Rie2,CR}.

\vspace{.2cm}

 In order to include disorder, a larger algebra is needed and this is defined next. Let $(\Omega,\tau,\ZM^d,\PM)$ be the dynamical system given in Definition~\ref{DisorderConfigSpace} (the measure $\PM$ plays no role here).  

\begin{definition} 
\label{def-rotalg}
The algebra of the bulk observables is defined as the universal C$^*$-algebra generated by $C(\Omega)$ and $u_1,\ldots,u_d$,
$$\mathcal A_d\;=\;C^\ast(C(\Omega),u_1,\cdots,u_d)\;,
$$
with the following additional commutation relations:
\begin{equation}
\label{BComm2}
\phi\, u_j\; =\; u_j\,(\phi\circ \tau_j)\,, 
\qquad
\forall \ \phi \in C(\Omega)\,, \;\; j=1,\ldots,d\,.
\end{equation}
If the dependence on the magnetic field is to be stressed, the notation $\Aa_{\BB,d}=\Aa_d$ will be used.
\end{definition}

As we shall see shortly, the disordered bulk Hamiltonians introduced in Chapter~\ref{Chap-Physics} can all be generated from $\Aa_d$. In the following, we offer the reader various ways to look at $\Aa_d$. Of central importance to the bulk-boundary analysis is the presentation of $\Aa_d$ as an iterated crossed product
\begin{equation}
\label{eq-algiter}
\Aa_d
\;=\;
C(\Omega)\rtimes_{\alpha_1}\ZM\ldots\rtimes_{\alpha_d}\ZM
\;,
\end{equation}
with adequate $\ZM$-actions $\alpha_j$, constructed below. This iterative construction immediately implies
\begin{equation}
\label{eq-algiter2}
\Aa_{d}
\;=\;
\Aa_{d-1}\rtimes_{\alpha_d}\ZM
\;,
\end{equation}
which will be used in the proof of the bulk-boundary correspondence. It will become apparent shortly that the iterated crossed product is connected to the Landau gauge. In the literature \cite{Bel,BES}, the bulk algebra is often introduced as a twisted crossed product
\begin{equation}
\label{eq-AlgSym}
\Aa_d
\;=\;
C(\Omega) \rtimes_{\alpha,\BB} \mathbb Z^d
\;,
\end{equation}
with a twist given by the magnetic field. This presentation has similarities to the symmetric gauge and will also be discussed below as well.

\vspace{.2cm}

The first step of the construction of \eqref{eq-algiter} is to consider the C$^\ast$-algebra $C(\Omega)$ of the continuous functions over $\Omega$ with the sup-norm. Then one considers $\Aa_1=C^\ast(C(\Omega),u_1)$ whose commutation relations define a $\ast$-automorphic action of $\ZM$ on $C(\Omega)$:
$$
\ZM \ni x_1\;\; \mapsto 
\;\;
\alpha_1^{x_1}(\phi) \;=\; (u_1)^{x_1} \, \phi \, (u_1^\ast)^{x_1} 
\;=\; \phi \circ \tau_{-x_1}
\;.
$$
Comparing {\it e.g.} with \cite{Dav}, one realizes that $\Aa_1$ is in fact the crossed product algebra associated to this action
$$
C^\ast(C(\Omega),u_1)\;=\;
C(\Omega) \rtimes_{\alpha_1} \ZM
\;.
$$
The next step is to consider $\Aa_2=C^\ast(\Aa_1,u_2)$, whose commutation relations define a $\ast$-automorphic action of $\ZM$ on $\Aa_1$:
$$
\ZM \ni x_2 \;\;\mapsto\;\; 
\alpha_2^{x_2}( \phi u_1) 
\;=\; (u_2)^{x_2} (\phi u_1) (u_2^\ast)^{x_2} 
\;=\; e^{\I x_2 B_{21}} \big(\phi \circ \tau_{-x_2}\big) u_1
\;,
$$
and for the same reasons
$$
C^\ast(\Aa_1,u_2)
\;=\;\Aa_1 \rtimes_{\alpha_2} \ZM
\;=\;C(\Omega) \rtimes_{\alpha_1} \ZM \rtimes_{\alpha_2} \ZM
\;.
$$
The steps can be iterated to finally obtain \eqref{eq-algiter}. The presentation as iterated crossed product is closely related to writing the dense set of non-commutative polynomials in $\mathcal A_d$ in the form
\begin{equation}
\label{eq-LandauPres}
p\;=\;
\sum_{x\in\ZM^d}
p(x)\,u^x,\,
\end{equation}
where $p(x)$ are continuous functions over $\Omega$ which are non-vanishing only for a finite number $x\in\ZM^d$, and $u^x$ are the monomials
$$
u^x\,=\,u_1^{x_1}\cdots u_d^{x_d}\;, 
\qquad x=(x_1,\ldots,x_d)\;.
$$
Note the particular ordering of the  $u_i$'s in $u^x$, which reflects the iterated nature of the crossed product and is connected to the Landau gauge. The monomials $u^x$ obey the following commutation relations:
\begin{equation}
\label{eq-Unmrel}
u^xu^y
\;=\;
e^{\I\langle x| \BB|y\rangle}u^yu^x
\;, 
\qquad   
u^xu^y
\;=\;
e^{\I\langle x| \BB_+|y\rangle} u^{x+y}
\;,
\end{equation}
and
$$
(u^x)^*
\;=\;
e^{\I\langle x|\BB_+|x\rangle }\, u^{-x}
\;.
$$
Care must be taken because $p(x)$ does not commute with $u^x$. Given a second polynomial $q=\sum_{y\in\ZM^d}q(y)\,u^y$, one has
\begin{align}
& 
p\,q\;=\;
\sum_{x\in \ZM^d}
\;\Big(
\sum_{y\in\ZM^d} p(y)\,\big (q(x-y)\circ \tau_{-y}\big )\,e^{\I \langle y|\BB_+| x-y\rangle}
\Big)
\,u^x
\;,
\label{eq-prodnodis}
\\
& p^*\;=\;
\sum_{x\in\ZM^d}
\overline{p(-x)\circ\tau_{-x}}\;e^{\I \langle x|\BB_+ | x\rangle }\,u^x\;.
\label{eq-adnodis}
\end{align}
The bulk algebra $\Aa_d$ is then just the closure of the set of these polynomials under the C$^*$-norm $\|p\|=\sup\,\|\pi(p)\|$, where the supremum is taken over all $\ast$-representations $\pi$. In particular, every element of $\Aa_d$ is a norm limit of non-commutative polynomials and can be written by the same formula \eqref{eq-LandauPres} with coefficients $p(x)$ having appropriate decay properties (instead of being of finite range, see Section~\ref{Sec-BulkFourier}). Many of the algebraic identities in the following will be stated for non-commutative polynomials, but they extend by continuity to the whole algebra $\Aa_d$. 

\vspace{.2cm}

For sake of concreteness, let us write out the dependence on $\omega$ explicitly as in
$$
p(\omega)
\;=\;
\sum_{x\in\ZM^d}
p(\omega,x)\,u^x
\;,
\qquad
p(\omega,x)\in\CM\;.
$$
Then the multiplication and adjunction rules become
\begin{align*}
& (p\,q)(\omega)
\;=\;
\sum_{x\in \ZM^d}
\;\Big(
\sum_{y\in\ZM^d} p(\omega,y) q(\tau_{-y}\omega,x-y)\,e^{\I \langle y|\BB_+| x-y\rangle}
\Big)
\,u^x
\;,
\\
& p^*(\omega)
\;=\;
\sum_{x\in\ZM^d}
\overline{p(\tau_{-x}\omega,-x)}\;e^{\I \langle x|\BB_+ | x\rangle }\,u^x\;.
\end{align*}
Furthermore, let us show how the iteration $\mathcal A_d = \mathcal A_{d-1} \rtimes_{\alpha_d} \mathbb Z$ of \eqref{eq-algiter2}  occurs naturally for the presentation in \eqref{eq-LandauPres}. Due to the particular ordering in the monomials $u^x$, one can write
\begin{equation}\label{P01}
p
\;=\;
\sum_{x_d \in \mathbb Z}\sum_{x \in \mathbb Z^{d-1}} p(x,x_d) u^x u_d^{x_d}
\;=\;
\sum_{x_d \in \mathbb Z} p_{d-1}(x_d) u_d^{x_d}
\;,
\end{equation}
where 
$$
p_{d-1}(x_d)
\;=\;
\sum_{x \in \mathbb Z^{d-1}} p(x,x_d) u^x \in \mathcal A_{d-1}
\;.
$$
Furthermore, if $q$ is another non-commutative polynomial from $\mathcal A_d$ which is decomposed in the same manner, then:
\begin{align*}
p \, q &\;=\; \sum_{y_d \in \mathbb Z} p_{d-1}(y_d) u_d^{y_d}\sum_{x_d \in \mathbb Z} q_{d-1}(x_d) u_d^{x_d} \nonumber \\
&\;=\; \sum_{x_d,y_d \in \mathbb Z} p_{d-1}(y_d) \big(u_d^{y_d} \ q_{d-1}(x_d) \ u_d^{-y_d}\big ) u_d^{x_d+y_d}
\;.
\end{align*}
By a change of variable $x_d \mapsto x_d -y_d$,
\begin{equation}
\label{eq-pqprod}
p \, q 
\;=\; 
\sum_{x_d\in \mathbb Z}\Big (\sum_{y_d \in \mathbb Z} p_{d-1}(y_d) \alpha_d^{y_d} \big( q_{d-1}(x_d-y_d)\big )\Big ) u_d^{x_d}
\;,
\end{equation}
and the r.h.s. is exactly the multiplication in $\mathcal A_{d-1} \rtimes_{\alpha_d} \mathbb Z$.

\vspace{.2cm}

Next let us explain how the bulk algebra can be viewed as the twisted crossed product \eqref{eq-AlgSym}.  For this, we consider the monomials
\begin{equation}
\label{eq-USymRel}
u_\sym^x 
\;=\; 
e^{\frac{\I}{2}\langle x |\BB_+|x \rangle } u^x
\;,
\end{equation}
which obey the relations
\begin{equation}
\label{eq-USymRel2}
u_\sym^x u_\sym^y 
\;=\; 
e^{\I \langle x|\BB |y \rangle} u_\sym^y u_\sym^x= e^{\frac{\I}{2}\langle x|\BB |y \rangle} u_\sym^{x+y}
\;,
\qquad
(u_\sym^x)^\ast\; =\;u_\sym^{-x}
\;.
\end{equation}
Instead of \eqref{eq-LandauPres}, one now decomposes the non-commutative polynomials as
\begin{equation}
\label{eq-psym}
p
\;=\; 
\sum_{x\in\ZM^d} p_\sym (x) u_\sym^x\,,
\end{equation}
where, as before, $p_\sym(x)\in C(\Omega)$ is non-vanishing only for a finite number $x\in\ZM^d$. Given a second polynomial $q=\sum_{x\in\ZM^d} q_\sym(x) u_\sym^x$, one has
\begin{align}
& 
p\, q
\;=\;
\sum_{x\in \ZM^d}
\;\Big(
\sum_{y\in\ZM^d} p_\sym(y) \, q_\sym(x-y)\circ\tau_{-y}
\;e^{\frac{\I}{2} \langle y|\BB| x\rangle}
\Big)
\,u_\sym^x
\;,
\label{eq-psymprop}
\\
& 
p^*\;=\;
\sum_{x\in\ZM^d}
\overline{p_\sym(-x)\circ\tau_{-x}} \ u_\sym^x.
\label{eq-psymstar}
\end{align}
By looking at the coefficients, one realizes that \eqref{eq-AlgSym} holds with the $\ZM^d$-action $\alpha$ on $C(\Omega)$ given by $\alpha_x(\phi)=\phi\circ\tau_{-x}$, and a twist given by the magnetic field.

\subsection{Covariant representations in the Landau gauge} 
\label{SubSec-URep}

Here we define the family of covariant representations which generate the Hamiltonians presented in Chapter~\ref{Chap-Physics} in the Landau gauge.

\begin{proposition} Recall the generators $e_j$ of $\ZM^d$, the right-shifts $S_j|x\rangle=|x+e_j\rangle$ and the position operator $X=(X_1,\ldots,X_d)$ on $\ell^2(\ZM^d)$. Then the following relations define a family $\{\pi_\omega\}_{\omega \in \Omega}$ of faithful $\ast$-representations of $\Aa_d$ on $\ell^2(\mathbb Z^d)$: 
\begin{equation}
\label{PiRep1}
\pi_\omega(u_j)
\;=\;
e^{\I\langle e_j| \BB_+|X\rangle} \,S_j
\;=\;
S_j \,e^{\I \langle e_j|\BB_+ | X \rangle }
\;, 
\qquad j=1,\ldots,d\;,
\end{equation}
and
\begin{equation}
\label{PiRep2}
\pi_\omega(\phi) 
\;= \;
\sum_{x \in \mathbb Z^d} \phi(\tau_x \omega)|x \rangle \langle x|
\;, 
\qquad \forall \ \phi \in C(\Omega)
\;.
\end{equation}
\end{proposition}

\noindent {\bf Proof.} We need to verify the commutation relations \eqref{BComm1} and \eqref{BComm2}. We have
$$
\pi_\omega(u_j) \pi_\omega(u_i) 
\;=\; 
e^{\I\langle e_j| \BB_+|X\rangle} \,S_j   \, e^{\I\langle e_i| \BB_+|X\rangle} \,S_i 
\;= \;
e^{\I\langle e_j| \BB_+|X\rangle}e^{\I\langle e_i| \BB_+|X-e_j\rangle} S_jS_i
\;.
$$
By switching $i \leftrightarrow j$, one immediately sees that
$$
\pi_\omega(u_i) \pi_\omega(u_j) 
\;=\; 
e^{\I(\langle e_i| \BB_+|e_j\rangle - \langle e_j| \BB_+|e_i\rangle) } \pi_\omega(u_j) \pi_\omega(u_i)
\;.
$$
The first part of the commutation relation \eqref{BComm1} is now established because of the identity $\langle e_i| \BB_+|e_j\rangle - \langle e_j| \BB_+|e_i\rangle = B_{i,j}$. The second part also follows because $\pi_\omega(u_j)$ are unitary.  For \eqref{BComm2}, we can use the following simple, but useful identity,
\begin{equation}\label{TransId}
|x \rangle \langle x|\, \pi_\omega(u_j) 
\;= \;
\pi_\omega(u_j)\, |x-e_j \rangle \langle x-e_j|
\;,
\end{equation}
which gives
$$
\pi_\omega(\phi)\, \pi_\omega(u_j)
\;=\; \pi_\omega(u_j) \, \sum_{x \in \mathbb Z^d} \phi(\tau_x \omega)|x -e_j \rangle \langle x -e_j |  
\;=\;  \pi_\omega(u_j) \,\pi_\omega(\phi \circ \tau_j)
\;.
$$
It is also clear that $\pi_\omega(\phi^\ast) = \pi_\omega(\phi)^\ast$. 
\hfill $\Box$

\vspace{0.2cm}

For the monomials $u^x$, we have 
$$
\pi_\omega(u_x) 
\;=\; 
\big( e^{\I\langle e_1| \BB_+|X\rangle} \,S_1 \big )^{x_1} \cdots \big( e^{\I\langle e_d| \BB_+|X\rangle} \,S_d \big )^{x_d}
\;.
$$
Since $\BB_+$ is a lower triangular matrix, all the phase factors commute with the shifts following them, in particular, $\pi_\omega(u^y)|0\rangle =|y\rangle$. Then  \eqref{eq-Unmrel} gives
$$
\pi_\omega(u^x)\,|y\rangle
\;=\;
\pi_\omega(u^x)\,\pi_\omega(u^y)\,|0\rangle
\;=\;
e^{\I \langle x|\BB_+|y\rangle}\,|x+y\rangle
\;.
$$
This shows that $\pi_\omega(u^x)$ are precisely the dual magnetic translations on $\ell^2(\ZM^d)$ in the Landau gauge, introduced in \eqref{DualMagTransLandau}. Hence, we can unify the notations 
$$
U^x\;=\;\pi_\omega(u^x)
\;=\;
S^x\,e^{\I \langle x|\BB_+|X\rangle}
\;=\;
\sum_{y\in\ZM^d}
e^{\I \langle x|\BB_+|y\rangle}\,
|y+x\rangle\langle y|
\;.
$$ 
Later, we will also use $U_j=U^{e_j}$. Note that $x\in\ZM^d\mapsto U^x$ provides a projective unitary representation of the translation group $\ZM^d$ on $\ell^2(\ZM^d)$. Lastly, let us write explicitly the representation of the non-commutative polynomials
\begin{equation}
\label{LandauRep}
\boxed{
\;
\Aa_d \ni p \;=\; \sum_{y \in \mathbb Z^d} p(\omega,y)u^y
\;\;\mapsto\;\;
\pi_\omega(p) \;=\; \sum_{x,y \in \mathbb Z^d} p(\tau_x \omega,y)|x \rangle \langle x | U^y.
\;}
\end{equation}

Now the covariance properties of these representations are investigated. The dual magnetic translations $U^x$ are invariant w.r.t. the magnetic translation $V^x$, introduced in \eqref{MagTransLandau}. Let us introduce the latter in a different way,
\begin{equation}
\label{eq-magtransLand}
\mathbb Z^d \ni x 
\;\;\mapsto \;\;
V^x\;=\;(V_1)^{x_1}\cdots (V_d)^{x_d}\;,
\qquad
V_j\;=\;
e^{\I\langle X|\BB_+|e_j\rangle}\,S_j
\;.
\end{equation}
Their commutation relations are
\begin{equation}
\label{Vs}
V^xV^y\;=\;
e^{-\I \langle x|\BB|y \rangle}\, V^{y}V^x\;=\;e^{-\I \langle x|\BB_+|y\rangle}\, V^{x+y}
\;,
\end{equation}
and
$$
(V^{x})^*\;=\;
e^{-\I \langle x|\BB_+|x\rangle }V^{-x}\;,
$$
hence they also form a projective unitary representation of $\ZM^d$ on $\ell^2(\ZM^d)$. Since $(V^{-y})^\ast |0\rangle = |y\rangle$,
$$
V^x|y\rangle 
\;=\; 
e^{\I \langle y|\BB_+|y\rangle } V^x V^{y}|0\rangle 
\;=\; 
e^{\I \langle x+ y|\BB_+|x\rangle}\,|y+x\rangle
\;,
$$ 
so that, indeed,
$$
V^x\;=\;e^{\I \langle X|\BB_+|x\rangle}\,S^x
\;,
$$
as in \eqref{MagTransLandau}. Recall that there is no Peierls phase factor when the magnetic translation is only along $x_d$ because $e^{\I \langle e_d+ y|\BB_+|e_d\rangle}=1$, which is a defining characteristic of the Landau gauge. The invariance relation is
$$
V^x\,U_j\,(V^x)^*\;=\;U_j
\;,
$$
which can be verified using the explicit actions of $U_j$ and $V^x$ on $\ell^2(\mathbb Z^d)$. Moreover, one has
$$
V^x\,\pi_\omega(\phi)\,(V^x)^*\;=\;
\pi_{\tau_x\omega}(\phi)
\;,
\qquad
\phi\in C(\Omega)
\;.
$$
Together this implies the covariant property
\begin{equation}
\label{eq-covarrel2}
\boxed{
V^x\, \pi_\omega(p) \,(V^x)^\ast 
\;=\; 
\pi_{\tau_x \omega} (p)
\;,
\qquad
p\in\Aa_d\;,\;\;x\in\ZM^d\;.
}
\end{equation}
Inversely, any finite range operator family $\{A_\omega\}_{\omega\in\Omega}$ on $\ell^2(\ZM^d)$ satisfying the covariance relation $V^xA_\omega(V^x)^*=A_{\tau_x \omega}$ for all $x\in\ZM^d$ is the representative $A_\omega=\pi_\omega(p)$ of a polynomial $p$ in $\Aa_d$. This property gives the physical meaning to the algebra $\mathcal A_d$, which now can be identified with the algebra of covariant operators on $\ell^2(\mathbb Z^d)$. Some of the special properties of these operators have been already highlighted in Section~\ref{Sec-UClassGeneral}.

\subsection{Covariant representations in the symmetric gauge}
\label{sec-SymmetricGauge}

Let us briefly comment here on the family $\{\pi_{\sym,\omega}\}_{\omega\in\Omega}$ of covariant representations, which generate the physical models in the symmetric gauge, as discussed in Chapter~\ref{Chap-Physics}. These representations were used in most prior works, {\it e.g.} \cite{Bel,BES}, but they will play little role in our analysis. They are obtained from $\pi_\omega$ via a gauge transformation
\begin{equation}
\label{SymRep}
\pi_{\sym,\omega}(p) 
\;=\; 
e^{-\frac{\I}{2} \langle X|\BB_+|X \rangle }\, \pi_\omega(p)\, e^{ \frac{\I}{2} \langle X|\BB_+|X \rangle }
\;,
\qquad
p\in\Aa_d
\;.
\end{equation}
One can immediately see that the dual magnetic translations in the symmetric gauge, introduced in Chapter~\ref{Chap-Physics} in \eqref{DualMagTransSymmetric}, are actually equal to
$$
U_\sym^y 
\;=\; 
\pi_{\sym,\omega}(u_\sym^y) 
\;=\; 
e^{\frac{\I}{2}\langle y |\BB_+|y \rangle } \, e^{-\frac{\I}{2} \langle X|\BB_+|X \rangle } \,U^y \,e^{ \frac{\I}{2} \langle X|\BB_+|X \rangle }
\;.
$$
Recalling the magnetic translations \eqref{MagTransSymmetric} in the symmetric gauge and the invariance relation $V^x_\sym \,U^y_\sym \,(V_\sym^x)^\ast =U^y_\sym$,  one obtains the covariance relation for the symmetric gauge
\begin{equation}
\boxed{
V^x_\sym \,\pi_{\sym,\omega}(p) \,(V_\sym^x)^\ast 
\;=\; 
\pi_{\sym,{\tau_x \omega}} (p)
\;,
\qquad
p\in\Aa_d\;,x\in\ZM^d
\;.
}
\end{equation}
Lastly, the representation of the symmetric non-commutative polynomials is given by
\begin{equation} 
\label{SymmetricRep}
\boxed{
\;\mathcal A_d \ni p\;=\;\sum_{y\in\ZM^d} p_\sym (y) u_\sym^y
\;\;
\mapsto \;\;
\pi_{\sym,\omega}(p) 
\;=\; 
\sum_{x,y\in\ZM^d}
p_\sym(\tau_x\omega,y) |x\rangle \langle x| \,U_\sym^y \;.
\;
}
\end{equation}

\subsection{The algebra elements representing the Hamiltonians}
\label{sec-examplesbulkalg}

As stressed several times before, the families of Hamiltonians $\{H_\omega\}_{\omega\in\Omega}$ introduced in Chapter~\ref{Chap-Physics} are representations of elements from $M_N(\mathbb C)\otimes\Aa_{d}$. This will shown explicitly here.  First, by comparing \eqref{UGenericModelBDSymmetric} and \eqref{SymmetricRep}, one can immediately conclude
\begin{equation}
\label{UCGenericGeneratorBulk}
\boxed{
\;
H_{\sym,\omega}
\;=\;
\pi_{\sym,\omega}(h_\sym)
\;,
\qquad 
h_\sym
\;=\;
\sum_{y\in \Rr} W_y \otimes u_\sym^y
\;\in\;  M_N(\mathbb C)\otimes\Aa_d 
\;.\;
}
\end{equation}
As for the Landau gauge, by comparing \eqref{UGenericModelBDLandau} and \eqref{LandauRep}, it follows that
\begin{equation}
\label{UCGenericGeneratorBulk2}
\boxed{
\;H_\omega
\;=\;
\pi_\omega(h)
\;,
\qquad
h
\;=\;
\sum_{y \in \Rr} e^{\frac{\I}{2} \langle y|\BB_+|y\rangle } \,W_y \otimes u^y
\;\in\;  M_N(\mathbb C)\otimes\Aa_d
\;.
\;
}
\end{equation}
By examining the relation \eqref{eq-USymRel} between the monomials $u^y$ and $u^y_\sym$, one immediately sees that $h_\sym = h$. All this applies equally well to the chiral symmetric Hamiltonians in both the symmetric gauge \eqref{UCGenericModelBDSymmetric} and the Landau gauge \eqref{UCGenericModelBDLandau}, the only difference is the particular form of $ W_y$.

\vspace{.2cm}

The algebra elements corresponding to the Fermi projection and the Fermi unitary operator can be obtained from the functional calculus on $\Aa_d$, provided the BGH holds. For example, in the Landau gauge, the Fermi projection $P_F = \{P_\omega\}_{\omega \in \Omega}$ is given by
$$
P_\omega \;=\; \pi_\omega (p_F)\;, 
\qquad 
p_F \;=\; \chi(h \leq \mu)
\;.
$$
If $h \in M_{2N}(\mathbb C) \otimes \Aa_d$ such the CH holds, then $J h J = -h$, hence
$$
\sgn(h) 
\;=\; 
\begin{pmatrix} 0 & u_F^* \\ u_F & 0 \end{pmatrix}
\;,
$$
which defines the element $u_F \in M_N(\mathbb C) \otimes \Aa_d$ representing the Fermi unitary operator $U_F = \{\pi_\omega(u_F) \}_{\omega \in \Omega}$. If only the MBGH holds, then $p_F$ is not in $\Aa_d$, but only lies in the non-commutative Sobolev spaces defined below.

\section{The algebras of half-space and boundary observables}
\label{Sec-HalfSpaceAlgebra}

The operator algebras for half-space and boundary observables are first introduced as mathematical objects. The canonical covariant representations are discussed in Sections~\ref{SubSec-HSRep} and the connection to the concrete models is established in Section~\ref{Sec-AlgElemHalfSpace}.

\subsection{Definition of the algebras and basic properties}

\begin{definition} 
\label{def-halfspacealg}
The algebra $\widehat{\Aa}_d$ of the half-space observables is defined as the universal C$^\ast$-algebra generated by $C(\Omega)$ and $\hat u_1,\ldots,\hat u_d$ satisfying the commutation relations:
\begin{equation}\label{HSComm1}
 \hat{u}_i\hat{u}_j\;=\;e^{\I B_{i,j}}\hat{u}_j\hat{u}_i\;,  
\qquad  i,j=1,\ldots,d\;,
\end{equation}
\begin{equation}\label{HSComm2}
 \hat u_j \hat u_j^\ast\; = \;\hat u_j^\ast \hat u_j\; = \;\one\,,
 \qquad  j=1,\ldots,d-1\;,
\end{equation}
and
\begin{equation}
\label{HSComm3}
\hat{u}_d^*\hat{u}_d\;=\;\one\;, \qquad \hat{u}_d\hat{u}_d^*\;=\;\one-\hat{e}\;,
\end{equation} 
for some projection $\hat{e}^2=\hat{e}^*=\hat{e}$ commuting with $\hat{u}_1,\ldots,\hat{u}_{d-1}$, as well as  the additional commutation relations
\begin{equation}\label{HSComm4}
\phi\,\hat{e}\;=\;
\hat{e}\,\phi\;,
\qquad
\phi\, \hat u_j\; =\; \hat u_j (\phi\circ \tau_j)\,, 
\qquad \phi\, \hat u_j^\ast \;=\; \hat u_j^\ast (\phi\circ \tau_j^{-1}) \;, 
\end{equation}
for all $\phi \in C(\Omega)$ and  $ j=1,\ldots,d$. The algebra 
$$
\widehat{\Aa}_d 
\;=\; 
C^*(C(\Omega),\hat u_1,\ldots,\hat u_d)
\;,
$$
will be called the half-space algebra or also the disordered non-commutative torus with a boundary. 
\end{definition}

The difference w.r.t. the bulk algebra $\Aa_d$ is that the last generator $\hat{u}_d$ is not unitary, but only a partial isometry. The label ``half-space" will become clear when the canonical representations are discussed below, as will the following terminology.

\begin{definition} 
The algebra of the boundary observables is the proper two-sided ideal $\Ee_d$ of $\widehat{\mathcal A}_d$ generated by the projection $\hat e$, namely, $\Ee_d=\widehat{\mathcal A}_d\, \hat e\, \widehat{\mathcal A}_d$.
\end{definition}

Let us point out that $\hat e$ is part of both the half-space and boundary algebras, but $\hat u_d$ is only in the half-space algebra. Hence $\Ee_{d}$ is indeed a proper ideal of $\widehat{\mathcal A}_{d}$. The element $\hat e$ commutes with all $\hat u_i$ for $i=1,\ldots,d-1$ and $\hat e \hat u_d = \hat u_d^\ast \hat e=0$. Therefore  $(\hat{u}_{d}^\ast)^n\hat{e}(\hat{u}_{d})^m=0$ whenever $n+m > 0$. Furthermore,
$$
\hat{u}_d^n (\hat{u}_d^*)^m
\;=\;
\left\{
\begin{array}{cc} 
\hat{u}_d^{n-m}\big(\one-\sum_{l=0}^{m-1} (\hat u_{d})^l\hat{e}(\hat{u}_{d}^*)^l\big)\;, & \;\;\;\;n\geq m\;,\medskip
\\
\big(\one-\sum_{l=0}^{n-1} (\hat u_{d})^l\hat{e}(\hat{u}_{d}^*)^l\big)(\hat{u}_d^*)^{m-n}\;, & \;\;\;\; n\leq m\;.
\end{array}
\right.
$$
hence dense subsets of the algebras $\widehat{\mathcal A}_d$ and $\Ee_d$ are linearly spanned by monomials of the form 
\begin{equation}
\label{eq-HSmonomials}
\phi\, \hat u_1^{x_1} \cdots \hat u_{d-1}^{x_{d-1}} \hat u_d^n (\hat u_d^\ast)^m 
\quad \mbox{and} \quad 
\phi\, \hat u_1^{x_1} \cdots \hat u_{d-1}^{x_{d-1} }(\hat u_{d})^n\hat{e}(\hat{u}_{d}^*)^m
\;,
\end{equation}
respectively, with $\phi\in C(\Omega)$, $x_i \in \mathbb Z$ and $m,n \in \mathbb N$. As such, elements of $\widehat{\mathcal A}_d$ and $\Ee_d$ can be presented in the form
\begin{equation}
\label{eq-phattilde}
\hat{p} 
\;=\; \sum_{n,m\geq 0} \hat{p}_{n,m}\,\hat{u}_d^n (\hat{u}_d^*)^m\;,
\qquad
\tilde{p}
\;=\;
\sum_{n,m\geq 0}\tilde{p}_{n,m}\,\hat{u}_d^n\,\hat{e}\,(\hat{u}_d^*)^m
\;,
\end{equation}
respectively, where both $\hat p_{n,m}$ and $\tilde{p}_{n,m}$ belong to $\Aa_{d-1}\cong C^\ast(C(\Omega),\hat u_1,\ldots,\hat u_{d-1})$. The algebraic operations in this presentation can be conveniently written out using the automorphism $\alpha_d:\Aa_{d-1}\to\Aa_{d-1}$ of Section~\ref{sec-rotalg}. By a similar calculation leading to \eqref{eq-pqprod}, one finds that in $\widehat{\Aa}_d$
\begin{align*}
& 
\hat{p}\,\hat{q}
\;=\;
\sum_{n,m\geq 0}
\left(
\sum_{k>l\geq 0}
\hat{p}_{n,k}\,\alpha^{n-k}_d\big(\hat{q}_{k,m+l-k}\big)
+
\sum_{l\geq k\geq 0}\hat{p}_{n+k-l,k}\,\alpha^{n-l}_d\big(\hat{q}_{l,m}\big)
\right)
\hat{u}_d^n (\hat{u}_d^*)^m
\;,
\\
& \hat{p}^*=\;
\sum_{n,m\geq 0}
\alpha_d^{m-n}(\hat{p}_{n,m}^*)\,
\hat{u}_d^m (\hat{u}_d^*)^n
\;.
\nonumber
\end{align*}
In $\Ee_d$, the expressions simplify to
\begin{align}
\label{eq-brules}
& 
\tilde {p}\,\tilde{q}
\;=\;
\sum_{n,m\geq 0}
\left(
\sum_{k\geq 0}
\tilde{p}_{n,k}\,\alpha^{n-k}_d\big(\tilde{q}_{k,m}\big)\,
\right)
\hat{u}_d^n\,\hat{e}\,(\hat{u}_d^*)^m
\;, \\
& \tilde{p}^*=\;
\sum_{n,m\geq 0}
\alpha_d^{m-n}(\tilde{p}_{n,m}^*)\,
\hat{u}_d^m\,\hat{e}\,(\hat{u}_d^*)^n
\;.\nonumber
\end{align}
%

\subsection{The exact sequence connecting bulk and boundary}
\label{sec-BBC}

Let us now consider the embedding $i:\Ee_{d}\hookrightarrow\widehat{\Aa}_{d}$ and the canonical surjective C$^\ast$-algebra homomorphism $\mbox{\rm ev}:\widehat{\Aa}_{d}\to {\Aa}_{d}$ defined by
$$
\mbox{\rm ev}(\phi)=\phi\;,
\qquad
\mbox{\rm ev}(\hat{u}_j)=u_j
\;, 
\qquad 
\mbox{\rm ev}(\hat{u}_j^\ast)=u_j^\ast\,,  
$$ 
for $j=1,\ldots d$. Then necessarily $\mbox{\rm ev}(\hat{e})=0$ so that:

\begin{proposition}
\label{prop-ExSeq}
The following sequence
\begin{equation}
\label{ExactSequence}
\boxed{
\;
\begin{diagram}
0 &\rTo &\mathcal E_{d}   &\rTo{i}  &\widehat{\mathcal A}_{d}  &\rTo{ \mathrm{ev}}  &\mathcal A_{d} &\rTo &0
\end{diagram}
\;
}
\end{equation}
is an exact sequence of C$^\ast$-algebras. 
\end{proposition}

The bulk-boundary correspondence for the topological invariants is rooted in this sequence, which is hence of major importance for what follows. Let us collect a number of basic implications of Proposition~\ref{prop-ExSeq}. First of all, $\widehat{\Aa}_{d}/\Ee_{d}\cong\Aa_{d}$. The sequence is never split-exact as a sequence between C$^\ast$-algebras, but it is split-exact as a sequence between linear spaces, with the split $i': {\Aa}_{d} \rightarrow \widehat{\Aa}_{d}$ given by:
\begin{equation}\label{iprime}
i'\big(\phi\, u_1^{x_1}\ldots u_d^{x_d} \big )
\;=\;
\left \{
\begin{array}{cc}
\phi\, \hat u_1^{x_1}\cdots \hat u_{d-1}^{x_{d-1}}\hat u_d^{x_d}, \;\; & \mathrm{if} \ x_d \geq 0,\medskip \\
\phi \,\hat u_1^{x_1}\cdots \hat u_{d-1}^{x_{d-1}}(\hat u_d^\ast)^{|x_d|},\;\; & \mathrm{if} \ x_d < 0.
\end{array} \right .
\end{equation}
The split is well defined because ${\Aa}_{d}$ is linearly spanned by the monomials considered above and the required property $\mathrm{ev}\circ i' =\mathrm{id}_{\Aa_{d}}$ can be verified through direct computation. Hence $\widehat{\Aa}_d=\Aa_d\oplus\Ee_d$ as linear spaces. We should emphasize that $i'$ is not an algebra homomorphism. For example, $u_d u_d^\ast = \one$, but $i'(u_d)i'(u_d^\ast) = \one - \hat e$. Nevertheless, the split-exact sequence between the linear spaces generates a useful presentation of the half-space algebra, since any element from $ \widehat{\Aa}_{d}$ can be written as a direct sum $ i'(p)+\tilde p$, with $p \in \Aa_{d}$ and $\tilde p \in \Ee_{d}$. Even more convenient, the elements of $\widehat{\Aa}_{d}$ can be represented as $\hat p = (p,\tilde p),$ where $p = \mathrm{ev}(\hat p) \in \Aa_{d}$ and $\tilde p = \hat p - i'(p) \in \Ee_{d}$. In this presentation, the multiplication in $ \widehat{\Aa}_{d}$ takes the form:
$$
(p,\tilde p) (q,\tilde q)\; =\; 
(pq, \widetilde{pq})
\;,
$$
where $\widetilde{pq} = \hat p \hat q - i'(pq)$. Furthermore, the $\ast$-operation becomes $(p,\tilde p)^\ast =(p^\ast, \tilde p^\ast)$.

\subsection{The Toeplitz extension of Pimsner and Voiculescu}
\label{Sec-PimsnerVoiculescu}

In this section, the exact sequence \eqref{ExactSequence} will be isomorphically mapped to the Toeplitz extension of Pimsner and Voiculescu \cite{PV} associated to the discrete time C$^*$-dynamical system $(\Aa_{d-1},\alpha_d,\ZM)$. Such a dynamical system always comes with the crossed product $\mathcal A_{d-1} \rtimes_{\alpha_d} \mathbb Z$ as well as its Toeplitz extension
\begin{equation}
\label{PVExactSequence}
\begin{diagram}
0 &\rTo &\mathcal A_{d-1}\otimes \mathcal K  &\rTo{ \psi }  &T(\mathcal A_{d-1})  &\rTo{\pi} &\mathcal A_{d-1} \rtimes_{\alpha_d} \mathbb Z &\rTo &0.
\end{diagram}
\end{equation}
Here, $T(\mathcal A_{d-1})$ is defined as the sub-algebra of $\Aa_{d} \otimes C^\ast \big (\widehat{S}\big )$ generated by $a\otimes \one$ and $u_d \otimes \widehat{S}$ with $a \in \Aa_{d-1}$ and $\widehat{S}$ a partial isometry on a separable Hilbert space $\mathcal H$, satisfying
$$\widehat{S}^*\; \widehat{S}\;=\;\one\,, \qquad \widehat{S}\;\widehat{S}^* = \one-\widetilde{P}
\;,
$$ 
with a non-trivial projection $\widetilde{P}$ on $\mathcal H$. Hence $T(\mathcal A_{d-1})$ is generated by the monomials
$$
a \, u_d^{n-m}\otimes \widehat{S}^{\; n} (\widehat{S}^\ast)^m
\,, \quad a \in \Aa_{d-1} \,.
$$
Modulo isometries, the Toeplitz extension is independent of $\mathcal H$, $\widehat{S}$ or $\widetilde{P}$ and therefore it is allowed to have the concrete realization of Chapter~\ref{Chap-Illustration} in mind. As before, $\mathcal K$ denotes the algebra of compact operators over $\mathcal H$ and thus only remains to define the maps appearing in \eqref{PVExactSequence}:
$$
\psi \big (a \otimes |n\rangle \langle m | \big )\; =\; 
u_d^n \, a \, (u_d^\ast)^m \otimes \widehat{S}^{\; n} \widetilde{P} (\widehat{S}^\ast)^m
$$
and 
$$
\pi \big(a \, u_d^{n-m}\otimes \widehat{S}^{\; n} (\widehat{S}^\ast)^m \big )\; =\; 
a \, u_d^{n-m}\,.
$$ 
Let us point out that if $d=1$ and $\Omega$ is just a point, then $\Aa_{d-1}=\CM$ and the exact sequence \eqref{PVExactSequence} reduces exactly to the sequence \eqref{ToeplitzExtension}, while with disorder it becomes \eqref{eq-toepextdisoredered}. This also justifies the terminology. The Toeplitz extension was used in \cite{PV} as a tool to calculate the $K$-theory of $\Aa_d=\mathcal A_{d-1} \rtimes_{\alpha_d} \mathbb Z$ in terms of the $K$-theory of $\mathcal A_{d-1}$, and this is precisely what will be done in Section~\ref{sec-KObsAlg}. Here we establish the connections of \eqref{PVExactSequence} to \eqref{ExactSequence}.

\begin{proposition} 
\label{PVLink2}
Let $\tilde{p}$ and $\hat{p}$ be decomposed as in \eqref{eq-phattilde}. 

\begin{enumerate}[\rm (i)]

\item The map $\widetilde \rho:\Ee_d \to\Aa_{d-1}\otimes \Kk$ given by
\begin{equation}\label{RhoEe}
\widetilde \rho(\tilde{p})
\;=\;
\sum_{n,m\geq 0}
\alpha_d^{-n}\big(\tilde{p}_{n,m}\big)
\otimes|n\rangle\langle m|
\;,
\end{equation}
is a C$^\ast$-algebra isomorphism.\newline

\item The map $\widehat \eta: \widehat{\mathcal A}_{d} \rightarrow T(\Aa_{d-1})$ given by
\begin{equation}\label{EtaAa}
\widehat \eta(\hat p) \;=\; \sum_{n,m\geq 0} \hat{p}_{n,m} \, u_d^{n-m} \otimes \widehat{S}^{\; n} (\widehat{S}^\ast)^m
\end{equation}
is a C$^\ast$-algebra isomorphism.\newline

\item The Pimsner-Voiculescu exact sequence and \eqref{ExactSequence} are isomorphic:
\begin{equation}\label{PVLink}
\begin{diagram}
0 &\rTo &\mathcal A_{d-1}\otimes \mathcal K  &\rTo{\psi} \ \  &T(\mathcal A_{d-1})  &\rTo{\pi} &\mathcal A_{d-1} \rtimes_{\alpha_d} \mathbb Z &\rTo &0\\
\ & &\uTo{\widetilde \rho}       &                                            &\uTo{\widehat \eta} & &\uCongruent                  &   \\
0 &\rTo &\mathcal E_d   &\rTo{i}  &\widehat{\mathcal A}_d  &\rTo{ \mathrm{ev}} \ \  &\mathcal A_d &\rTo &0
\end{diagram}
\end{equation}
This diagram is commutative.
\end{enumerate}
\end{proposition}

\noindent {\bf Proof.}  All affirmations follow from straightforward computations.\hfill $\Box$

\vspace{0.2cm}

This isomorphism will be used in Section~\ref{sec-PairingsDuality}, where the bulk-edge correspondence principle is established.

\subsection{Half-space representations}
\label{SubSec-HSRep}

The half-space algebra $\widehat{\Aa}_{d}$, and thus its sub-algebra $\Ee_d$ have canonical faithful $\ast$-representations on the Hilbert space $\ell^2(\ZM^{d-1}\times\NM)$ which are constructed and described in this section. Recall that $\PI_d:\ell^2(\ZM^{d})\to\ell^2(\ZM^{d-1}\times\NM)$ is the surjective partial isometry satisfying $\PI_d^*|x\rangle=|x\rangle$ for $x\in\ZM^{d-1}\times\NM\subset\ZM^d$.

\begin{proposition} The following relations
$$
\widehat \pi_\omega(\hat u_j)
\;=\;
\PI_d\pi_\omega(u_j)\PI_d^*
\;=\; 
\PI_d U_j\PI_d^*
\;=\;
e^{\I\langle e_j| \BB_+|X\rangle} \PI_d S_j\PI_d^*\; ,
\;
$$
for $j=1,\ldots,d$,
$$
\widehat \pi_\omega(\phi) 
\;=\; 
\PI_d\pi_\omega(\phi)\PI_d^*
\;=\; 
\sum_{n\in \mathbb N} \sum_{x\in \mathbb Z^{d-1}} \phi(\tau_{x,n}\omega)|x,n\rangle \langle x,n |
\;,
$$
for $\phi \in C(\Omega)$, and
$$
\widehat \pi_\omega(\hat e)\; =\; P_{\hat e}\;=\;\sum_{y\in\ZM^{d-1}}|y,0\rangle\langle y,0|
\;,
$$
define a family of faithful $\ast$-representations of $\widehat \Aa_d$ on $\ell^2(\mathbb Z^{d-1}\times \mathbb N)$.
\end{proposition}

\noindent {\bf Proof.} We need to verify the commutation relations in Definition~\ref{def-halfspacealg}. Since $U_j$ commutes with $\PI_d^\ast\PI_d$ for $j=1,\ldots,d-1$, and since $\PI_d \PI_d^\ast = \one_{\ell^2(\ZM^{d-1}\times\NM)}$, 
$$
\widehat \pi_\omega(\hat u_i) \widehat \pi_\omega(\hat u_j) 
\;=\; \PI_d U_i U_j \PI_d^\ast
\;=\;
e^{iB_{i,j}}\PI_d U_j U_i \PI_d^\ast 
\;=\;
e^{iB_{i,j}}\widehat \pi_\omega(\hat u_j) \widehat \pi_\omega(\hat u_i)
$$
for all $i=1,\ldots,d$, hence the first set of commutation relations \eqref{HSComm1} are automatically satisfied. For the same reason,
$$
\widehat \pi_\omega(\hat u_j)\widehat \pi_\omega(\hat u_j^\ast) 
\;=\; 
\PI_d U_j U_j^\ast \PI_d^\ast 
\;=\; 
\one_{\ell^2(\ZM^{d-1}\times\NM)}
\;,
$$
and the second set of commutation relations \eqref{HSComm2} follows. Next, from \eqref{TransId} one finds
$$
\PI_d^\ast\PI_d U_d 
\;=\; 
U_d \big (\PI_d^\ast\PI_d + \sum_{y\in\ZM^{d-1}}|y,-1\rangle\langle y,-1| \big )
\;, 
\quad  
U_d \PI_d^\ast\PI_d \; =\;  
(\PI_d^\ast\PI_d - P_{e})U_d
\;,
$$
hence
$$
\widehat \pi_\omega(\hat u_d)^\ast \widehat \pi_\omega (\hat u_d) 
\;= \; \PI_d U_d^\ast U_d \big (\PI_d^\ast\PI_d + \sum_{y\in\ZM^{d-1}}|y,-1\rangle\langle y,-1|\big ) \PI_d^\ast 
\;=\; 
\one_{\ell^2(\ZM^{d-1}\times\NM)}
$$
and
$$
\widehat \pi_\omega(\hat u_d) \widehat \pi_\omega (\hat u_d)^\ast 
\;= \; 
\PI_d U_d^\ast U_d \big (\PI_d^\ast\PI_d -P_{\hat e} ) \PI_d^\ast 
\;=\; 
\one_{\ell^2(\ZM^{d-1}\times\NM)}-\widehat \pi_\omega(\hat e)
\;,
$$
which confirm the commutation relations \eqref{HSComm3}. Lastly, since $\pi_\omega(\phi)$ commutes with  $\PI_d^\ast \PI_d$, 
$$
\widehat \pi_\omega(\phi) \widehat \pi_\omega(\hat u_j) 
\;=\; 
\PI_d \pi_\omega(\phi) U_j \PI^\ast 
\;=\;
\PI_d U_j \pi_\omega(\phi \circ \tau_j) \PI_d^\ast 
\;=\; 
\pi_\omega(\hat u_j) \widehat \pi_\omega(\phi \circ \tau_j)
\;,
$$
for all $j=1,\ldots,d$, and the commutation relations \eqref{HSComm4} follow.\hfill $\Box$

\vspace{0.2cm}

Let us write the representation explicitly for the non-commutative polynomials. It is useful to decompose $\hat{p}=i'(p)+\tilde{p}$ with $p=\ev(\hat{p})$, since then
\begin{equation}
\label{RepHS2}
\widehat \pi_\omega (\hat p) 
\;=\; 
\PI_d \,\pi_\omega(p)\, \PI_d^* 
\;+\; 
\widehat \pi_\omega (\tilde p)
\;,
\end{equation}
which shows that essentially only the representation of the boundary algebra is new. Therefore, we also write $\widetilde \pi_\omega (\tilde p)=
\widehat \pi_\omega (\tilde p)$ for $\tilde{p}\in\Ee_d$. If we decompose as in \eqref{eq-phattilde}, then
$$
\widetilde{ \pi}_\omega(\tilde p) 
\, = \,
\sum_{n,m,k \in \mathbb N} \sum_{x,y,z \in \mathbb Z^{d-1}} \tilde p_{n,m}(\tau_{x,k}\omega,y) |x,k\rangle \langle x,k| U^{y} U_d^{n} |z,0\rangle \langle z,0|  U_d^{-m},
$$
and we can use \eqref{TransId} to transfer all the $U$'s to the end, to conclude:
\begin{equation}\label{HSRep}
\boxed{
\;
\begin{array}{c}
\tilde{p}
\;=\;
\sum\limits_{n,m \in \mathbb N} \sum\limits_{y \in \mathbb Z^{d-1}}\tilde{p}_{n,m}(\omega,y) \, u^y \,\hat{u}_d^n\,\hat{e}\,(\hat{u}_d^*)^m 
\;\in\; \Ee_d \\
 \downarrow \\
 \widetilde{ \pi}_\omega(\tilde p) 
\; = \;
\sum\limits_{n,m \in \mathbb N} \sum\limits _{x,y \in \mathbb Z^{d-1}} \tilde p_{n,m}(\tau_{x,n}\omega,y) |x,n\rangle \langle x,n| U^{(y,n-m)}\;.
\end{array}
\;
}
\end{equation}
Because both $\PI_d\PI_d^*$ and $P_{\hat{e}}$ are invariant under translations in $\ZM^{d-1}\times\{0\}$, the representations $\widehat \pi_\omega $ (hence also $\widetilde \pi_\omega$) inherit from \eqref{eq-covarrel2} the covariance property
$$
\widehat V^{(x,0)}\,\widehat \pi_\omega (\hat p)\,\big (\widehat V^{(x,0)}\big )^*\;=\;\widehat{\pi}_{\tau_x\omega}(\hat p)
\;,
\qquad
\hat{p}\in\widehat{\Aa}_{d}
\;,
\;\;
x\in \ZM^{d-1}
\;,
$$
where the magnetic translations are given by $\widehat V^{(x,n)}=\PI_d V^{(x,n)} \PI_d^*$ for $(x,n)\in\ZM^{d-1}\times \NM$. We also mention the following property w.r.t. the magnetic translations in the $d$th direction
\begin{align}\label{eq-perptrans}
\widehat V^{(0,k)} \widetilde{\pi}_{\tau_d^k\omega}(\tilde p) \big (\widehat V^{(0,k)} \big)^\ast 
= 
\sum\limits_{n,m \in \NM} \sum\limits _{x,y \in \mathbb Z^{d-1}}\!  \tilde p_{n,m}(\tau_{x,n+k}\omega,y) |x,n+k\rangle \langle x,n+k| U^{(y,n-m)}
\,, 
\end{align}
which effectively translates the boundary by $k\geq 0$ units. As one can see, if the lattice sites are relabelled such that $(x,n+k)$ becomes $(x,n)$, then $\widehat V^{(0,k)} \widetilde{\pi}_{\tau_d^k\omega}(\tilde p) \big (\widehat V^{(0,k)} \big)^\ast$ becomes identical with $\widetilde{\pi}_\omega (\tilde p)$.

\subsection{Algebra elements representing half-space Hamiltonians}
\label{Sec-AlgElemHalfSpace}

The generic half-space Hamiltonians were introduced in Section~\ref{Sec-HSHamilt}. The generic form of a covariant family $\{\widehat H_\omega \}_{\omega\in\Omega}$ of half-space Hamiltonians  was given in \eqref{HalfSpaceGenericH}. Here, we want to show explicitly that every half-space Hamiltonian can be represented uniquely as $\widehat H_\omega = \widehat \pi_\omega(\hat h)$ with some adequate $\hat{h}\in\widehat{\Aa}_d$. Using the decomposition from \eqref{HalfSpaceGenericH} and by comparing with \eqref{RepHS2}, we see that 
$$
\widehat H_\omega 
\;=\;
\PI_d \pi_\omega(h) \PI_d^\ast \;+\; \widetilde \pi_\omega(\tilde h)\;,
$$
with $h$ given in \eqref{UCGenericGeneratorBulk2}. Our task was reduced to finding $\tilde h\in\Ee_d$ for the generic $\widetilde H_\omega$ in \eqref{UHalfSpaceGenericHLandau}. By comparing with \eqref{HSRep}, one immediately finds
\begin{equation}
\boxed{
\;
\widetilde H_\omega = \widetilde \pi_\omega(\tilde h)
\;, 
\quad 
\tilde h 
\;=\; 
\sum_{n,m\leq R}\;\sum_{y \in \Rr'}\;\sum_{x\in \mathbb Z^{d-1}} e^{\frac{\I}{2}\langle y,n-m |\BB_+|y,n-m \rangle } \, \widetilde W^{y}_{n,m}\,  u^y\,\hat{u}_d^n\,\hat{e}\,(\hat{u}_d^*)^m
\;.
\;}
\end{equation}
All the above applies to the unitary as well as the chiral unitary class, the only difference being the special form of the hopping matrices in the latter case.

\section{The non-commutative analysis tools}
\label{Section-Calculus}

\subsection{The Fourier calculus}
\label{Sec-BulkFourier}

The Fourier calculus on $\Aa_d$ is defined by the following $\ast$-action of the $U(1)^{\times d}$ group on $\mathcal A_d$
$$ 
u_j \;\;\mapsto\;\; e^{-\I k_j} u_j\;, 
\qquad  
k_j\in [0,2\pi]\;, 
\;\; 
j=1,\ldots,d\;.
$$
This action generates a $d$-parameter group $k\in \mathbb T^d\mapsto\rho_k$ of continuous $*$-automorphisms \cite{Dav}, acting as
$$
\rho_k (p) \;=\; \sum_{x \in \mathbb Z^d} e^{-\I \langle x|k\rangle} p(x)u^x
$$
on the non-commutative polynomials. Given a generic element $a\in \Aa_d$, one can define its Fourier coefficients
$$
\Phi_x(a) 
\;=\; 
\left [ \int_{\mathbb T^d} \frac{dk}{(2\pi)^d} \ e^{\I \langle x|k \rangle } \rho_k (a) \right ] (u^x)^\ast \;, 
\qquad x \in \mathbb Z^d
\;.
$$
The Fourier coefficients are ordinary functions over the space of disorder configurations $\Omega$. For $x=0$, $\Phi_0$ is actually an expectation of $\mathcal A_d$ onto $C(\Omega)$ (see \cite{Dav} pp. 222 for details). For a non-commutative polynomial $p=\sum_x p(x) u^x$, we have $\Phi_x(p) = p(x)$. For a generic element $a \in \Aa_d$, the Ces\`{a}ro sums
\begin{equation}\label{Cesaro}
a^{(n)}
\;=\;
\sum_{x \in V_n} \prod_{j=1}^d \left (1-\frac{|x_j|}{n+1} \right ) \Phi_{x}(a) u^x
\;,
\end{equation}
with $V_n = [-n,\ldots,n]^d$, converge in norm to $a$ as $n \rightarrow \infty$ \cite{Dav}. While we already knew that the algebra of non-commutative polynomials is dense in $\Aa_d$, \eqref{Cesaro} provides an explicit approximation of $a$ in terms of such non-commutative polynomials. It also tells us that two elements with the same Fourier coefficients are identical. Hence various actions on $\Aa_d$, such as the derivations below, can be defined by specifying their action on the Fourier coefficients or on the non-commutative polynomials.

\vspace{.2cm}

Now we consider the algebra of boundary observables. The Fourier calculus over $\Ee_d$ is defined by the following $\ast$-action of the $U(1)^{\times (d-1)}$ group
$$ 
\hat u_j \;\;\mapsto \;\;e^{-\I k_j} \hat u_j
\;, \qquad k_j \in [0,2\pi]
\;, \;\;
j=1,\ldots,d-1
\;.
$$
The remaining generator $\hat u_d$, hence also the projector $\hat e$, remain unchanged. This generates a $(d-1)$-parameter group $k\in \mathbb T^{d-1}\mapsto \hat \rho_k$ of continuous $*$-automorphisms, acting as
$$
\widetilde \rho_k (\hat p) 
\;=\; 
\sum_{n,m \in \mathbb N} \sum_{x \in \mathbb Z^{d-1} } e^{-\I \langle x|k\rangle} \tilde p_{n,m}(x)u^x \hat u_d^n (\hat u_d^\ast)^m
$$
on the non-commutative polynomials from $\Ee_d$. Given a generic element $\tilde a\in \Ee_d$, one can define its Fourier coefficients
$$
\widetilde \Phi_x(\tilde a) 
\;=\; \left [ \int_{\mathbb T^{d-1}} \frac{dk}{(2\pi)^{d-1}} \ e^{\I \langle x|k \rangle } \widetilde \rho_k (\tilde a) \right ] (\hat u^x)^\ast
\;, 
\qquad x \in \mathbb Z^{d-1}.
$$
These Fourier coefficients belong to the algebra $\Ee_1$.

\subsection{Non-commutative derivations and integrals}
\label{Sec-DiffCalculus}

The Fourier calculus over the algebra of bulk observables generates a system of unbounded closed $*$-derivations $\partial =(\partial_1,\ldots,\partial_d)$. Indeed, let $C^n(\Aa_d)$ be the linear subspace spanned by those elements $a\in \Aa$ for which $\rho_k(a)$ is an $n$-times differentiable function of $k$. Then the derivations are defined over $C^1(\Aa_d)$ as the generators of the automorphisms $\rho_k$. Their actions on the non-commutative polynomials are given explicitly by
\begin{equation}
\label{BulkDerivation}
\boxed{
\; \partial_j \sum_{x \in \mathbb Z^d} p(x) u^x 
\;=\; 
-\,\I \sum_{x \in \mathbb Z^d} x_j\, p (x)u^x
\;.\;
 }
\end{equation}
The derivations satisfy the Leibniz rule
$$
\partial (ab)\;=\; (\partial a)b\;+\;a(\partial b)
\;,
\qquad
a,b\in C^1(\Aa_d)
\;,
$$
and for the representations on $\ell^2(\ZM^d)$
\begin{equation}
\label{BulkDerivationRep}
\boxed{
\;
\pi_{\omega}(\partial a)
\;=\;
\I \big [ \pi_{\omega}(a),X \big ]
\;.
\;}
\end{equation}

The Fourier calculus also defines a faithful continuous trace over $\Aa_d$. Indeed, the map $a \rightarrow \Phi_0(a)$ generates a faithful and continuous expectation from $\Aa_d$ to $C(\Omega)$ \cite{Dav}. Combined with the continuous and normalized trace over $C(\Omega)$ given by $\int_\Omega d \PM(\omega) \, \phi(\omega)$, it defines the canonical trace on $\Aa_d$
\begin{equation}
\label{BulkTrace}
\boxed{
\;
\Tt(a) \;=\; \int_\Omega \mathbb P(d\omega) \, \Phi_0(a)
\;. \;
}
\end{equation}
For the non-commutative polynomials, the trace can be computed as 
$$
\Tt(p) 
\;=\; 
 \int_\Omega \mathbb P(d\omega) \, p(\omega,0)
 \;.
$$
The trace $\Tt$ is continuous, normalized, $\Tt(\one)=1$, and invariant w.r.t. the automorphisms $\rho_k$. Its physical meaning can be understood from   Birkhoff's ergodic theorem. As shown in \eqref{eq-BrikhoffTV}, $\Tt$ is actually equal to the trace per unit volume on $\ell^2(\mathbb Z^d)$
\begin{equation}\label{TracePerVolume}
\Tt(a)
\;=\;
\lim_{V \rightarrow \ZM^d} \frac{1}{|V|}\mathrm{Tr}\big (\PI_V \, \pi_\omega (a) \, \PI_V^\ast \big )
\;,
\end{equation}
where $V$ is a cube in $\mathbb Z^d$ and $|V|$ is its cardinality and the equality holds for $\PM$-almost all $\omega$.

\begin{remark}
When $\Omega$ consists of just one point and $\BB=0$, then the operators in the algebra $\Aa_d$ are periodic and actually $\Aa_d$ is isomorphic to the continuous functions $C(\TM^d)$ on the $d$-dimensional torus $\TM^d=(-\pi,\pi]^d$. This can be explicitly seen via the discrete Fourier transform $\Ff:\ell^2(\ZM^d)\to L^2(\TM^d)$ defined by
$$
(\Ff\phi)(k)
\;=\;
(2\pi)^{-\frac{d}{2}}
\;\sum_{n\in\ZM^d}\phi_{n}\;e^{-\I\langle n| k\rangle}
\;,
\qquad
k\in\TM^d
\;.
$$
If now $a\in\Aa_d$, then 
$$
\Ff\pi(a)\Ff^*
\;=\;
\int^\oplus_{\TM^d}dk\;a(k)
\;,
$$
where the r.h.s. is a multiplication operator with $a(k)= \sum_{n\in\ZM^d}a(n)e^{-\I\langle n| k\rangle}$. Now the trace per unit volume becomes
$$
\Tt(a)
\;=\;
\int_{\TM^d}\frac{dk}{(2\pi)^d}\;a(k)
\;,
$$
and the derivations satisfy for $a\in C^1(\Aa_d)$
$$
\Ff\,\pi(\partial a)\,\Ff^*
\;=\;
\Ff\,\I[\pi(a),X]\,\Ff^*
\;=\;
\int^\oplus_{\TM^d}dk\;\partial_k a(k)
\;.
$$
This establishes the connection between the familiar calculus over the Brillouin zone and the non-commutative analysis tools.
\hfill $\diamond$
\end{remark}

Now we turn our attention to the algebra of boundary observables where we can define again a non-commutative differential calculus. As before, the system of derivations $\widetilde \partial=(\widetilde \partial_1,\ldots,\widetilde \partial_{d-1})$ is defined over $C^1(\Ee_d)$ by the generators of the automorphisms $\widetilde \rho_k$. On the non-commutative polynomials,
\begin{equation}
\label{HSDerivation}
\boxed{
\;\widetilde \partial_j \tilde p 
\;=\; 
-\,\I \sum_{n,m \in \mathbb N}\;\sum_{x \in \mathbb Z^{d-1}}\; x_j \,\tilde p_{n,m}(\omega, x) \,u^x \, u_d^n \, \hat e \, (\hat u_d^\ast)^m
\;.\;
}
\end{equation}
The derivations $\widetilde \partial$ obey the Leibniz rule and
\begin{equation}
\label{eq-HSderivrep}
\boxed{
\;\widetilde \pi_\omega (\widetilde \partial_j \tilde a) 
\;=\; 
\I \big [ \widetilde \pi_\omega (a), X_j \big ]
\;
}
\end{equation}
on $\ell^2(\mathbb Z^{d-1} \times \mathbb N)$, for $j=1,\ldots,d-1$. A system of derivations can be also introduced over the algebra of half-space physical observables,
$$
\widehat \partial_j \hat a 
\;=\; 
i'(\partial_j a)\; + \;\widetilde \partial_j \tilde a\;, 
\qquad j = 1, \ldots, d-1
\;,
$$
for any $\hat a = (a, \tilde a) \in \widehat \Aa_d$. On the non-commutative polynomials,
\begin{equation}
\label{FHSDerivation}
\boxed{\;
\widehat \partial_j \hat p 
\;=\; -\,\I \sum_{n,m \in \mathbb N}\;\sum_{x \in \mathbb Z^{d-1}}\; x_j\, \hat p_{n,m}(\omega, x)\, u^x \,u_d^n (\hat u_d^\ast)^m
\;.\;
}
\end{equation}

The Fourier calculus  over $\Ee_d$ presented in the previous section provided us with $\widetilde \Phi_0$, which is a continuous expectation from $\Ee_d$ to $\Ee_1$ \cite{Dav}. Therefore, as already seen in the bulk case, a canonical trace can be introduced over $\Ee_d$ once we define a canonical trace over $\Ee_1$. The non-commutative polynomials from $\Ee_1$ have the form
$$
\tilde p \;=\; 
\sum_{n,m \in \mathbb N} \tilde p_{n,m} \, \hat u_d^n\, \hat e \,(\hat u_d^\ast)^m
\;,
$$
with $\tilde p_{n,m}$ ordinary continuous functions over $\Omega$. These coefficients can also be seen as functions over $\Omega$ with values in $\Kk$, the algebra of compact operators on $\ell^2(\NM)$. Hence a trace can be canonically defined by  
$\sum_{n} \int_\Omega \PM(d\omega) \tilde p_{n,n}(\omega)$ which then can be promoted to a lower-semicontinuous trace $\widetilde\Tt_1$ over $\Ee_1$ \cite{Bla}. The trace over $\Ee_d$ is then defined as
\begin{equation}
\label{HSTrace}
\boxed{
\;\widetilde \Tt(\tilde a) \;=\; \widetilde \Tt_1 \Big ( \widetilde \Phi_0(\tilde a) \Big )
\;.\;
}
\end{equation}
For the non-commutative polynomials from $\Ee_d$, it takes the explicit form
\begin{equation}
\widetilde \Tt(\tilde p) 
\;=\; 
\sum_{n \in \mathbb N} \int_\Omega \mathbb P(d\omega) \ \tilde p_{n,n}(\omega,0)
\;.
\end{equation}
The trace $\widetilde \Tt$ is lower semicontinuous and invariant w.r.t. the automorphisms $\widetilde \rho_k$. Using Birkhoff's ergodic theorem together with an average over the position of the boundary, one can write $\widetilde \Tt$ as the trace per area on $\ell^2(\mathbb Z^{d-1} \times \mathbb N)$:
\begin{equation}\label{TracePerArea}
\widetilde \Tt(\tilde a) 
\;= \;
\lim_{A \rightarrow \mathbb Z^{d-1}} \lim_{K \rightarrow \infty} \frac{1}{K|A|}
\;
\sum_{k=1}^K\Tr \Big ( \PI_{A\times\mathbb N} \, \widetilde \pi_{\tau_d^k\omega} (\tilde a) \, \PI_{A\times \mathbb N}^\ast \Big )
\;,
\end{equation}
where $A$ is a cube from $\mathbb Z^{d-1}$ and $|A|$ is its cardinality.

\vspace{.2cm}

The pairs $(\partial, \Tt)$ and $(\tilde \partial,\widetilde \Tt)$ define the non-commutative differential calculus over the algebras of bulk  and boundary observables, respectively. As we've already seen, the physical models are rather generated from $M_N(\mathbb C) \otimes \Aa_d$ and $M_N(\mathbb C) \otimes \Ee_d$, but the non-commutative calculi extend naturally over these algebras as $(\one \otimes \partial, {\rm Tr} \otimes \Tt)$ and $(\one \otimes \widetilde \partial, {\rm Tr} \otimes \widetilde \Tt)$, respectively. To ease the notation, we will continue to use $(\partial, \Tt)$ and $(\tilde \partial,\widetilde \Tt)$ for these situations too. Next let us list a few useful identities.

\begin{proposition}\label{Indenties1}
The following holds for $(\partial, \Tt)$, and analogously for $(\widetilde \partial, \widetilde \Tt)$:
\begin{enumerate}[\rm (i)]

\item Let $e$ be a projection from $C^1(\Aa_d)$. Then
\begin{align*}
 e (\partial_i e) & \;= \;(\partial_i e)(1-e)\;, \\ 
 (1-e)(\partial_i e) &\; =\; (\partial_i e ) e \;, \\ 
 (1-2e) (\partial_i e) & \;=\; - (\partial_i e ) (1-2e) \;.
 \end{align*}

\item Let $e$ be a projection from $C^2(\Aa_d)$. Then
\begin{align*}
 e(\partial_i \partial_j e) e & \;=\; - e \{\partial_i e, \partial_j e\}\; =\; - \{\partial_i e, \partial_j e\} e\;, \\
 (1-e)(\partial_i \partial_j e) (1-e) & \;=\;  (1-e) \{\partial_i e, \partial_j e\} \;=\; \{\partial_i e, \partial_j e\}(1-e)\;,
\end{align*}
where $\{\, ,\,\}$ denotes the anti-commutator.

\item Let $a \in C^1(\Aa_d)$ be invertible. Then $a^{-1} \in C^1(\Aa_d)$ and
$$
\partial_i a^{-1}\; = \;- a^{-1}(\partial_i a) a^{-1}\;.
$$

\item Let $a, b \in C^1(\Aa_d)$. Then
$$
\Tt\big ( \partial_i a \big ) \;=\;0\;, 
\qquad
\Tt \big (a (\partial_i b) \big) \;= \;- \Tt \big((\partial_i a) b \big)
\;.
$$
\end{enumerate}
\end{proposition}

\subsection{The smooth sub-algebras and the Sobolev spaces}
\label{Sec-Sobolev} 

Due to the result below, the spaces of infinitely differentiable elements
$$
\mathscr A_d
\;=\;
C^\infty(\Aa_d)
\;=\;
\bigcap_{n\geq 1} C^n(\Aa_d)
\;,
\qquad
\mathscr E_d
\;=\;
C^\infty(\Ee_d)
\;=\;
\bigcap_{n\geq 1} C^n(\Ee_d)
\;,
$$
can be endowed with a locally convex topology so as to become Fr\'{e}chet algebras. In particular, this implies that they are metrizable and  complete. Before proceeding, we want to elaborate briefly on why these technical structures are important in what follows. The $K$-groups of operator algebras (see Chapter~\ref{Chap-KTheory}) can be introduced in a topological or purely algebraic fashion, and the non-commutative geometry program can be carried in both settings \cite{Var}. However, for applications in solid state physics we eventually would like to come back to the topological $K$-groups. While the algebraic $K$-groups can be defined for generic $\ast$-algebras, the topological $K$-groups are natural for $C^\ast$-algebras and, at most, can be defined for Fr\'{e}chet algebras. As such, the smooth sub-algebras $\mathscr A_d$ and $\mathscr E_d$ will posses well-defined topological $K$-groups. Since the topological invariants can only be defined over the smooth sub-algebras, they can provide information only about these $K$-groups. If one is only interested in defining topological invariants, this aspect will be marginally relevant, but if one cares about the classification of the topological insulators, then it is imperative to make sure that the $K$-groups of the smooth sub-algebras coincide with the ones of the original algebras. An important technical result in non-commutative geometry states that this is indeed the case if the smooth sub-algebras are dense and invariant w.r.t. the holomorphic calculus \cite{Con,GVF}. The good news is that all these issues have a simple resolution for the present context. The statements below follow from the work by Rennie \cite{Ren} (see, particularly, the examples provided at page 131) which covers both the unital and non-unital algebras. This is relevant here since the algebra $\Ee_d$ has no unit.

\begin{proposition}[The smooth sub-algebras defined] \label{SmoothAlgebras}
\begin{enumerate}[\rm (i)]
\item The $\ast$-algebra $\mathscr A_d$ endowed with the topology induced by the seminorms 
$$
\|a \|_\alpha \;=\; \|\partial^\alpha a \|\;, 
\qquad 
\partial^\alpha = \partial_1^{\alpha_1} \cdots \partial_d^{\alpha_d}
\;,\;\;
\alpha \;=\; 
(\alpha_1, \ldots \alpha_d)
\;,
$$ 
is a dense Fr\'{e}chet sub-algebra of $\Aa_d$ which is stable under holomorphic calculus. The norm appearing on the right, above, is the $C^\ast$-norm of $\Aa_d$.

\item The completion of the dense sub-algebra of non-commutative polynomials from $\Ee_d$ in the topology induced by the seminorms 
$$
\|\tilde p \|_{\alpha,\beta} 
\;=\; 
\sup_{n,m\in \mathbb N} n^{\beta_1} m^{\beta_2} \|(\tilde \partial^\alpha \tilde p)_{n,m} \|\;, 
\qquad 
\tilde \partial^\alpha 
\;=\; 
\tilde \partial_1^{\alpha_1} \cdots \tilde \partial_{d-1}^{\alpha_{d-1}}
\;,
$$ 
is a Fr\'{e}chet algebra $\mathscr E_d$ which is stable under holomorphic calculus. The norm appearing on the right, above, is the $C^\ast$-norm of $\Aa_{d-1}$.
\end{enumerate}
\end{proposition}

The smooth algebras can be characterized as the sub-algebras of elements with rapidly decaying Fourier coefficients, more precisely:

\begin{proposition}
\label{RapidDecay}
 If $a \in \mathscr A_d \subset \Aa_d$, then for any $d$-index $\alpha$:
\begin{equation}
\label{Decay1}
x^{\alpha}\|\Phi_x(a)\|_{C(\Omega)} 
\;\leq\; 
\|\partial^{\alpha} a\| 
\;<\; \infty\;, 
\qquad 
x^{\alpha} = x_1^{\alpha_1} \cdots x_d^{\alpha_d}
\;.
\end{equation}
Conversely, if for every $d$-index $\alpha$
$$
x^{\alpha} \|\Phi_x(a)\|_{C(\Omega)} 
\; < \; \infty
$$ 
uniformly in $x$, then $a$ belongs to $\mathscr A_d$. Similarly for the boundary algebra, if $\tilde a \in \mathscr E_d$ then for any indices $\beta_1$, $\beta_2$ and $(d-1)$-index $\alpha$
\begin{equation}
\label{Decay2}
x^\alpha n^{\beta_1} m^{\beta_2} \|\widetilde \Phi_x(\tilde a)_{n,m}\|_{C(\Omega)}  
\;\leq\; 
\|\widetilde \partial^{\alpha} \tilde a\| 
\;<\; \infty\;, 
\qquad 
x^{\alpha} = x_1^{\alpha_1} \cdots x_{d-1}^{\alpha_{d-1}}
\;.
\end{equation}
\end{proposition}

\noindent {\bf Proof.} One has
$$
|x^{\alpha}\Phi_x(a)(\omega)| 
\;=\; 
| \langle  0 |\pi_\omega(\partial^{\alpha} a) | - x \rangle |
\;\leq\;  
\sup_{\omega \in \Omega} \|\pi_\omega(\partial^{\alpha} a)\| 
\;=\; 
\|\partial^{\alpha} a\| 
\;<\; \infty
\;.
$$
The other cases are treated similarly.
\hfill $\Box$

\vspace{0.2cm} 

It follows from the estimates presented in Proposition~\ref{FermiProjection1} that the Fermi projections (and the Fermi unitary operators) of finite-range bulk Hamiltonians belong to the smooth algebra $\mathscr A_d$, provided BGH (and CH) holds. Furthermore, from the estimates presented in Proposition~\ref{HFSmoothFunc1}, it follows that under BGH any smooth function of finite-range half-space Hamiltonians belongs to the smooth algebra $\mathscr E_d$.

\vspace{0.2cm}

Another important issue is to find the maximal domains of the linear functionals defined using the non-commutative differential calculus. For the cases of interest in Chapter~\ref{Chap-TopologicalInvariants}, they are given by certain Sobolev spaces \cite{BES,PLB,PS}.  To define these spaces we need first to introduce and characterize the so-called non-commutative $L^s$-spaces. Let us consider first the bulk case. Denoting the absolute value of an element $a\in \Aa_d$ as usual by $|a|=\sqrt{a^\ast a}$, we set
\begin{equation}
\label{LPNorm}
\|p\|_s \;=\; 
{\cal T}\left ( |p|^s\right)^{\frac{1}{s}} \; , 
\qquad s\in [1,\infty]\;. 
\end{equation}
This defines a norm on $\Aa_d$. The completion of $\Aa_d$ under this norm defines the non-commutative $L^s$-space, which as usual are denoted by $L^s({\Aa_d},{\cal T})$. Of particular importance is the space $L^\infty({\Aa_d},{\cal T})$ which represents the weak von Neumann closure of $\Aa_d$. This space can be also viewed \cite{Con3} as the closure of the non-commutative polynomials from $\Aa_d$ under the norm 
$$
\|p\|_\infty\; =\; 
\mathbb P\!-\!\operatorname*{\mathrm{esssup}}\limits_{\omega \in \Omega} \|\pi_\omega (p)\|
\;.
$$
Since von Neumann algebras are stable under the Borel functional calculus, the Fermi projections and Fermi unitary operators belong to $L^\infty({\Aa_d},{\cal T})$, regardless of the existence of spectral or mobility gaps at the Fermi level. The topology of $L^\infty({\Aa_d},{\cal T})$ is, however, too strong to be useful in the strong disorder regime. For example, the Fermi projections do not vary continuously w.r.t. $\|\cdot \|_{L^\infty}$ when the models are deformed continuously, even when the Fermi level is located in a region of Anderson localized spectrum. This is another reason for introducing the non-commutative Sobolev spaces. We continue, however, first with the characterization of the $L^s$-spaces.

\vspace{.2cm}

The non-commutative version of H\"older's inequality is a useful tool for the characterization of the $L^s$-spaces,
\begin{equation}
\label{Holder}
 \|f_1 \cdots f_k\|_s 
 \;\leq \;
 \|f_1\|_{s_1} \cdots \| f_k \|_{s_k}
\;, 
\qquad \frac{1}{s_1} \,+\, \ldots \,+ \,\frac{1}{s_k} \,=\, \frac{1}{s} 
\;.
\end{equation}
Note that, in general, the $L^s$-spaces are not closed under multiplication, hence they are not algebras, but only Banach spaces. Nevertheless, \eqref{Holder} enables one to make sense of the products of elements from different $L^s$-spaces, such as the one on the l.h.s. of  \eqref{Holder}, as elements of lower $L^s$-spaces. Taking some of the $f$'s in \eqref{Holder} to be the identity in $\Aa_d$, the following sequence of inclusions can be derived 
\begin{equation}
\label{eq-sequence}
L^\infty ({\Aa_d},{\cal T})  \subset \ldots \subset L^s({\Aa_d},{\cal T}) \subset \ldots \subset L^1({\Aa_d},{\cal T})
\;.
\end{equation}
Dense subspaces of elements for each of the $L^s$-spaces are furnished by the non-commutative polynomials with coefficients in $L^s(\Omega, \PM)$. In fact, the maps $\Phi_x$ which give the Fourier coefficients can be extended by continuity on the non-commutative $L^s$-spaces and, using the classical H\"older inequality, one sees that the Fourier coefficients take values in $L^s(\Omega, \PM)$. Lastly, let us state a useful upper bound on the $L^s$-norms.

\begin{proposition} 
\label{prop-lsbulk}
Let $a \in L^s(\Aa_d, \Tt)$ and assume $s$ is integer. Then
\begin{equation}
\label{eq-lsbound0}
\|a\|_s 
\;\leq \;
2\, \sum_{x \in \ZM^d} \left [ \int_\Omega \PM(d\omega) \, |a(\omega,x)|^s \right ]^\frac{1}{s}
\;,
\end{equation}
where $a(\omega,x)$ it the Fourier coefficient at $x$.
\end{proposition}

\noindent {\bf Proof.} It is enough to establish the estimate on a dense subspace, which we take to be the algebra $\Aa_d$. Let us first assume that $a$ is self-adjoint, in which case $|a|^s = |a^s|$. Furthermore, the absolute value of any self-adjoint operator $f$ can be computed by applying the continuous function $t \, \sgn(t)$ on $f$. By approximating the sign function by the smooth function $\sgn_\varepsilon(t) = \tanh(t/\varepsilon)$, one can write $|f| = \lim_{\varepsilon \rightarrow 0} f \, \sgn_\varepsilon (f)$ with a limit taken inside $\Aa_d$ w.r.t. the C$^*$-norm. The point of the last expression is that $\sgn_{\epsilon}(f) \in \Aa_d$, while $\sgn(f)$ is not. Also, note that $f \, \sgn_{\epsilon}(f)$ is an increasing sequence of positive operators as $\varepsilon \rightarrow 0$. Since the trace $\Tt$ is continuous,
$$
\Tt(|f|) 
\;=\; 
\lim_{\varepsilon \rightarrow 0} \;\Tt\big ( f \, \sgn_\varepsilon (f) \big ) 
\;=\; 
\lim_{\varepsilon \rightarrow 0} \;\int_\Omega \PM(d \omega) \, ( f \, \sgn_\varepsilon (f) \big )(\omega,0)
\;.
$$
We now take $f=a^s$ and denote $v_\varepsilon = \sgn_\varepsilon (a^s)$.
We will exploit the fact that
\begin{equation}
\label{eq-umat}
\sup_{\omega \in \Omega}|v_\varepsilon(\omega,x)| 
\; \leq \;
1,
\end{equation}
which follows from
$$
|v_\varepsilon(\omega,x)|
\;=\;
|\langle 0|\pi_\omega (v_\varepsilon) |-x \rangle| 
\;\leq\;  
\|\sgn_\varepsilon\big (\pi_\omega (a)^s)\| 
\;\leq \;
1 \;.
$$
Let us evaluate the product $a^s v_\epsilon$ at $x=0$, explicitly,
$$
\big (a^s v_\varepsilon\big ) (\omega,0) 
\,= \!
\sum_{x_1,\ldots,x_s} \!\!
e^{\I \phi} a(\tau_{-x_{s+1}}\omega, x_s-x_{s+1}) \cdots a(\tau_{-x_2}\omega,x_1-x_2) v_\varepsilon(\tau_{-x_1} \omega,-x_1),
$$
where $e^{i\phi}$ is a phase factor containing all the Peierls factors, $x_1,\ldots,x_s$ run over $\ZM^d$ and $x_{s+1}=0$. With the notation $\omega_i= \tau_{-x_{i+1}}\omega$ and change of variable $y_i = x_i - x_{i+1}$, and from \eqref{eq-umat}, it follows
$$
\Tt\big(a^s v_\varepsilon\big ) 
\; \leq \;
\int_\Omega \PM(d \omega) \, \sum_{y_1,\ldots,y_s \in \ZM^d} \; \prod_{i=1}^s |a(\omega_i,y_i)|
\;.
$$
The terms above are all positive hence one can interchange the integral and the summations and, after applying the classical H\"older inequality,
$$
\Tt\big(a^s v_\varepsilon \big ) 
\; \leq \;
 \sum_{y_1,\ldots,y_s \in \ZM^d} \ \prod_{i=1}^s \left [\int\limits_\Omega \PM(d \omega) \, |a(\omega,y_i)|^s \right ]^\frac{1}{s}
\;=\;
\left ( \sum_{x \in \ZM} \|a(\cdot,x)\|_{L^s(\Omega,\PM)} \right )^s
\;,
$$
where we also used the invariance of $\PM$ against lattice translations. This is a uniform upper bound in $\varepsilon$. Hence it applies to the limit, too. The statement now follows for self-adjoint $a$, even without the factor $2$ in front. The result can be extended to the generic case by using the decomposition $a = a_r + \I a_i$ into the real and imaginary parts, $a_r = \frac{1}{2}(a+ a^\ast)$ and $a_i = -\frac{\I}{2} (a - a^\ast)$. Indeed
$$
\|a\|_{L^s} 
\;\leq\; 
\|a_r\|_{L^s}+\|a_i\|_{L^s} 
\;\leq\; 
\sum_{x \in \ZM^d} \Big (\|a_r(\cdot,x)\|_{L^s(\Omega,\PM)} + \|a_i(\cdot,x)\|_{L^s(\Omega,\PM)} \Big ) 
\;,
$$
and the statement follows from the definition of the real and imaginary parts and after applying the triangle inequality once again.
\hfill $\Box$

\vspace{.2cm}

We are now ready to introduce the non-commutative Sobolev spaces for the bulk algebra. Let $\alpha =(\alpha_1,\ldots, \alpha_d)$ be a $d$-index, as above, and $|\alpha|=\alpha_1+\ldots \alpha_d$. Then
\begin{equation}
\|p\|_{s,k}
\;=\;
\sum_{0\leq |\alpha| \leq k} \|\partial^\alpha p\|_s
\; ,
\qquad 
s \in [1,\infty]\;, \;\; k\in \mathbb N
\end{equation}
defines a norm on the algebra of non-commutative polynomials from $\Aa_d$. The completions under these norms define the first class of non-commutative Sobolev spaces, which will be denoted by $\mathcal W_{s,k}({\Aa_d},{\cal T})$. These spaces represent the maximal domain for the multilinear forms defined in Chapter~\ref{Chap-TopologicalInvariants}. The use of $\mathcal W_{p,r}(\mathcal A_d,\mathcal T)$ for the index theorems in \cite{BES,PLB} depends critically on the computation of a certain Dixmier trace, which is highly technical and is further complicated when the dimensionality of the space is odd. However, as in \cite{PS}, we can avoid all this by introducing a second class of Sobolev spaces, generated by the closure of the algebra of non-commutative polynomials from $\Aa_d$ under the norm
$$
\|p\|'_{s,k} 
\;=\; 
\sum_{x \in \mathbb Z^d} (1+|x|)^k \left [ \int_\Omega \mathbb P(d \omega) |p(\omega,x)|^s \right ]^\frac{1}{s}
\;, 
\qquad 
s \in [1,\infty]\;, \;\; k\in \mathbb N
\;. 
$$
These Banach spaces will be denoted by $\mathcal W'_{s,k}(\mathcal A_d,\mathbb P)$. Some of their important properties are listed below.

\begin{proposition}\label{SobolevBulk} The two classes of Sobolev spaces satisfy the following relations:

\begin{enumerate}[\rm (i)]

\item $\mathcal W'_{s,k}(\mathcal A_d,\mathbb P)$ is invariant to the $\ast$-operation.

\item $\mathcal W'_{s,k}(\mathcal A_d,\mathbb P) \subset \mathcal W'_{s',k'}(\mathcal A,\mathbb P)$, whenever $s \leq s'$ and $k \leq k'$.

\item $\mathcal W'_{s,k}(\mathcal A_d,\mathbb P)\subset \mathcal W_{s,k}(\mathcal A_d,\mathcal T)$ for $s$ integer.

\end{enumerate}

\end{proposition}

\noindent {\bf Proof.} (i) Since the norms $\| \cdot \|'_{s,k}$ are invariant to the transformations $p(\omega, x) \mapsto p(\tau_{y}\omega,- x)$ and to the complex conjugation of $p(\omega, x)$, the equality $\|p^\ast \|_{s,k} = \|p\|_{s,k}$ holds. (ii) From the very definition, 
\begin{equation}
\label{Comp1}
\|p\|'_{s,k} 
\;\leq\; 
\|p\|'_{s,k'}
\;, 
\qquad {\rm for} \; k \leq k'
\;,
\end{equation}
and H\"older's inequality gives $\|p\|'_{s,k} \leq \|p\|'_{s',k}$ whenever $s \leq s'$. (iii) We will show
\begin{equation}
\label{NormInequality}
\|p\|_{s,k} 
\;\leq\; 
2\,\mathcal N_k \,\|p\|'_{s,k}
\;,
\end{equation}
with a constant $\mathcal N_k$ specified below. Indeed, from \eqref{eq-lsbound0},
$$ 
\|\partial^{\alpha} p\|_s  \; \leq \;
2\sum_{x \in \ZM} \|\partial^{\alpha} p(\cdot,x)\|_{L^s(\Omega,\PM)}
 \; \leq \;
2\,\sum_{x \in \ZM^d} |x|^{|\alpha|}\| p(\cdot,x)\|_{L^s(\Omega,\PM)}
\;,
$$
hence $\|\partial^{\alpha} p\|_s  \leq 2\,\|p\|'_{s,|\alpha|}.$ Then
$$
\|p\|_{s,k} 
\;=\;
\sum_{|\alpha|\leq k} \|\partial^{\alpha}p\|_s 
\;\leq\; 
2\sum_{|\alpha|\leq k} \|p\|'_{s,|\alpha|} 
\;\leq \;
2\mathcal N_k \|p\|'_{s,k}
\;,
$$
where $\mathcal N_k$ is the cardinality of the set $\{\alpha \in \mathbb N^d\,:\,  |\alpha| \leq k\}$.  \hfill $\Box$

\vspace{.2cm}

Let us now turn our attention to the boundary algebras and spaces. The definition of the non-commutative $L^s$-spaces is universal, hence $L^s(\Ee_d, \widetilde \Tt)$ is defined in the same way as the closure of the algebra $\Ee_d$ under the norm
$$
\|\tilde a \|_s 
\;=\; 
\widetilde \Tt \big ( |\tilde a|^s \big )^\frac{1}{s}
\;.
$$ 
Again, a special role is played by $L^\infty(\Ee_d,\widetilde \Tt)$ which can be also characterized as $L^\infty(\Ee_d,\widetilde \Tt) \simeq L^\infty(\Aa_{d-1},\Tt) \otimes \mathcal B$, where $\mathcal B$ is the algebra of bounded operators on a separable Hilbert space. In particular, this implies that $L^\infty(\Ee_d,\widetilde \Tt)$ has a unit. Since $\Ee_d$ does not have a unit, there are no inclusions as in \eqref{eq-sequence} for the boundary $L^s$-spaces. Recall that the elements from $\mathscr E_d$ are of the form 
$$
\tilde a 
\;=\;
\sum_{n,m \in \mathbb N}\; \sum_{x \in \mathbb Z^{d-1}} \;\tilde a_{n,m} (\omega,x) \,\hat u^x \,(\hat u_d)^n \,\hat e\, (\hat u_d^\ast)^m
\;,
$$ 
where the Fourier coefficients $\tilde a_{n,m}(\omega,x)$ are continuous functions in $\omega$. It will be convenient to denote the matrix with entries $\tilde a_{n,m}(\omega,y)$ by $\tilde a(\omega,x)$.

\begin{proposition}
\label{prop-lsboundary}
 Let $\tilde a \in L^s(\Ee_d, \widetilde \Tt)$ and assume $s$ integer. Then
\begin{equation}
\label{eq-lsbound}
\|\tilde a\|_s 
\;\leq\; 
2 \,\sum_{x \in \ZM^{d-1}} \left [ \int_\Omega \PM(d\omega) \, \|\tilde a(\omega,x)\|_{(s)}^s \right ]^\frac{1}{s}
\;,
\end{equation}
where $\|b\|_{(s)}=\tr(|b|^s)^{\frac{1}{s}}$ represents $s$-Schatten norm of a matrix $b$.  
\end{proposition}

\noindent {\bf Proof.} As in Proposition~\ref{prop-lsbulk}, it is enough to establish the inequality for self-adjoints from $\Ee_d$. Since $\widetilde \Tt$ is lower semicontinuous, we can still apply
$$
\widetilde \Tt(|\tilde a|^s) 
\;=\; 
\lim_{\varepsilon \rightarrow 0}\;\widetilde \Tt\big ( \tilde a^s \tilde v_\varepsilon \big ) 
\;=\; 
\lim_{\varepsilon \rightarrow 0}\; \int_\Omega \PM( d \omega) \, \tr \big ( (\tilde a^s \tilde v_\varepsilon)(\omega,0)\big )
\;,
$$
where $\tilde v_\varepsilon = \sgn_\varepsilon (\tilde a^s)$. From \eqref{eq-brules}, the product $\tilde a^s \tilde v_\epsilon$ at $x=0$ is given by,
$$
\big (\tilde a^s \tilde v_\varepsilon\big ) (\omega,0) 
\;= \sum_{x_1,\ldots,x_s \in \ZM^{d-1}} e^{\I \phi} \tilde a(\omega_s, x_s-x_{s+1}) \cdots \tilde a(\omega_1,x_1-x_2) \tilde v_\varepsilon(\omega_0,-x_1)
\,,
$$
where $e^{i\phi}$ is a phase factor containing all the Peierls factors, $\omega_i$ are translates of $\omega$ and $x_{s+1}=0$. By taking the trace norm, factoring out the matrix norm of $\tilde v_\varepsilon(\omega_0,-x_1)$ which is bounded by $1$, and by applying H\"older's inequality \refeq{Holder}, we obtain
$$
\tr \big ( (\tilde a^s \tilde v_\varepsilon)(\omega,0)\big ) 
\; \leq \;
\sum_{y_1,\ldots,y_s \in \ZM^{d-1}} \;\prod_{i=1}^s\; \|\tilde a(\omega_i, y_i)\|_{(s)}
\;,
$$
where $y_i = x_i - x_{i+1}$ for $i=1,\ldots,s$. Finally, by taking the integrals w.r.t. $\omega$ and applying H\"older's inequality once again,
$$
\int_\Omega \PM(d \omega) \, \tr \big ( (\tilde a^s \tilde v_\varepsilon)(\omega,0)\big ) 
\; \leq \;
\sum_{y_1,\ldots,y_s \in \ZM^{d-1}} \;\prod_{i=1}^s \left [\int_\Omega \PM(d \omega) 
\; \|\tilde a(\omega_i, y_i)\|_{(s)}^s \right ]^\frac{1}{s}\; ,
$$
Since $\PM$ is invariant against translations, the above inequality can be cast in the form presented in the statement.
\hfill $\Box$

\vspace{.2cm}

The first class of non-commutative Sobolev spaces corresponding to the algebra $\Ee_d$ of the boundary observables can be defined in the same way as ${\mathcal W}_{s,k}(\Ee_d,\widetilde \Tt)$, using the obvious substitutions. Given the isomorphism $\Ee_d \simeq \Kk \otimes \Aa_{d-1}$, the second class of Sobolev spaces ${\mathcal W}'_{s,k}(\Ee_d,\mathbb P)$ can also be canonically defined for the boundary algebra, as the closure of the algebra of non-commutative polynomials from $\Ee_d$ under the norm 
$$
\|\tilde p\|'_{s,k} 
\;=\; 
\sum_{x \in \mathbb Z^{d-1}} (1+|x|)^k \left [ \int_\Omega \mathbb P(d \omega)\; \|\tilde p(\omega,x)\|_{(s)}^s \right ]^\frac{1}{s}
\;, 
\qquad 
s \in [1,\infty]\;, \;\; k\in \mathbb N
\;. 
$$
Note that the quantity appearing between the  square brackets is just the $L^s$-norm for the trace $\int_\Omega \PM(d\omega) {\rm tr}(\cdot)$. The following properties of these spaces are proved as in Proposition~\ref{SobolevBulk}.

\begin{proposition}\label{SobolevBoundary} The two classes of Sobolev spaces satisfy the following relations:

\begin{enumerate}[\rm (i)]

\item $\mathcal W'_{s,k}(\Ee_d,\mathbb P)$ is invariant to the $\ast$-operation.

\item $\mathcal W'_{s,k}(\Ee_d,\mathbb P) \subset \mathcal W'_{s,k'}(\Ee,\mathbb P)$, whenever $k \leq k'$.

\item $\mathcal W'_{s,k}(\Ee_d,\mathbb P)\subset\mathcal W_{s,k}(\Ee_d,\widetilde \Tt)$ for $s$ integer.

\end{enumerate}

\end{proposition}


\subsection{Derivatives with respect to the magnetic field}

One further element of analysis will be used below, namely the derivatives w.r.t. the components of the magnetic field. These so-called Ito-derivatives (due to similarities with SPDE's) were introduced by Rammal and Bellissard \cite{RB} and further developed and used in \cite{Bel3,ST,LP}. Recall that the magnetic field $\BB=(B_{i,j})_{1\leq i,j\leq d}$ is an antisymmetric real matrix with entries in $[0,2\pi)$, hence a point in the torus of dimension $\frac{d(d-1)}{2}$ which will be denoted by $\Xi$. As this section is about the dependence on $\BB$, the algebras will carry a supplementary index $\Aa_{\BB,d}$. Togehter they form a continuous field of C$^*$-algebras which will be denoted by $\Ff_d=(\Aa_{\BB,d})_{\BB\in\Xi}$. A dense set inside this algebra are the non-commutative polynomials which now carry a supplmentary index $\BB$. In the symmetric presentation similar to \eqref{eq-psym}, they are given by
\begin{equation}
\label{eq-psym2}
p(\BB,\omega)
\;=\; 
\sum_{x\in\ZM^d} p_\sym (\BB,\omega,x)\, u_\sym^x\,,
\end{equation}
with complex coefficients $p_\sym (\BB,\omega,x)\in\CM$. The C$^*$-norm on $\Ff_d$ is then given by $\|p\|=\sup_{\BB\in\Xi}\|p(\BB)\|$ where $p(\BB)=p(\BB,\,.\,)\in\Aa_{\BB,d}$. Let now C$^1_K(\Ff_d)$ be the dense set of non-commutative polynomials having coefficients $p_\sym (\BB,\omega,x)$ which depend in a differentiable manner on $\BB$. For $p\in$C$^1_K(\Ff_d)$, the $\tfrac{1}{2}d(d-1)$ Ito-derivatives are introduced by
\begin{equation}\label{ItoDer}
\boxed{
\;
(\delta_{i,j} p)(\BB,\omega)
\;=\;
\sum_{x\in\ZM^d} \big ( \partial_{B_{i,j}}\,p_\sym(\BB,\omega,x)\big ) \; u_\sym^x\; , 
\qquad 1\leq i <j \leq d
\;.
\;}
\end{equation}
A norm on C$^1_K(\Ff_d)$ can then be defined by
$$
\| p\|_{C^1(\Ff_d)} 
\;=\; 
\|p\|\;+\; \sum_{j=1}^d \|\partial_j p\|
\;+\; \sum_{i,j=1}^d \|\delta_{i,j}p \|
\;.
$$
The closure of C$^1_K(\Ff_d)$ w.r.t. this norm is a Banach space denoted by C$^1(\Ff_d)$. Let us collect the most important facts about the Ito-derivatives.

\begin{proposition}\label{prop-deltajs}
The following identities hold:

\begin{enumerate}[\rm (i)] 

\item Let $p\in C^1(\Ff_d)$. Then $\Tt(p)$ is a differential function of $\BB$ and
$$
\partial_{B_{i,j}}\Tt \big (p \big )
\; =\;\Tt \big ( \delta_{i,j} p \big ) \; .
$$

\item Let $p\in C^1(\Ff_d)$. Then
$$
\delta_{i,j}(p^*) \;=\;(\delta_{i,j} p)^*\;, \qquad
\delta_{i,j}\partial_k p \;=\;\partial_k\delta_{i,j} p \; .
$$

\item If $p=p^*$ and $f\in C^2(\RM)$, then $f(p)\in C^1(\Ff_d)$.

\item Let $p,q \in C^1(\Ff_d)$. Then $pq \in C^1(\Ff_d)$ and
$$
\delta_{i,j}(pq)
\;=\;
(\delta_{i,j}p)q\;+\;p(\delta_{i,j}q)
\;-\;
\tfrac{\I}{2} \bigl(\partial_{i}p\,\partial_{j}q\,-\,\partial_{j}p\,\partial_{i}q\bigr)
\;.
$$

\item The Ito-derivative obeys the Leibniz rule under the trace, 
$$
\Tt \big ( \delta_{i,j}(pq) \big )
\;=\;
\Tt \big ( (\delta_{i,j}p)q\;+\;p(\delta_{i,j}q) \big ) \;.
$$

\item If $p$ is invertible from $C^1(\Ff_d)$, then $p^{-1} \in C^1(\Ff_d)$ and
$$
\delta_{i,j} (p^{-1})
\;=\;
p^{-1}
\left(
-\delta_{i,j} p
\,-\frac{\I}{2}\partial_{i}p\, p^{-1}\partial_{j}p
+\frac{\I}{2}\, \partial_{j}p\, p^{-1}\partial_{i}p
\right)
p^{-1}
\;.
$$

\item If $e$ is a projection from $C^1(\Ff_d)$, then
\begin{align*}
& e(\delta_{i,j} e) e \;=\; \tfrac{\I}{2} e[\partial_i e , \partial_j e ] \;=\;\tfrac{\I}{2} [\partial_i e , \partial_j e ]e\;,
\\
& (1-e)(\delta_{i,j} e) (1-e)\;=\; -\, \tfrac{\I}{2} (1-e)[\partial_i e , \partial_j e ] \;=\;-\,\tfrac{\I}{2} [\partial_i e , \partial_j e ](1-e) \; .
\end{align*}

\end{enumerate}

\end{proposition}

\noindent {\bf Proof.} The identity (i) follows from the definition \eqref{ItoDer} and that of the trace. (ii) is a direct consequence of the definition \eqref{ItoDer}. (iii) can be checked via Laplace  transform and a generalized DuHamel formula,  (iv) is obtained by taking the Ito derivative on \eqref{eq-psymprop}. (v) follows from (iv) by observing that, by using the cyclic property of the trace, the third term of (iv) can be written as a total derivation. (vi) is obtained by observing that $\delta_{i,j} (p p^{-1}) =0$ and using (iv). As for (vii) see \cite{ST} for details.
\hfill $\Box$

\section{The exact sequence of periodically driven systems}
\label{sec-PerDriven}

As already stressed in the preface, the bulk-boundary correspondence is just one instance where exact sequences of C$^*$-algebras are useful. As a second example, let us sketch in this section how to associate an exact sequence to periodically driven systems. In such a system, the Hamiltonian depends continuously and periodically on a time parameter $t$. We choose $t$ to vary in the interval $[0,2\pi)$. Hence is given a path $t\in[0,2\pi)\mapsto h_t=h_t^*\in M_N(\mathbb C)\otimes \Aa_d$ with $h_0=h_{2\pi}$. Each matrix element of this path defines an element in the C$^*$-algebra $C(\SM^1,\Aa_d)$. If furthermore is given a loop $t\in[0,2\pi)\mapsto \mu_t\in\RM$ such that $\mu_t$ lies in a gap of $h_t$, then there are also associated the (instantaneous) adiabatic projections $p_{A,t}=\chi(h_t\leq \mu_t)\in M_N(\mathbb C)\otimes \Aa_d$. Again $p_A=\{p_{A,t}\}_{t\in\SM^1}$ is an element in $M_N(\mathbb C)\otimes C(\SM^1,\Aa_d)$ which is actually a projection. As will be shown in Section~\ref{sec-polar} the orbital polarization is expressed in terms of this projection and is of topological nature. This topology can also be read off certain unitary elements in $M_N(\mathbb C)\otimes \Aa_d$ (see the stroboscopic interpretation in Section~\ref{sec-polar}). To make this connection, the following exact sequence will be used: 
\begin{equation}
\label{eq-PerDrivenSeq}
\begin{diagram}
0 &\rTo & S\Aa_d   &\rTo{i}  & C(\SM^1,\Aa_d)  &\rTo{ \mathrm{ev}}  &\mathcal A_d &\rTo &0
\end{diagram}
\end{equation}
Here $S\Aa_d=C_0((0,2\pi),\Aa_d)$ is the so-called suspension of $\Aa_d$ which is embedded as a subalgebra in $C(\SM^1,\Aa_d)$, and $\mbox{\rm ev}$ is the evaluation at $0\cong 2\pi$. A follow-up on how this sequence is used is given in Sections~\ref{sec-adiabatic} and \ref{sec-polar}. The natural extension of the analysis tools to $C(\SM^1,\Aa_d)$ is described in Section~\ref{Sec-Suspension}.

\chapter{$K$-theory for topological solid state systems}
\label{Chap-KTheory}

\abstract{The first part of the chapter reviews the $K$-theory of unital and non-unital C$^\ast$-algebras, particularly, the $K$-groups and their standard characterization, the six-term exact sequences and their connecting maps as well as the suspensions and Bott periodicity. In the second part, the analysis is specialized to the observable algebras defined in Chapter~\ref{Chap-Observables}. Using the Pimsner-Voiculescu sequence, this allows to present the generators of the $K$-groups in detail.  In the third part, various connecting maps for solid state systems are computed explicitly. }

\section{Review of key elements of $K$-theory}
\label{sec-Ktheory}

This section, which is intended for non-specialists, collects in a highly condensed form, {\it e.g.} from \cite{RLL,WO,GVF,Bla0}, the essential facts from $K$-theory of operator algebras needed in the sequel. To give a head-start, recall from Chapter~\ref{Chap-Physics} that the topologies of the solid state systems from the complex classes are encoded in the Fermi projection or Fermi unitary operator. The complex $K$-groups are of central importance because they deal precisely with the projections and the unitary elements of an algebra, for which they provide a classification by stable homotopy. As elaborated on many occasions,  {\it e.g.} in \cite{Kit,SCR,Pro3}, this stable homotopy criterion is exactly the one sought when classifying the topological condensed matter phases. The $K$-groups not only allow to distinguish the topological phases but also to identify the generators of the entire sequences of topological phases. Recall also from Chapter~\ref{Chap-Physics} that one of the main conjectures is that the topology of the solid state systems can be recovered from the boundary physics. $K$-theory will enable us to identify a projection and a unitary operator from the algebra of boundary observables, which encode the same topological information as the Fermi projection and Fermi unitary operator. This is accomplished with the so-called six-term exact sequence of $K$-theory associated to \eqref{ExactSequence}.  

\subsection{Definition and characterization of $K_0$ group}
\label{sec-K0props}

Let $\Aa$ be a C$^\ast$-algebra with or without the unit. The following definition and characterization of $K_0(\Aa)$ is borrowed from \cite{RLL}. Recall that a projection is an element of the algebra which obeys $e^2=e$ and $e^* = e$. Let $\mathcal P_N(\mathcal A)$ denote the set of projections from $M_N(\mathbb C) \otimes \mathcal A$ and consider the infinite union 
\begin{equation}
\mathcal P_\infty (\mathcal A) 
\;=\; 
\cup_{N=1}^\infty \mathcal P_N(\mathcal A)
\;,
\end{equation}
where $\mathcal P_N(\mathcal A)$'s are considered pairwise disjoint. On $\mathcal P_\infty (\mathcal A)$, one introduces the addition operation
$$
e \oplus e'  
\;=\; 
\begin{pmatrix} e\; & 0 \\ 0 & \;e' \end{pmatrix}
\;=\; 
\mathrm{diag}(e,e')
\;,
$$
so that $e \oplus e' \in \mathcal P_{N+M}(\mathcal A)$ when $e \in \mathcal P_N(\mathcal A)$ and $e' \in \mathcal P_M(\mathcal A)$. The following defines an equivalence relation on $\mathcal P_\infty (\mathcal A)$, which is compatible with the addition,
\begin{equation}\label{Equivalence1}
\mathcal P_N(\mathcal A) \;\ni\; e \;\sim_0 \;e' \;\in \;\mathcal P_M(\mathcal A) 
\quad \Longleftrightarrow \quad 
\left \{
\begin{array}{l}
e \,=\, vv^* \\
 e'\,=\,v^*v
 \end{array}
 \right.
\end{equation}
for some $v$ from $M_{N\times M}(\mathbb C) \otimes \mathcal A$. This is a slight extension of the Murray-von Neumann equivalence relation because the projections can belong to different matrix algebras. Let $[ \,.\, ]_0$ denote the equivalence classes corresponding to \eqref{Equivalence1}. Then $(\mathcal P_\infty (\mathcal A)/\sim_0,+)$ with
$$
[e]_0 \;+\; [e']_0 
\;=\; 
[e \oplus e']_0
\;=\;
[e' \oplus e]_0
$$
becomes an Abelian semi-group.  

\begin{definition} The group $K_{00}(\Aa)$ is defined as the enveloping abelian Grothen\-dieck group of $\mathcal P_\infty (\mathcal A)/\sim_0$.
\end{definition}

For C$^\ast$-algebras with a unit, the $K_{00}$-group coincides with the actual $K_0$-group, but this is not the case for algebras without a unit. For example, the algebra $\Ee_d$ of the boundary observables does not have a unit, hence we need the definition of $K_0$-group for non-unital algebras. The construction is, however, useful even for the characterization of the $K_0$-groups of unital algebras. It starts from the observation that any C$^\ast$-algebra, unital or not, accepts a unique extension $\Aa^+$ such that the following long diagram is split-exact
\begin{equation}
\label{eq-s411}
\begin{diagram}
0 & \rTo  & \Aa & \rTo{i} &\Aa^+  & \pile{\rTo{\pi} \\ \lTo_{\lambda}}  & \mathbb C & \rTo & 0
\;.
\end{diagram}
\end{equation}
The algebra $\Aa^+$ can be presented as 
$$
\{(a,t)\ : \ a\in\Aa, \ t\in\mathbb C\}
\;,
$$
with $i(a) = (a,0)$, $\pi(a,t)=t$ and splitting $\lambda(t) = (0,t)$, and with the multiplication rule and adjunction:
$$
(a,t)(a',t')\; =\; (aa'+t'a+ta',tt')
\;,
\qquad
(a,t)^*\;=\;(a^*,\overline{t})
\;.$$
The algebra $\Aa^+$ can be endowed with a C$^\ast$-norm and $\Aa^+$ has a unit 
\begin{equation}
\label{eq-one+}
\one^+\, =\, (0,1)
\;.
\end{equation}
Hence, this is a canonical way to adjoin a unit to a non-unital C$^\ast$-algebra. In terms of this unit, $\Aa^+$ can be represented as 
$$
\Aa^+
\;=\;
\{a+t\,\one^+ \ : \ a\in \Aa, \ t\in \mathbb C\}
\;,
$$
where we dropped the inclusion and wrote $i(a)$ as $a$. If $\varphi : \Aa \rightarrow \Bb$ is a C$^\ast$-algebra homomorphism, then $\varphi$ extends canonically to a homomorphism $\varphi^+ : \Aa^+ \rightarrow \Bb^+$, by declaring 
$$
\varphi^+(a + t \, \one^+) = \varphi(a) + t \, \one^+.
$$
If $\Aa$ has a unit $\one$, then $\one$ is a projection in $\Aa^+$ and each element can be represented uniquely as $a+t(\one^+ - \one)$. In turn, this gives an C$^\ast$-algebra isomorphism $\psi:\Aa^+\to\Aa \oplus \mathbb C$ by $\psi(a,t)=(a+t\,\one,t)$. This isomorphism does not exist if $\Aa$ is not unital. 

\vspace{.2cm}

Throughout, we will use a uniform definition and characterization of the $K_0$ group, regardless of lack or presence of a unit element. Note that any homomorphism $\varphi:\Aa \rightarrow \Bb$ between C$^\ast$-algebras induces a homomorphism between the $K_{00}$-groups:
$$
K_{00}(\Aa) \;\ni\; [e]_0 \;\;\mapsto\;\;\varphi_*[e]_0 
\;=\; 
[\varphi(e)]_0 \;\in\; K_{00}(\Bb)
\;.
$$
In particular, $\pi$ in \eqref{eq-s411} induces a map from $K_{00}(\Aa^+)$ to $K_{00}(\mathbb C)=\mathbb Z$.

\begin{definition}\label{K0Group} 
The $K_0$ group of the C$^\ast$-algebra $\Aa$ is defined as
$$
K_0(\Aa) \;=\;
 \mathrm{Ker}\{\pi_\ast : K_{00}(\Aa^+) \rightarrow K_{00}(\mathbb C)\}
 \;.
 $$
\end{definition} 

If $\Aa$ is unital, then $K_0(\Aa) \simeq K_{00}(\Aa)$ and, apparently, in this case we do not really need $\Aa^+$ but, as we shall see below, the characterization of the $K_0$-group is greatly simplified if we follow Definition~\ref{K0Group}. The use of $\Aa^+$ also comes handy when morphisms between unital and non-unital algebras are considered. Below, we provide the standard picture of the $K_0$ group. Throughout the book, the unit elements of the matrix algebras $M_N(\mathbb C) \otimes \Aa^+$ will be denoted by $\one_N$. They should not be confused with the unit elements of $M_N(\mathbb C) \otimes \Aa$ when $\Aa$ has a unit.

\begin{proposition}[Standard picture of the $K_0$ group.]
\label{prop-standardK0}

\begin{enumerate}[\rm (i)]

\item The group can be presented as:
\begin{align*}
K_0(\Aa)\;&=\;\big \{[e]_0 - [s(e)]_0 \ : \ e \in \mathcal P_\infty(\Aa^+) \big \} \\
&=\;\big \{[e]_0 - [s(e)]_0 \ : \ e \in \mathcal P_N(\Aa^+), \ N\in \mathbb N \big \}
\;,
\end{align*}
where $s = \lambda \circ \pi$ is the morphism which identifies the scalar part of elements.

\item The addition is given by
$$
\big ([e]_0 - [s(e)]_0 \big ) + \big ([e']_0 - [s(e')]_0 \big) 
\;=\;
[e\oplus e']_0 - [s(e) \oplus s(e')]_0
\;.
$$

\item If $e$ and $e'$ belong to same $M_N(\CM)\otimes \Aa^+$ and $ee'=0$, then 
$$
[e\oplus e']_0 \;=\; [e+e']_0, \quad [s(e) \oplus s(e')]_0 \;=\; [s(e+e')]_0.
$$

\item The inverse of an element $e \in \mathcal P_N(\Aa^+)$ is
$$
-\big ([e]_0 - [s(e)]_0 \big )\; =\; [\one_N - e]_0 - [\one_N-s(e)]_0\;.
$$

\item The zero element in $K_0(\Aa)$ can be characterized as 
$$
[e]_0 - [s(e)]_0 
\;=\; 
0 \;\; \Longleftrightarrow \;\;  e \oplus \one_M \sim_0 s(e) \oplus \one_M
$$ 
for some finite integer $M$. Consequently
$$
[e]_0 - [s(e)]_0
\;=\;
[e']_0 - [s(e')]_0 
\;\;\Longleftrightarrow \;\;
e \oplus \one_M \sim_0 e' \oplus \one_{M'}
$$
for some finite integers $M$ and $M'$.

\item Any homomorphism $\varphi : \Aa \rightarrow \Bb$ between C$^\ast$-algebras, unital or not, induces a group homomorphism
$$
\varphi_* : K_0(\Aa) \rightarrow K_0(\Bb)\;, 
\qquad 
\varphi_*([e]_0 - [s(e)]_0) 
\;=\; [\varphi^+(e)]_0 - [s(\varphi^+(e))]_0
\;.
$$ 
If $\varphi^1$ and $\varphi^2$ are homotopic, then the group morphisms $\varphi^1_\ast$ and $\varphi^2_\ast$ coincide.

\item If $e$ is a projection from $\Pp_\infty(\Aa)$, then $e$ is also a projection from $\Pp_\infty(\Aa^+)$ and furthermore $s(e)=0$. As such, $[e]_0$ just by itself is an element of $K_0(\Aa)$.

\end{enumerate}

\end{proposition}

\noindent {\bf Proof.} See \cite{RLL}.
\hfill $\Box$

\subsection{Definition and characterization of $K_1$ group}

The following definition and characterization of $K_1(\mathcal A)$ group is also borrowed from \cite{RLL}. Let $\Aa$ be a C$^\ast$-algebra, unital or not. Let $\mathcal U_N (\Aa^+)$ denote the group of unitary elements of $M_N(\mathbb C) \otimes \Aa^+$ and let
\begin{equation}
\mathcal U_\infty (\Aa^+) 
\;=\; 
\cup_{N=1}^\infty \mathcal U_N(\Aa^+)
\;,
\end{equation}
where $\mathcal U_N(\mathcal A^+)$'s are again considered pairwise disjoint. Define the following binary operation on $\mathcal U_\infty (\Aa^+)$:
$$
u \oplus v 
\;=\; 
\mathrm{diag}(u,v)
\;,
$$
so that $u\oplus v \in \mathcal U_{N+M}(\Aa^+)$ when $u \in \mathcal U_N(\Aa^+)$ and $v \in \mathcal U_M(\Aa^+)$.
Define the equivalence relation, which is compatible with the binary operation,
$$ 
\mathcal U_N(\Aa^+) \;\ni\; u \sim_1 v \in \mathcal U_M(\Aa^+) 
\;\;\Longleftrightarrow \;\;
u\oplus \one_{K-N} \sim_h v\oplus \one_{K-M}
$$
for some $K\geq \max(N,M)$. Here, $\sim_h$ denotes the homotopy equivalence inside $\mathcal U_K(\Aa^+)$, namely, $u \sim_h v$ if $u$ can be continuously deformed into $v$, with respect to the topology of $M_K(\mathbb C) \otimes \Aa^+$, without ever leaving the unitary group $\mathcal U_K(\Aa^+)$. 

\begin{definition} For any C$^\ast$-algebra $\Aa$, the $K_1$-group is defined as
$$
K_1(\Aa)\;=\; 
\mathcal U_\infty(\Aa^+)/\sim_1
$$
equipped with the commutative addition:
$$
[u]_1 + [v]_1 \;=\; [u\oplus v]_1
\;,
$$
where $[ \,.\, ]_1$ denotes the equivalence classes w.r.t. $\sim_1$.
\end{definition}

\begin{proposition}[Standard picture of the $K_1$ group] 
\label{prop-standardK1}
\begin{enumerate}[\rm (i)]

\item The group can be presented as:
\begin{align*}
K_1(\Aa) & = \{[u]_1 \ : \ u \in \mathcal U_\infty(\Aa^+)\} \\
& = \{[u]_1 \ : \ u \in \mathcal U_N(\Aa^+), \ N \in \mathbb N\}
\;.
\end{align*}

\item In general, $u\oplus \one_N \sim_1 u$. In particular, the units are all equivalent: $\one_N \sim_1 \one_M$ for all natural numbers $N$ and $M$. Their common equivalence class give the zero element:
$$
[\one]_1\; =\; 0\;.
$$

\item If $u,v \in \mathcal U_N(\Aa^+)$ and $u \sim_h v$, then $[u]_1 = [v]_1$. 

\item If $u,v \in \mathcal U_N(\Aa^+)$, then
$$
[uv]_1 
\;=\; 
[vu]_1 
\;=\; 
[u]_1 + [v]_1
\;.
$$
According to the second point, we can always place the representatives $u$ and $v$ in a common $\mathcal U_K(\Aa^+)$. As such, $K_1(\Aa)$ can be also presented as a multiplicative group.

\item If $u \in \mathcal U_N(\Aa^+)$, then the inverse of $[u]_1$ is given by
$$ 
- \,[u]_1 \;=\; [u^{-1}]_1 \; = \;  [u^\ast].
$$

\item If $\Aa$ is a unital algebra and $u \in \Aa$ is a unitary element, then $u$ can be promoted to a unitary in $\Aa^+$ by
$$
u^+\;=\; u \;+\;(\one^+ - \one).
$$
This is a group homomorphism which can be automatically extended to a homomorphism between $\mathcal U_\infty(\Aa)$ and $\mathcal U_\infty(\Aa^+)$. Then the equality $\mathcal U_\infty(\Aa^+)/\sim_1 = \mathcal U_\infty(\Aa)/\sim_1$ follow, or, put it differently, $K_1(\Aa) = K_1(\Aa^+)$. 

\item Any C$^\ast$-algebra  homomorphism $\varphi : \Aa \rightarrow \Bb$ between C$^\ast$ algebras, unital or not, induces a group homomorphism
\begin{equation}
\varphi_\ast : K_1(\Aa) \rightarrow K_1(\Bb)\;, 
\qquad 
\varphi_\ast [v]_1 = [\varphi^+(v)]_1
\;.
\end{equation}
If two homomorphisms $\varphi^1$ and $\varphi^2$ are homotopic, then the induced maps $\varphi^1_\ast$ and $\varphi^2_\ast$ coincide.

\end{enumerate}

\end{proposition}

\noindent {\bf Proof.} See \cite{RLL}.
\hfill $\Box$


\subsection{The six-term exact sequence}\label{SixTermSequence}

The central result of $K$-theory states that the $K$-groups of three C$^\ast$-algebras (unital or not) in a short exact sequence 
\begin{equation}
\label{FirstDiagram}
\begin{diagram}
0 & \rTo  & \Ee & \rTo{i} &\widehat \Aa  & \rTo{\mathrm{ev}}  & \Aa & \rTo & 0
\end{diagram}
\end{equation}
form a six-term exact sequence of Abelian groups
\begin{equation}
\label{6TermDiagram}
\begin{diagram}
K_0(\Ee) &\rTo^{\ \ i_\ast \ \ } & K_0(\widehat \Aa) & \rTo{\ \ \mathrm{ev}_\ast \ \ } & K_0(\Aa) \\
\uTo{\mathrm{Ind}}                 &               \     \                        &     \     \                                 &       \     \                   &\dTo{\mathrm{Exp}} \\
K_1(\Aa)        & \lTo{\ \ \mathrm{ev}_\ast \ \ } & K_1(\widehat \Aa) & \lTo{\ \ i_\ast \ \ }    & K_1(\Ee)
\end{diagram}
\end{equation}
As already discussed in our introductory remarks, this diagram is the key to the bulk-boundary correspondence. Here we define the connecting maps in \eqref{6TermDiagram}, called the index and the exponential maps. A proof that these definitions indeed lead to an exact sequence can be found in \cite{Bla0,WO,RLL}. The physical implications and applications are presented in the following sections.

\vspace{.2cm}

The index map  is defined as follows (see \cite{RLL}, pp. 153). First, note that the evaluation map ${\rm ev}$ in \eqref{FirstDiagram} is surjective, so  ${\rm id}\otimes \, \mathrm{ev}^+: M_N(\mathbb C) \otimes \widehat \Aa^+ \rightarrow M_N(\mathbb C) \otimes \Aa^+$ is also surjective for any $N\in \mathbb N$. In such circumstances, there is the elementary but profound observation that
$$
({\rm id} \otimes \mathrm{ev}^+)\big(\mathcal U_{N}(\widehat \Aa^+)_0\big ) 
\;=\;
\mathcal U_{N}(\Aa^+)_0\;, 
\qquad \forall \ N\in \mathbb N
\;,
$$
where $( \ )_0$ denotes the connected component of the unity. This tells that any element from the connect component of the unity of $\mathcal U_{N}(\Aa^+)$ has a preimage in the connected component of the unity of $\mathcal U_{N}(\widehat \Aa^+)$. If $v$ is a unitary from $\mathcal U_{N}(\Aa^+)$, then $\mathrm{diag}(v,v^\ast)$ belongs to $\mathcal U_{2N}(\Aa^+)_0$, hence there exists a unitary $\hat w \in \mathcal U_{2N}(\widehat \Aa^+)_0$ such that 
$$
({\rm id} \otimes \mathrm{ev}^+)(\hat w)\; =\;\mathrm{diag}(v,v^\ast)
\;.
$$
The element $\hat w$ so defined is called a lift and the standard notation is
$$
\hat w
\;=\; 
\mathrm{Lift}\big (\mathrm{diag}(v,v^\ast)\big )
\;.
$$
The lift is unique up to homotopies. Next, one considers the projection
\begin{equation}\label{IndProj}
\hat w \; \mathrm{diag}(\one_N, 0_N) \; \hat w^\ast 
\;\in\; 
\mathcal P_{2N}(\widehat \Aa^+)
\;,
\end{equation}
whose homotopy class is  entirely determined by $v$ and, moreover,
$$
s (\hat w \, \mathrm{diag}(\one_N, 0_N) \, \hat w^\ast \big)\;\sim_h \;\mathrm{diag}(\one_N,0_N)\;.
$$ 
Since 
$$
({\rm id}\otimes {\rm ev}^+)\big(\hat w \, \mathrm{diag}(\one_N, 0_N) \, \hat w^\ast \big )
\;=\;
{\rm diag}(\one_N,0_N)
\;,
$$ 
and the short sequence \eqref{FirstDiagram} is exact, the projector \eqref{IndProj} is in the image of $\Ee^+$ in $\widehat \Aa^+$ under $i^+$. Throughout, we will identify $i^+(\Ee^+)$ and $\Ee^+$. Then the index map is defined as
\begin{equation}\label{IndexMapDefinition}
\boxed{
\;
\mathrm{Ind}\big ([v]_1 \big ) \; =\; [\hat w \, \mathrm{diag}(\one_N, 0_N) \, \hat w^\ast]_0 \;-\;  [\mathrm{diag}(\one_N, 0_N)]_0\; \in\; K_0(\Ee)
\;.
\;
}
\end{equation} 

The lift $\hat{w}$ is not constructively defined above, but, as we shall see in Section~\ref{sec-expBBC}, for the Fermi unitary operator, the lift can be generated using the functional calculus with the Hamiltonian, through an explicit procedure which also has a certain physical interpretation. Note that the projection \eqref{IndProj} belongs to $[0]_0$ class of $K_0(\widehat \Aa\;)$ because $\hat w$ can be continuously deformed to $\one_{2N}$. But as an element of $K_0(\Ee\;)$, this is not the case because, in general, $\hat w$ can not be continuously deformed to $\one_{2N}$ and keep the projection \eqref{IndProj} inside $i^+(\Ee^+)$. If $v=\mathrm{ev}(\hat v)$, however, then
$$
\mathrm{Lift}\big(\mathrm{diag}(v,v^\ast) \big )\; = \;\mathrm{diag}(\hat v,\hat v^\ast)\; ,
$$
in which case the projector \eqref{IndProj} is just ${\rm diag}(\one_N,0_N)$, hence its class in $K_0(\Ee)$ is trivial. This shows that the sequence \ref{6TermDiagram} is exact at $K_1(\Aa)$. Lastly, note that, for  $v \in \mathcal U_{N}(\Aa^+)$ and $v' \in \mathcal U_{M}(\Aa^+)$,
$$
\mathrm{Lift}\big(\mathrm{diag}(v\oplus v' ,(v\oplus v')^\ast) \big ) \;\sim_1 \;\mathrm{Lift}\big(\mathrm{diag}(v,v^\ast) \big ) \oplus \mathrm{Lift}\big(\mathrm{diag}(v',v'^\ast) \big )\;,
$$
which in turn gives
$$
\Ind ([v \oplus v']_1) = \Ind([v]_1) + \Ind([v']_1)
\;. 
$$
In other words, the index map is indeed a group homomorphism.

\vspace{.2cm}

The exponential map is defined as follows (see \cite{RLL}, pp. 209). Consider an element from $K_0(\Aa)$ which, according to the standard characterization, can always be represented as $[e]_0 -[s(e)]_0$ with $e \in \mathcal P_N(\Aa^+)$ for some $N$. Since the evaluation map is surjective we can always find a lift for $e$
$$
\hat g \;=\; \mathrm{Lift}(e) \; \in\; M_N(\widehat \Aa^+)\;,
 \qquad 
 ({\rm id} \otimes \mathrm{ev}^+)(\hat g) \;=\; e
 \;.
$$ 
The lift is unique up to homotopies and it can always be chosen self-adjoint, in which case one can define the unitary element
\begin{equation}\label{UE}
\exp(2\pi \I \hat g) \;\in\; \Uu_N(\widehat \Aa^+)
\;,
\end{equation}
such that
$$
({\rm id}\otimes {\rm ev}^+ )\big ( \exp(2\pi \I \hat g) \big ) \;=\; \one_N\;.
$$
Since the sequence \eqref{FirstDiagram} is exact, the unitary element \eqref{UE} has a pre-image in $\Uu_N(\Ee^+)$. Following the same convention as for the index map, {\it i.e.} $\Ee^+ = i^+(\Ee^+)$, the exponential map is defined as
\begin{equation}\label{ExpMapDefinition}
\boxed{
\;\mathrm{Exp}\big ([e]_0 - [s(e)]_0 \big ) \;=\;-\, [\exp(2\pi \I \hat g)]_1 \;=\; [ \exp(-2\pi \I \hat g)]_1 \;\in \;K_1(\Ee)\;.\;
}
\end{equation} 

As for the index map, we shall see that the lift $\hat g$ corresponding to the Fermi projection can be constructed explicitly using the functional calculus with the Hamiltonian. Note also that, if $e \in \mathcal P_N(\Aa^+)$ and $e' \in \mathcal P_M(\Aa^+)$, then
$$
\mathrm{Lift}(e \oplus e' ) 
\;=\; 
\mathrm{Lift}(e) \oplus \mathrm{Lift}(e')\; =\;\mathrm{diag}(\hat g, \hat g')
\;,
$$ 
hence the exponential map is indeed a group homomorphism. Lastly, if $e = \mathrm{ev}( \hat e)$ for some $\hat e \in \mathcal P_N(\widehat \Aa^+)$, then $\mathrm{Lift}(e)=\hat e$ and $\exp(2\pi \I \hat e)=\one_N$, hence trivial. This shows that sequence \eqref{6TermDiagram} is exact at $K_0(\Aa)$.

\subsection{Suspension and Bott periodicity}
\label{Sec-SuspensionBottPer}

The suspension $S\Aa$ and cone $C\Aa$ of a C$^*$-algebra $\Aa$ are defined as
$$
S\Aa\;=\;C_0((0,2\pi),\Aa)
\;,
\qquad
C\Aa\;=\;C_0([0,2\pi),\Aa)
\;,
$$
with the C$^*$-norm given by the supremum over the intervals of the C$^*$-norm on $\Aa$. Of course, the boundary $2\pi$ can be replaced by any other positive number or $\infty$.  The suspension can alternatively be thought of as the algebra of the loops over $\Aa$ which are pinned at one point
\begin{equation}\label{Suspension}
S\Aa \;\cong\; 
\{f\in C(\mathbb T,\Aa) \ : \ f(0)=0\}
\;.
\end{equation}
Suspension and cone are connected by an exact sequence
\begin{equation}
\label{eq-SuspensionSeq}
\begin{diagram}
0 & \rTo  & S\Aa & \rTo{i} & C\Aa  & \rTo{\mathrm{ev}}  & \Aa & \rTo & 0
\end{diagram}
\;,
\end{equation}
where $i$ is the obvious inclusion and $\mbox{\rm ev}$ the evolution at $0$. A special case of this exact sequence was already discussed in Section~\ref{sec-PerDriven}. Note that neither $S\Aa$ nor $C\Aa$ have a unit, hence their $K$-groups are necessarily defined through $(S\Aa)^+$ and $(C\Aa)^+$. Now the cone $C\Aa$ is contractible and therefore has trivial $K$-groups. Therefore the six-term exact sequence \eqref{6TermDiagram} associated to \eqref{eq-SuspensionSeq} decouples and index and exponential become isomorphisms.

\begin{theorem} 
\label{theo-Ksuspension}
$K_1(\Aa) \simeq K_0(S \Aa)$.
\end{theorem}

\begin{theorem} 
\label{theo-KBott}
$K_1(S \Aa) \simeq K_0(\Aa)$. 
\end{theorem}

As pointed out above, index and exponential map provide the isomorphism. As these isomorphisms can be made more explicit and play a role in what follows, we will write them out below. But first, let us mention that, together, the two theorems from above provide the Bott periodicity
$$
K_0(S S \Aa) 
\;\simeq \;
K_1(S \Aa) 
\;\simeq\; 
K_0(\Aa)
\;.
$$
We have already discussed, at the beginning of Chapter~\ref{Chap-Physics}, the relevance of the Bott periodicity  to the classification of the topological band insulators. In the present work, the suspensions and the above isomorphisms will be used in a different direction, namely, to give alternative representations of the topological invariants. 

\vspace{.2cm}

The $\theta$-map, which gives the isomorphism in Theorem~\ref{theo-Ksuspension}, is defined as follows. First, note that, for any $v \in \mathcal U_N(\Aa^+)$, $s(v) \in M_N(\mathbb C)$ which is a simply connected space. Hence, $s(v)$ is always a homotopy of the identity and $v \sim_1 s(v)^\ast \, v$, and the latter has the useful property that $s(s(v)^\ast \, v)=\one_N$. Therefore, any class from $K_1(\Aa)$ can be represented by $v's$ with $s(v)=\one_N$, for some integer $N$. Another general fact is that $\one_{2N}$ and $\mathrm{diag}(v,v^\ast)$ are homotopic in $\mathcal U_{2N}(\Aa^+)$. As such, there exists a continuous interpolation $w_t$ inside $\mathcal U_{2N}(\Aa^+)$ such that $w_0=\one_{2N}$ and $w_{2\pi}=\mathrm{diag}(v,v^\ast)$, which can always be normalized such that $s(w_t) =\bm \one_{2N}$, for all $t \in [0,2 \pi]$. Setting $w=\{w_t\}_{t \in [0,2 \pi]} \in M_{2N}(\CM) \otimes (C\Aa)^+$, the $\theta$-isomorphism is defined as
\begin{equation}\label{ThetaMap}
\boxed{
\; K_1(\Aa) \;\ni\; [v]_1 \;\overset{\theta}{\mapsto}\; [w \, \mathrm{diag}(\one_N,0_N) \, w^\ast]_0 
\;-\;
[{\rm diag}(\one_N,0_N)]_0 \;\in\; K_0(S\Aa)\;.\;
}
\end{equation}
The construction is even more flexible, namely it is sufficient to choose a path $w_t$ from $\one_{2N}$ to $\mathrm{diag}(v,v')$ for a given unitary $v'$ (chosen such that a path exists, which is the case for $v'=v^*$). Note that the projection
$$\{e_t\}_{t \in [0,2\pi]} \;=\; 
\{w_t \, \mathrm{diag}(\one_N,0_N) \, w_t^\ast \}_{t \in [0,2\pi]}
$$
is in the image of $M_{2N}(\CM) \otimes (S\Aa)^+$ under $i^+$, hence it belongs to $\Pp_{2N}\big((S\Aa)^+ \big )$. Also, note that the proper normalization of $w_t$ ensures that $s(e_t)={\rm diag}(\one_N,0_N)$, hence the $\theta$-map is conform with the standard picture of the $K_0$-group. A common choice \cite{RLL,Par} for $w_t$ is 
\begin{equation}
\label{TetaMapChoice}
w_t\;=\;
r_t \, \mathrm{diag}(v^\ast,\one_N) \, r_t^\ast \, \mathrm{diag}(v,\bm \one_N)
\;,
\end{equation}
with
\begin{equation}\label{RChoice}
r_t
\;=\;
\begin{pmatrix}
\cos\big(\frac{t}{4}\big) \one_N & \sin\big(\frac{t}{4}\big) \one_N \medskip \\
-\sin\big(\frac{t}{4}\big) \one_N & \cos\big(\frac{t}{4}\big) \one_N
\end{pmatrix}
\;.
\end{equation}
With this choice
\begin{equation}\label{ThetaEt}
e_t 
\;=\; 
r_t
\begin{pmatrix}
\cos^2( \tfrac{t}{4}) \one_N & \cos( \tfrac{t}{4}) \, \sin (\tfrac{t}{4}) \, v^\ast \\
\cos(\tfrac{t}{4}) \, \sin (\tfrac{t}{4}) \, v & \sin^2 (\tfrac{t}{4})\one_N
\end{pmatrix}
r_t^\ast
\;.
\end{equation}

The isomorphism  in Theorem~\ref{theo-KBott} is given by the Bott map
\begin{align}\label{BetaMap}
\boxed{
\;
K_0(\Aa) \;\ni\; [e]_0\;-\;[s(e)]_0 \;\overset{\beta}{\mapsto}\;
 [(\one_N-e)\,+\, \exp(\I t) e]_1 \;\in \;K_1(S\Aa)
 \;,
}
\end{align}
for $e \in \mathcal P_N(\Aa^+)$. Note that the unitary $(\one_N-e)+ \exp(\I t) e$ indeed belongs to $\Uu_{N}\big ((S\Aa)^+)$.

\subsection{The inverse of the suspension map}
\label{sec-inverseTheta}

Let $t\in[0,2\pi)\mapsto e_t\in M_N(\mathbb C)\otimes \Aa$ be a closed smooth loop of projections in $\Aa$ defining a projection $e$ in $M_N(\mathbb C)\otimes C(\SM^1,\Aa)$. With $e_0$ viewed as a constant loop, then the class $[e]_1-[e_0]_1\in K_0(C(\SM^1,\Aa))$ is an element in $K_0(S\Aa)$ because its image under evaluation in the split exact sequence (different from \eqref{eq-SuspensionSeq}!)
\begin{equation}
\label{eq-InvThetaSeq}
\begin{diagram}
0 &\rTo & S\Aa   &\rTo{i}  & C(\SM^1,\Aa)  &\rTo{ \mathrm{ev}}  &\mathcal A &\rTo &0
\end{diagram}
\end{equation}
vanishes. Hence $i_*^{-1}([e]_0-[e_0]_0)\in K_0(S\Aa)$ is well-defined and any element of $K_0(S\Aa)$ is of this form. The aim is to determine a preimage of $i_*^{-1}([e]_0-[e_0]_0)\in K_0(S\Aa)$ in $K_1(\Aa)$ under the suspension map $\theta$. Our answer in Theorem~\ref{theo-adiabatic0} will involve the adiabatic time evolution as introduced by Kato \cite{Kat}, and this is an alternative argument showing the surjectivity of the $\theta$-map ({\it e.g.} Section 7.2 of \cite{WO}). As we are not aware of a reference, we provide a detailed proof.

\begin{proposition}
\label{prop-AdiabaticEvol}
Let $h_t=h_t^*$ be a path of self-adjoints in $M_N(\mathbb C)\otimes \Aa$ satisfying $[h_t,e_t]=0$. Then the solution $v_t\in\Aa^+$ of the adiabatic evolution
\begin{equation}
\label{eq-AdiabaticEv0}
\I\,\partial_t v_{t}\;=\;
\big(h_t+\I[\partial_t e_{t},e_{t}]\big)v_{t}
\;,
\qquad
v_{0}=\one_N
\;,
\end{equation}
is unitary and satsifies
$$
e_{t}
\;=\;
v_{t}\,e_0\,v_{t}^*
\;.
$$
\end{proposition}

\noindent {\bf Proof.} First of all, as $h_t+\I[\partial_t e_{t},e_{t}]$ is self-adjoint the solution $v_t$ is indeed unitary. Furthermore,
\begin{align*}
\partial_t(
v_{t}^*\,e_t\,v_{t})
\;
& =\;
-v_{t}^*(\partial_tv_{t})v_{t}^*\,e_t\,v_{t}
+
v_{t}^* (\partial_t\,e_t)\,v_{t}
+
v_{t}^*\,e_t\,\partial_tv_{t}\,
\\
& =\;
v_t^*\Big(
\I(h_t+\I[\partial_te_t,e_t])e_t+\partial_te_t-e_t\I(h_t+\I[\partial_te_t,e_t])
\Big) v_t
\;,
\end{align*}
which now vanishes by hypothesis and $\partial_te_t=e_t\partial_te_t+\partial_te_te_t$. As $v_{0}^*\,e_0\,v_{0}=e_0$ the proof is completed.
\hfill $\Box$

\vspace{.2cm}

Let us point out that $h_t=0$ is a possible choice. Furthermore, $[v_{t}]_1=0$ in $K_1(\Aa)$ because it is $v_t$ is path connected to the identity. The Poincar\'e map $v_{2\pi}$ of the adiabatic evolution is in general different from the identity, but $e_{2\pi}=e_{0}$ implies
\begin{equation}
\label{eq-adprops0}
v_{2\pi}\,e_0\,v_{2\pi}^*
\;=\;
e_0
\;.
\end{equation}
Therefore the range of $e_0$ is invariant under $v_{2\pi}$ and hence $e_0 v_{2\pi} e_0+\one_N-e_0$ is a unitary in $\Aa^+$. The following result now determines the inverse of $\theta$ and shows that it has some structural similarity with the Bott map. The freedom of choice of $h_t$ in \eqref{eq-AdiabaticEv0} reflects that many $v_{2\pi}$'s define the same $K_1$-class. 

\begin{theorem}
\label{theo-adiabatic0}
Let $t\in[0,2\pi)\mapsto e_t\in M_N(\mathbb C)\otimes \Aa$ be a closed smooth loop of projections in $\Aa$ and $t\in[0,2\pi)\mapsto v_t$ an associated adiabatic evolution. Then
\begin{equation}
\label{eq-AdiabaticK}
\theta^{-1} \big(i_*^{-1}([e]_0-[e_0]_0)\big)
\;=\;
[e_0 v_{2\pi} e_0\,+\,\one_N\,-\,e_0]_1
\;.
\end{equation}
\end{theorem}

\noindent {\bf Proof.} Using the rules of Section~\ref{sec-K0props}, one finds in $K_1(C(\SM^1,\Aa))$
\begin{align*}
[e]_0-[e_0]_0
\;&=\;
[e]_0\;+\;[\one_N-e_0]_0\;-\;[\one_N]_0
\\
&
\;=\;
\left[
\begin{pmatrix}
e & 0 \\ 0 & \one_N-e_0
\end{pmatrix}
\right]_0
\;-\;
\left[
\begin{pmatrix}
e_0 & 0 \\ 0 & \one_N-e_0
\end{pmatrix}
\right]_0
\;.
\end{align*}
Note that also the r.h.s. is in the image of $i_*$ and hence represents an element of $K_0(S\Aa)$. Now let us introduce the path of unitaries
$$
s\in[0,\tfrac{\pi}{2}]\;\mapsto\;
r_s
\;=\;
\begin{pmatrix}
e_0+\cos(s)(\one_N-e_0) & -\sin(s)(\one_N-e_0) \\
\sin(s)(\one_N-e_0) & e_0+\cos(s)(\one_N-e_0) 
\end{pmatrix}
\;.
$$
It allows to write, still in a form lying in the image of $i_*$ for every $s$,
$$
[e]_0-[e_0]_0
\;=\;
\left[
r_s
\begin{pmatrix}
e & 0 \\ 0 & \one_N-e_0
\end{pmatrix}
r_s^*\right]_0
\;-\;
\left[
r_s\begin{pmatrix}
e_0 & 0 \\ 0 & \one_N-e_0
\end{pmatrix}
r_s^*
\right]_0
\;.
$$
As 
$$
r_s\begin{pmatrix}
e_0 & 0 \\ 0 & \one_N-e_0
\end{pmatrix}
r_s^*
\;=\;
\begin{pmatrix}
e_0 +\sin^2(s)(\one_N-e_0) & -\cos(s)\sin(s)(\one_N-e_0) \\ -\cos(s)\sin(s)(\one_N-e_0) & \cos^2(s)(\one_N-e_0)
\end{pmatrix}
\;,
$$
we will choose $s=\frac{\pi}{2}$. Then
$$
r_{\frac{\pi}{2}}
\begin{pmatrix}
e & 0 \\ 0 & \one_N-e_0
\end{pmatrix}
r_{\frac{\pi}{2}}^*
\;=\;
w
\begin{pmatrix}
\one_N & 0 \\ 0 & 0
\end{pmatrix}
w^*
\;,
$$
with $w=\{w_t\}_{t\in[0,2\pi)}$ given by
$$
w_t
\;=\;
r_{\frac{\pi}{2}}
\begin{pmatrix}
v_{t} & 0 \\ 0 & \one_N
\end{pmatrix}
r_{\frac{\pi}{2}}^*
\;=\;
\begin{pmatrix}
e_0v_{t}e_0+\one_N- e_0& e_0 v_{t}(\one_N-e_0)\\ (\one_N-e_0) v_{t}e_0 & (\one_N-e_0)v_{t}(\one_N-e_0)+e_0
\end{pmatrix}
\;.
$$
Thus
$$
[e]_0-[e_0]_0
\;=\;
\left[w
\begin{pmatrix}
\one_N & 0 \\ 0 & 0
\end{pmatrix}
w^*
\right]_0
\,-\,
\left[
\begin{pmatrix}
\one_N & 0 \\ 0 & 0
\end{pmatrix}
\right]_0
\;.
$$
Now $w_{2\pi}$ is diagonal due to \eqref{eq-adprops0} with upper left entry $e_0v_{2\pi}e_0+\one_N- e_0$. 
Comparing with the definition \eqref{ThetaMap} of $\theta$, the result follows.
\hfill $\Box$

\section{The $K$-groups of the algebras of physical observables}
\label{sec-KObsAlg}

The $K$-theory of the algebras of bulk, half-space and boundary observables can be determined from the six-term exact sequence
\begin{equation}\label{PSixTermDiagram}
\begin{diagram}
& K_0(\Ee_d) & \rTo{\ \ i_\ast \ \ } & K_0(\widehat{\Aa}_d) & \rTo{\ \ {\rm ev}_\ast \ \ } & K_0(\Aa_d) &\\
& \uTo{\rm Ind} & \  &  \ & \ & \dTo{\rm Exp} & \\
& K_1(\Aa_d)  & \lTo{{\rm ev}_\ast} & K_1(\widehat{\Aa}_d) & \lTo{i_\ast} & K_1(\Ee_d) &
\end{diagram}
\end{equation}
associated to \eqref{ExactSequence}. Since the algebra $\Aa_d$ of the bulk observables can be presented as an iterated crossed product by $\mathbb Z$, the computation of the $K$-groups reduces to a standard application of the Pimsner-Voiculescu machinery  \cite{PV}. Furthermore, the generators of $K_{0,1}(\Aa_d)$ groups can be explicitly identified based on the work of Elliott \cite{Ell} and \cite{Rie1} on the non-commutative torus. All these generators are presented in this section.

\subsection{The Pimsner-Voiculescu sequence and its implications}
\label{Sec-KPimsnerVoiculescu}

It was shown in Section~\ref{Sec-PimsnerVoiculescu} that the short exact sequence 
$$
\begin{diagram}
0 & \rTo  & \Ee_d & \rTo{i} & \widehat{\mathcal A}_{d}  & \rTo{\mathrm{ev}}  & \Aa_d & \rTo & 0
\end{diagram}
$$ 
between the algebras of physical observables is isomorphic to the Toeplitz extension of Pimsner and Voiculescu \cite{PV}.  Here we collect the $K$-theoretic consequence of this fact, namely that \eqref{PSixTermDiagram} is identical to the Pimsner-Voiculescu $6$-term sequence. Throughout, the identifying maps of Section~\ref{Sec-PimsnerVoiculescu} will be freely used and $K_{0,1}(\Aa_{d-1})$ will be identified with $K_{0,1}(\Aa_{d-1}\otimes \Kk)$ via the isomorphism induced by the imbedding $a \rightarrow a \otimes |1\rangle \langle 1|$. As a straightforward consequence of \eqref{RhoEe}, we have the isomorphisms
$$
\widetilde \rho_\ast : K_j(\Ee_d) \rightarrow K_j(\Aa_{d-1})\;, 
\qquad 
j=0,1\;.
$$
To go further, two additional natural maps can be defined, the inclusion $ \mathfrak{i}(a) = a \otimes \one$ of $\Aa_{d-1}$ into its Toeplitz extension $T(\Aa_{d-1})$ and the inclusion $\mathfrak{j}$ of $\Aa_{d-1}$ into $\Aa_d = \Aa_{d-1} \rtimes_{\alpha_d} \mathbb Z$. One important result of \cite{PV} is that the inclusion $\mathfrak{i}$ generates group isomorphisms:
$$
\mathfrak{i}_\ast : K_j (\Aa_{d-1}) \rightarrow K_j(T(\Aa_{d-1}))
\;, 
\qquad 
j=0,1
\;.
$$
Hence the natural inclusion
\begin{equation}
\label{eq-natincl}
\mathfrak{i}': \Aa_{d-1} \rightarrow \widehat \Aa_d, \qquad \mathfrak{i}'=\widehat \eta^{-1} \circ \mathfrak{i}
\;,
\end{equation}
generates the group isomorphisms
$$ 
\mathfrak{i}'_\ast : K_j(\Aa_{d-1}) \rightarrow K_j(\widehat \Aa_d)\;, 
\qquad j=0,1\;.
$$
Moreover, the following identity holds
$$
\psi_\ast \,= \,\mathfrak{i}'_* \circ (\one-\alpha_d^{-1})_\ast \circ \widetilde \rho_\ast \,.
$$
As such, \eqref{PSixTermDiagram} can be rewritten as:
\begin{equation}\label{GSixTermDiagram}
\begin{diagram}
& K_0(\Ee_d) & \rTo{\mathfrak{i}'_* \circ (\one-\alpha_d^{-1})_\ast \circ \tilde \rho_\ast \ } & K_0(\widehat \Aa_d) & \rTo{\ \  \rm ev_\ast \ \ } & K_0(\Aa_d) &\\
& \uTo{\rm Ind} & \  &  \ & \ & \dTo{\rm Exp} & \\
& K_1(\Aa_d)  & \lTo{\ \  \rm ev_\ast \ } & K_1(\widehat \Aa_d) & \lTo{\ \ \mathfrak{i}'_* \circ (\one-\alpha_d^{-1})_\ast \circ \tilde \rho_\ast} & K_1(\Ee_d) &
\end{diagram}
\end{equation}
Using the isomorphisms listed above, this diagram can be seen to be completely equivalent to the standard six-term exact sequence of \cite{PV}:
\begin{equation}\label{PVSixTermDiagram}
\begin{diagram}
& K_0(\Aa_{d-1}) & \rTo{(\one-\alpha_d^{-1})_\ast \ \ } & K_0(\Aa_{d-1}) & \rTo{\ \  \mathfrak{j}_\ast \ \ } & K_0(\Aa_d) &\\
& \uTo{\rm Ind} & \  &  \ & \ & \dTo{\rm Exp} & \\
& K_1(\Aa_d)  & \lTo{\ \  \mathfrak{j}_\ast \ \ } & K_1(\Aa_{d-1}) & \lTo{\ \ (\one-\alpha_d^{-1})_\ast} & K_1(\Aa_{d-1}) &
\end{diagram}
\end{equation}
One insight that came out of this diagram is that the $K$-groups of the crossed product by $\mathbb Z$ depend on the action $\alpha_d$, or better said on the homotopy class of $\alpha_d$. On the other hand, if the homotopy class of $\alpha_d$ is trivial, then the six-term diagram becomes a straightforward tool for the computations of the $K$-groups, and this is the case in the present application. For the following, it will be crucial that the disorder space $\Omega$ is contractible (in fact convex), resulting from the assumed contractibility of the local disorder spaces $\Omega_0^y$.

\begin{proposition}\label{Prop-Homotopy}
The map $\alpha_d :\Aa_{d-1} \rightarrow \Aa_{d-1}$ defined as before, $\alpha_d(p) = u_d \, p \, u_d^\ast$, $p \in \Aa_{d-1}$, is homotopic to the identity.
\end{proposition} 

\noindent {\bf Proof.}  The action of $\alpha_d$ on the generators of $\Aa_{d-1}$ is
$$\alpha_d(u_j \phi) 
\;=\; 
e^{\I B_{d,j}}\, u_j(\phi\circ \tau_d^{-1})
\;.
$$
For $t\in [0,1]$,
$$
\xi_t : \Omega \rightarrow \Omega\;, 
\qquad 
\xi_t(\omega) = t \omega + (1-t) \tau_d^{-1}(\omega)
\;,
$$
is a homotopy between $\tau_d$ and the identity, which commutes with the action of $\mathbb Z^{d-1}$ for all $t$'s. Then
$$ 
\alpha_d^t(u_j \phi) \;=\; e^{\I (1-t) B_{d,j}} u_j(\phi\circ \xi_t)
$$
defines the family of $\ast$-endomorphisms which interpolates continuously between $\alpha_d$ and the identity. By definition ({\it cf.} p.~43 in \cite{RLL}), this is the desired homotopy equivalence. \hfill $\Box$

\vspace{.2cm}

There is an important direct consequence of the above statement, which will be essential at several points of our presentation.

\begin{proposition}\label{InnerAuto} 
Let $e \in \mathcal P_N(\Aa_{d-1})$ be a projection. Then there exist the unitary elements $\overline{w}_e$ and $w_e$ from $\mathcal U_N(\Aa_{d-1})_0$, the connected component of the unity, such that $\overline{w}_e=u_d w_eu_d^*$ and 
$$
({\rm id} \otimes \alpha_d)(e)\;=\; (\one_N \otimes u_d) e (\one_N \otimes u_d^\ast) \;=\; \overline{w}_e \, e\, (\overline{w}_e)^\ast
\;,
$$ 
and for the inverse action,
$$
({\rm id} \otimes \alpha_d^{-1})(e)\;=\; (\one_N \otimes u_d^\ast) e (\one_N \otimes u_d) \;=\; w_e^\ast \, e\, w_e
\;.
$$ 

\end{proposition}

\begin{remark} One should be aware that the above statement does not imply that $\alpha_d$ is an inner automorphisms because, as the notation suggests, $w_e$ and $\overline{w}_e$ both depend on $e$.\hfill $\diamond$  \end{remark}

\noindent {\bf Proof.}  Since $\alpha_d^t$ are $\ast$-endomorphisms, $({\rm id} \otimes \alpha_d^t)(e)$ are projections for all $t \in [0,1]$. As such, there is a homotopy of projections between $({\rm id} \otimes\alpha_d)(e)$ and $e$, in which case the construction of the unitary element $\overline{w}_e$ can be accomplished by many methods, in particular by Proposition~\ref{prop-AdiabaticEvol}. Inverting the action readily leads to the second identity.  \hfill $\Box$ 

\begin{proposition}\label{PropKGroups} 
For $d\geq 1$, the $K$-groups of the observable algebras are given by
$$
K_j(\Aa_d)\;=\;
K_j(\Ee_{d+1})\;=\;K_j(\widehat{\Aa}_{d+1})\;=\;\ZM^{2^{d-1}}
\;, \qquad  
j=0,1\;.
$$
\end{proposition}

\noindent {\bf Proof.}  We have already seen above that:
$$
K_j (\Ee_d)\; \simeq \;K_j(\widehat \Aa_d)\; \simeq \;K_j(\widehat \Aa_{d-1})\;.
$$
Since the homotopy class of $\alpha_d$ is trivial by Proposition~\ref{Prop-Homotopy}, the upper-left corner of the six-term diagram \eqref{PVSixTermDiagram} becomes the following short exact sequence of Abelian groups
\begin{equation}\label{ExactSeq1}
\begin{diagram}
0 & \rTo & K_0(\Aa_{d-1}) & \rTo{\rm ev_\ast \ } &  K_0(\Aa_{d}) & \rTo{\rm Exp \ }& K_1(\Aa_{d-1}) &\rTo & 0
\;.
\end{diagram}
\end{equation}
Similarly, the lower-right corner of \eqref{PVSixTermDiagram} gives
\begin{equation}\label{ExactSeq2}
\begin{diagram}
0 & \rTo & K_1(\Aa_{d-1}) & \rTo{\rm ev_\ast \ } &  K_1(\Aa_d) & \rTo{\rm Ind \ }& K_0(\Aa_{d-1}) &\rTo & 0
\end{diagram}
\end{equation}
The $K$-groups can now be derived iteratively, starting from the $K$-groups of $\Aa_1$ which has $C(\SM^1)$ as a retract so that $K_0 (\Aa_1) =K_1 (\Aa_1)= \mathbb Z$. Indeed, the only abelian group extension of $\ZM$ by $\ZM$ is $\ZM^2$, so that \eqref{ExactSeq1} and \eqref{ExactSeq2} for $d=2$ imply $K_0 (\Aa_2) =K_1 (\Aa_2)= \mathbb Z^2$. This procedure can now be iterated to complete the proof.
\hfill $\Box$

\vspace{.2cm}

Let us point out that the above argument also shows
$$
K_0(\Aa_d) \; \simeq \;K_0(\Aa_{d-1}) \oplus K_1(\Aa_{d-1})\;\simeq \;K_0(\widehat \Aa_d) \oplus K_1(\Ee_d)
\;,
$$
and
$$
K_1(\Aa_d) \;\simeq\;
 K_1(\Aa_{d-1}) \oplus K_0(\Aa_{d-1})
\;\simeq \;
K_1(\widehat \Aa_d) \oplus K_0(\Ee_d) 
 \;.
 $$
This holds for $d\geq 2$. The case $d=1$ is described in \eqref{SixTermDiagram}.

\subsection{The inverse of the index map}

The following explicit construction of the index map is reproduced from \cite{KRS} (see Proposition~A.1) and it also follows from \cite{PV}. This result is instrumental for the construction of the generators of the $K$-group, presented in the following section, as well as for Section~\ref{sec-PairingsDuality}

\begin{proposition}
\label{InverseIndexMap} 
Consider the Pimsner-Voiculescu exact sequence and let $e$ be a projection from $\Pp_N(\Aa_{d-1})$.  With the unitary $ w_e \in \Aa_{d-1}$ from Proposition~\ref{InnerAuto} and $u_d$ identified with $\one_N \otimes u_d$, let us set
\begin{equation}
v 
\;=\; 
(\bm \one_N - e) + e\, w_e \, u_d^\ast
\; \in \;\Uu_N(\Aa_d)
\;.
\end{equation}
This is a pre-image of $e$ for the index map: 
$$
\Ind[v]_1 \;=\; [e]_0
\;.
$$
\end{proposition}

\noindent {\bf Proof.} First of all, checking the unitarity of $v$ is elementary if one observes that $w_e \, u_d^\ast$ and $e$ commute. Next, with the partial isometry $\widehat{S}$ and projection $\widetilde{P}$ as in the construction of the the Pimsner-Voiculescu sequence in Section~\ref{Sec-PimsnerVoiculescu}, we can generate the following lift
$$
{\rm Lift}\big({\rm diag}(v,v^\ast)\big )
\;=\; \begin{pmatrix}
(\bm \one_N - e)\otimes 1 + e \, w_e \, u_d^\ast \otimes \widehat{S}\,^\ast  & 0 \\
e \otimes \widetilde{P} & (\bm \one_N - e)\otimes 1 + \overline{w}_e^\ast \,u_d \,e \otimes \widehat{S} 
\end{pmatrix}
\;.
$$
Indeed, by recalling that $\widehat S \, ^\ast \, \widehat S=\one$ and $\widehat S \; \widehat S \, ^\ast = \one - \widetilde P$, one can easily verify that the r.h.s. is a unitary element from $\mathcal U_{2N}(T(\Aa_d))$. We also recall that $\pi (1 \otimes \widetilde{P}) =0$, hence $\pi (e \otimes \widetilde{P}) =0$, and $\pi(u_d \otimes \widehat S) = u_d$ while $\pi(u_d^\ast \otimes \widehat S\, ^\ast) = u_d^\ast$. Thus
$$
\pi\big( ( \one_N - e)\otimes 1 + e \, w_e \, u_d^\ast \otimes \widehat{S}\,^\ast\big) 
\;=\; 
(\one_N - e) + e \, w_e \, u_d^\ast \; =\; v
\;,
$$
and similarly for the other diagonal term, using  that $\overline{w}_e^\ast \,u_d \, e \, = \, e \, \overline{w}_e^\ast \,u_d$. This proves the second claim. Now, a direct computation gives
\begin{align*}
{\rm Lift}\big({\rm diag}(v,v^\ast)\big ) {\rm diag}(\bm \one_N,0_N) {\rm Lift}\big({\rm diag}(v,v^\ast)\big )^\ast 
\;=\;
{\rm diag}(\bm \one_N, e \otimes \widetilde{P})
\;,
\end{align*}
hence
\begin{align*}
{\rm Ind}([v]_1)  \;=\; [e]_0 ,
\end{align*}
from the definition \eqref{IndexMapDefinition} of the index map.
\hfill $\Box$

\subsection{The generators of the $K$-groups}
\label{Sec-GeneratorsKGroups}

Proposition~\ref{PropKGroups} shows that $K$-groups of $\Aa_d$ and the $d$-dimensional rotation algebra coincide. This is routed in the fact that the contractability of $\Omega$ implies that the rotation algebra (with same magnetic field) is a retract of $\Aa_d$ (see also Proposition~\ref{prop-1dhomotop} for a direct argument). Therefore, also the generators of the $K$-groups $K_{0,1}(\Aa_d)$  can be chosen to be elements of the rotation algebra, which is also denoted by $\Aa_d$ in this section (as it corresponds to the special case of a set $\Omega$ having only one point). These generators have been analyzed in detail by Elliott \cite{Ell} and Rieffel \cite{Rie1}. Here we present an iterative construction in increasing dimension $d$ based on the Pimsner-Voiculescu exact sequence  \eqref{PVSixTermDiagram} just as the proof of Proposition~\ref{PropKGroups} . Supplementary information on the generators and their pairing with the cyclic cohomology can then be found in Section~\ref{sec-PairingsRange}. 

\vspace{.2cm}

Let us begin with several well-known explicit computations. The group $K_0(\Aa_1) \cong \ZM$ is generated by the identity and $K_1(\Aa_1) \cong \ZM$ by $[u_1]_1$. The group  $K_1(\Aa_2) \cong \ZM^2$ is generated by $[u_1]_1$ and $[u_2]_1$, while $K_0(\Aa_2) \cong \ZM^2$ is generated by the identity and by the Powers-Rieffel projection $[e_{\{1,2\}}]_0$ \cite{Rie0,WO}, given by
\begin{equation}
\label{eq-Rieffel}
e_{\{1,2\}} \;=\; u_2^\ast \, g(u_1) \;+\; f(u_1) \;+ \;g(u_1) \, u_2\;,
\end{equation}
where $f$ and $g$ are properly chosen functions such that $e_{\{1,2\}}$ is indeed a projection and $\Tt(e_{\{1,2\}})=\frac{1}{2\pi}B_{1,2}=\theta$. Through a direct computation ({\it cf.} p.~116 in \cite{PV}), one can show that
\begin{equation}\label{eq-ExpRiefel}
\Exp[e_{\{1,2\}}]_0 
\;=\; 
[\exp (-2\pi \I \, \hat e_{\{1,2\}})]_1 
\;=\; 
[\one-\hat e + \hat u_1 \hat e]_1
\;=\;
[u_1]_1\;,
\end{equation}
where the last two elements lie in $K_1(\Ee_1)$ and $K_1(\Aa_1)$ respectively, which are identified as in Proposition~\ref{PropKGroups}. The identity \eqref{eq-ExpRiefel} is the $K$-theoretic essence of the bulk-edge correspondence for two-dimensional quantum Hall effect. Let us go one step further to $d=3$ before describing the general structure of the $K$-groups.  The group $K_0(\Aa_3) \cong \ZM^4$ is generated by the identity and the three Powers-Rieffel projections $e_{\{i,j\}}$ corresponding to the two-dimensional tori $C^\ast(u_i,u_j)$, $i \neq j \in \{1,2,3\}$. On the other side, $K_1(\Aa_3) \cong \ZM^4$ is generated by $u_1, u_2, u_3$ and the additional unitary operator \cite{PV,Wal}
$$
v_{\{1,2,3\}} 
\;=\;
\one\;-\;e_{\{1,2\}} \;+\;e_{\{1,2\}} \, \breve u_3\, u_3^*
\;,
$$
where $\breve u_3\in\Aa_{2}$ implements the action of $u_3$ on the Powers-Rieffel projection by an inner automorphism of $\Aa_{2}$ (see Proposition~\ref{InnerAuto}). Due Proposition~\ref{InverseIndexMap} one then has
\begin{equation}\label{eq-3dtopgen}
\mathrm{Ind}[v_{\{1,2,3\}}]_1\; = \;[e_{\{1,2\}}]_0
\;.
\end{equation}

We now provide the general iterative procedure for constructing the generators of the $K$-groups. For this, suppose the $K$-theory of $\Aa_{d-1}$ has been already computed. Then the inclusion maps $j:\Aa_{d-1}\to\Aa_d$ in \eqref{PVSixTermDiagram} induce injections
\begin{equation}\label{eq-KEmbed}
j_*:K_{0,1}(\Aa_{d-1})\rightarrow K_{0,1}(\Aa_{d})
\;,
\end{equation}
so that the generators of $K_{0,1}(\Aa_{d-1})$ are naturally identified with generators of $K_{0,1}(\Aa_{d})$. By doing so, we already identified half of the generators in dimension $d$. Still by \eqref{PVSixTermDiagram}, the index and exponential maps are surjections and can therefore be inverted to injective maps
\begin{equation}
\label{eq-KIncrease}
\Exp_d^{-1}: K_{1}(\Aa_{d-1})\rightarrow K_{0}(\Aa_d)
\;,
\qquad
\Ind_d^{-1}: K_{0}(\Aa_{d-1})\rightarrow K_{1}(\Aa_d)
\;,
\end{equation}
which supply the other half of the generators. As it will become apparent below, it is convenient to work with $-\Ind_d^{-1}$, where minus means inversion in the $K_1$ group. The index $d$ on $\Exp$ and $\Ind$ indicate that they correspond to \eqref{PVSixTermDiagram}. The inverse of the index map is written down explicitly in Proposition~\ref{InverseIndexMap}. To our best knowledge, a similar simple construction of the inverse of the exponential map is not known, except in the case $d=2$ already mentioned above. 

\vspace{0.2cm}

If we apply the above iteration, starting from $\Aa_0\cong\CM$, we can compute the $K$-theory of $\Aa_d$ for arbitrary $d$. In particular, this will reproduce the explicit computations of the $K$-theories $\Aa_1$, $\Aa_2$ and $\Aa_3$ provided above. The generators of $K_0(\Aa_d)$ provided by the iteration will be uniquely labeled as $[e_I]_0$ by the increasingly ordered subsets $I \subset \{1,\ldots,d\}$ of even cardinality $|I|$. As a convention,  the empty set $I=\emptyset$ of zero cardinality is a valid choice and $[e_\emptyset]_0$ represents the class of the unit element. Likewise, the generators of $K_1(\Aa_d)$ are uniquely labeled as $[v_J]_1$ by the increasingly ordered subsets $J \subset \{1,\ldots,d\}$ of odd cardinality $|J|$. Due to \eqref{eq-KEmbed} applied iteratively, if $I$ and $J$ are subsets of $\{1,\ldots,d'\}$, the corresponding generators $[e_I]_0$ and $[v_J]_1$ can be seen as generators of $K_0(\Aa_d)$ and $K_1(\Aa_d)$ for any $d\geq d'$. The iteration is started by choosing $e_\emptyset=\one$ as representative for the generator of $K_0(\Aa_0)$.  New generators in $K_{0,1}(\Aa_d)$, namely not inherited from $K_{0,1}(\Aa_{d-1})$ via \eqref{eq-KEmbed}, correspond to labels $I$ and $J$ containing the index $d$ and are obtained using \eqref{eq-KIncrease}. They are defined by the equations
\begin{equation}
\label{eq-ExpIndLink}
\boxed{
\Ind^{-1}_d[e_I]_0\;=\; - \,[v_{I\cup\{d\}}]_1
\;,
\qquad
\Exp^{-1}_d[v_J]_1\;=\; [e_{J\cup\{d\}}]_0
\;.
}
\end{equation}
The labelling by the subsets provides the following decomposition of the $K$-groups
\begin{equation}
\label{K0Structure}
K_0(\Aa_d)\; =\; \sum_{I\subset \{1,\ldots,d\}} \mathbb Z
\;, 
\qquad
K_1(\Aa_d)\; =\;\sum_{J\subset \{1,\ldots,d\}} \mathbb Z
\;, 
\end{equation}
where the sums run over $|I|$ even and $|J|$ odd respectively. Accordingly, one can count again the dimensionality of the $K$-groups,
$$
\sum_{k=0}^{[d/2]}  \binom {d} {2k}\; =\; 2^{d-1}\;,
\qquad
\sum_{k=0}^{[d/2]}  \binom {d} {2k+1} \;=\; 2^{d-1}\;,
$$
in agreement with Proposition~\ref{PropKGroups}.  

\vspace{.2cm}

Starting from the generator $[e_\emptyset]_0\in K_0(\Aa_0)$, one first infers from \eqref{eq-ExpIndLink} that $\nu_{\{1\}}=u_1$ specifies the generator of $K_1(\Aa_1)$. Applying \eqref{eq-ExpIndLink} for $d=2$, leads to $v_{\{2\}}=u_2$ and, due to \eqref{eq-ExpRiefel}, to the Powers-Rieffel projection $e_{\{1,2\}}$. For $d=3$, the new generators in $K_1(\Aa_3)$ are $v_{\{3\}}=u_3$ and minus $v_{\{1,2,3\}}$ defined above, see \eqref{eq-3dtopgen}, while the new generators for $K_0(\Aa_3)$ are the Powers-Rieffel projections $e_{\{1,3\}}$ and $e_{\{2,3\}}$ defined as in \eqref{eq-Rieffel}.

\vspace{.2cm}

The generators of the $K$-groups of the algebras of half-space and boundary observables can be derived from the generators of the bulk algebra and the isomorphisms established in Section~\ref{Sec-KPimsnerVoiculescu}. In particular, the isomorphisms between the $K$-groups induced by $\mathfrak{i}'$ defined in \eqref{eq-natincl} provide the generators of $K_{0,1}(\widehat \Aa_d)$ groups:
\begin{equation}\label{GeneratorHatK0}
\hat e_{I} \;=\; \mathfrak{i}'_*(e_I)\;,
\qquad
\hat v_{J} \;=\; \mathfrak{i}'_*(v_J)\;,
\end{equation}
with $I,J\subset\{1,\ldots,d-1\}$ of even and odd cardinality respectively. The evaluation map sends these generators into
$$
{\rm ev} (\hat e_{I}) \;=\; e_{I}\;, 
\qquad {\rm ev} (\hat v_J) \;=\; v_J
\;.
$$
Finally, the generators of $K_0(\Ee_d)$ consist of $\hat e$ and
\begin{equation}\label{GeneratorTildeK0}
\tilde e_I
\; =\; 
\hat e_I \,\hat e
\;,
\end{equation}
still with $I\subset \{1,\ldots,d-1\}$. The $K_1(\Ee_d)$ group is generated by 
\begin{equation}
\label{GeneratorTildeK1}
\tilde v_J\;=\;\one\,-\,\hat e \,-\, \hat v_J\, \hat e\, 
\;, 
\end{equation}
where $J\subset \{1,\ldots,d-1\}$ has odd cardinality. If we recall that the projection $\hat e$ is sent to $0$ by the imbedding $i$, then we see explicitly how $i_*$ sends the entire groups $K_0(\Ee_d)$ and $K_1(\Ee_d)$ into the trivial classes $0$ and $[\one]_1$ repectively. Further let us note that
$$
\widetilde \rho (\tilde e_I)\;  =\; e_I\;,
\qquad
\widetilde \rho (\tilde v_J) \;=\; v_J
\;.
$$

Next let us discuss how the generators are mapped by the connecting maps of the six-term exact sequence \eqref{GSixTermDiagram} rather than \eqref{PVSixTermDiagram}, which given various isomorphisms of Section~\ref{Sec-KPimsnerVoiculescu} is merely a rewriting of \eqref{eq-ExpIndLink}. First of all,
$$
\mathrm{Ind}[\, v_J\,]_1 
\;=\; 
[\tilde e_I]_0\,, 
\qquad 
J=I\cup\{d\}\;,\;\;I\subset\{1,\ldots,d-1\}
\;.
$$
All the other generators, namely $v_J$ with $d\not\in J$, are sent to the trivial class by the index map. As for the exponential map, note that $\hat e_I$ with $d\not\in I$ provides a lift of $e_I$ in $\widehat \Aa_d$. As $d\not\in I$, this lift is again a projection and
$$
{\rm Exp}[e_I]_0 \;= \;[\exp(-2 \pi \I \hat e_I)]_1\; =\; [0]_1\;,
$$
which merely confirms that ${\rm Exp} \circ {\rm ev}_\ast=0$. On the other hand, if $d\in I$, the lift is no longer a projection, and one has
$$
{\rm Exp} [e_{I}]_0
\; = \;
[\tilde v_{I\setminus\{d\}}]_1\;.
$$
A particular case of this is \eqref{eq-ExpRiefel}.

\vspace{.2cm}

To round up this section, let us briefly place the classification of the unitary and chiral unitary classes of topological insulators into this $K$-theoretic context. According to  \eqref{K0Structure} and the discussion before, the $K_0$-group has a generator which involves all space dimensions only if the space dimension is even. This top generator $e_{[1, d]}$ generates all the strong phases of topological insulators appearing in the first row of the classification  (Table \ref{Table1}) under dimension $d$. In other words, modulo lower generators, the Fermi projector of any strong topological insulator from the unitary class belonging to the $n$-th phase in $d$ space dimensions, is stably homotopic to
$$
p_F \;\sim_0 \;{\rm diag}(e_{[1, d]}, \ldots, e_{[1, d]})\;,
$$
with precisely $n$ copies of $e_{[1, d]}$ appearing inside the diagonal. It will be shown in the next chapter, see in particular Section~\ref{sec-PairingsRange}, that the strong even Chern character pairs non-trivially with the generator $e_{[1, d]}$, but its pairings with all other  generators vanish. These comments transpose to the $K_1$-group in connection with the chiral unitary topological insulators. According to \eqref{K0Structure}, $K_1(\Aa_d)$  has a generator which involves all space dimensions only when $d$ is odd. This top generator is $v_{[1,d]}$ and, as we shall see, this generator pairs non-trivially with the strong odd Chern character, hence it generates all the strong phases of topological insulators appearing in the second row of the classification table.

\section{The connecting maps for solid state systems}
\label{Sec-ConnectingMaps}

Various connecting maps between the $K$-groups were defined and discussed in Sections~\ref{SixTermSequence} and \ref{Sec-SuspensionBottPer}. Here the general theory is applied to different exact sequences associated to solid state systems.

\subsection{The exponential map for the bulk-boundary correspondence}
\label{sec-expBBC}

We begin by considering the exact sequence
\begin{diagram}
0 &\rTo &\mathcal E_d   &\rTo{i}  &\widehat{\mathcal A}_d  &\rTo{ \mathrm{ev}}  &\mathcal A_d &\rTo &0
\end{diagram}
constructed in Section~\ref{sec-BBC}. As we have already seen, associated is the induced six-term exact sequence between the $K$-groups
\begin{equation}
\label{diag-6K}
\begin{diagram}
K_0(\Ee_d) &\rTo^{\ \ i_\ast \ \ } & K_0(\widehat \Aa_d) & \rTo{\ \ \mathrm{ev}_\ast \ \ } & K_0(\Aa_d) \\
\uTo{\mathrm{Ind}}                 &               \     \                        &     \     \                                 &       \     \                   &\dTo{\mathrm{Exp}} \\
K_1(\Aa_d)        & \lTo{\ \ \mathrm{ev}_\ast \ \ } & K_1(\widehat \Aa_d) & \lTo{\ \ i_\ast \ \ }    & K_1(\Ee_d)
\;.
\end{diagram}
\end{equation}
Here we are interested in the class $[p_F]_0 \in K_0(\Aa_d)$ of the Fermi projection  $p_F = \chi(h\leq \mu)$ of a bulk Hamiltonian satisfying the BGH, with the aim of expressing its image under the exponential map in terms of the finite range half-space Hamiltonian $\hat h = (h,\tilde h) \in M_N(\mathbb C)\otimes \widehat \Aa_d$ (which is a particular lift of $h$ in the above exact sequence). This is achieved by the following result.  

\begin{proposition}[\cite{SKR}] 
\label{ExpMap}
The class of the Fermi projection in $K_0(\Aa_d)$ is mapped under the exponential into
\begin{equation}
\label{ExpMapFormula}
\boxed{
\;
\Exp([p_F]_0)
\;=\;
\left[\exp(2\pi\I\, \FFunc(\hat h))\right]_1
\;\in\;
K_1(\Ee_d)
\;,
\;
}
\end{equation}
where $\FFunc :\RM\to [0,1]$ is a non-decreasing continuous function equal to $0$ below the insulating gap $\Delta$ and to $1$ above $\Delta$. Above, the functional calculus involving $\FFunc$ is carried out in $M_N(\CM) \otimes \widehat \Aa_d$ algebra, while the one involving the $\exp$ function in $M_N(\CM) \otimes \widehat \Aa_d^+$ algebra.
\end{proposition}

\noindent {\bf Proof.} In the light of statement (vii) of Proposition~\ref{prop-standardK0}, $p_F$ can be viewed as an element of $\Pp_N(\Aa_d^+)$. The statement follows directly from definition \eqref{ExpMapDefinition} of the exponential map, provided we can show that $(1-\FFunc) (\hat h)$, viewed as an element of $M_N(\mathbb C) \otimes \widehat{\Aa}_d^+$, is a lift of $p_F$. We have
$$
({\rm id} \otimes {\rm ev}^+)\Big ((1-\FFunc) \big (\hat h \big ) \Big ) 
\;=\; 
({\rm id} \otimes {\rm ev})\Big ((1-\FFunc) \big (\hat h \big ) \Big ) 
$$
and, since ${\rm id} \otimes {\rm ev}$ is a bounded homomorphism of C$^\ast$-algebras, it commutes with the continuous functional calculus 
$$
({\rm id} \otimes {\rm ev})\Big ((1-\FFunc) \big (\hat h \big ) \Big ) 
\;=\; 
(1-\FFunc ) \big (({\rm id} \otimes {\rm ev})(\hat h) \big ) 
\;= \;
(1-\FFunc ) (h) 
\;= \;
p_F
\;,
$$  
where the last equality follows because $1-\FFunc$ is equal to $1$ below the bulk gap and to $0$ above the bulk gap.\hfill $\Box$

\vspace{.2cm}

The exponential connecting map provides a unitary element in the boundary algebra $M_N(\mathbb C) \otimes \Ee_d^+$ which encodes the topology of the system. Due to its central importance for the bulk-boundary problem we call the image the boundary unitary element and used the notation:
\begin{equation}\label{TildeU}
\boxed{
\;
\tilde u_\Delta \;=\; \exp \big (2\pi\I\, \FFunc (\hat h) \big ) \in M_N(\mathbb C) \otimes \Ee_d^+\;.
\;
}
\end{equation}
We used the label $\Delta$ because $\tilde u_\Delta - \one_N$ can be constructed entirely from the spectral subspace of $\hat h$ corresponding to the bulk insulating gap $\Delta$.  Indeed, $\exp(2\pi \I \FFunc ) - 1$ is a smooth function with support in the insulating gap. Furthermore, according to Proposition~\ref{HFSmoothFunc1}, $\tilde u_\Delta-\one_{N}$ belongs to the smooth algebra $M_N(\mathbb C) \otimes \mathscr E_d$ and to any Sobolev space $W^{s,k}({\Ee_d},\widetilde \Tt)$, which is an important technicality playing a role in the definition of the boundary topological invariant.

\subsection{The index map  for the bulk-boundary correspondence}
\label{sec-indBBC}

This section deals with the other connecting map of the exact sequence \eqref{diag-6K}, namely the index map. The unitary specifying an element of $K_1(\Aa_d)$ is the Fermi unitary associated to a chiral symmetric Hamiltonian $\hat h = (h,\tilde h) \in M_{2N}(\mathbb C) \otimes \widehat \Aa_d$ via
$$
\sgn(h)
\;=\;
\begin{pmatrix}
0 & u_F^\ast \\ u_F & 0
\end{pmatrix}
\;,
\qquad
J\;=\;
\begin{pmatrix}
\one_N & 0 \\ 0 & -\one_N
\end{pmatrix}
\;.
$$
The present task is to compute the element $\Ind([u_F]_1 )$ of $K_0(\Ee_d)$. Obviously, this will be relevant for the bulk-boundary problem in the chiral unitary class of topological insulators.

\begin{proposition}
\label{IndMap} 
Suppose BGH and CH hold for $\hat h = (h,\tilde h) \in M_{2N}(\mathbb C) \otimes \widehat \Aa_d$ and let $u_F$ be the Fermi unitary operator associated to $h\in M_{2N}(\mathbb C) \otimes \Aa_d$. Let $\GFunc :\RM\to [-1,1]$ be a non-decreasing smooth function, equal to $\pm 1$ above/below the bulk insulating gap, respectively, and odd under inversion, $\GFunc(-x) = -\GFunc(x)$. Then 

\begin{enumerate}[\rm (i)]

\item The class $[u_F]_1 \in K_1(\Aa_d)$ is mapped by the index map into
\begin{equation}\label{IndMapFormula}
\Ind \big ([u_F]_1 \big )
\;=\;
\left[
e^{-\I\frac{\pi}{2} \GFunc (\hat h)}
\,
{\rm diag}(\one_N,0_N)
\,e^{\I\frac{\pi}{2} \GFunc (\hat h)}
\right]_0
\;-\;
\left[{\rm diag}(\one_N, 0_N)
\right]_0
\;.
\end{equation}
Above, the functional calculus involving $\GFunc$ is carried out in $M_{2N}(\CM) \otimes \widehat \Aa_d$ algebra, while the one involving the exponential function in $M_{2N}(\CM) \otimes  \widehat \Aa_d^+$ algebra.

\item The projection provided by the index map and explicitly written above belongs to the smooth sub-algebra $M_N(\CM) \otimes \mathscr{E}_d^+$.

\end{enumerate}

\end{proposition}

\noindent {\bf Proof.} (i) Recall statement (vi) of Proposition~\ref{prop-standardK1}, which says that the Fermi unitary element $u_F$ from $\Uu_{N}(\Aa_d)$ can be promoted to a unitary element $u_F^+$ from $\Uu_{N}(\Aa_d^+)$. According to definition \eqref{IndexMapDefinition} of the index map, we need to find an explicit lift in $\Uu_{2N}(\widehat \Aa_d^+)$ of 
$$
{\rm diag}\big (u_F^+,(u_F^\ast)^+\big )
\;=\; 
{\rm diag}(u_F,u_F^\ast)^+ 
\;\in\; 
\Uu_{2N}(\Aa_d^+) \; .
$$ 
But if we find a lift of ${\rm diag}(u_F,u_F^\ast)$ to $\Uu_{2N}(\widehat \Aa_d)$, then this lift can be automatically promoted to a lift in $\Uu_{2N}(\widehat \Aa_d^+)$ of ${\rm diag}(u_F,u_F^\ast)^+$. Furthermore, since
$$
{\rm diag}(u_F,u_F^\ast) 
\;= \;
\begin{pmatrix} 0_N & \bm 1_N \\ \bm 1_N & 0_N \end{pmatrix}\begin{pmatrix} 0_N & u_F^\ast \\ u_F & 0_N \end{pmatrix}
\;=\; 
\begin{pmatrix}0_N & \bm 1_N \\ \bm 1_N & 0_N \end{pmatrix} \sgn(h)
\;,
$$
the problem is reduced to finding an appropriate unitary lift for ${\rm sgn}(h)$ in $\Uu_{2N}(\widehat \Aa_d)$. Following the same strategy as for the exponential map, we can consider
$$
{\rm Lift}\big ({\rm sgn}(h) \big ) 
\;=\; 
\I \,e^{- \I \frac{\pi}{2} \GFunc (\hat h)}
\;,
$$
and we can verify that
$$
({\rm id}\otimes {\rm ev})\Big (\I e^{ - \I \frac{\pi}{2} \GFunc(\hat h)} \Big ) 
\;=\; 
\I \,e^{ -\I \frac{\pi}{2} \GFunc(({\rm id}\otimes {\rm ev}^+)(\hat h))}
\;=\; 
\I \,e^{- \I \frac{\pi}{2} \GFunc(h)} 
\;=\; 
\sgn(h)\;.
$$ 
Then the unitary $\hat w$ in the definition \eqref{IndexMapDefinition} of the index map becomes
$$
\hat w 
\;=\; 
\I \, \begin{pmatrix}0_N & \bm 1_N \\ \bm 1_N & 0_N \end{pmatrix}e^{- \I \frac{\pi}{2} \GFunc (\hat h)} 
\;\in\; 
\Uu_{2N}(\widehat \Aa_d^+)
\;, 
$$
where the exponentiation is considered inside $M_{2N}(\CM) \otimes \widehat \Aa_d^+$. The statement then follows from the definition of the index map and the homotopy argument used in Proposition~\ref{prop-1d}.

\vspace{.1cm}

(ii) We have
\begin{align*}
e^{-\I\frac{\pi}{2}\GFunc(\hat h)}
\,
{\rm diag}(\one_N, 0_N)
\,e^{\I\frac{\pi}{2}\GFunc(\hat h)} & 
\;=\; \tfrac{1}{2} \,e^{-\I\frac{\pi}{2}\GFunc(\hat h)}
\,
(\bm 1_{2N} + \widehat J \; )
\,e^{\I\frac{\pi}{2}\GFunc(\hat h)} \\
& 
\;=\; \tfrac{1}{2}\,\bm 1_{2N}  \;+\; \tfrac{1}{2}\, \widehat J \; e^{\I \frac{\pi}{2} (\GFunc(\hat h) - \GFunc(- \hat h))} \\
&
\;=\; \tfrac{1}{2} \,\widehat J \; ( e^{\I \pi \GFunc(\hat h)} +\one_{2N} )
\;+\; {\rm diag}(0_N,\one_N) \; .
\end{align*}
The function $e^{\I \pi \GFunc(x)}+1$ is smooth and with support inside the bulk insulating gap $\Delta$. Hence, the non-scalar part of the projection is a function of $\hat h$ which satisfies all conditions of Proposition~\ref{HFSmoothFunc2}, hence in $M_N(\CM)\otimes \mathscr{E}_d$. 
\hfill $\Box$

\vspace{0.2cm}

The conclusion is that the index map provides a projection from the smooth boundary sub-algebra $M_{2N}(\CM)\otimes \mathscr{E}_d^+$, which can be used to encode the topology of the boundary and will be of central importance for the bulk-boundary problem. We call it the chiral boundary projection and use the notation
\begin{equation}\label{BoundaryProjection1}
\boxed{
\;
\tilde p_\Delta 
\;=\;
e^{-\I\frac{\pi}{2}\GFunc(\hat h)}
{\rm diag}(\one_N, 0_N)
\,e^{\I\frac{\pi}{2}\GFunc(\hat h)}\; \in \;M_{2N}(\CM)\otimes \mathscr{E}_d^+
\;.
\;
}
\end{equation}
We used the label $\Delta$ because, as we have seen above, $\tilde p_\Delta - s(\tilde p_\Delta)$ can be constructed entirely from the spectral subspace of $\hat h$ corresponding to the bulk insulating gap $\Delta$. Also, note that, since $\diag(\one_N,0_N)$ and $\diag(0_N,\one_N)$ are homotopic, the index map can be also written as:
\begin{equation}\label{IndMapFormula2}
\boxed{
\;
\Ind \big ([u_F]_1 \big )
\;=\;
\left[\tilde p_\Delta \right]_0
\;-\;
\left[s(\tilde p_\Delta)
\right]_0
\;.
}
\end{equation}

Finally, let us consider the particular case when the bundary spectrum has gaps, which due to the chiral symmetry must occur symmetrically relative to the origin. This allows to further simplify the image of the index map and will be of particular physical relevance in Section~\ref{sec-SurfacQHE}, where we analyze the situation when a magnetic field perpendicular to the boundary of a chiral topological insulator opens gaps in the boundary spectrum.

\begin{proposition}
\label{prop-SurfaceStateGap}
If $[-\delta,\delta] \subset \Delta$ such that $\pm \delta$ lie in a spectral gap of $\hat{h}$, then the spectral projector $\tilde{p}(\delta)=\chi(-\delta\leq\hat{h}\leq\delta)$ belongs to $\mathscr{E}_d$ and the chiral boundary projection can be chosen as
$$
\tilde p_\Delta \;=\; \widehat J \; \tilde{p}(\delta)\; +\; \diag(0_N,\one_N) \; .
$$
Furthermore, $\tilde{p}(\delta)$ can be simultaneously diagonalized with $\widehat J$, namely there exist the mutually orthogonal projections $\tilde{p}_{\pm}(\delta) \in \mathscr{E}_d$ with $\tilde{p}(\delta)=\tilde{p}_{+}(\delta)+\tilde{p}_{-}(\delta)$ and $\widehat J \, \tilde{p}_{\pm}(\delta)=\pm \tilde{p}_{\pm}(\delta)$. Then
$$
\tilde p_\Delta -s(\tilde p_\Delta)
\;=\;
\tilde{p}_{+}(\delta) - \tilde{p}_{-}(\delta)
\;,
$$
and
$$
[\tilde p_\Delta]_0 -[s(\tilde p_\Delta]_0
\;=\;
\big [\tilde{p}_{+}(\delta) \big ]_0-\big [\tilde{p}_{-}(\delta) \big ]_0
\;.
$$
\end{proposition}

\noindent {\bf Proof.} All the statements follow as in Proposition~\ref{prop-1dbis}, by choosing a smooth function $\GFunc$ and such that $\GFunc=0$ inside the interval $[-\delta,\delta]$, $\GFunc=-1$ on $(-\infty, -\delta)\cap \sigma(\hat h)$ and $\GFunc=+1$ on $(\delta, + \infty) \cap \sigma(\hat h)$. \hfill $\Box$

\vspace{0.2cm}

The following is a trivial consequence of the above, but nevertheless it is important to state it explicitly.

\begin{corollary}
\label{coro-GappedEf}
 If the spectrum of $\hat h$ is gapped at $E=0$, then the chiral boundary projection can be chosen to be trivial:
$$
\tilde p_\Delta \;=\;  \diag(0_N,\one_N) \;,
$$
and $\tilde p_\Delta - s(\tilde p_\Delta)=0$.
\end{corollary}

\subsection{The Bott map of the Fermi projection}

The topology of the solid systems from the unitary class is encoded in the Fermi projection $p_F = \chi(h\leq \mu)$. Using the Bott connecting map \eqref{BetaMap} for suspensions, we show here that, equivalently, the topology can be encoded using the resolvent function of the Hamiltonian. As shown in Section~\ref{Sec-Suspension}, this can be used to reformulate the bulk invariants in terms of the resolvent function. Such expressions are well-known in the physics literature \cite{Vol,QHZ,EG1,EG2}.

\begin{proposition}
\label{prop-Greenlink}
Consider a finite range bulk Hamiltonian $h \in M_N(\mathbb C)\otimes \Aa_d$ obeying BGH. Let $\Gamma_F$ be a negatively oriented curve in the resolvent set of $h$ such that the Fermi projection is given by $p_F=\oint_{\Gamma_F}\frac{dz}{2\pi \I}\,g_z$ where $g_z=(h-z)^{-1}$ is the resolvent function. Then the Bott isomorphism $\beta:K_0(\Aa_d)\mapsto K_1(S \Aa_d)$ satisfies
\begin{equation}
\label{eq-Bottmap}
\beta [p_F \oplus p_F]_0
\;=\;
\big[z\in\Gamma_F\mapsto (h-\bar z)g_z \big]_1
\;.
\end{equation}
\end{proposition}

\noindent {\bf Proof.} 
First, note that $(h-\bar z)g_z$ is a unitary element from $\Aa_d$ such that $(h-\bar z_0)g_{z_0}=\one_{N}$, where $z_0$ is one of the points where $\Gamma_F$ traverses the real axis. As such, by a proper parametrization of $\Gamma_F$,  $z\in\Gamma_F\mapsto (h-\bar z)g_z$ becomes an element of $\Uu_{N}\big ((S\Aa_d)^+\big )$, as required.
In the path given in \eqref{eq-Bottmap}, the Hamiltonian $h$ can be homotopically deformed to the flat band Hamiltonian $\one_N-2p_F$, while at the same time $\Gamma_F$ is deformed to $t\in[0,2\pi]\mapsto \exp(-\I t)-1$ and $z_0$ can be taken as $0$. Then
$$
(h-\bar z)g_z 
\;=\; 
\frac{2-\exp(\I t)}{2-\exp(-\I t)}(1-p_F) \;+\; \exp(2 \I t) p_F
\;.
$$
The r.h.s. can be continuously deformed to $(\one-p_F) + \exp(2 \I t) p_F$ without leaving $\Uu_{N}\big ((S\Aa_d)^+\big )$. In other words
$$
(h-\bar z)g_z 
\;\sim_1 \;
\big ((\one-p_F) \;+ \;\exp(\I t) p_F
\big)^2
\;.
$$
One can recognize inside the square the unitary element 
$$
u\; =\; 
\big ( t\in [0,2\pi] \rightarrow (\one-p_F) \;+\; \exp(\I t) p_F \big )
$$ 
from the definition of the Bott map \eqref{BetaMap}. Then
\begin{align*}
\big[z\in\Gamma_F\mapsto  & (h-\bar z)g_z  \big]_1 
\;=\; 
[u^2]_1 
\;=\;
[{\rm diag} (u,u)]_1
\;=\;
\beta [{\rm diag}(p_F,p_F)]_0
\;.
\end{align*}
The statement is proved.  
\hfill $\Box$

\begin{remark}
It is possible to restate  \eqref{eq-Bottmap} as
$$
\beta [p_F ]_0
\;=\;
\big[z\in\Gamma_F\mapsto (g_{z_0})^{-1}g_z \big]_1
\;.
$$
where $z_0$ is some fixed point on the loop which is choose to correspond to $0$ so that the loop on the r.h.s. is in $(S\Aa)^+$. Its values are not in the unitaries, but only in the invertibles which by polar composition can be retracted to the unitaries. While this second formula looks more compact, \eqref{eq-Bottmap} has the advantage of remaining valid in the regime of the MBGH. When the pairings are calculated under a BGH also the second path can be used, see Theorem~\ref{theo-VEG}.
\hfill $\diamond$
\end{remark}

\vspace{-.4cm}

\subsection{The $K$-theory of periodically driven systems}
\label{sec-adiabatic}

Here we suppose given the set-up described in Section~\ref{sec-PerDriven}, namely let $t\in[0,2\pi)\mapsto h_t\in M_N(\mathbb C)\otimes \Aa_d$ be a closed smooth loop attached to $h=h_0$ and suppose that there exists a loop $t\in[0,2\pi)\mapsto \mu_t\in\RM$ is such that $\mu_t$ lies in a gap of $h_t$.  Associated are then the (instantaneous) adiabatic projections $p_{A,t}=\chi(h_t\leq \mu_t)\in M_N(\mathbb C)\otimes \Aa_d$. They specify an element in $M_N(\mathbb C)\otimes C(\SM^1,\Aa_d)$ which is denoted by $p_A=\{p_{A,t}\}_{t\in\SM^1}$. We also set $p_{A,0}=p_F=\chi(h_0\leq \mu_0)$, and write also $p_F$ for the constant smooth loop. Then the class $[p_A]_1-[p_F]_1\in K_0(C(\SM^1,\Aa_d))$ can be viewed as an element in $K_0(S\Aa_d)$, if the exact sequence \eqref{eq-PerDrivenSeq}, which is a special case of \eqref{eq-InvThetaSeq}, is invoked. Based on the results of Section~\ref{sec-inverseTheta}, we can now determine the preimage of $[p_A]_0-[p_F]_0\in K_0(S\Aa_d)$ in $K_1(\Aa_d)$ under the suspension map $\theta$. The adiabatic time evolution $v_{A,t}$ is given by 
\begin{equation}
\label{eq-AdiabaticEv}
\I\,\partial_t v_{A,t}\;=\;
\big(h_t\;+\;\I[\partial_t p_{A,t},p_{A,t}]\big)v_{A,t}
\;,
\qquad
v_{A,0}=\one_N
\;.
\end{equation}
Hence Theorem~\ref{theo-adiabatic0} implies the following result which is at the heart of the stroboscopic interpretation of the polarization as discussed in Section~\ref{sec-polar}.

\begin{proposition}
\label{prop-adiabatic}
Assume all the above. Then
\begin{equation}
\label{eq-AdiabaticK2}
\theta [p_F v_{A,2\pi} p_F\;+\;\one_N\;-\;p_F]_1
\;=\;
[p_A]_0\;-\;[p_F]_0
\;,
\end{equation}
where on the r.h.s. $p_F$ denotes the constant path so that $[p_A]_0-[p_F]_0\in K_0(S\Aa_d)$.
\end{proposition}

\chapter{The topological invariants and their interrelations}
\label{Chap-TopologicalInvariants}

\abstract{This chapter first reviews the cyclic cohomology for general C$^\ast$-algebras and its pairing with the $K$-theory, which produces numerical topological invariants. The discussion is then specialized to the algebras of physical observables. The strong and the weak topological invariants, for both bulk and boundary, are defined as pairings of specific cyclic cocycles with the elements of the $K$-groups encoding the topology of the solid state systems. The duality of the pairings with respect to the connecting maps is proved and the equality between the bulk and boundary invariants is established. Lastly, generalized Streda formulas are derived and used to determine the range of the topological invariants.}

\section{Notions of cyclic cohomology}
\label{sec-CycCoh}

The cyclic (co)homology \cite{Con0,Tsy} is a theory for both commutative and non-commutative C$^\ast$-algebras, which can be regarded as a natural extension of the classical de Rham theory (see \cite{Kha} for insightful discussion). Of key importance for the invariants of solid state systems are the explicit pairing formulas between the cyclic cocycles and the elements of the $K_0$ and $K_1$-groups. As we shall see, the numerical topological invariants used in the classification of the unitary and chiral unitary classes of topological insulators can be obtained this way.  

\vspace{0.2cm}

Below, we present some key aspects of the cyclic cohomology which are instrumental for our goals. For example, as already pointed out in \cite{KRS}, the proof of equality between the bulk and the boundary invariants relies on the invariance of the pairings against the deformations of {\it both} the cyclic cocycles and the $K$-group elements. It would be very difficult, if not impossible, to prove this equality by brute computation, yet an elegant argument is possible when taking full advantage of the cyclic cohomology theory. As such, we feel that a brisk introduction to this theory is absolutely necessary. Below we will make references to the de Rham cohomology because physicists are familiar with this theory, but the reader should be aware that cyclic cohomology is in fact a generalization of the de Rham homology \cite{Con0}. 

\vspace{0.2cm}

The setting is that of a C$^\ast$-algebra $\Aa$. One considers densely defined $(n+1)$-multilinear functionals $\varphi$ on $\Aa$ satisfying the cyclicity relation 
\begin{equation}
\varphi(a_1,\ldots,a_n,a_0)
\;=\;
(-1)^n \varphi(a_0,a_1,\ldots,a_n)
\;,
\end{equation}
which play the same role as the differential forms in the classical de Rham theory. The equivalent of the exterior derivative is played by the Hochschild coboundary map 
\begin{align*}
b \varphi(a_0,a_1,\ldots,a_{n+1}) 
\;=\; & \sum_{j=0}^n (-1)^j \varphi(a_0,\ldots, a_j a_{j+1}, \ldots a_{n+1}) \bigskip \nonumber \\
& +(-1)^{n+1} \varphi(a_{n+1}a_0, \ldots, a_n) \;. 
\end{align*}
Note that indeed $b \circ b =0$. The cyclic cohomology of $\Aa$ is defined as the cohomology of the complex
$$
\ldots \;\stackrel{b}{\longrightarrow} \;\Cc ^{n-1}(\Aa) \;\stackrel{b}{\longrightarrow}\; \Cc^n(\Aa) \;\stackrel{b}{\longrightarrow}\; \ldots
\;,
$$
where $\Cc^n(\Aa)$ are the linear spaces of the cyclic $(n+1)$-linear functionals. The objects of the cyclic cohomology are the cyclic cocycles defined by
\begin{equation}
\ b \, \varphi \;=\;0\;, 
\qquad 
\varphi \in \Cc^n(\Aa)
\;, \;\;n\geq 0
\; .
 \end{equation}
They play the same role as the closed differential forms in the classical de~Rham cohomology. The cohomology class $[\varphi]$ contains all $\varphi' \in \Cc^n(\mathcal A)$ with $ \varphi' = \varphi + b \phi$ for some $\phi \in \Cc^{n-1}(\mathcal A)$. A cyclic cocycle $\varphi$ from $\Cc^n(\Aa)$ will be called $n$-cyclic cocycle and $\varphi$ will be called odd (even) if $n$ is an odd (even) integer. 

\vspace{0.2cm}

The domains of the cyclic cocycles need not be the entire algebra $\Aa$, but they must all include a dense Fr\'{e}chet sub-algebra $\mathscr{A}$ of $\mathcal A$ which is invariant under the holomorphic functional calculus. The terminology and its significance was already explained in Section~\ref{Sec-Sobolev} where this sub-algebra was called smooth, in analogy with the classical case (see \cite{Ren} for a detailed discussion). 

\begin{example}[Standard cocycles for the unital case]\label{UnitalStandardCocycle}
Let $\partial_1,\ldots,\partial_k$ be commuting derivations on a unital algebra $\Aa$ and let $\mathscr A$ be the smooth sub-algebra of Proposition~\ref{SmoothAlgebras}~(i). Assume the existence of a continuous trace such that $\Tt(\partial_j a)=0$ for all $a \in \mathscr A$ and $j=1,\ldots,k$. Then
\begin{equation}
\varphi(a_0,a_1,\ldots,a_k) 
\;=\; 
\sum_{\rho \in \Ss_k} (-1)^\rho \;\Tt\Big ( a_0 \prod_{i=1}^k \partial_{\rho_i} a_i \Big )
\end{equation}
satisfies $b\varphi =0$, hence it is a cyclic cocycle over $\Aa$ with domain $\mathscr A$. Above, $\Ss_k$ denotes the group of permutations and $(-1)^\rho$ the signature of the permutation. \hfill $\diamond$
\end{example}

\begin{example}[Standard cocycles for the non-unital case] \label{NonUnitalStandardCocycle}
Let $\partial_1,\ldots,\partial_k$ be commuting derivations on a non-unital algebra $\Aa$ and let $\mathscr A$ be a smooth sub-algebra, as in Proposition~\ref{SmoothAlgebras}~(ii). Assume the existence of a lower semicontinuous trace $\Tt$ such that $\Tt(\partial_j a)=0$ for all $a \in \mathscr A$ and $j=1,\ldots,k$. Extend the derivations and the trace over $\Aa^+$ by declaring $\partial(\one^+)=0$ and $\Tt(\one^+) = 1$. Then
\begin{equation}
\varphi(a_0,a_1,\ldots,a_k) 
\;=\; 
\sum_{\rho \in \Ss_k} (-1)^\rho \;\Tt\Big ( a_0 \prod_{i=1}^k \partial_{\rho_i} a_i \Big )
\end{equation}
is a cyclic cocycle over $\Aa^+$ with domain $\mathscr A^+$. Furthermore,
$$
\varphi(a_0,a_1,\ldots,a_k) 
\;=\; 
\varphi\big(a_0-s(a_0),a_1-s(a_1),\ldots,a_k-s(a_k)\big)
\;.
$$
Note that the scalar part of $a_1,\ldots,a_k$ can be dropped because $\partial(\one^+)=0$, and the scalar part of $a_0$ can be dropped for the same reason due to the cyclicity of $\varphi$. \hfill $\diamond$
\end{example}

\begin{remark} According to the work by Nest \cite{Nes, Nes2}, the above cyclic cocycles generate the entire (periodic) cyclic cohomology of the smooth non-commutative torus, hence of the smooth algebras of bulk and boundary observables.
\hfill $\diamond$
\end{remark} 

We now introduce the concept of pairing. In the classical de~Rham theory, a differential form defined over a smooth manifold can be integrated over a closed sub-manifold of appropriate dimension. If the form is closed, then the integral is invariant to smooth deformations of both the closed sub-manifold and of the closed differential form. As a result, the integral defines a paring between the cohomology class of the closed differential form and the homotopy class of the closed sub-manifold. The equivalent of all these in the non-commutative setting is the pairing between the cyclic cocycles and the classes of the $K$-groups. 
 
 \begin{theorem}[Pairing even cocycles with $K_0$-classes \cite{Con0}]\label{CyclicCocycles1}
Let $\varphi$ be an even cyclic cocycle over $\Aa^+$ with domain $\mathscr A^+$, and let $\tr \, \# \, \varphi$ be its natural extension over $\Kk \otimes \mathscr A^+$, where $\Kk$ is the algebra of compact operators. Then the map 
\begin{equation}\label{Cocycle2}
\mathcal P_\infty(\mathscr{A}^+) \ni e \;\;\mapsto \;\;(\tr  \, \# \,  \varphi) (e,\ldots,e) \in \mathbb C
\end{equation}
is constant on the equivalence class of $e$ in $K_0(\mathscr{A})$ $(=K_0(\Aa))$ and on the equivalence class of $\varphi$ in the cyclic cohomology. As such, there exists a pairing between  $K_0 (\mathcal A)$ and the even cyclic cohomology of $\mathcal A$,
\begin{equation}\label{Pairing1}
\boxed{
\;
\big \langle [\varphi],[e]_0-[s(e)]_0 \big \rangle 
\;=\;
(\tr  \, \# \,  \varphi) (e,\ldots,e)
\;,\;
}
\end{equation}
where on the r.h.s. it is understood that the representative for the class $[e]_0$ was chosen from the smooth sub-algebra $\mathscr{A}^+$. Moreover, the map 
$$
[e]_0-[s(e)]_0 \in K_0(\Aa)\mapsto \big \langle [\varphi],[e]_0-[s(e)]_0 \big \rangle\in\CM
$$
is a homomorphism of abelian groups.
\end{theorem} 

\begin{theorem}[Pairing odd cocycles with $K_1$-classes \cite{Con0}]\label{CyclicCocycles2}
Let $\varphi$ be an odd cyclic cocycle over $\Aa^+$ with domain $\mathscr A^+$, and let $\tr  \, \# \,  \varphi$ be its natural extension over $\Kk \otimes \mathscr{A}^+$. Then the map 
\begin{equation}\label{Cocycle1}
\mathcal U_\infty(\mathscr{A}^+) \ni v \;\;\mapsto\;\; 
(\tr  \, \# \,  \varphi) (v^\ast-\one,v-\one,\ldots,v^\ast-\one,v-\one)\; \in\; \mathbb C
\end{equation}
is constant on the equivalence class of $v$ in $K_1(\mathscr{A})$ $(=K_1(\Aa))$ and on the equivalence class of $\varphi$ in the cyclic cohomology. As such, there exists a natural pairing between  $K_1 (\mathcal A)$ and the odd cyclic cohomology of $\mathcal A$,
\begin{equation}\label{Pairing2}
\boxed{
\;
\big \langle [\varphi],[v]_1 \big \rangle 
\;=\;
(\tr  \, \# \,  \varphi) (v^\ast-\one,v-\one,\ldots,v^\ast-\one,v-\one) \;\in\; \mathbb C
\;,\;
}
\end{equation}
where on the r.h.s. it is understood that the representative for the class $[v]_1$ was chosen from the smooth sub-algebra $\mathscr{A}^+$. Moreover, the map 
$$
[v]_1\in K_1(\Aa)\mapsto \big \langle [\varphi],[v]_1 \big \rangle \in\CM
$$
is a homomorphism of abelian groups.
\end{theorem} 

As shown in Examples \ref{UnitalStandardCocycle} and \ref{NonUnitalStandardCocycle}, cyclic cocycles can be straightforwardly defined for both algebras of bulk and of boundary observables, using the non-commutative calculus presented in Section~\ref{Sec-DiffCalculus}. Let us point out that the above statements give no information about the range of the pairings in \eqref{Pairing1} and \eqref{Pairing2}, except that they are some countable subgroup of $\mathbb C$. In Section~\ref{sec-PairingsRange} these ranges will be determined explicitly.

\section{Bulk topological invariants defined}
\label{Sec-BulkTopInvariants}

Consider an ordered subset $I=\{i_1, \ldots,i_n\} \subset \{1,\ldots,d\}$, with order not necessarily the one induced by $\ZM$. We define $(n+1)$-cyclic cocycles $\xi_I: \mathcal W_{n,1}(\Aa_d,\Tt)^{\times{n+1}}\to \CM$ by
\begin{equation}
\xi_I(a_0,\ldots,a_n)
\; = \; \Lambda_n \sum_{\rho\in \Ss_n}(-1)^{\rho}
\;\Tt \Big (a_0\,\partial_{\rho_1}a_1\cdots \partial_{\rho_n}a_n\Big)
\;,
\label{XiBulkDef}
\end{equation}
where elements $\rho\in \Ss_n$ of the symmetric group are viewed as a bijective map from $\{1,\ldots,n\}$ onto $I$ with signature $(-1)^\rho$, and the normalization constants chosen as
\begin{equation}
\label{eq-normconst}
\Lambda_n
\;=\;
\frac{(2\I \pi)^{\frac{n}{2}}}{\frac{n}{2}!}
\quad \mbox{for }n\;\mbox{even}\,,
\qquad
\Lambda_n
\;=\;
\frac{\I\,(\I \pi)^\frac{n-1}{2}}{n!!}
\quad \mbox{for }n\;\mbox{odd}\,.
\end{equation}
The associated pairing of $\xi_I$ with $|I|$ even and odd respectively define the bulk Chern numbers of the projections and unitary elements, respectively:
$$
\Ch_I(e)
\;=\;
\;\big \langle [\xi_{I}],[e]_0-[s(e)]_0 \big \rangle 
\;,
\qquad
\Ch_I(v)
\;=\;
\big \langle [\xi_{I}],[v]_1 \big \rangle
\;.
$$
In previous works \cite{Con,Pim,KRS}, other normalization coefficients were used. Our present choices are as in \cite{PLB,PS} and assure that the top pairings are integer-valued in any dimension. 

\begin{remark} 
The cocycles $\xi_I$ can be shown to be continuous over the Sobolev space $\mathcal W_{n,1}(\Aa_d,\Tt)$, $n=|I|$, by using the non-commutative Holder inequality \eqref{Holder}.  Hence, we chose to define the cocycles from the beginning over their maximal domain of continuity, but we recall that the smooth sub-algebra $\mathscr A_d$ is contained in $\mathcal W_{n,1}(\Aa_d,\Tt)$. 
\hfill $\diamond$
\end{remark}

\begin{remark} 
When pairing the cocycles with the $K$-groups, the cocycles are extended over the matrix algebras as $\tr  \, \# \,  \xi_I$ (see \cite{Con0} for the standard procedure in the generic case). For the particular cocycles considered here, this amounts to replacing $\Tt$ by $\tr\otimes\Tt$ in the above definitions. This will tacitly be assumed for all observables algebras and suppressed in the notations, as already done in Section~\ref{Sec-DiffCalculus}. 
\hfill $\diamond$
\end{remark}

We now combine the above cyclic cocycles with the Fermi projections and the Fermi unitary operators. 

\begin{theorem}[The bulk invariants defined \cite{PLB,PS}]\label{Th-BulkInv}

\begin{enumerate}[\rm (i)]

\item  Let $h\in M_N(\mathbb C) \otimes \Aa_d$ be a finite hopping range bulk Hamiltonian and assume that BGH holds. If $p_F=\chi(h \leq \mu)$ denotes the Fermi projection and $I \subset \{1,\ldots,d\}$ is an ordered subset with $|I|$ even, then
\begin{equation}
\label{EvenBulkChernNumbers1}
\boxed{
\;
{\rm Ch}_I (p_F) \;=\; \Lambda_{|I|} \sum_{\rho\in \Ss_{|I|}}(-1)^{\rho}
\;\Tt \Big (p_F \prod_{j=1}^{|I|}\partial_{\rho_j}p_F \Big)
\;
}
\end{equation}
is a real number which remains constant under the continuous deformations of $h$ defined in Definition~\ref{ContDeformation}, as long as BGH holds.

\item Let $h \in M_{2N}(\mathbb C) \otimes \Aa_d$ be a finite hopping range bulk Hamiltonian and assume  that BGH and CH hold.  If $u_F$ is the Fermi unitary operator and $I \subset \{1,\ldots,d\}$ is an ordered subset with $|I|$ odd, then
\begin{equation}\label{OddBulkChernNumbers1}
\boxed{
\;
{\rm Ch}_I (u_F)  \;=\;  \Lambda_{|I|} \sum_{\rho\in \Ss_{|I|}}(-1)^{\rho}
\;\Tt \Big ( (u_F^\ast - \bm 1_{N}) \prod_{j=1}^{|I|} \partial_{\rho_j}u_F^{\ast_{j-1}} \Big)
\;
}
\end{equation}
is a real number which remains constant under the continuous deformations of $h$ defined in Defintion~\ref{ContDeformation}, as long as BGH holds.
\end{enumerate}
\end{theorem} 

\begin{remark} 
To simplify the notations, we used above $\ast_j$ for the $j$-fold convolution $\ast_j= \ast \circ \ldots \circ \ast$, equal to $\ast$ if $j$ is odd and to the identity map if $j$ is even. The notation will prove useful in several other places.
\hfill $\diamond$
\end{remark}

\noindent {\bf Proof.}  (i) Under BGH, the Fermi projection $p_F$ can be computed as a smooth function of $h$.  As a consequence, $p_F$ is an element of the C$^\ast$-algebra $\Aa_d$ and it defines a class in $K_0(\Aa_d)$. Furthermore, the Hamiltonian has a finite  range hopping. Hence, according to Proposition~\ref{FermiProjection1}, $p_F$ belongs to the smooth sub-algebra $\mathscr A_d$, and in fact to any of the Sobolev spaces $\mathcal W_{s,k}({\Aa_d},\Tt)$. Then Eq.~\eqref{EvenBulkChernNumbers1} is just the pairing $\big \langle [\xi_I],[p_F]_0 - [s(p_F)]_0 \big \rangle$ because $s(p_F)=0$. According to Proposition~\ref{FermiProjection2}, a continuous deformation of $h$ which does not violate BGH generates a homotopy of $p_F$ inside $\mathscr A_d$, hence inside a class of $K_0(\mathscr A_d)$. Then the statement follows from Theorem~\ref{CyclicCocycles1}. The proof of (ii) parallels the above, except that at the end one invokes Theorem~\ref{CyclicCocycles2}. 
\hfill $\Box$

\begin{remark} We stressed above the smoothening process because the C$^\ast$-algebras are stable only under the functional calculus with continuous functions. For example, if the Fermi level lies inside the energy spectrum, then the smoothening argument can no longer be applied and $p_F$ and $u_F$ are no longer elements of $\Aa_d$. However, if the Fermi level lies in a region of Anderson localized spectrum, then $p_F$ and $u_F$ remain inside the Sobolev spaces $\mathcal W_{n,1}({\Aa_d},{\cal T})$ according to Propositions~\ref{FermiProjection11} and \ref{FermiUnitary1}, and hence lie in the maximal domain of the cyclic cocycles $\xi_I$. This regime will be analyzed in Chapter~\ref{Chap-IndexTheorems}.
\hfill $\diamond$
\end{remark}

\begin{remark}
Let us recall some further aspects already stressed in Sections~\ref{Sec-UClassGeneral} and \ref{Sec-CUClassGeneral}. First of all, for periodic systems the invariants \ref{EvenBulkChernNumbers1} and \ref{OddBulkChernNumbers1} reduce to the expressions \eqref{EvenChernK} and \eqref{OddChernK} already used for solid state systems in prior works \cite{ASSS,RSFL}, and known from differential topology (over the torus). Furthermore, the invariants \ref{EvenBulkChernNumbers1} and \ref{OddBulkChernNumbers1} can be calculated from the covariant physical representations  $P_\omega = \pi_\omega(p_F)$ and $U_\omega = \pi_\omega(u_F)$ if one interprets the trace $\Tt$ as the trace per unit volume. The outcome is then $\PM$-almost surely constant, as already stressed in \eqref{EvenChernR} and \eqref{OddChernR}.
\hfill $\diamond$
\end{remark}

\section{Boundary topological invariants defined}
\label{Sec-BoundaryTopInvariants}

Let $\tilde I\subset \{1,\ldots,d-1\}$ be an ordered subset with the order not necessarily the one induced by $\ZM$. Extend the differential calculus $(\widetilde \partial, \widetilde \Tt)$ as in Example~\ref{NonUnitalStandardCocycle}. Then the $(n+1)$-linear maps $\widetilde \xi_{\tilde I}: \Ww_{n,1}(\Ee^+_d,\widetilde \Tt)^{\times{n+1}}\to \CM$ defined by
\begin{equation}\label{XiBoundaryDef}
\widetilde \xi_{\tilde I}(\tilde a_0,\ldots,\tilde a_n)
\; = \; 
\Lambda_{|\tilde I|}\;
\sum_{\rho\in \Ss_n}(-1)^{\rho}
\;\widetilde \Tt (\tilde a_0\widetilde \partial_{\rho_1} \tilde a_1\cdots \widetilde \partial_{\rho_n} \tilde a
_n)
\end{equation}
are cyclic cocycles over the algebra of boundary observables. The associated pairings with $K$-group elements then define the even and odd Chern numbers:
$$
\widetilde{\Ch}_{\tilde{I}}(e)
\;=\;
\;\big \langle [\widetilde \xi_{\tilde{I}}],[\tilde{e}]_0-[s(\tilde{e})]_0 \big \rangle 
\;,
\qquad
\widetilde{\Ch}_{\tilde{I}}(\tilde{v})
\;=\;
\big \langle [\widetilde{\xi}_{\tilde{I}}],[\tilde{v}]_1 \big \rangle
\;.
$$
As before, the cocycles and hence the pairings are defined over their maximal domains and we recall that $\mathscr E_d \subset \Ww_{n,1}(\Ee_d,\widetilde \Tt)$ hence $\mathscr E^+_d \subset \Ww_{n,1}(\Ee^+_d,\widetilde \Tt)$. We now combine these cocycles with the boundary unitary operator $\tilde u_\Delta$ of \eqref{TildeU} and with the chiral boundary projection $\tilde p_\Delta$ of \eqref{BoundaryProjection1}.

\begin{theorem}[The boundary invariants defined]\label{Th-BoundaryInvariants}
\begin{enumerate}[\rm (i)]

\item Let $\hat h = (h,\tilde h)\in M_N(\mathbb C) \otimes \widehat \Aa_d$ be a half-space Hamiltonian of finite hopping range such that BGH holds.  If $\tilde I \subset \{1,\ldots,d-1\}$ is an ordered subset with $|\tilde I|$ odd, then
\begin{equation}\label{OddBoundaryChernNumbers1}
\boxed{\;
\widetilde {\rm Ch}_{\tilde I} (\tilde u_\Delta)  \;= \; 
\Lambda_{|\tilde I|} \sum_{\rho\in \Ss_{|\tilde I|}}(-1)^{\rho}
\;\widetilde \Tt \Big ( (\tilde u_\Delta^\ast - \bm 1_{N}) \prod_{j=1}^{|\tilde I|} \widetilde \partial_{\rho_j}\tilde u_\Delta^{\ast_{j-1}} \Big)
\;}
\end{equation}
is a real number which remains constant under the continuous deformations of $\hat h$ defined in Definition~\ref{ContDeformation}, provided BGH continues to hold.

\item Let $\hat h = (h,\tilde h) \in M_{2N} (\mathbb C) \otimes \widehat \Aa_d $ be a half-space Hamiltonian of finite hopping range such that BGH and CH hold.  If $\tilde I \subset \{1,\ldots,d-1\}$ is an ordered subset with $|\tilde I|$ even, then
\begin{equation}\label{EvenBoundaryChernNumbers1}
\boxed{
\;\widetilde {\rm Ch}_{\tilde I} (\tilde p_\Delta)  \;= \; \Lambda_{|\tilde I|} \sum_{\rho\in \Ss_{|\tilde I|}}(-1)^{\rho}
\;\widetilde \Tt \Big (\tilde p_\Delta \prod_{j=1}^{|\tilde I|} \widetilde \partial_{\rho_j}\tilde p_\Delta \Big)
\;}
\end{equation}
is a real number which remains constant under the continuous deformations of $\hat h$ defined in Definition~\ref{ContDeformation}, provided BGH and CH continue to hold.

\end{enumerate}
\end{theorem} 

\noindent {\bf Proof.}  We recall the discussion below Eq.~\eqref{TildeU} where it was shown that $\tilde u_\Delta - \one_N$ belongs to the smooth algebra $M_N(\mathbb C) \otimes \mathscr E_d$. A similar conclusion was achieved in Proposition~\ref{IndMap} for the chiral boundary projection. Therefore Eqs.~\eqref{OddBoundaryChernNumbers1} and \eqref{EvenBoundaryChernNumbers1} 
 are just the pairings $\big \langle [\widetilde \xi_{\tilde I}],[\tilde u_\Delta]_1 \big \rangle$ and $\big \langle [\widetilde \xi_{\tilde I}],[\tilde p_\Delta]_0 -[s(\tilde p_\Delta)]_0 \big \rangle$ for $|\tilde I|$ odd or even, respectively. According to Proposition~\ref{HFSmoothFunc2}, any continuous deformation of $\hat h$ generates a homotopy of $\tilde u_\Delta - \bm 1_N$ and $\tilde p_\Delta$ inside the smooth algebra. Then the statements follow from Theorems~\ref{CyclicCocycles1} and \ref{CyclicCocycles2}.\hfill $\Box$

\begin{remark}
The boundary invariants can also be expressed in terms of the physical observables. This follows directly from the above definitions, the canonical representation defined in Sections~\ref{SubSec-HSRep} and the representation \eqref{TracePerArea} of the trace per unit area $\widetilde \Tt$. For example, in the case of odd $|\tilde I|$, let  $\widetilde U_\omega = \widetilde \pi_\omega(\tilde u_\Delta)$ be the physical representations at a disorder configuration $\omega$. Then 
$$
\widetilde {\rm Ch}_{\tilde I} (\tilde u_\Delta) \;=\;  \Lambda_{|\tilde I|} \sum_{\rho\in \Ss_{|\tilde I|}}(-1)^{\rho}
\;\widetilde \Tt \Big ((\widetilde U_\omega^\ast - \one) \prod_{j=1}^{|\tilde I|}\I [\widetilde U_\omega^{\ast_{j-1}},X_{\rho_j}] \Big)
\;,
$$
$\PM$-almost surely.
\hfill $\diamond$
\end{remark}

\begin{example} Let $d$ be even. We demonstrate here that, in the periodic case, the odd $\widetilde{\rm Ch}_{d-1}(\tilde u_\Delta)$  for $d$ even reduces to the quantity $\chi \sum \nu_W$ defined in \eqref{BB2} of Section~\ref{Sec-UClassGeneral} in terms of the chiralities of the Weyl points:
\begin{equation}\label{BI1}
\widetilde{\rm Ch}_{d-1}(\tilde u_\Delta) 
\;=\; 
\chi\,\sum_W \nu_W.
\end{equation}
The computation is restricted to $d=4$ (but the generalization is possible, see \cite{Kub}) and translation invariance is assumed in the directions parallel to the boundary. In this case, $\chi = -1$. Let there be a Weyl point in the boundary spectrum, assumed to be isolated from the other possible Weyl singularities. We may assume $E^W=0$ and $k^W=0$ without loss of generality. Further let us choose $\FFunc$ such that $\FFunc(E) =0$ for $E \leq - \delta$ and $\FFunc(E)=1$ for $E \geq \delta$, with $\delta$ arbitrarily small.  The computation of the boundary invariant involves only the band spectrum inside $[-\delta, \delta]$ and, since $\delta$ is arbitrarily small, we can use the Weyl Hamiltonian \eqref{eq-Weyl} to describe the bands connected at the Weyl point. By a change of variables $k_j \rightarrow \frac{k_j}{v_j}$, we can reduce the Weyl Hamiltonian to $\langle k, \sigma \rangle$, $k \in \mathbb R^3$. In the process, a sign factor appears, equal precisely to $\nu_W$. After a Fourier transform, $\widetilde{\rm Ch}_{d-1}(\tilde u_\Delta)$ becomes
$$
\widetilde{\mathrm{Ch}}_{d-1} (\tilde u_\Delta)
\;=\;
-\,\frac{\nu_W}{24\pi^2} \int d^3 k \sum_{\rho \in \Ss_3}(-1)^\rho \, \mathrm{tr} \Big ( \prod_{j=1}^3 \widetilde U(k)^{-1}\partial_{\rho_j} \widetilde U(k) \Big )
\;,
$$
with $\widetilde U(k) = - e^{2\pi \I \FFunc (\langle k,\sigma\rangle )}$, where we inserted a harmless minus sign. It is convenient to make the change of variable $\kappa = \frac{k}{\delta}$. Since $\widetilde U(\kappa)=\one$ for $|\kappa| \geq 1$, one can view $\widetilde U(\kappa)$ as a map from the three-dimensional unit ball with its boundary $|\kappa |= 1$ identified with a point, to the SU$(2)$ group which is parametrized precisely by this space. Then, as noted in \cite{WQZ} in a different context, the degree of this map is well defined and is equal to the r.h.s. of the above equation, times $\nu_W$. As the degree is defined for any continuous function, one can deform $\FFunc $ from a smooth to a continuous map which we choose to be $\FFunc (E) = \tfrac{1}{2}(1 + \frac{E}{\delta})$ for $|E|<\delta$, and $\FFunc (E)=0$ for $E \leq - \delta$ and $\FFunc (E) = 1$ for $E \geq \delta$. Using the new variable, the computation reduces to finding the degree of the map
$$
\widetilde U(\kappa) 
\;=\; 
-\,e^{\I \pi (1+ \langle \kappa ,\sigma \rangle )} 
\;=\; 
\cos \big (\pi |\kappa | \big )\, + \,\I \sin \big (\pi |\kappa| \big) \left \langle \frac{\kappa}{|\kappa|} ,\sigma \right \rangle
\;.
$$
But this is just the inverse of the standard parametrization of SU$(2)$, hence a homeomorphism of degree $-1$. The calculation can then be repeated for the rest of the Weyl points and the conclusion will be $\widetilde{\rm Ch}_d (\tilde u_\Delta)= - \sum \nu_W$. This shows that the l.h.s. of \eqref{BB2} is indeed minus the boundary invariant of \eqref{OddBoundaryChernNumbers1}. That it is indeed connected to the bulk invariant will follow from the results of Section~\ref{sec-PairingsDuality}.
\hfill $\diamond$
\end{example}

\begin{example} Let $d$ be odd. We demonstrate here that, for the periodic case, the even $\widetilde{\rm Ch}_{d-1}(\tilde p_\Delta)$ reduces (up to a sign) to the one defined in Section~\ref{Sec-CUClassGeneral} in terms of the chiralities of the Dirac points:
\begin{equation}
\label{BI2}
\widetilde{\rm Ch}_{d-1}(\tilde p_\Delta) \;=\; -\,\chi \, \sum_D \nu_D.
\end{equation}
The computation is restricted to $d=3$ (hence $\chi = - 1$) and translation invariance is assumed in the directions parallel to the boundary. The assumptions and the settings are the same as in the previous example and we start directly with the computation of $\widetilde{\rm Ch}_{d-1}(\tilde p_\Delta)$ for the chiral boundary projection
$$
\widetilde P(k) \;=\;  e^{-\I \frac{\pi}{2} \GFunc(\widehat H_k)}{\rm diag}(\one,0)e^{\I \frac{\pi}{2} \GFunc(\widehat H_k)}
\;,
$$
with $\widehat H_k$ a Dirac Hamiltonian of positive chirality
$$
\widehat H_k \;=\; \begin{pmatrix} 0 & k_1-\I k_2 \\  k_1+\I k_2 & 0 \end{pmatrix},
$$
and $\GFunc$ an odd function such that $\GFunc = \pm 1$ above/below an interval $[-\delta, \delta]$. We have
$$
\widehat H_k \psi_\pm (k) 
\;=\; 
\pm r \, \psi_\pm(k)
\;, 
\qquad 
\psi_\pm(k) \;=\; \tfrac{1}{\sqrt{2}}\begin{pmatrix}
1 \\ \pm \ e^{-\I \alpha} \end{pmatrix}
\;,
$$
where $k_1+\I k_2 = r e^{\I \alpha}$, and note that $\widehat J \psi_\pm (k) = \psi_\mp (k)$. Then
$$
\widetilde P(k) \;=\;  
|\varphi(k)\rangle \langle \varphi(k) |
\;,
\qquad
\varphi(k)\;=\;\tfrac{1}{\sqrt{2}}\Big ( \psi_+(k) + e^{\I \pi \GFunc(r)} \psi_-(k) \Big )
\;,
$$
which can be verified by a direct computation. More explicitly,
$$
\varphi(k)
\; =\; 
 \tfrac{1}{2}\begin{pmatrix} 1 + e^{\I \pi \GFunc(r)} \\
 \big( 1-e^{\I \pi \GFunc(r)} \big ) \ e^{-\I \alpha} 
\end{pmatrix}
\;,
\qquad  
\varphi'(k) 
\;=\;
\tfrac{1}{2}\begin{pmatrix} \big ( 1 + e^{\I \pi \GFunc(r)} \big ) \ e^{\I \alpha} \\
1-e^{\I \pi \GFunc(r)}  
\end{pmatrix}
\;,
$$
where we displayed two expressions which differ by just a gauge factor. The first expression has a limit as $k\rightarrow 0$ and the second one has a limit as $k\rightarrow \infty$. We now proceed as
$$
\mathrm{Ch}_2(\widetilde P) 
\;=\; 
-\tfrac{1}{2 \pi \I}\left(\int_{|k| \, \leq \, R} +\int_{|k| \, \geq \, R}\right)  \mathrm{tr}\Big ( \widetilde P(k) \ {\bf  d} \widetilde P(k) \wedge {\bf d} \widetilde P(k) \Big )
\;,
$$
where $\bf d$ is the exterior derivative, and we apply Stokes' theorem to continue
\begin{align*}
\mathrm{Ch}_2(\widetilde P) & \;=\; -\tfrac{1}{2 \pi \I} \; \int\limits_{|k|=R} \Big [ \big \langle \varphi(k) | {\bf d}  \varphi(k) \big \rangle -  \big \langle \varphi'(k) | {\bf d}  \varphi'(k) \big \rangle \Big ] \\
& \;=\; -\tfrac{1}{2\pi \I} \; \int_0^{2\pi} d \alpha \Big ( \big \langle \varphi(R e^{\I \alpha}) | \partial_\alpha  \varphi(R e^{\I \alpha}) \big \rangle - \big \langle \varphi'(R e^{\I \alpha}) | \partial_\alpha  \varphi'(R e^{\I \alpha}) \big \rangle \Big) .
 \end{align*}
Using the explicit expressions of $\varphi$ and $\varphi'$, the integrant can be seen to be independent of $\alpha$ and
$$
\mathrm{Ch}_2(\widetilde P) 
\;=\;
-\tfrac{1}{4}\Big ( - \big |1-e^{\I \pi \GFunc(R)} \big |^2 - \big |1 + e^{\I \pi \GFunc(R)}\big |^2\Big ) 
\;=\; 1 \;=\; -\chi\, \nu_D\;.
$$
For a Dirac Hamiltonian of negative chirality,  we only need to change $\alpha$ into $-\alpha$ everywhere in the above calculations, whose effect will be a change of sign for the Chern number, as expected. 
\hfill $\diamond$
\end{example}

We end this section by introducing more precise terminology and conventions, which apply to both bulk and boundary invariants:

\begin{enumerate}[1)]

\item In analogy to the classical differential geometry (compare \eqref{EvenChernK} and \eqref{OddChernK} and the classical expressions from \cite{Par}), we call the invariants the even or odd Chern numbers. 

\item The cyclic cocycles \eqref{XiBulkDef} and \eqref{XiBoundaryDef} will be called the bulk and boundary Chern cocycles.

\item If $|I|$ or $|\tilde I|$ equals the maximum even (odd) number allowed by the space dimension, then we call the pairings with ${\rm Ch}_I$ or $\widetilde {\rm Ch}_{\tilde I}$ a top even (odd) Chern number. The rest of the invariants will be called lower Chern numbers. 

\item If $I=\{1,\ldots,d\}$ or $\tilde I = \{1,\ldots,d-1\}$ with order induced by $\ZM$, then parings with ${\rm Ch}_I$ and $\widetilde {\rm Ch}_{\tilde I}$ will be called the strong Chern numbers and will simply be denoted by ${\rm Ch}_d$ and $\widetilde {\rm Ch}_{d-1}$, respectively. The rest of the invariants are called the weak Chern numbers.
\end{enumerate}

\begin{remark}
The weak Chern numbers are not integer valued except if $|I|=d-1$ and $|\tilde{I}|=d-2$, see Section~\ref{sec-PairingsRange}. The strong Chern numbers are the only invariants for which an index theorem holds (which is proved in Chapter~\ref{Chap-IndexTheorems}). This allows to define the strong invariants also in the more general conditions of MBGH, which is not the case for the weak invariants.
\hfill $\diamond$
\end{remark}

\section{Suspensions and the Volovik-Essin-Gurarie invariants}
\label{Sec-Suspension}

In this section we use the connecting maps for suspensions to derive equivalent expressions of the bulk topological invariants, in terms of the resolvent functions of the Hamiltonians. These expressions are of interest because the resolvent function, hence the topological invariants, can be generalized via the one-particle Green's function \cite{Mah} to models which include the electron-electron interaction \cite{Gur}. One should be aware, though, that the one-particle Green's functions can display singularities which are not well understood at this time, and that the problem of defining the topological invariants in the presence of electron-electron interaction remains an open problem.   

\vspace{.2cm}

Let us now consider the $n$-cocycle $\xi_I$ given in \eqref{XiBulkDef}, corresponding to a set of indices $I \subset \{1,\ldots,d\}$ with $|I| = n$. Its suspension $\xi_I^s:\Ww^{n,1}(S \Aa_d,\Tt^s)^{\times{(n+2)}}\to \CM $ is the following cyclic $(n+2)$-multilinear map over the suspension algebra $S \Aa_d$
$$
\xi_I^s(a_{0}, \ldots,a_{n+1}) 
\;=\;  
\Lambda_{n+1} \sum_{\sigma \in \Ss_{n+1}}(-1)^{\sigma}  \Tt^s \Big (a_{0,t} \prod_{i=1}^{n+1} \partial_{\rho_i}a_{i,t} \Big )\;, 
\qquad  
a_{i}\,=\,\{a_{i,t}\}_{t\in \TM}
\;.
$$
Here, elements $\sigma\in \Ss_{n+1}$ are viewed as maps from $\{1,\ldots, n+1\}$ to $\{0\} \cup I$, furthermore $\partial_0 = \partial_t$, and $\Tt^s$ denotes the following trace on $S \Aa_d$
$$\Tt^s (a) 
\;=\; 
\int^{2\pi}_0\frac{dt}{2\pi}\,\Tt ( a_t )
\;,
\qquad
a\,=\,\{a_t\}_{t\in(0,2\pi)}\in S\Aa_d
\;.
$$
Let us write the suspended cocycle more explicitly:
\begin{align*}
\xi_I^s(a_0, \ldots,a_{n+1}) 
\; = & \; \Lambda_{n+1} \sum_{j=1}^{n+1}(-1)^{j-1} \sum_{\rho\in \Ss_n}(-1)^{\rho} 
\\
& \;\;\;\;
\Tt^s \Big (a_{0,t}\Big (\prod_{i=1}^{j-1} \partial_{\rho_i}a_{i,t} \Big) \partial_t a_{j,t} \Big ( \prod_{i=j}^{n}\partial_{\rho_i} a_{i+1,t} \Big ) \Big )
\;.
\end{align*} 
The following result, due to Pimsner \cite{Pim}, shows that the suspension of cocycles is dual to the suspension maps $\theta$ and $\beta$ of $K$-theory. The proof given below is considerably streamlined. It also extends to suspensions of pairings over the edge algebra, if adequate trace class conditions are imposed. Actually, the same proof carries over to any cocycle over some C$^*$-algebra that is obtained as in Examples \ref{UnitalStandardCocycle} or \ref{NonUnitalStandardCocycle}.

\begin{theorem}
\label{theo-suspension} 
For $|I|$ even,
\begin{equation}
\label{eq-pairsuspeven}
\boxed{
\;\big \langle [\xi_{I}],[e]_0-[s(e)]_0 \big \rangle \;=\; -\,\big \langle {[\xi}_I^s],\beta [e]_0 \big \rangle
\;,
\;}
\end{equation}
while for $|I|$ odd,
\begin{equation}
\label{eq-pairsuspodd}
\boxed{
\;\big \langle [\xi_{I}],[v]_1 \big \rangle
\;=\;
\big \langle {[\xi}_I^s],\theta [v]_1 \big \rangle
\;.
\;}
\end{equation}
\end{theorem}

\noindent {\bf Proof.} Let $|I|=2k$ and let $e \in \mathcal P_N(\Aa_d^+)$. Further let 
$$
v\;=\;(\one-e)\;+\;{u}e \in (S\Aa_d)^+
\;,
$$ 
with $u=\{e^{\I t}\}_{t\in [0,2\pi]}$, so that $\beta[e]_0=[v]_1$. Then $\partial_0 v =\I u e$ and $e \partial_i e=e\partial_i e (\one_N-e)$ for $i \geq 1$, so that $e(\prod_{i=1}^{j-1}\partial_{\rho_i}e)e=0$ for even $j$, while for odd $j$ the $e$ on the r.h.s. can be dropped. Using these facts and $e u=u e$, one can evaluate 
\begin{align*}
\big \langle [\xi_I^s],[v]_1 \big \rangle
& \;=\;
\I (k+1)\Lambda_{2k+1} 
\sum_{\rho\in \Ss_{2k}}(-1)^{\rho}
 \\
& \; \; \;\;\; \;\;\;
\int^{2\pi}_0\frac{dt}{2\pi} \, \Tt\left(
\,u(u-\one)^k(u^*-\one)^{k+1}
\,e\prod_{j=1}^{2k}\partial_{\rho_j}e\right)
\\
& \;=\;
\I\,\Lambda_{2k+1}
\,(k+1)\,\binom{2k+1}{k}\;
\sum_{\rho\in \Ss_{2k}}(-1)^{\rho}
\;\Tt\left(
e\prod_{j=1}^{2k}\partial_{\rho_j}e\right)
\\
& \;=\;
\I\,\Lambda_{2k+1}
\,(k+1)\,\binom{2k+1}{k}\;
\frac{1}{\Lambda_{2k}}\;
\big \langle [\xi_{I}],[e]_0 - [s(e)]_0 \big \rangle
\;,
\end{align*}
where we used
$$
\int^{2\pi}_0\frac{dt}{2\pi} \;u(u-\one)^k(u^*-\one)^{k+1}
\;=\;\binom{2k+1}{k}
\;.
$$
The first statement then follows from \eqref{eq-normconst} which implies
\begin{equation}
\label{eq-ConstId}
\I (k+1)  \Lambda_{2k+1} \begin{pmatrix} 2k+1 \\ k \end{pmatrix} 
\;=\; 
-\,\Lambda_{2k}
\;,
\end{equation}

\vspace{.1cm}

For \eqref{eq-pairsuspodd}, let $|I|=n=2k-1$ and $v\in \mathcal U_N(\Aa_d^+)$. Recall the $\theta$-map from Section~\ref{Chap-KTheory}:
$$
\theta [v]_1 \;=\;[e_t]_0-[{\rm diag}(\bm \one_N,0_N)]_0
\;, 
\qquad 
e_t \;=\; r_t \, p_t \, r_t^\ast
\;,
$$
with 
$$
p_t
\;=\; 
\begin{pmatrix}
c^2 \bm \one_N & c s v^\ast \\
c s v & s^2 \bm \one_N
\end{pmatrix}
\;, 
\qquad 
c=\cos (\tfrac{t}{4})\;, \; \; s = \sin( \tfrac{t}{4})\;,
$$
and $r_t$ is the rotation matrix given in \eqref{RChoice}. In the following we will drop the explicit dependence of $e$, $r$ and $p$ on $t$, as already done for $c$ and $s$. Then
$$
\big \langle [\xi_I^s],\theta [v]_1 \big \rangle  \;=\;  \Lambda_{2k}\sum_{j=1}^{2k}(-1)^{j-1} 
\!\!
\sum_{\rho\in \Ss_{2k-1}}
\!\!
(-1)^{\rho} \;
\Tt^s \Big (e \Big ( \prod_{i=1}^{j-1} \partial_{\rho_i}e \Big ) \partial_t e\Big ( \prod_{i=j}^{2k-1}\partial_{\rho_i}e \Big )\Big ).
$$
Next let us collect some useful identities
$$
\partial_{\rho_i} e \;=\; r (\partial_{\rho_i} p) r^\ast\;, 
\qquad 
i=1,\ldots,2k-1
\;,
$$ 
and
$$
\partial_t e
\; =\; 
r \big ( \partial_t p + r^\ast  (\partial_t r) \, p - p \, r^\ast \partial_t r \big ) r^\ast
\;.
$$
Note that the flanking by $r$ and $r^\ast$ can be dropped above because of the cyclic property of the trace. Another identity is
$$
r^\ast \partial_t r \;=\;
-\,\tfrac{1}{4}\begin{pmatrix}
0_N & - \bm \one_N \\
\bm \one_N & 0_N
\end{pmatrix}
\;.
$$
As $p$ is a projection itself, we can use the identities mentioned in the proof of \eqref{eq-pairsuspeven} to move $p$ around inside the trace, and to conclude that 
\begin{align*}
\big \langle [\xi_I^s], \theta [v]_1 \big \rangle
\; = & \; \Lambda_{2k}\sum_{j=1}^{2k}(-1)^{j-1} \sum_{\rho\in \Ss_{2k-1}}(-1)^{\rho} \\
& \;\;\;\;
\Tt^s  \left  (  p \Big ( \prod_{i=1}^{j-1} \partial_{\rho_i}p \Big ) \left  ( \partial_t p - \tfrac{(-1)^j}{4}
\begin{pmatrix}
0_N & - \bm \one_N \\
\bm \one_N & 0_N
\end{pmatrix} \right )
   \Big ( \prod_{i=j}^{2k-1} \partial_{\rho_i}p \Big ) \right )
\;.
\end{align*}
Let us split the above pairing into two summands
$$   
T_1 
\;=\;
\Lambda_{2k}\;\sum_{j=1}^{2k}(-1)^{j-1} \sum_{\rho\in \Ss_{2k-1}}(-1)^{\rho}
\, \Tt^s \left (p \Big ( \prod_{i=1}^{j-1} \partial_{\rho_i}p \Big )  \partial_t p \Big ( \prod_{i=j}^{2k-1} \partial_{\rho_i}p \Big ) \right )
$$
and
$$
T_2 
\;=\; 
\tfrac{1}{4}\,\Lambda_{2k}\;
\sum_{j=1}^{2k} \sum_{\rho\in \Ss_n}(-1)^{\rho}  \,\Tt^s \left  ( p \Big ( \prod_{i=1}^{j-1} \partial_{\rho_i}p \Big )
\begin{pmatrix}
0_N & - \bm \one_N \\
\bm \one_N & 0_n
\end{pmatrix}
\Big ( \prod_{i=j}^{2k-1} \partial_{\rho_i}e_t \Big ) \right ) 
\;.
$$

The second term $T_2$ is identically zero. Indeed, note first that
$$
\partial_{\rho_i} p 
\;=\; 
cs \begin{pmatrix}
0_N  &  \partial_{\rho_i} v^\ast \\
\partial_{\rho_i} v & 0_N
\end{pmatrix}.
$$
Then inside the last trace we have the projection $p$ followed by $2k$ off-diagonal matrices, hence the off-diagonal part of $p$ leads to an overall trace-less term and only the diagonal part of  $p$ needs to be taken into account. But the diagonal part of $p$ is $p_{\rm diag}= {\rm diag}(c^2 \one_N, s^2 \one_N)$, which commutes with the derivations $\partial_i$. Then
\begin{align*}
T_2\;=\;
& 
\tfrac{1}{4}\,\Lambda_{2k}\sum_{j=1}^{2k}(-1)^{j-1} \sum_{\rho\in \Ss_n}(-1)^{\rho} \int^{2\pi}_0\! \frac{dt}{2\pi} \cdot\\
&  \;\;\;\cdot\Tt  \left  ( \partial_{\rho_1} \left ( 
p_{\rm diag} \, p \, \partial_{\rho_2}p \cdots  \partial_{\rho_{j-1}}p
\begin{pmatrix}
0_N & - \bm \one_N \\
\bm \one_N & 0_N
\end{pmatrix}
\partial_{\rho_j} p  \cdots \partial_{\rho_{2k-1}}p \right )\right )
\;,
\end{align*}
because, when applying the Leibniz rule for $\partial_{\rho_1}$, all the terms containing $\partial_{\rho_1} \partial_{\rho_i} p$ cancel out identically due to the anti-symmetrizing factor $(-1)^\rho$. The trace of a total derivation vanishes by Proposition~\ref{Indenties1}, so $T_2=0$.  As for $T_1$, let us first note 
$$
\partial_t p
\;=\;
\tfrac{1}{4}\,\begin{pmatrix}
-2cs \, \one_N & (c^2-s^2)\,v^* \\ (c^2-s^2)\,v & 2cs  \, \one_N
\end{pmatrix}
\;,
$$
so that
\begin{align*}
T_1\;
& = \; \Lambda_{2k}
\sum_{j=1}^{2k}  (-1)^{j-1}
\sum_{\rho\in \Ss_{2k-1}}(-1)^{\rho}
\,\int^{2 \pi}_0\frac{dt}{8\pi} \,(cs)^{2k-1}\cdot
\\
& \;\;\;\;\;\;\; \cdot
\Tt
\left(
p\prod_{i=1}^{j-1}
\begin{pmatrix}
0_N & \partial_{\rho_i}v^*\\
\partial_{\rho_i}v & 0_N
\end{pmatrix}
\begin{pmatrix}
-2cs \, \one_N & (c^2-s^2)\,v^* \\ 
(c^2-s^2)\,v & 2cs \, \one_N
\end{pmatrix}
\prod_{i=j}^{2k-1}
\begin{pmatrix}
0_N & \partial_{\rho_i}v^*\\
\partial_{\rho_i}v & 0_N
\end{pmatrix}
\right)
\;.
\end{align*}
Now one can use the following anti-commutation relations
$$
\left\{
\begin{pmatrix}
0_N & \partial_i v^*\\
\partial_i v & 0_N
\end{pmatrix}
,
\begin{pmatrix}
\one_N & 0_N\\
0_N & -\one_N
\end{pmatrix}
\right\}
\;=\;0\;,
\qquad
\left\{
\begin{pmatrix}
0_N & \partial_{i}v^*\\
\partial_{i}v & 0_N
\end{pmatrix}
,
\begin{pmatrix}
0_N & v^*\\
v & 0_N
\end{pmatrix}
\right\}
\;=\;0\;,
$$
together with the explicit expression for $p$ to further simplify
\begin{align*}
T_1 \;  =\;
\Lambda_{2k}\; &
\sum_{j=1}^{2k}
\sum_{\rho\in \Ss_{2k-1}}(-1)^{\rho}
\,\int^{2 \pi}_0\frac{dt}{8\pi} \,(cs)^{2k-1}\cdot \\
& \cdot\Tt
\left(
\begin{pmatrix}
-cs \, \one_N & c^2\,v^* \\ -s^2\,v & cs \, \one_N
\end{pmatrix}
\prod_{i=1}^{2k-1}
\begin{pmatrix}
0_N & \partial_{\rho_i}v^*\\
\partial_{\rho_i}v & 0_N
\end{pmatrix}
\right)
\;.
\end{align*}
The product of matrices inside the trace is off-diagonal and therefore the diagonal term of the first factor inside the trace does not contribute (namely the one with a factor $cs$). Then
\begin{align*}
T_1
& \; =\;
k\,\Lambda_{2k}
\sum_{\rho\in \Ss_{2k-1}}(-1)^{\rho}
\,\int^{2 \pi}_0\frac{dt}{4\pi} \,(cs)^{2k-1}\cdot\, \\
& 
\;\;\;\;\;\;
\cdot\left[
\,c^2\,\Tt\left(
v^*\partial_{\rho_1}v
\prod_{j=1}^{k}
\partial_{\rho_{2j}}v^*\partial_{\rho_{2j+1}}v\right)
\,-\,s^2
\,\Tt\left(
v\partial_{\rho_1}v^*
\prod_{j=1}^{k}
\partial_{\rho_{2j}}v\partial_{\rho_{2j+1}}v^*\right)
\right]
\\
&
\;=\;
\frac{k\,\Lambda_{2k}}{\Lambda_{2k-1}}
\int^{2 \pi}_0\frac{dt}{4\pi} \,(cs)^{2k-1}\,
\left[\,c^2\,
\big \langle [{\xi}_I],[v]_1 \big \rangle
\;-\,s^2\,\big \langle [{\xi}_I],[v^*]_1 \big \rangle
\right]
\;.
\end{align*}
The pairing is a group homomorphism from the $K_1$-group, hence
$$
0\;=\; 
\big \langle [{\xi}_I],[\bm \one_N]_1 \big \rangle 
\;=\; 
\big \langle {[\xi}_I],[v v^*]_1 \big \rangle
\;=\;
\big \langle {[\xi}_I],[v]_1 \big \rangle \;+\; \big \langle [{\xi}_I],[v^\ast]_1 \big \rangle
\;,
$$
so that $\big \langle [{\xi}_I],[v^*]_1 \big \rangle=- \big \langle [{\xi}_I],[v]_1 \big \rangle$. Now using this fact and the integral 
$$
\int^{2\pi}_0\frac{dt}{4\pi}\;(cs)^{2k-1}\;=\;\frac{1}{2^{2k-1}}\,\frac{1}{\pi}\,\frac{(2k-2)!!}{(2k-1)!!}
\;=\;\frac{1}{2^{k}}\,\frac{1}{\pi}\,\frac{(k-1)!}{(2k-1)!!}
\;,
$$
one can conclude the proof by using the expressions for $\Lambda_n$ in \eqref{eq-normconst}. \hfill $\Box$

\vspace{.2cm}

As an application of the above result, we provide two expressions of the bulk invariants in even dimensions in terms of the resolvent function. This will establish a link with the invariants introduced by Volovik \cite{Vol} for $d=2$ and arbitrary $d$ by Essin and Gurarie \cite{EG1,EG2}.  

\begin{theorem}
\label{theo-VEG} 
Consider the settings and the notations of Proposition~\ref{prop-Greenlink} and let $t\in[0,2\pi]\mapsto z(t)$ be a parametrization of the loop $\Gamma_F$ such that $z(0)$ belongs to the bulk spectral gap. Assume $|I|=2k$. Below, each $\sigma$ is seen as a bijection from $\{1,\ldots,2k+1\}$ onto $\{0\} \cup I$.

\begin{enumerate}[\rm (i)]
\item Consider the unitary operator
$$
q \;=\; \big [t \in [0,2\pi] \rightarrow q_z=(h-\bar z)(h-z)^{-1}\big ]\; \in\; \Uu_N\big ((S \Aa_d)^+\big ).
$$
Then
$$
\Ch_{I}(p_F)
\;=\;
-\,\tfrac{1}{2}\, \Lambda_{2k+1}
\,
\sum_{\sigma\in \Ss_{2k+1}}(-1)^{\sigma}
\,\Tt^s(q^\ast \partial_{\sigma_1}q\,\partial_{\sigma_2}q^\ast\cdots\partial_{\sigma_{2k+1}}q)
\;.
$$
\item Consider the invertible element from $(S \Aa_d)^+$
$$
g = \big [t \in [0,2\pi] \rightarrow g_z=(h-z)^{-1}\big ].
$$
Then
$$
\Ch_{I}(p_F)
\;=\;
-\,\Lambda_{2k+1}
\,
\sum_{\sigma\in \Ss_{2k+1}}(-1)^{\sigma}
\,\Tt^s(g^{-1}\partial_{\sigma_1}g\,\partial_{\sigma_2}g^{-1}\cdots\partial_{\sigma_{2k+1}}g)
\;.
$$
\end{enumerate}
\end{theorem}

\noindent {\bf Proof.} (i) This follows directly by combining Proposion~\ref{prop-Greenlink} with \eqref{eq-pairsuspeven} in Theorem~\ref{theo-suspension} and using the additivity of the even Chern number. (ii) Note that the r.h.s. is just the odd Chern number applied on the invertible element $g$. Then the identity follows from the first statement by observing that $q = (g^\ast)^{-1} g$ and using the factorization of the odd Chern numbers with respect to the multiplication
$$
{\rm Ch}_{2k+1}(q) \;=\; {\rm Ch}_{2k+1}\big ((g^\ast)^{-1}\big )\; +\;  {\rm Ch}_{2k+1}(g)\;.
$$
Hence only remains to note that ${\rm Ch}_{2k+1}\big ((g^\ast)^{-1}\big ) = - {\rm Ch}_{2k+1}\big (g^\ast \big ) = {\rm Ch}_{2k+1} (g )$.
\hfill $\Box$

\begin{remark} The first expression is new and has the advantage that $q_z$ is differentiable in $z$ even when the bulk gap is filled with the dense point spectrum, hence this expression works even under MBGH. We have carried out numerical calculations which indicate that the second expression fails in this regime. 
\hfill $\diamond$
\end{remark}

Another application of Theorem~\ref{theo-suspension}, following from \eqref{eq-pairsuspodd}, is the following expression of the strong invariants in odd space dimensions in terms of the resolvent, similar as in \cite{EG1,EG2}.

\begin{theorem}
\label{theo-VEGodd} Suppose $h$ satisfies the BGH and CH with Fermi unitary operator $u_F$. Set $g_0=h^{-1}$. Then
$$
\Ch_d(u_F)
\;=\;
\frac{\Lambda_{d}}{2}
\,
\sum_{\sigma\in \Ss_{d}}(-1)^{\sigma}
\,\Tt(J g_0\partial_{\sigma_1}g_0^{-1}\,\partial_{\sigma_2}g_0\cdots\partial_{\sigma_d}g_0^{-1})
\;.
$$
\end{theorem}

\section{Duality of pairings and bulk-boundary correspondence}
\label{sec-PairingsDuality}

In this section we focus on the short exact sequence \eqref{ExactSequence} and on the induced six-term exact sequence \eqref{PSixTermDiagram} between the $K$-groups. The connecting maps in this exact sequence link $K_0$ and $K_1$-classes, both of which have pairings with cyclic cocycles. Duality theory studies the equalities between such pairings. A general theorem of this kind was proved in \cite[Theorem~A.10]{KRS}, based on various prior works \cite{Pim,EN,ENN,Nes}. Since it plays a central role in the following, not only the statement but also the proof are reproduced below. The theorem is then adapted to present context and the equality between the bulk and the boundary invariants  is established. By doing so, we attain one of the main aims of the book.

\begin{theorem} \label{theo-duality} {\rm \cite{KRS}} 
Consider the Pimsner-Voiculescu exact sequence:
\begin{diagram}
0 &\rTo &\mathcal A_{d-1}\otimes \mathcal K  &\rTo{\psi}  &T(\mathcal A_{d-1})  &\rTo{\pi} &\mathcal A_{d-1} \rtimes_{\alpha_d} \mathbb Z &\rTo &0.
\end{diagram}
and its corresponding six-term diagram
\begin{diagram}
& K_0(\Aa_{d-1}) & \rTo{(\one-\alpha_d^{-1})_\ast \ \ } & K_0(\Aa_{d-1}) & \rTo{\ \  i'_\ast \ \ } & K_0(\Aa_d) &\\
& \uTo{\rm Ind} & \  &  \ & \ & \dTo{\rm Exp} & \\
& K_1(\Aa_d)  & \lTo{\ \  i'_\ast \ \ } & K_1(\Aa_{d-1}) & \lTo{\ \ (\one-\alpha_d^{-1})_\ast} & K_1(\Aa_{d-1}) &
\end{diagram}
Then, for a set of indices $I$ such that $d\not\in I$ and $|I|$ odd,
\begin{equation}
\label{eq-pairdualeven1}
\boxed{
\;\big \langle [\xi_{I\cup\{d\}}],[e]_0 - [s(e)]_0 \big \rangle
\;=\;
\big \langle \xi _I,\Exp [e]_0\big \rangle
\;,
\;}
\end{equation}
while for $|I|$ even,
\begin{equation}
\label{eq-pairdualodd1}
\boxed{
\;
\big \langle [\xi_{I\cup\{d\}}],[v]_1\big \rangle
\;=\;
-\, \big \langle [ \xi_I],\Ind [v]_1 \big \rangle
\;.
\;}
\end{equation}
\end{theorem}

\noindent {\bf Proof.}  Let us consider first the second identity. Throughout the notations from Sections~\ref{Sec-PimsnerVoiculescu} and \ref{Sec-KPimsnerVoiculescu} will be used. We also recall that $\Aa_d = \Aa_{d-1} \rtimes_{\alpha_d} \mathbb Z$ and that $\Aa_{d-1}$ is viewed as a sub-algebra of $\Aa_d$. Let $e \in \mathcal P_{N}(\Aa_{d-1}^+)$ be such that $[e]_0 - [s( e)]_0$ is contained in the image of the index map. In Proposition~\ref{InverseIndexMap} we found a pre-image of $e$ under the index map, given by
\begin{equation}
v 
\;=\; 
(\bm \one_N - e) + e\, w_e u_d^\ast 
\;.
\end{equation}
Recall that $w_e u_d^\ast$ commutes with $e$. We will proceed with the computation of the r.h.s. of Eq.~\eqref{eq-pairdualodd1} for this $v$. Let us use the notation $u=w_e \, u_d^\ast$ such that 
$$
v \;=\; 
(\bm \one_N - e) \,+\, e\, u
\;=\;
\bm \one_N \,+\, e\,(u-\bm \one_N)
\;.
$$ 
We have:
$$
\big \langle [\xi_{I\cup\{d\}}],[v]_1\big \rangle 
\;=\; 
\Lambda_{2k+1} \! \sum_{\sigma\in \Ss_{2k+1}}
\!
(-1)^{\rho}  \Tt \Big ((u^\ast -\bm 1_N) e  \prod_{j=1}^{2k+1}\, \partial_{\sigma_j} \big ((u^{*_{j-1}} -\bm 1_N) e \big ) \Big )\,,
$$
where $\sigma$ is seen as a map from $\{1,\ldots,2k+1\}$ to $I \cup \{d\}$ and we rearranged the factors slightly, using the fact that $e$ and $u$ commute. At this point, one cannot proceed directly. Instead, one should note how much the calculation would simplify if $u$ was replaced by $u_d^\ast$, because of the simplification of the derivations. This can be achieved by replacing the action $\alpha_d$ by $\alpha'_d = \alpha_d \circ {\rm Ad}_{w_e^\ast}$. Indeed, the action of $\alpha'_d$ on $\Aa_{d-1}$ is implemented on $\Aa_{d-1}$ by $u'_d = u_d \, w_e^\ast$, which is precisely $u^\ast$. Let us then complete the calculation for the simpler case of $\Aa_{d-1} \rtimes_{\alpha'_d} \mathbb Z$. All objects will carry a prime for this algebra. First of all,
\begin{align*}
\big \langle [\xi'_{I\cup\{d\}}], & [v']_1\big \rangle
\;=\; \Lambda_{2k+1} \sum_{j=1}^{2k+1}(-1)^{j-1} \sum_{\rho\in \Ss_{2k}}(-1)^{\rho} \ \Tt' \Big ( (u'_d -\bm 1_N) e 
\cdot
\\
& \cdot\Big (\prod_{i=1}^{j-1}\, \partial'_{\rho_i} \big ( (u'_d)^{*_i} e- e \big ) \Big)\partial'_{\rho_d} \big ( (u'_d)^{*_j} e- e \big ) \Big(\prod_{i=j}^{2k}\, \partial'_{\rho_i} \big ( (u'_d)^{*_i} e- e \big ) \Big ) \Big )
\;.
\end{align*}
where $\rho$ maps $\{1,\ldots,2k\}$ onto $I$, this time. Noticing that 
$$
\partial'_{\rho_i}\big ( (u'_d)^{*_i} e - e \big )\;=\; \big ((u'_d)^{*_i} -\bm 1_N) \partial'_{\rho_i} e
\;,
\qquad
\partial'_d \big ( (u'_d)^{*_j} e - e\big ) \;=\;-\I (-1)^j (u'_d)^{*_j} e
\;,
$$
and recalling that $u'_d$ commutes with $e$, we can write:
\begin{align*}
\big \langle [\xi'_{I\cup\{d\}}],&[v']_1\big \rangle
\;=\;-\I  \Lambda_{2k+1} \sum_{j=1}^{2k+1}(-1)^{2j-1} \sum_{\rho\in \Ss_{2k}}(-1)^{\rho}\cdot \\
& \cdot\Tt' \Big ((u'_d -\bm 1_N)^{k+1} \big ( (u'_d)^\ast -\bm 1_N)^k (u'_d)^{*_j} e \Big (\prod_{i=1}^{j-1}\partial'_{\rho_i} e \Big ) e \Big (\prod_{i=j}^{2k+1}\, \partial'_{\rho_i} e\Big ) \Big )
\;.
\end{align*}
Now $e \Big (\prod_{i=1}^{j-1}\, \partial'_{\rho_i} e \Big ) e =0$ for $j-1$ odd, while for $j-1$ even one  can erase $e$ on the l.h.s. Hence
\begin{align*}
\big \langle [\xi'_{I\cup\{d\}}],[v']_1\big \rangle
\;=\; & \I (k+1)  \Lambda_{2k+1} \sum_{\rho\in \Ss_{2k}}(-1)^{\rho}\cdot \\
& \;\cdot
\Tt' \Big ((u'_d -\bm 1_N)^{k+1} \big ( (u'_d)^\ast -\bm 1_N)^k (u'_d)^* e \prod_{i=1}^{2k}\, \partial'_{\rho_i} e \Big )
\;.
\end{align*}
From the definition of the trace for crossed products, only the terms not containing $u'_d$ contribute. There are
$$
\sum_{l=0}^k \binom{k+1}{l} \binom{k}{l} 
\;=\; \binom{2k+1}{k} 
$$
such terms, hence
\begin{align*}
\big \langle [\xi'_{I\cup\{d\}}],[v']_1\big \rangle
\;=\;  \,\I (k+1)  \Lambda_{2k+1} \binom{2k+1}{k}\sum_{\rho\in \Ss_{2k}}(-1)^{\rho} 
 \Tt' \Big ( e\prod_{i=1}^{2k}\, \partial'_{\rho_i} e\Big )
 \;.
\end{align*}
Note that $\Tt$ and $\Tt'$ as well as $\partial'$ and $\partial$ coincide on $\mathcal A_{d-1}^+$, hence we can erase the primes. 
With \eqref{eq-ConstId} one deduces
$$
\big \langle [\xi'_{I\cup\{d\}}],[v']_1\big \rangle 
\;=\; 
-\, \big \langle [\xi_I],[e]_0 - [s(e)]_0\big \rangle
\;.
$$

Our next task is to show the equality of pairings:
\begin{equation}
\label{BigEq}
\big \langle [\xi_{I\cup\{d\}}],[v]_1\big \rangle 
\;=\; 
\big \langle [\xi'_{I\cup\{d\}}],[v']_1\big \rangle
\;.
\end{equation}
Following \cite{KRS}, we imbed the crossed products $\Aa_{d-1} \rtimes_{\alpha_d} \mathbb Z$ and $\Aa_{d-1} \rtimes_{\alpha'_d} \mathbb Z$ in a common crossed product $(M_2(\mathbb C) \otimes \Aa_{d-1}) \rtimes_{\alpha''_d} \mathbb Z$, where
$$
\alpha''_d \begin{pmatrix} a & b \\ c & d \end{pmatrix} 
\;=\; 
\begin{pmatrix} \alpha_d(a) & \alpha_d(b) \overline{w}_e \\ \overline{w}_e^\ast \alpha_d(c) & \alpha'_d(d) \end{pmatrix}
\;,
$$
where $\overline{w}_e=u_d w_eu_d^*$ as in Proposition~\ref{InnerAuto}. If $u''_d$ is the element of $(M_2(\mathbb C) \otimes \Aa_{d-1}) \rtimes_{\alpha''_d} \mathbb Z$ implementing the action $\alpha''_d$, then the two imbeddings are explicitly given by:
$$ 
R\Big (\sum_{n \in \mathbb Z^{d-1}} p_n u_d^n \Big ) 
\;=\; 
\sum_{n \in \mathbb Z^{d-1}} \begin{pmatrix} p_n & 0 \\ 0 & 0 \end{pmatrix} (u''_d)^n
$$
and 
$$ 
R'\Big (\sum_{n \in \mathbb Z^{d-1}} p'_n (u'_d)^n \Big ) 
\;=\; 
\sum_{n \in \mathbb Z^{d-1}} \begin{pmatrix} 0 & 0 \\ 0 & p'_n \end{pmatrix} (u''_d)^n
\;.
$$
Obviously, $R$ and $R'$ are homomorphisms. The next ingredient is the family of smooth inner automorphisms:
$$ 
{\rm Ad}_{V_t} : (M_2(\mathbb C) \otimes \Aa_{d-1}) \rtimes_{\alpha''_d} \mathbb Z 
\;\rightarrow \;
(M_2(\mathbb C) \otimes \Aa_{d-1}) \rtimes_{\alpha''_d} \mathbb Z
\;, 
\qquad 
t \in [0,\pi]
\;,
$$
corresponding to the unitary elements 
$$
V_t 
\;=\; 
\begin{pmatrix} 
\cos (\tfrac{t}{2} )& - \sin (\tfrac{t}{2}) \\
\sin (\tfrac{t}{2}) & \cos (\tfrac{t}{2})
\end{pmatrix} \otimes \bm 1 \;\in\; M_2(\mathbb C) \otimes \Aa_{d-1}
\;.
$$ 
Let us focus for the moment at $t=\pi$. The action of ${\rm Ad}_{V_\pi}$ on $M_2(\mathbb C) \otimes \Aa_{d-1}$ is:
$$
{\rm Ad}_{V_\pi} \begin{pmatrix} a & b \\ c & d \end{pmatrix} 
\;=\; 
\begin{pmatrix} d & -c \\ -b & a \end{pmatrix}
$$
and
$$
{\rm Ad}_{V_\pi}(u''_d) 
\;=\; 
\begin{pmatrix} 0 & -1 \\ 1 & 0 \end{pmatrix} u''_d \begin{pmatrix} 0 & 1 \\ -1 & 0 \end{pmatrix} 
\;=\; 
\begin{pmatrix} 0 & -1 \\ 1 & 0 \end{pmatrix} \alpha''_d \begin{pmatrix} 0 & 1 \\ -1 & 0 \end{pmatrix} u''_d
\;,
$$
and, using the definition of $\alpha''_d$,
$$
{\rm Ad}_{V_\pi}(u''_d) 
\;=\; 
\begin{pmatrix}  \overline{w}_e^\ast & 0 \\ 0 & \overline{w}_e \end{pmatrix} u''_d
\;.
$$
Then it can be checked that the following diagram commutes:
\begin{diagram}
&\Aa_{d-1} \rtimes_{\alpha_d} \mathbb Z  &\rTo{ \ \ R \ \ \ } \ \  & (M_2(\mathbb C) \otimes \Aa_{d-1}) \rtimes_{\alpha''_d} \mathbb Z \\
&\dTo{Q}       &                                            &\dTo{{\rm Ad}_{V_\pi} }                  \\
&\Aa_{d-1} \rtimes_{\alpha'_d} \mathbb Z   &\rTo{\ \ R' \ }  & (M_2(\mathbb C) \otimes \Aa_{d-1}) \rtimes_{\alpha''_d} \mathbb Z  
\end{diagram}
with $Q$ the $\ast$-isomorphism defined by $Q(a) =a$ for a $a \in \Aa_{d-1}$ and $Q(u_d) = \overline{w}_e u'_d$. In particular, $Q(v)=v'$. We now start the ascent towards \eqref{BigEq}. The first step is to recognize that
$$ 
\xi_{ I \cup \{d\} } \;=\; \xi''_{I \cup \{d\} } \circ R
\;, 
\qquad 
\xi'_{ I \cup \{d\} } 
\;=\; 
\xi''_{I \cup \{d\} } \circ R'
\;,
$$
that is, the two cocycles involved in \eqref{BigEq} are pullbacks of the same cocycle $\xi''_{I \cup \{d\} }$ defined over $(M_2(\mathbb C) \otimes \Aa_{d-1}) \rtimes_{\alpha''_d} \mathbb Z$. This gives:
$$
\big \langle [\xi_{I\cup\{d\}}],[v]_1\big \rangle 
\;=\;  
\big \langle [\xi''_{I \cup \{d\} } \circ R],[v]_1 \big \rangle
\;.
$$
The second step is to realize that the cocycles 
$$
\xi''_{I \cup \{d\} } \circ R \quad {\rm and} \quad  \xi''_{I \cup \{d\} } \circ {\rm Ad}_{V_\pi} \circ R
$$
are connected by the homotopy $\xi''_{I \cup \{d\} } \circ {\rm Ad}_{V_t} \circ R$. Hence:
$$
\big \langle [\xi''_{I \cup \{d\} } \circ R],[v]_1 \big \rangle 
\;=\; 
\big \langle [\xi''_{I \cup \{d\} } \circ {\rm Ad}_{V_\pi} \circ R],[v]_1 \big \rangle
\;.
$$
For the last step, we use ${\rm Ad}_{V_\pi} \circ R = R' \circ Q$ implying that
\begin{align*}
\big \langle [\xi''_{I \cup \{d\} } \circ {\rm Ad}_{V_\pi} \circ R],[v]_1 \big \rangle 
& \;=\; \big \langle [\xi''_{I \cup \{d\} } \circ R'\circ Q],[v]_1 \big \rangle \\
& \;=\; \big \langle [\xi''_{I \cup \{d\} } \circ R'],[Q(v)]_1 \big \rangle \\
& \;=\; \big \langle [\xi'_{I \cup \{d\} }],[v']_1 \big \rangle.
\end{align*}
The statements \eqref{BigEq} and thus also \eqref{eq-pairdualodd1} are now proved.

\vspace{0.2cm}

The first identity \eqref{eq-pairdualeven1} follows from \eqref{eq-pairdualodd1}  if  the following commutative diagram is used:
\begin{diagram}
&K_0(\Aa_d)  &\rTo{ \ \ \beta \ \ } \ \  & K_1(S \Aa_d) \\
&\dTo{{\rm Exp} }      &                                            &\dTo{{\rm Ind} }                  \\
&K_1(\Aa_{d-1})   &\rTo{\ \ \theta \ \ }  & K_1(S \Aa_{d-1}) 
\end{diagram}
namely $\Exp=\theta^{-1}\circ\Ind\circ\beta$. Indeed, one finds
\begin{align*}
\big \langle\xi_{I\cup\{d\}},[e]_0 -[s(e)]_0\big \rangle 
& \;=\; -\,\big \langle \xi_{I\cup\{d\}}^s,\beta [e]_0\big \rangle 
\; =\; \,\big \langle \xi_I^s,\Ind \circ\beta [e]_0\big \rangle \\
&
\;=\; \,\big \langle \xi_I^s,\theta \circ \Exp  [e]_0\big \rangle  \;=\;
\,\big \langle \xi_I,\Exp [e]_0\big \rangle
\;,
\end{align*}
where we used both equalities of Theorem~\ref{theo-suspension}.
\hfill $\Box$

\vspace{.2cm}

The above result can be extended to the algebras of physical observables using the isomorphisms established in Sections~\ref{Sec-PimsnerVoiculescu} and \ref{Sec-KPimsnerVoiculescu}.

\begin{corollary}\label{coro-duality}
Consider the exact sequence \eqref{ExactSequence} between the algebras of observables
\begin{diagram}
0 &\rTo &\mathcal E_d   &\rTo{i}  &\widehat{\mathcal A}_d  &\rTo{ \mathrm{ev}}  &\mathcal A_d &\rTo &0
\end{diagram}
and the associated six-term exact sequence \eqref{PSixTermDiagram} between the $K$-groups
\begin{diagram}
& K_0(\Ee_d) & \rTo{\ \ i_\ast \ \ } & K_0(\widehat{\Aa}_d) & \rTo{\ \ {\rm ev}_\ast \ \ } & K_0(\Aa_d) &\\
& \uTo{\rm Ind} & \  &  \ & \ & \dTo{\rm Exp} & \\
& K_1(\Aa_d)  & \lTo{{\rm ev}_\ast} & K_1(\widehat{\Aa}_d) & \lTo{i_\ast} & K_1(\Ee_d) &
\end{diagram}
Then, for a set $I$ of indices such that $d\not\in I$ and $|I|$ odd,
\begin{equation}
\label{eq-pairdualeven2}
\big \langle [\xi_{I\cup\{d\}}],[e]_0 - [s(e)]_0 \big \rangle
\;=\;
\big \langle [\widetilde{\xi}_I],\Exp [e]_0\big \rangle
\;,
\end{equation}
while for $|I|$ even and $v \in \mathcal U_N(\Aa_d^+)$,
\begin{equation}
\label{eq-pairdualodd2}
\big \langle [\xi_{I\cup\{d\}}],[v]_1\big \rangle
\;=\;
-\,\big \langle [ \widetilde{\xi}_I],\Ind [v]_1 \big \rangle
\;.
\end{equation}
\end{corollary}

\noindent {\bf Proof.} We will use the isomorphism \eqref{PVLink} between the exact sequence of observables algebras and the Pimsner-Voiculescu sequence. Let $e \in \mathcal P_N(\Aa_d^+)$. Then
$$
\big \langle [\widetilde \xi_I], \Exp [e] _0  \big \rangle  
\;=\; 
\big \langle \widetilde \rho^\ast [\widetilde \xi_I], \widetilde \rho_\ast\Exp [e] _0 \big \rangle
\;.
$$
Note that $\widetilde \rho_\ast\Exp [e] _0 = \Exp[e]_0$, where the latter exponential map is the one appearing in the Pimsner-Voiculescu sequence. Also, $\widetilde \rho^\ast [\widetilde \xi_I]= [\xi_I]$. Then 
$$
\big \langle [\widetilde \xi_I], \Exp [e] _0  \big \rangle 
\;=\; 
\big \langle [\xi_I], \Exp [e] _0 \big \rangle 
\;=\; 
\big \langle [\xi_{I \cup \{d\} }], [e] _0 -[s(e)]_0\big \rangle
\;.
$$
The second identity follows in a similar way.\hfill $\Box$

\vspace{.2cm}

\begin{theorem}[Equality between the bulk and boundary invariants]
\label{th-BBEquality}
\begin{enumerate}[\rm (i)]
\item Let $\hat h = (h,\tilde h) \in M_N(\mathbb C) \otimes \widehat \Aa_d$ satisfying BGH, and let $I$ be a set of indices such that $|I|=2k-1<d$ and $d \notin I$. Then:
\begin{equation}
\label{EQ1}
\boxed{
\;{\rm Ch}_{I \cup \{d\}}(p_F) \;=\; \widetilde{\rm Ch}_I(\tilde u_\Delta)
\;.
\;}
\end{equation}
\item Let $\hat h = (h,\tilde h) \in M_N(\mathbb C) \otimes \widehat \Aa_d$ satisfying BGH and CH, and let $I$ be a set of indices such that $|I|=2k<d$ and $d \notin I$. Then:
\begin{equation}
\label{EQ2}
\boxed{
\;{\rm Ch}_{I \cup \{d\}}(u_F) \;=\; -\;\widetilde{\rm Ch}_I(\tilde p_\Delta)\;.
\;}
\end{equation}
\end{enumerate}
\end{theorem}

\noindent {\bf Proof.} The bulk and boundary invariants are defined in Theorems~\ref{Th-BulkInv} and \ref{Th-BoundaryInvariants}. For (i), we now combine Proposition~\ref{ExpMap} and \eqref{TildeU}  with \eqref{eq-pairdualeven2}, we have
\begin{align*}
{\rm Ch}_{I \cup \{d\}}(p_F) 
& \;= \;\big \langle [\xi_{I \cup \{d\}}],[p_F]_0 \big \rangle\; =\; \big \langle [\widetilde \xi_I ],\Exp [p_F]_0 \big \rangle \\
& \;=\; \big \langle [\widetilde \xi_I ], [\tilde u_\Delta]_1 \big \rangle 
\;= \; \widetilde {\rm Ch}_{I}(\tilde u_\Delta)
\;.
\end{align*}

\noindent Similarly, (ii) follows by combining Proposition~\ref{IndMap} and \ref{BoundaryProjection1} with \eqref{eq-pairdualodd2}:
\begin{align*}
{\rm Ch}_{I \cup \{d\}}(u_F) &\, =\, \big \langle [\xi_{I \cup \{d\}}],[u_F]_1 \big \rangle \,=\, -\,\big \langle [\widetilde \xi_I ],\Ind [u_F]_1 \big \rangle  \\
& \, =\, -\,\big \langle [\widetilde \xi_I ], [\tilde p_\Delta]_0 \big \rangle\, = \,-\,\widetilde {\rm Ch}_{I}(\tilde p_\Delta)\,,
\end{align*}
completing the proof.
\hfill $\Box$

\vspace{0.2cm}

\begin{remark} The bulk-boundary correspondence relations are fully compatible with the ones stated in Chapter~\ref{Chap-Physics}. Indeed, by combining \eqref{EQ1} with \eqref{BI1} we can reproduce \eqref{BB1} for the unitary class A, and by combining \eqref{EQ2} with \eqref{BI2} we can reproduce \eqref{BB2} for the chiral unitary class AIII.
\hfill $\diamond$
\end{remark}

\section{Generalized Streda formulas}

The results in this section are inspired by the Streda formula \cite{Str} for quantum Hall effect in $d=2$, connecting the derivative of the electron density $n$ and the Hall conductance, $\partial_B n = \sigma_H$. As we shall see in Section~\ref{sec-transport}, this translates into
\begin{equation}
\label{eq-Streda}
\partial_{B_{12}}\,\Tt(p_F)
\;=\;
\frac{1}{2\pi}\;\Ch_2(p_F)
\;,
\end{equation}
where $p_F\in\Aa_d$ is the Fermi projection. An algebraic proof of this identity which is not based on Bloch theory was provided by Rammal and Bellissard \cite{RB}. More recently, it was shown that the Streda formula holds also when the Fermi level lies in a region of dynamical Anderson localization as well as in higher dimensions \cite[Theorem~7]{ST}. It is useful to read this formula in the following manner: on the l.h.s. is the magnetic field derivative of a pairing with a $0$-cocycle, while on the r.h.s. is a pairing with a $2$-cocycle. Likewise, the generalized Streda formulas derived below connect derivatives of pairings with $n$ cocycles w.r.t. the magnetic field to pairings with $n+2$ cocycles.  Due to the importance of the Streda formula in condensed matter physics, let us begin by providing a short proof separately before going to more complicated algebraic manipulations.

\begin{proposition}
\label{prop-derizero}
Suppose that a projection $e\in\Pp_N(\Ff_d)$ is Ito-differentiable. Then for given $i,j\in\{1,\ldots,d\}$ with $i<j$,
\begin{equation}
\label{eq-Streda2}
\partial_{B_{i,j}}\,\big \langle [\xi_{\emptyset}],[e]_0-[s(e)]_0 \big \rangle
\;=\;
\frac{1}{2\pi}\;
\big \langle [\xi_{\{i,j\}}],[e]_0-[s(e)]_0 \big \rangle
\;.
\end{equation}
\end{proposition}

\noindent {\bf Proof.} The proof uses rules (i) and (vii) of Proposition~\ref{prop-deltajs},
\begin{align*}
\partial_{B_{i,j}}\,\Tt(e)
\; &=\;
\Tt \big (\delta_{i,j}e \big )
\;=\;
\Tt \big (e(\delta_{i,j}) e + (\one-e)(\delta_{i,j}e)(\one-e) \big) \\
& = \;\tfrac{\I}{2} \Tt \big (e[\partial_i e , \partial_j e] \big )\; -\; \tfrac{\I}{2} \Tt \big ((\one-e)[\partial_i e , \partial_j e] \big ) \\
& = \;\tfrac{\I}{2} \tfrac{1}{2\pi \I} \big ( \xi_{\{i,j\}}(e,e,e) \;-\; \xi_{\{i,j\}}(\one-e,\one-e,\one-e)\big ) \; ,
\end{align*}
and the statement follows because $e$ and $\one-e$ are orthogonal and they add up to identity, hence 
$$
\big \langle [\xi_{\{i,j\}}],[\one-e]_0-[s(\one-e)]_0\big \rangle 
\;=\;
- \,\big \langle [\xi_{\{i,j\}}],[e]_0 - [s(e)]_0\big \rangle \; .
$$
The pairing with the scalar part $s(e)$ vanishes.
\hfill $\Box$

\vspace{.2cm}

For $d=2$, Propostion~\ref{prop-derizero} reduces to the Streda formula \eqref{eq-Streda}. Propostion~\ref{prop-derizero} is also particularly interesting for dimension $d=3$, because the pairing $\big \langle \xi_{\{i,j\}},[e]_0\big \rangle$ on the r.h.s. of \eqref{eq-Streda2} is integer valued, see Section~\ref{sec-PairingsRange}. Hence, in $d=3$, the relation \eqref{eq-Streda2} can be regarded as the Streda formula for the three-dimensional quantum Hall effect. Next let us turn to higher cocycles. As a next step, we derive a paring with a $1$-cocyle w.r.t. a magnetic field.

\begin{proposition}
\label{prop-deri1coc}
Suppose that $a\in M_N(C^1(\Ff_d))$ is invertible. Then for given $i,j\in\{1,\ldots,d\}$ and $k\not = i,j$,
\begin{equation}
\label{eq-deri1co}
\partial_{B_{i,j}}\,\big \langle [\xi_{\{k\}}],[a]_1\big \rangle
\;=\;
\frac{1}{2\pi}\;
\big \langle [\xi_{\{i,j,k\}}],[a]_1\big \rangle
\;.
\end{equation}
If $k\in\{i,j\}$, then the l.h.s. of \eqref{eq-deri1co} vanishes.
\end{proposition}

\noindent {\bf Proof.}
The definitions and rules (i) and (v) of Proposition~\ref{prop-deltajs} imply
\begin{align*}
 \partial_{B_{i,j}}\,
\big \langle [\xi_{\{k\}}],[a]_1\big \rangle
\; & =\;
\Lambda_1\;
\partial_{B_{i,j}}\,\Tt \big (a^{-1}\partial_k a\big )
\;=\;
\I\;
\Tt \big (\delta_{i,j}(a^{-1}\partial_k a) \big ) \\
& =\;
\I\,
\Tt \big (\delta_{i,j}(a^{-1})\partial_k a +  a^{-1}\delta_{i,j}(\partial_k a) \big )\\
& =\;
\I\,
\Tt \big (\delta_{i,j}(a^{-1})\partial_k a  - (\partial_k a^{-1})\delta_{i,j}a \big )
\;.
\end{align*}
Using rule (iii) of Proposition~\ref{Indenties1} and a cyclic permutation on the second term,
\begin{align*}
 \partial_{B_{i,j}}\,
\big \langle [\xi_{\{k\}}],[a]_1\big \rangle
\; & =\;
\I\,
\Tt \big (\delta_{i,j}(a^{-1})\partial_k a  + a^{-1}(\delta_{i,j}a) a^{-1} (\partial_k a)\big )
\;.
\end{align*}
Then rule (vi) of Proposition~\ref{prop-deltajs} gives
\begin{align*}
 \partial_{B_{i,j}}\,
\big \langle [\xi_{\{k\}}],[a]_1\big \rangle
\; =\;
-\,\tfrac{1}{2}\;
\Tt
\big(a^{-1}
(\partial_{i}a)\, (\partial_{j}a^{-1})
(\partial_k a)
\,-\, a^{-1}(\partial_{j}a)\, (\partial_i
a^{-1})
(\partial_k a)
\big )
\;.
\end{align*}
If say $i=k$, then one can use the identity $a^{-1}\partial_j a=-(\partial_j a^{-1})\,a$ twice on the second term to verify that both terms cancel.
Otherwise, using the cyclicity of the trace, and with $\sigma\in \Ss_3$ viewed as map from $\{1,2,3\}$ to $\{i,j,k\}$, the derivative can be written as
\begin{align*}
\partial_{B_{i,j}}\,
\big \langle [\xi_{\{k\}}],[a]_1\big \rangle
& =\;
-\,\tfrac{1}{2} \tfrac{1}{3}
\sum_{\sigma\in \Ss_3}
(-1)^\sigma\;
\Tt
\big(a^{-1}
(\partial_{\sigma_1}a)\,a^{-1} (\partial_{\sigma_2}a) \,a^{-1}
(\partial_{\sigma_3}a)
\big )
\\
& =\;
\frac{1}{2\pi}\,\big \langle [\xi_{\{i,j,k\}}],[a]_1\big \rangle
\;,
\end{align*}
where in the last step the constant $\Lambda_3=  - \, \frac{\pi}{3}$ was restored.
\hfill $\Box$

\vspace{.2cm}

The identity in Proposition~\ref{prop-deri1coc} is particularly interesting in dimensions $d=3$ and $d=4$ where the pairing on the r.h.s. is integer valued. This will be exploited in Section~\ref{sec-ChiralPol}.
Up to now, only derivatives of pairings with $0$ and $1$-cocycles were considered. The generalized Streda formulas concern pairings with higher cocycles.

\begin{theorem}
\label{theo-StredaGen}
Let $I \subset \{1,\ldots,d\}$ be an ordered subset and $i,j\not\in I$, where the ordering of $I$ is not necessarily the one induced by $\ZM$. Then:

\begin{enumerate}[\rm (i)]
\item If $|I|$ is even and $e$ is a projection from $C^1(\Ff_d)$,
\begin{equation}
\label{eq-Stredaeven}
\boxed{
\;\partial_{B_{i,j}}\,\big \langle [\xi_{I}],[e]_0 - [s(e)]_0\big \rangle\;=\; \frac{1}{2\pi}\;\big \langle [\xi_{\{i,j\} \cup I}],[e]_0 -[s(e)]_0 \big \rangle\;.
\;}
\end{equation}

\item  If $|I|$ is odd and $v$ is a unitary from $C^1(\Ff_d)$,
\begin{equation}
\label{eq-Stredaodd}
\boxed{
\;\partial_{B_{i,j}}\,\big \langle [\xi_{I}],[v]_1\big \rangle\;=\; \frac{1}{2\pi}\;\big \langle [\xi_{\{i,j\} \cup I}],[v]_1\big \rangle\;.
\;}
\end{equation}
\end{enumerate}
The ordering of the indices on the r.h.s is as the notation $\{i,j\} \cup I$ implies. If $i\in I$ or $j\in I$, the derivatives on the l.h.s. of \eqref{eq-Stredaeven} and \eqref{eq-Stredaodd} vanish.
\end{theorem}

\noindent {\bf Proof.} (i) Let $|I|=n$ be even. Using rules (i) and (v) of Proposition~\ref{prop-deltajs},
\begin{align*}
\partial_{B_{i,j}}\xi_I(e,\ldots,e) \;=\;  \xi_I(\delta_{i,j}e,e,\ldots,e)  + \Lambda_n \sum_{\sigma \in \Ss_n} (-1)^\sigma \Tt \big ( e \delta_{i,j} (\partial_{\sigma_1}e \cdots \partial_{\sigma_n}e ) \big ) \; .
\end{align*}
The second term can be further processed using the identity
\begin{align*}
\delta_{i,j}(p_1 \cdots p_n) \;=\; & \sum_{k=1}^n p_1 \cdots (\delta_{i,j} p_k ) \cdots p_n \\
& - \tfrac{\I}{2} \sum_{\rho \in \Ss_2} \sum_{k<l} (-1)^\rho p_1 \cdots (\partial_{\rho_1}p_k) \cdots  (\partial_{\rho_2} p_l) \cdots p_n \; ,
\end{align*}
obtained by iterating rule (iv) of Proposition~\ref{prop-deltajs}. The result is
\begin{align*}
\partial_{B_{i,j}} & \xi_I(e,\ldots,e) \;=\; \sum_{k=1}^{n+1} \xi_I(e,\ldots, \delta_{i,j}e,\ldots,e) 
\\
& \; - \;
\tfrac{\I}{2} \Lambda_n \sum_{\sigma \in \Ss_n} \sum_{\rho \in \Ss_2} \sum_{k<l}(-1)^{\sigma +\rho} 
\Tt \big ( e \partial_{\sigma_1}e \cdots \partial_{\rho_1}\partial_{\sigma_k}e \cdots  \partial_{\rho_2}\partial_{\sigma_l}e \cdots \partial_{\sigma_n} e\big ) \; ,
\end{align*}
where $\delta_{i,j} e$ is at the $k$-th position in the first summand and $(-1)^{\sigma + \rho}=(-1)^\sigma (-1)^\rho$. We will compute separately the first and second terms, called $T_1$ and $T_2$ in the following.  Using the cyclic property of the cocycle
$$
T_1  \;=\; (n+1) \xi_I((\delta_{i,j} e),e,\ldots, e)  \;=\; (n+1) \Lambda_n \sum_{\sigma \in \Ss_n} (-1)^\sigma \Tt \big ( \delta_{i,j} (e) \partial_{\sigma_1} e \cdots \partial_{\sigma_n} e \big ) \; .
$$
Since $n$ is even, one can replace $\delta_{i,j}e$ inside of the trace by its diagonal part 
$$
e(\delta_{i,j}e) e + (1-e)(\delta_{i,j} e)(1-e) \; .
$$ 
Now applying rules of Proposition~\ref{prop-deltajs}(vii), one obtains
\begin{align*}
T_1  & \;=\; -\tfrac{\I}{2}(n+1) \Lambda_n \sum_{\sigma \in \Ss_n} \sum_{\rho \in \Ss_2} (-1)^{\sigma +\rho} \,  \Tt \big ( (1-2e) \partial_{\rho_1}e \partial_{\rho_2}e \partial_{\sigma_1}e \cdots \partial_{\sigma_n}e \big ) \\
&\;=\; -\tfrac{\I}{2}(n+1) \Lambda_n \sum_{\sigma \in \Ss_n} \sum_{\rho \in \Ss_2} (-1)^{\sigma +\rho} \,  \Tt \big ( (1-2e) \partial_{\sigma_1}e \cdots \partial_{\sigma_n}e \partial_{\rho_1}e \partial_{\rho_2}e \big ) \; ,
\end{align*} 
where in the second line we used the last of the rules in Proposition~\ref{Indenties1}(i)  and the cyclic property of the trace.  As for the second term, we first perform a partial integration with respect to $\partial_{\sigma_k}$ and use the anti-symmetrizing factor $(-1)^\sigma$ to cancel all but one term produced when applying Leibniz rule. The result is
\begin{align*}
T_2 \;=\; \tfrac{\I}{2} \Lambda_n \sum_{\sigma \in \Ss_n}  \sum_{\rho \in \Ss_2} \sum_{k<l}(-1)^{\sigma +\rho} \,  \Tt \big ( \partial_{\sigma_k} e \partial_{\sigma_1}e \cdots \partial_{\rho_1}e \cdots  \partial_{\rho_2}\partial_{\sigma_l}e \cdots \partial_{\sigma_n} e\big ) \; .
\end{align*}
We reorder the $\sigma_i$'s up to $i=k$, which brings a sign factor $(-1)^{k-1}$. Also, note that $\partial_{\rho_2}\partial_{\sigma_l}e$ can be replaced by its diagonal part 
$$
e(\partial_{\rho_2}\partial_{\sigma_l}e)e + (1-e) (\partial_{\rho_2}\partial_{\sigma_l}e) (1-e) \; ,
$$
which enables us to apply the rules (ii) of Proposition~\ref{Indenties1}. We obtain
\begin{align*}
T_2 \;=\;  \tfrac{\I}{2} \Lambda_n & \sum_{\sigma \in \Ss_n} \sum_{\rho \in \Ss_2} \sum_{k<l}(-1)^{\sigma +\rho} \,  (-1)^{k-1}  \\
& \Tt \big ( \partial_{\sigma_1} e \cdots \partial_{\sigma_k}e \partial_{\rho_1}e \cdots \partial_{\sigma_{l-1}}e \, (1-2e)\{\partial_{\rho_2}e,\partial_{\sigma_l} e\}\partial_{\sigma_{l+1}}e \cdots \partial_{\sigma_n} e\big ) \; .
\end{align*}
Now Proposition~\ref{Indenties1}(i) allows to bring $1-2e$ in front:
\begin{align*}
T_2 \;=\;  -\tfrac{\I}{2} \Lambda_n & \sum_{\sigma \in \Ss_n} \sum_{\rho \in \Ss_2} (-1)^{\sigma +\rho} \,   \\ 
& \sum_{k<l} (-1)^{k+l} \; \Tt \big ( (1-2e) \partial_{\sigma_1} e \cdots \partial_{\sigma_k}e \partial_{\rho_1}e  \cdots \{\partial_{\rho_2}e,\partial_{\sigma_l} e\} \cdots \partial_{\sigma_n} e\big ) \; .
\end{align*}
After writing out the anti-commutators explicitly,  all terms in the sum over $l$ cancel out due to the sign factor $(-1)^l$, except for the first and last ones:
\begin{align*}
T_2 \;=\;  -\tfrac{\I}{2} \Lambda_n & \sum_{\sigma \in \Ss_n} \sum_{\rho \in \Ss_2} (-1)^{\sigma +\rho} \,    \\
 \sum_{k=1}^{n-1} \Big [ & (-1)^{k+k+1} \; \Tt \big ( (1-2e) \partial_{\sigma_1} e \cdots \partial_{\sigma_k}e \partial_{\rho_1}e\partial_{\rho_2}e\partial_{\sigma_{k+1}}e \cdots \partial_{\sigma_n} e\big ) \\
&+ (-1)^{k+n} \; \Tt \big ( (1-2e) \partial_{\sigma_1} e \cdots \partial_{\sigma_k}e \partial_{\rho_1}e\partial_{\sigma_{k+1}}e \cdots \partial_{\sigma_n} e \partial_{\rho_2}e \big ) \Big ] \; .
\end{align*}
At this point we use again the last rules of Proposition~\ref{Indenties1}(i) and the cyclic property of the trace to move all $\partial_{\rho_2}e$ to the right. After reordering $\partial_{\sigma_i}$'s, 
\begin{align*}
T_2 \;=\;  -\tfrac{\I}{2} \Lambda_n  \sum_{\sigma \in \Ss_n} \sum_{\rho \in \Ss_2} & (-1)^{\sigma +\rho} \,     \Big [ -(n-1) \, \Tt \big ( (1-2e) \partial_{\sigma_1} e \cdots \partial_{\sigma_n}e \partial_{\rho_1}e\partial_{\rho_2} e \big ) \\
&+ \sum_{k=1}^{n-1} (-1)^{k} \; \Tt \big ( (1-2e) \partial_{\sigma_1} e \cdots \partial_{\sigma_k}e \partial_{\rho_1}e \cdots \partial_{\sigma_n} e \partial_{\rho_2}e \big ) \Big ] \; .
\end{align*}
Combining with $T_1$, one concludes
\begin{align*}
& \partial_{B_{i,j}} \xi_I(e,\ldots,e) 
\\
&\;\;=\; -\tfrac{\I}{2} \Lambda_n  \sum_{\sigma \in \Ss_n} \sum_{\rho \in \Ss_2} (-1)^{\sigma +\rho} \,  
\sum_{k=0}^{n} (-1)^{k} \; \Tt \big ( (1-2e) \partial_{\sigma_1} e \cdots \partial_{\sigma_k}e \partial_{\rho_1}e \cdots \partial_{\sigma_n} e \partial_{\rho_2}e \big ) \; ,
\end{align*}
where we used the fact that the terms $k=0$ and $k=n$ are both equal to 
$$
\Tt \big ( (1-2e) \partial_{\sigma_1} e \cdots \partial_{\sigma_n}e \partial_{\rho_1}e\partial_{\rho_2} e \big ).
$$ 
The result can be further processed to
\begin{align*}
\partial_{B_{i,j}} \xi_I(e,\ldots,e) &  \;=\;  -\, \tfrac{\I}{n+2}\, \Lambda_n\,\tfrac{1}{2}  \sum_{\sigma \in \Ss_n} \sum_{\rho \in \Ss_2}   (-1)^{\sigma +\rho} \,  \\
& \sum_{0 \leq k \neq l \leq n}\!\! (-1)^{k+l+1}  \tfrac{1}{2}\; \Tt \big ( (1-2e) \partial_{\sigma_1} e \cdots \partial_{\sigma_k}e \partial_{\rho_1} e \cdots \partial_{\sigma_l} e \partial_{\rho_2}e \cdots \partial_{\sigma_n} e\big ) 
\end{align*}
or, since $k+l$ is even,
\begin{align*}
\partial_{B_{i,j}} \xi_I(e,\ldots,e) \;=\; \tfrac{1}{2\pi}\, \Lambda_{n+2} \sum_{\sigma \in \Ss_{n+2}} (-1)^\sigma   \tfrac{1}{2}\; \Tt \big ( (1-2e) \partial_{\sigma_1} e \cdots  \partial_{\sigma_{n+2}} \big ) \; ,
\end{align*}
with $\sigma$ viewed as a bijection from $\{1,\ldots,n+2\}$ to the ordered set $I \cup \{i,j\}$. This is precisely the r.h.s. of \eqref{eq-Stredaeven}, because the pair $\{i,j\}$ can be brought in front of $I$ without changing the signatures. 

\vspace{0.2cm}

(ii) For the proof of \eqref{eq-Stredaodd} one could in principle proceed in a similar manner as above, but it is more simple to use the suspension of pairings proved in Theorem~\ref{theo-suspension}. In particular, the identity \eqref{eq-pairsuspodd} shows that for odd $|I|$
$$
\big \langle [\xi_{I}],[v]_1\big \rangle
\;=\;
\big \langle [\xi^s_{I}],\theta [v]_1\big \rangle
\;=\;
\big \langle [\xi_{I\cup\{d+1\}}],\theta [v]_1\big \rangle
\;,
$$
where the last pairing is seen as a pairing over the algebra $\Aa_{d+1}$ with vanishing magnetic field components $B_{i,d+1}$, $i=1,\ldots,d$, which contains $S\Aa_d$ as a subalgebra. Now one can use \eqref{eq-Stredaeven} to deduce
$$
\partial_{B_{i,j}}\,\big \langle [\xi_{I}],[v]_1\big \rangle
\;=\;
\frac{1}{2\pi}\,
\big \langle [\xi_{ \{i,j\} \cup I \cup \{d+1\}}],\theta [v]_1\big \rangle
\;=\;
\,\frac{1}{2\pi}\,
\big \langle [\xi_{\{i,j\} \cup I}],[v]_1\big \rangle
\;,
$$
where in the last step the suspension \eqref{eq-pairsuspodd} was applied a second time.
\hfill $\Box$

\vspace{0.2cm}

Below we specialize Theorem~\ref{theo-StredaGen} to condensed matter systems. Using the bulk-boundary duality, generalized Streda formulas can be automatically formulated for the boundary pairings too.

\begin{corollary}\label{coro-Streda} Let $\hat h = (h,\tilde h) \in \widehat \Aa_d$ and assume that BGH holds. Then the Fermi projection (hence the Fermi unitary element too, if CH applies) are Ito-differentiable \cite{ST}, and:
\begin{enumerate}[\rm (i)]
\item If $I \subset \{1, \ldots, d\}$ with $|I|$ even and $\{i,j\} \notin I$, 
\begin{equation}
\partial_{B_{i,j}} {\rm Ch}_I(p_F) 
\;=\; 
\frac{1}{2\pi}\;  {\rm Ch}_{\{i,j\} \cup I}(p_F) \; .
\end{equation}

\item If $\tilde I \subset \{1, \ldots, d-1 \}$ with $|\tilde I|$ odd and $\{i,j\} \notin \tilde I$,
\begin{equation}
\partial_{B_{i,j}} \widetilde{\rm Ch}_{\tilde I}(\tilde u_\Delta) 
\;=\; 
\frac{1}{2\pi}\;  \widetilde{\rm Ch}_{\{i,j\} \cup \tilde I  }(\tilde u_\Delta) \;.
\end{equation}

\item If CH holds, too, and $|I|$ is odd,
\begin{equation}
\partial_{B_{i,j}} {\rm Ch}_{I}(u_F) 
\;=\; 
\frac{1}{2\pi} \; {\rm Ch}_{\{i,j\} \cup I}(u_F).
\end{equation}

\item If $\tilde I \subset \{1, \ldots, d-1 \}$ with $|\tilde I|$ even and $\{i,j\} \notin \tilde I$,
\begin{equation}
\partial_{B_{i,j}} \widetilde{\rm Ch}_{\tilde I}(\tilde p_\Delta) 
\;=\; 
\frac{1}{2\pi} \; \widetilde{\rm Ch}_{\{i,j\} \cup \tilde I  }(\tilde p_\Delta) \;.
\end{equation}
\end{enumerate}
\end{corollary}

As we shall see in Chapter~\ref{Chap-Conclusions}, the above formulas will be instrumental for the physical interpretation of the bulk and boundary invariants of topological insulators.

\section{The range of the pairings and higher gap labelling}
\label{sec-PairingsRange}

Let us begin by recalling from Section~\ref{sec-CycCoh} that the map
$$
[e]_0  \in K_0(\Aa_{\BB,d})\;\mapsto\;\;
\Tt(e)
\;=\;
\big \langle [\xi_{\emptyset}],[e]_0\big \rangle \in \RM
\;,
$$
is a homomorphism of abelian groups. As $K_0(\Aa_{\BB,d})\cong \ZM^{2^{d-1}}$, the range $\Ran(\xi_\emptyset)$ is a discrete subset of $\RM$ which can be calculated once the image of the generators is determined. In dimension $d=2$, it was accomplished by Pimsner \cite{Pim} who established $\Ran(\xi_\emptyset)=\ZM+\frac{B_{1,2}}{2\pi}\,\ZM$. For higher dimensional non-commutative tori, the range of the trace was computed in \cite{Ell} (see \cite{BM} for extensions and new conjectures). In order to understand the link with the generalized Streda formulas, let us take a closer look at the well understood case of $d=2$. If $e_{\{1,2\}}$ is the Powers-Rieffel projection given in \eqref{eq-Rieffel}, then
$$
\big \langle [\xi_{\emptyset}],[\one]_0\big \rangle\;=\;1\;,
\qquad
\big \langle [\xi_{\emptyset}],[e_{\{1,2\}}]_0\big \rangle
\;=\;\tfrac{1}{2\pi}\,B_{1,2}
\;.
$$ 
One way to verify this is to use the Streda formula \eqref{eq-Streda2}, which implies that $\big \langle [\xi_{\emptyset}],[e_{\{1,2\}}]_0\big \rangle$ is linear in $B_{1,2}$. The slope is given by $\frac{1}{2\pi}$ times the Chern number of $e_{\{1,2\}}$ which is $1$, and the constant term vanishes because the $\Tt(e_{\{1,2\}})$ converges to $0$ as $B_{1,2}$ goes to $0$. All these facts apply, in particular, to the Fermi projection $p_F$ of a Hamiltonian in $\Aa_{\BB,2}$ if the Fermi level lies in a gap, and hence the integrated density of states $\Tt(p_F)$ takes values in $\ZM+\frac{B_{1,2}}{2\pi}\,\ZM$. This fact is referred to as gap labelling \cite{Bel}.

\vspace{.2cm}

It is the object of this section to determine the range of pairings with higher cocycles from the generalized Streda formulas and thereby to establish a gap labelling by higher cocycles. The essential inputs for this are:

\begin{itemize}


\item The duality of pairings established in Section~\ref{sec-PairingsDuality}, together with the behavior of the generators $[e_I]_0$ and $[v_J]_1$ under the connecting maps, summarized in \eqref{eq-ExpIndLink}.

\item The fact that $\lim_{\BB\to 0}\Tt(e_I)=0$ unless $I=\emptyset$. For example, the Powers-Rieffel projection has this property. The general case follows from Elliott's work \cite{Ell}. 

\end{itemize}

We will adopt the notation from \cite{BM} and define $\BB_I$ to be the skew-symmetric matrix obtained by restricting the indices to the set $I$. Furthermore, ${\rm Pf}(\BB_I)$ will denote its Pfaffian. Then, with the input from above, we can prove the following result. 

\begin{theorem}
\label{theo-PairingRange}
Let $I,J\subset\{1,\ldots,d\}$ be increasingly ordered sets of even cardinality. Then the generators $e_J$ of $K_0(\Aa_{\BB,d})$ presented in Section~\ref{Sec-GeneratorsKGroups} and the cocycles $\xi_I$ pair as follows:
\begin{align}
\label{eq-paringrangeeven1}
& \big \langle [\xi_{I}],[e_J]_0\big \rangle
\;=\;
0\;,
& I\setminus J\not=\emptyset\;,
\\
\label{eq-paringrangeeven2}
& \big \langle[ \xi_{I}],[e_J]_0\big \rangle
\;=\;
1 \;, & I=J\;,
\\
\label{eq-paringrangeeven3}
& \big \langle [\xi_{I}],[e_J]_0\big \rangle
\;=\;
(2\pi)^{-\frac{1}{2}| J\setminus I|} \; {\rm Pf}(B_{J\setminus I})\;,
& I\subset J\;.
\end{align}
Similarly, for index sets $I$ and $J$ of odd cardinality and generators $v_J$ of $K_1(\Aa_{\BB,d})$ defined in Section~\ref{Sec-GeneratorsKGroups}, one has
\begin{align}
\label{eq-paringrangeodd1}
& \big \langle [\xi_{I}],[v_J]_1\big \rangle
\;=\;
0\;,
& I \setminus J\not=\emptyset\;,
\\
\label{eq-paringrangeodd2}
& \big \langle [\xi_{I}],[v_J]_1\big \rangle
\;=\;
1 \;, & I=J\;,
\\
\label{eq-paringrangeodd3}
& \big \langle [\xi_{I}],[v_J]_1\big \rangle
\;=\;
 (2\pi)^{-\frac{1}{2}| J\setminus I|} \; {\rm Pf}(B_{J\setminus I})\;,
& I\subset J\;
\end{align}
\end{theorem}

\noindent {\bf Proof.} First of all, the generators $e_J$ and $v_J$ are all in $C^1(\Ff_d)$, which follows automatically from the their explicit construction. Then, \eqref{eq-paringrangeeven1} follows from the fact that when $i\in I\setminus J$, the cocycle involves a derivative $\partial_i$ for which $\partial_i e_J=0$ because $e_J$ only lies in the algebra generated by $u_j$ with $j\in J$, see Section~\ref{Sec-GeneratorsKGroups}. Similarly, one argues for \eqref{eq-paringrangeodd1}. 

Next, let us show \eqref{eq-paringrangeeven2} and \eqref{eq-paringrangeodd2}. These pairings are the strong pairings over the algebra $\Aa_{\BB,|I|}$. For $J=\emptyset$, $\langle \xi_\emptyset, e_\emptyset \rangle =1$ because the trace is normalized. Now \eqref{eq-paringrangeodd2} for $J=\{1\}$ can be studied as a paring over $\Aa_1$. As $[v_{\{1\}}]_1=[u_1]_1$ and $\partial_1 u_1=-\I u_1$ by \eqref{BulkDerivation}, one finds from the definition \eqref{XiBulkDef}
$$
\big \langle [\xi_{\{1\}}],[v_{\{1\}}]_1\big \rangle
\;=\;
\I\; \Tt(u_1^\ast \partial_1u_1)
\;=\;
1
\;.
$$
The pairing \eqref{eq-paringrangeeven2} for $J=\{1,2\}$ can be deduced via the bulk-boundary correspondence \eqref{eq-pairdualeven1} combined with the defining relation \eqref{eq-ExpIndLink} of the generators
$$
\big \langle [\xi_{\{1,2\}}],[e_J]_0\big \rangle
\;=\;
\big \langle [\xi_{\{1,2\}}],\Exp^{-1}[v_{\{1\}}]_1\big \rangle
\;=\;
\;\big \langle [\xi_{\{1\}}],[v_{\{1\}}]_1\big \rangle
\;=\;1
\;.
$$
If we now assume that \eqref{eq-paringrangeodd2} is true for some $I$ with $|I|$ odd and $j$ is larger then any index in $I$, then one infers from \eqref{eq-ExpIndLink} and the bulk-boundary principle \eqref{eq-pairdualeven2}
$$
\big \langle [\xi_{I\cup \{j\}}],[e_{I \cup \{j\}}]_0\big \rangle
\;=\;
\big \langle [\xi_{I\cup \{j\}}],\Exp_j^{-1}[v_{I}]_1\big \rangle
\;=\;
\;\big \langle [\xi_{I}],[v_{I}]_1\big \rangle
\;=\;1
\;.
$$
Likewise, if \eqref{eq-paringrangeeven2} holds for some $I$ with $|I|$ even, then from \eqref{eq-ExpIndLink} and  \eqref{eq-pairdualodd2},
$$
\big \langle [\xi_{I\cup \{j\}}],[v_{I \cup \{j\}}]_0\big \rangle
\;=\;
\big \langle [\xi_{I\cup \{j\}}],-\Ind_j^{-1}[v_{I}]_1\big \rangle
\;=\;
\;\big \langle [\xi_{I}],[e_{I}]_1\big \rangle
\;=\;1
\;.
$$

\vspace{.1cm}

Now let us consider \eqref{eq-paringrangeeven3} as a weak pairing in $\Aa_{\BB,|J|}$. Then by Theorem~\ref{theo-StredaGen} one has for any pair partition $(p_1,\ldots,p_k)$ of $J\setminus I$
$$
\partial_{B_{p_1}}
\!\cdots\,
\partial_{B_{p_k}}
\big \langle [\xi_{I}],[e_J]_0\big \rangle
\;=\;
\tfrac{1}{(2\pi)^k}
\;\big \langle [\xi_{\{p_1,\ldots,p_k\} \cup I}],[e_J]_0\big \rangle
\;.
$$
After re-ordering the indices in the increasing order,
\begin{equation}
\label{eq-toporder}
\partial_{B_{p_1}}
\!\cdots\,
\partial_{B_{p_k}}
\big \langle [\xi_{I}],[e_J]_0\big \rangle
\;=\;
\tfrac{1}{(2\pi)^k} \;\eta_{p_1,\ldots,p_k}
\big \langle [\xi_{J}],[e_J]_0\big \rangle 
\;,
\end{equation}
where $\eta_{p_1,\ldots,p_k}$ is the sign of the permutation which arranges $\{p_1,\ldots,p_k\}$ in an increasing order. Furthermore, derivatives w.r.t. $B_{i,j}$ with $i$ and $j$ either in $I$ or appearing twice vanish. In order to avoid redundancies,  we only consider pairs $p_l=(i_l,j_l)$ satisfying $i_l<i_{l+1}$ and $i_l<j_l$.  All these facts imply that $\big \langle [\xi_{I}],[e_J]_0\big \rangle$ is a polynomial in the magnetic field components of the form
$$
\big \langle [\xi_{I}],[e_J]_0\big \rangle
\;=\;
\sum_{l=0}^k
\sum_{(p_1,\ldots,p_l)}\alpha^{(l)}_{p_1,\ldots,p_l}\;B_{p_1}\cdots B_{p_l}
\;,
$$
where the sum runs over pair partitions of subsets of $J\setminus I$ of cardinality $2l$ with pairs $p_l=(i_l,j_l)$ satisfying $i_l<i_{l+1}$ and $i_l<j_l$, and the $\alpha^{(l)}_{p_1,\ldots,p_l}$ are complex coefficients.  Let us first show that all lower coefficients with $l<k$ actually vanish. They are given by
$$
\alpha^{(l)}_{p_1,\ldots,p_l}=\left . \partial_{B_{p_1}}
\!\cdots\,
\partial_{B_{p_l}}
\big \langle [\xi_{I}],[e_J]_0\big \rangle \right |_{\BB=\bm 0}
\;=\;
\left . \tfrac{1}{(2\pi)^k}
\;\big \langle [\xi_{\{p_1,\ldots,p_l\} \cup I}],[e_J]_0\big \rangle \right |_{\BB=\bm 0}
\;.
$$
Using the duality of the pairings and the iteration used to construct the generators in Section~\ref{Sec-GeneratorsKGroups}, the last pairing is given by
$$
\big \langle [\xi_{\{p_1,\ldots,p_l\} \cup I}],[e_J]_0\big \rangle
\;=\;
\big \langle [\xi_{\emptyset}],[e_{J \setminus (\{p_1,\ldots,p_l\} \cup I)}]_0\big \rangle
\;.
$$
But as $l<k$, this is a pairing of the trace with a non-trivial class, hence it vanishes as $\BB$ goes to zero by Elliott's result \cite{Ell}. The top coefficients $\alpha^{(k)}_{p_1,\ldots,p_k}$ can be read off from \eqref{eq-toporder} so that, as the sum runs over the ordered pair partitions as above,
\begin{align*}
\big \langle [\xi_{I}],[e_J]_0\big \rangle
 \;=\;
\tfrac{1}{(2\pi)^k} \;\big \langle [\xi_{J}],[e_J]_0\big \rangle 
\;\sum_{(p_1,\ldots,p_k)}
\eta_{p_1,\ldots,p_k}
\;B_{p_1}\cdots B_{p_l}
 \;=\;
\tfrac{1}{(2\pi)^k} \;{\rm Pf}(\BB_{J \setminus I})
\;.
\end{align*}
For \eqref{eq-paringrangeodd3} one can proceed similarly.
\hfill $\Box$

\vspace{.2cm}

Theorem~\ref{theo-PairingRange} immediately implies the following generalization of the results from \cite{Pim,Ell,BM} which deal with the case $I=\emptyset$. Our new result, applied to solid state systems, provides a gap labelling by higher invariants. 

\begin{corollary}
\label{coro-Ranges}
The image $\Ran(\xi_I)$ of the index pairings with a cocycle $\xi_I$ on $\Aa_{\BB,d}$ is given by
$$
\boxed{\;
\Ran(\xi_I)
\;=\;
\ZM\;+\;
\sum_{I\subset J} (2\pi)^{-\frac{1}{2}| J\setminus I|}
\,{\rm Pf}(\BB_{J\setminus I}) \,\ZM
\;,
\;}
$$
where the sum goes only over $J$'s with $|J|$ even.
\end{corollary}

\chapter{Index theorems for solid state systems}
\label{Chap-IndexTheorems}

\abstract{The values of the parings between cyclic cohomology and $K$-theory have already been determined in Section~\ref{sec-PairingsRange} and only the strong topological invariants take integer values in general. In this chapter index theorems are proved for these strong invariants which allow to extend this integrality to the regime with mobility gap. For this purpose it is first shown how cyclic cocycles obtained from  Fredholm modules pair integrally with $K$-theory. When combined with a local index formula, this allows to prove the integrality of the strong topological invariants and to establish their stability under MBGH. Furthermore, the delocalized nature of surface states is proved for non-trivial topological insulators.}

\section{Pairing $K$-theory with Fredholm modules}   
\label{sec-KFredPair}

This section, essentially taken from \cite{Con}, reviews the definition of (bounded) Fredholm modules and how they pair with $K$-theory. 

\begin{definition}
\label{FredModDef} 
Let $\mathcal A$ be a C$^\ast$-algebra. An even Fredholm module $(\pi,\mathcal{H},F,\gamma)$ over  $\mathcal A$ consists of a representation $\pi$ of $\mathcal A$ on a Hilbert space $\mathcal{H}$, a bounded operator $F$ and a grading operator $\gamma$ on $\mathcal H$ satisfying $\gamma^* = \gamma$ and $\gamma^2={\bf 1}$, such that:
\begin{enumerate}[\rm (i)]
\item $F^* =F\,,$ 
\item $F^2=\one\,,$ 
\item For all $a \in \mathcal A$, $[F,\pi(a)] \in \mathcal K(\mathcal H)$ is compact,
\item $\gamma \pi(a)= \pi(a) \gamma\;\;$ for all $ a \in \mathcal A$, 
\item $\gamma F = -F \gamma\,.$
\end{enumerate}
An odd Fredholm module $(\pi, \mathcal{H},F)$ over a C$^*$-algebra $\mathcal A$ is defined by just the first three properties {\rm (i)-(iii)}. For an even Fredholm module, the operator $F$ is off-diagonal in the spectral representation of the grading $\gamma$ and the off-diagonal is a unitary operator $G$ which will also be called the Dirac unitary:
\begin{equation}
\label{eq-GFdef}
F
\;=\;
\begin{pmatrix}
0 & G^\ast \\ G & 0
\end{pmatrix}
\;,
\qquad
\gamma
\;=\;
\begin{pmatrix}
\one & 0 \\ 0 & -\one
\end{pmatrix}
\;.
\end{equation}
The restrictions of $\pi$ to the spectral subspaces of $\gamma$ are also denoted by $\pi$.  
\end{definition}

Let us point out the similarities of the operator $F$, $G$ and $\gamma$ with the flat band Hamiltonian $\sgn(H)$, the Fermi unitary $U_F$ an the chiral symmetry $J$ respectively. Thus in the terminology of symmetries, one could refer to an even Fredholm module also as an odd Fredholm module with chiral symmetry $\gamma$. For the index calculation below, it is important to impose further traceclass conditions on the Fredholm modules.

\begin{definition} A Fredholm module is said to be $n$-summable over a dense Fr\'{e}chet sub-algebra $\mathscr A \in \mathcal A$ if $[F,\pi(a)]$ belongs to the $n$-th Schatten class for all $a$ in $\mathscr A$:
\begin{equation}
\label{eq-Schattenbound}
{\rm Tr}\Big( \big |[F,\pi(a)] \big |^n\Big) 
\;<\; \infty
\;.
\end{equation}
\end{definition}

Given an $(n+1)$-summable Fredholm module one can build up a quantized calculus consisting of a graded algebra 
$$
{\bm \Omega} \;=\; 
\bigoplus {\bm \Omega}^k\;, 
\qquad  
{\bm \Omega}^k 
\;=\; \mathrm{span}\{ a_0[F,a_1] \cdots [F,a_k]\,:\, \ a_j \in \mathscr{A}\}
\;,
$$
furnished with an external differentiation
$$ 
{\bm \Omega}^k \ni \eta \;\;\mapsto\;\; d \eta \;=\; F \eta -(-1)^k \eta F \in {\bm \Omega}^{k+1}
\;,
$$
and a closed graded trace:
$${\bm \Omega}^{n}\ni \eta 
\;\;\mapsto\;\;
\mathrm{Tr}'\{\eta\}
\;=\; 
\left \{
\begin{array}{cc}
\tfrac{1}{2} \ \mathrm{Tr}(\gamma F d\eta )\;, & \mbox{ for an even Fredholm module}\;, \\
\tfrac{1}{2} \ \mathrm{Tr}(F d\eta)\;, & \mbox{ for an odd Freholm module}\;.
\end{array} \right. 
$$

\begin{proposition}[\cite{Con}, p.~295-296]
\label{ChernPairing} 
Given an $(n+1)$-summable Fredholm module over $\Aa$ which is even for $n$ even and odd for $n$ odd, define
\begin{align}
\label{EvenChernCharacter}
\zeta_{n}(a_0,a_1, & \ldots,a_{n}) 
\;=\; 
{\rm \Gamma}_n \ \mathrm{Tr}'\big(\pi(a_0) [F,\pi(a_1)] \cdots [F,\pi(a_{n})]\big)
\;,
\end{align}
where 
$$
{\rm \Gamma}_n\;=\;(-1)^\frac{n}{2} \;\;\;\;\mbox{ for }n \mbox{ even },
\qquad 
{\rm \Gamma}_n \;=\;(-1)^\frac{n+1}{2} 2^{-n}\;\;\;\;\mbox{ for }n \mbox{ odd }.
$$
Then $\zeta_n$ is a cyclic $n$-cocycle over $\mathscr A$. Its cohomology class is called the $n$-th even Connes-Chern character of the Fredholm module and will be denoted by $\mathrm{Ch}_n(\mathcal H,F)$. The Connes-Chern characters pair integrally with the $K$-groups via indices of Fredholm operators on $\Hh$:

\begin{enumerate}[\rm (i)]
\item Even Connes-Chern characters pair with $K_0(\mathcal A)$:
\begin{equation}
\label{ChernPairing1}
\langle \mathrm{Ch}_n(\mathcal H,F),[e]_0-[s(e)]_0 \rangle 
\;=\; 
\Ind \Big ( \pi(e)  \, G \, \pi(e) \Big) \; \in\; \mathbb{Z}
\;,
\end{equation} 
where $G$ is defined by \eqref{eq-GFdef} and it is understood that, on the r.h.s., the representative $e$ for the class $[e]_0$ was taken from $\mathscr A$ and $\pi(e) G \pi(e)$ is an operator on the range of $\pi(e)$.

\item Odd Connes-Chern characters pair with $K_1(\mathcal A)$:
\begin{equation}\label{ChernPairing2}
\langle \mathrm{Ch}_n(\mathcal H,F),[v]_1 \rangle 
\;=\;
\Ind \Big ( E \, \pi(v) \, E \Big ) \ \in \mathbb{Z}
\;,
\end{equation}
where $E$ is the projection $E=\frac{1}{2}(\one+F)$. It is understood that, on the r.h.s., the representative $v$ of the class $[v]_1$ was taken from $\mathscr A$ and $E \pi(v) E$ is an operator on the range of $E$.
\end{enumerate}
\end{proposition}

\begin{remark} The constants $\Gamma_n$ are chosen as in \cite{Kha} (see pp. 174 and 176), but note that our super-trace $\Tr'$ differs by a factor $\frac{1}{2}$. 
\hfill $\diamond$ 
\end{remark}

The $(n+1)$-summability condition implies the Calderon-Fedosov conditions \cite{Cal,Fed}, namely for even $n$
\begin{align*}
&{\rm Tr}\Big ( \pi(e) - \big (\pi(e)  \, G^{*_j} \, \pi(e)\big ) \big (\pi(e)  \, G^{*_{j+1}} \, \pi(e)\big )^\ast \Big)^{m} 
\;<\; \infty\;, 
\qquad
j=0,1
\;,
\end{align*}
and for odd $n$
\begin{align*}
{\rm Tr}\Big (E - \big( E \, \pi(v)^{*_j} \, E \big) \big(E \, \pi(v)^{*_{j+1}} \, E\big ) \Big)^{m} \;<\; \infty \;,
\qquad
j=0,1
\;,
\end{align*}
where $m$ is any integer larger or equal to $\tfrac{1}{2}(n+1)$. The summability condition ensures that these operators belong to the Fredholm class and, moreover, allows to calculate the Fredholm index using the Calderon-Fedosov formula \cite{Cal,Fed} which leads to the proof of Proposition~\ref{ChernPairing} (see also  \cite{Kha} for a detailed derivation).  These traceclass estimates are relevant for Section~\ref{SecIndexFormulas}.

\section{Fredholm modules for solid state systems}

Fredholm modules in the sense of Definition~\ref{FredModDef} are typically obtained from so-called $K$-cycles consisting of a representation and an unbounded Dirac operator satisfying certain properties. $K$-cycles are also called spectral triples or unbounded Fredholm modules. We will not build up the full formalism in this section \cite{Con,GVF}, but rather directly present the natural Dirac operators for condensed matter systems and show how they lead to summable even and odd Fredholm modules over the algebras $\Aa_d$ and $\Ee_d$ of bulk and boundary observables. This implements Connes' program for solid state systems.

\vspace{.2cm}

Let us first comment on how these structures came to be. In the non-commutative geometry of the IQHE in two space dimensions \cite{Bel2,BES,Con}, the construction of the (even) Fredholm module starts from the unitary transformation 
$$
G \;=\; 
\frac{X_1 + \I X_2}{|X_1 + \I X_2|}
\;,
$$ 
where $X_1$ and $X_2$ are the self-adjoint position operators on $\ell^2(\ZM^2)$ and $G|0\rangle=|0\rangle$. This unitary describes the effect of a magnetic tube flux threaded through the lattice at the origin \cite{ASS,BES}, see also \cite{DS}. The operator $F$ and the grading are then defined as in \eqref{eq-GFdef}. The Fredholm module described above can be recast in a form which pertains to higher dimensional extensions:
$$
F\; =\; 
\frac{ \sigma_1 X_1 + \sigma_2 X_2  }{ |\sigma_1 X_1 + \sigma_2 X_2| }
\;,
$$
and $\gamma = \sigma_3$, with $\sigma_i$'s being the Pauli matrices. Hence the operator $F$ is nothing else but the sign of the self-adjoint Dirac operator $D=\sigma_1X_1+\sigma_2 X_2$ on $\CM^2\otimes\ell^2(\ZM^2)$. After Fourier transform, $D$ takes the more familiar form $\langle (\sigma_1,\sigma_2), \nabla_k \rangle$ of the Dirac operator on the two-torus of the momentum space, and the grading is the natural grading of the complex Clifford algebra $Cl_{2}$ given by $\sigma_3 = -\I \gamma_1 \gamma_2$. It is now quite apparent how to extend the Dirac operator in the position representation and the associated the Fredholm module to higher space dimensions, both even and odd.

\vspace{0.2cm}

Let us first focus on the Fredholm modules for the bulk algebras. In the following, $\gamma$ and $\sigma$ represent the generators of the irreducible representations of the even and odd Clifford algebras, respectively, with the CCR strictly enforced. Also, we recall that the sign function is defined as $\sign(t) = \pm 1$ for strictly positive/negative $t$ and $\sign(0) = 0$.

\begin{proposition}
\label{FredMod}  
Let $x_0$   be a shift taking values in the open cube $\Cc_d = (0,1)^d$.
\begin{enumerate}[\rm (i)] 
\item For even space dimensions $d=2k$, let 
$$
D_{x_0}: \mathbb C^{2^k} \otimes \ell^2(\mathbb Z^d) \rightarrow \mathbb C^{2^k} \otimes \ell^2(\mathbb Z^d) 
\;,
\qquad
D_{x_0} \;=\; \sum_{i=1}^d\gamma_i \otimes (X_i + x_{0,i})
\;,
$$
be the shifted self-adjoint Dirac operator. Then
$$
\Ff_{\rm even} 
\;=\; 
\Big ({\rm id}\otimes \pi_\omega, \mathbb C^{2^k} \otimes \ell^2(\mathbb Z^d), \sgn (D_{x_0}),\gamma_0 \otimes \one \Big)_{(x_0,\omega) \in \Cc_d \times \Omega}
$$
defines a field of even Fredholm modules over $\Aa_d$. 

\item For odd space dimensions $d=2k+1$, let 
$$ 
D_{x_0}: \mathbb C^{2^k} \otimes \ell^2(\mathbb Z^d) \rightarrow \mathbb C^{2^k} \otimes \ell^2(\mathbb Z^d)
\;,
\qquad  
D_{x_0} \;=\; \sum_{i=1}^d\sigma_i \otimes (X_i + x_{0,i})
\;,
$$
be the shifted self-adjoint Weyl operator. Then
$$
\Ff_{\rm odd} 
\;=\; 
\Big ({\rm id}\otimes \pi_\omega, \mathbb C^{2^k} \otimes \ell^2(\mathbb Z^d), \sgn (D_{x_0}) \Big)_{(x_0,\omega) \in \Cc_d \times \Omega}
$$
defines a field of odd Fredholm modules over $\Aa_d$. 

\end{enumerate}
\end{proposition}

In the following, we will use the conventions
\begin{equation}
\label{eq-conventions}
\sgn(D_{x_0})\;=\;F_{x_0}\;, 
\qquad
{\rm id} \otimes \pi_\omega \;=\; \pi_\omega \;,
\qquad 
\gamma_0 \otimes \one \;=\; 
\gamma_0
\;,
\end{equation}
as well as $E_{x_0}=\frac{1}{2}(\one+F_{x_0})$ and $G_{x_0}$ as in \eqref{eq-GFdef}. To further compactify notations, we will use $\nu = (\gamma_1, \ldots,\gamma_d)$ for $d$ even, and $\nu = (\sigma_1, \ldots,\sigma_d)$ for $d$ odd. Furthermore, the standard basis for $\mathbb C^{2^k} \otimes \ell^2(\mathbb Z^d)$ will be denoted by $|\alpha\rangle \otimes |x\rangle$. If $A$ is an operator on this Hilbert space, then $\langle x |A | y \rangle $ will be understood as the $2^k \times 2^k$ matrix and $|\langle x |A | y \rangle |$ will represent the operator norm of the matrix.

\vspace{.2cm}

\noindent {\bf Proof.} Items (i), (ii), (iv) and (v) of Definition~\ref{FredModDef} are satisfied by construction because the spectra of the Dirac and Weyl operators do not contain the origin due to the shift $x_0$. Let us now concentrate at point (iii) of Definition~\ref{FredModDef}. Since the algebra of compact operators is closed in the operator norm, it is enough to consider only polynomials $\sum_y p(y) u^y$ from $\Aa_d$. Then
$$
[F_{x_0}, \pi_\omega(u^y)]  
\;=\; 
\big (F_{x_0} - F_{x_0-y} \big ) \pi_\omega(u^y)
\;.
$$
The matrix elements of $F_{x_0} - F_{x_0-y}$ are diagonal,
$$
\langle x |F_{x_0} - F_{x_0-y} | x' \rangle 
\;=\; 
\delta_{x,x'}\big ( \sgn\langle \nu,x + x_0 \rangle - \sgn \langle \nu, x -y + x_0 \rangle \big )
\;,
$$
where the sign functions act on Hermitean matrices as usual. As
\begin{equation}\label{Asy}
\sgn\langle \nu,x  \rangle - \sgn \langle \nu, x -y \rangle 
\;\sim \;
|x|^{-1} \big \langle \nu,y +|x|^{-2}\langle x, y  \rangle x \big \rangle
\;,
\qquad
\mbox{\rm for }\;|x| \rightarrow \infty
\;,
\end{equation}
the diagonal elements of the diagonal operator $F_{x_0} - F_{x_0-y}$ decay to zero. Hence it is compact. 
\hfill $\Box$

\vspace{0.2cm}

In order to define the Connes-Chern characters, the Fredholm modules must be finitely summable. Hence, the following somewhat technical statement is important. 

\begin{lemma}
\label{BulkSummability} 
The families of Fredholm modules $\Ff_{\rm even}$ and $\Ff_{\rm odd}$ are $n$-summable over the Fr\'echet algebra $\mathscr A_d$ for any $n \geq d+1$.
\end{lemma}

\noindent {\bf Proof.} Let $a = \sum_{y} a(y) u^y \in \mathscr A_d$ and recall that $a(y)$ are continuous functions of $\omega$. Due to the Minkovski inequality for the Schatten norms and the usual decomposition $a=a_r +\I a_i$, where $a_r = \frac{1}{2}(a+a^\ast)$ and $a_i = \frac{1}{2 \I}(a-a^\ast)$ are the real and imaginary parts of $a$, respectively,  it is sufficient to consider the case $a=a^*$. Recall that $\langle x |\cdot |x' \rangle$ is viewed as a matrix, which in this context is from $M_{Q}(\CM)$ with $Q$ the dimension of the space for the irreducible representations of the Clifford algebras. We will first show that
\begin{equation}
\label{S1}
\sum_{x' \in \ZM^d}\| \langle x | [F_{x_0}, \pi_\omega(a)]^k| x' \rangle \|_{(1)} 
\;\leq\; 
\mbox{\rm const}\cdot |x + x_0|^{-k}
\;,
\end{equation}
for any positive integer k, where we recall that $\| \cdot \|_{(n)}$ represents the $n$-Schatten norm. Let us denote the l.h.s. of \eqref{S1} by $Y$. Let us write out the commutator
$$
\langle x|   [F_{x_0}, \pi_\omega(a)]|y\rangle  
\;=\;  
e^{\I \langle x-y|\BB_+ | y \rangle } a(\tau_x \omega, x-y) \big ( \sgn\langle \nu, x + x_0 \rangle - \sgn \langle \nu,y + x_0 \rangle \big )
\,.
$$
When raising the commutator to power $k$, one generates convolutions over the lattice sites accompanied by shifts of $\omega$ and Peierls factors. As such, the following upper bound can be readily derived
\begin{align*}
Y \;\leq\;\, 
& Q \, \sum_{x_1,\ldots,x_{k+1} \in \mathbb Z^d}\, \delta_{x_1, 0}\,  \prod_{i=1}^k \;|a(\tau_{x_i} \omega, x_i-x_{i+1})| \\
& \cdot \big | \sgn \langle \nu,x_i + x + x_0 \rangle - \sgn \langle \nu,x_{i+1} + x + x_0 \rangle \big |
\;,
\end{align*}
where $|\cdot|$ represents the matrix norm (same as absolute value if the matrix is one-dimensional). Due to the asymptotic behavior in Eq.~\eqref{Asy}, the supremum 
$$
S(y,y')\;=\;\sup_{x \in \mathbb R^d} \;| x|\, \big |\sgn \langle \nu,y + x \rangle - \sgn \langle \nu, y' + x \rangle \big |
$$
is finite. Since the sign function is invariant to scaling by scalars, the homogeneity $ S(s y, s y') = s\,S(y, y')$ holds for all $s\in\RM_+$. Taking $s = (|y|+|y'|)^{-1}$, one concludes
$$
S(y,y') 
\;\leq \;
(|y|+|y'|) \sup_{|x|+|x'| = 1} \;S(x,x')
\;.
$$
Returning to $Y$, this gives
$$
Y 
\;\leq \;
 \mbox{\rm const}\cdot |x + x_0|^{-k} 
\,\sum_{x_1,\ldots,x_{k+1} \in \mathbb Z^d}\, \delta_{x_1, 0} 
\prod_{i=1}^k \;(| x_i|+|x_{i+1}|)\;  |a(\tau_{x_i}\omega, x_i-x_{i+1})|
\;.
$$ 
We now make the change of variables $y_i = x_{i+1}-x_i $, $i=1,\ldots,k$, and since $ x_1=0$,
$$
 x_{i+1} 
\;=\;
y_i + \ldots + y_1
\;\;\; \Longrightarrow \;\;\;
|x_{i+1}|\; \leq\; \prod_{j=1}^k (1+| y_j|)
\;.
$$
Then let us continue with
\begin{align}\label{KPower}
Y 
\; \leq \;
 \mbox{\rm const}\cdot |x + x_0|^{-k} 
\,\sum_{y_1,\ldots,y_k \in \mathbb Z^d}  
\prod_{i=1}^k \;(1+|y_i|)^k\; |a(\tau_{x_i}\omega, y_i)|
\;,
\end{align} 
and furthermore
\begin{align*}
Y 
\; \leq \;
 \mbox{\rm const}\cdot |x + x_0|^{-k} \Big (\sum_{y \in \mathbb Z^d}(1+| y|)^k\; \sup_{\omega \in \Omega} |a(\omega, y)| \Big )^k.
\end{align*} 
The sum is finite due to \eqref{Decay1} and this concludes the proof of \eqref{S1}. Now, since $a$ is self-adjoint, $\I [F_{x_0}, \pi_\omega(a)]$ is self-adjoint and
$$
\big |[F_{x_0}, \pi_\omega(a)] \big |^{d+1}
\;=\; 
\big (\I [F_{x_0}, \pi_\omega(a)]\big )^{d+1} U
\;,
$$
with $U = \sgn \big ( \I [F_{x_0}, \pi_\omega(a)]\big )^{d+1}$ a unitary operator. Of course, for $d$ odd, $U$ is just the identity. Then
\begin{align*}
{\rm Tr}\left (\big |[F_{x_0}, \pi_\omega(a)]\big |^{d+1} \right ) 
\; & \leq \;
\sum_{x \in \ZM^d} \sum_{x' \in \ZM^d} \|\langle x | \big (\I [F_{x_0}, \pi_\omega(a)]\big )^{d+1} |x' \rangle \langle x' |U|x \rangle \|_{(1)}
\end{align*}
and, by using $|\langle x' |U|x \rangle|\leq 1$ together with \eqref{S1},
$$
{\rm Tr}\left (\big |[F_{x_0}, \pi_\omega(a)]\big |^{d+1} \right ) \;\leq \;
\mbox{\rm const}\cdot \sum_{x \in \ZM^d} |x + x_0|^{-d-1} 
\;<\; \infty
\;,
$$
and $(d+1)$-summability follows. \hfill $\Box$

\vspace{0.2cm}

The proof shows that the summability condition for both $\Ff_{\rm even}$ and $\Ff_{\rm odd}$ will not hold for integers lower than $d+1$. Hence, the lowest Connes-Chern characters that can be defined for  $\Ff_{\rm even}$ and $\Ff_{\rm odd}$ are ${\rm Ch}_d$. This makes out the difference between the strong and the weak Chern numbers. Let us now collect the main conclusion for the topological invariants following from the pairing of the Fredholm modules and $K$ groups of $\Aa_d$. 

\begin{theorem}
\label{BulkChernCharacters} 
The $d$-th Connes-Chern characters
$$
\zeta_{d}( a_0, a_1, \ldots,a_{d}) 
\;=\;  
{\rm \Gamma}_d \, \mathrm{Tr}'\Big (\pi_\omega(a_0) [F_{x_0},\pi_\omega (a_1)], \ldots, [F_{x_0},\pi_\omega (a_d)] \Big ) 
$$
are well defined over $\mathscr A_d$ for both $\Ff_{\rm even}$ and $\Ff_{\rm odd}$. They pair integrally with the $K$-groups of $\Aa_d$ as described in Proposition~\ref{ChernPairing}. In particular:

\begin{enumerate}[\rm (i)]

\item If $d$ is even and $h \in M_N(\mathbb C) \otimes \Aa_d$ is a short-range Hamiltonian for which BGH applies, then
$$
({\rm tr} \otimes \zeta_d) \big (p_F,\ldots,p_F \big) 
\;=\; 
\Ind\big ( P_\omega G_{x_0} P_\omega \big ) 
\;,
$$
where $P_\omega= \pi_{\omega}(p_F) = \chi(H_\omega \leq \mu)$ is the physical representation of the Fermi projection $p_F$ at disorder configuration $\omega$. 

\item If $d$ is odd and $h \in M_{2N}(\mathbb C) \otimes \Aa_d$ is a short-range Hamiltonian for which BGH and CH apply, then
$$
({\rm tr} \otimes \zeta_d ) \big (u_F^\ast-\bm 1_{N},\ldots,u_F-\bm 1_{N} \big ) 
\;=\; 
\Ind \big (E_{x_0}  U_\omega E_{x_0} \big )
\;,
$$
where $U_\omega = \pi_\omega (u_F)$ is the physical representation of the Fermi unitary $u_F$ at disorder configuration $\omega$.
\end{enumerate}
In both odd and even cases, the Fredholm indices are independent of $\omega$ or $x_0$. 
\end{theorem}

We now turn our attention to the Fredholm modules over the boundary algebra $\Ee_d$. Since $\Ee_d \simeq \Aa_{d-1} \otimes \Kk$, the analysis and the results and proofs are quite similar to that for the bulk algebra except for the shift from even to odd and odd to even. In accordance with the above, we use tildes on the notations such as \eqref{eq-conventions}. In an analogous manner to Proposition~\ref{FredMod} one first checks the following.

\begin{proposition}
\label{FredModEdge}
Let $\tilde x_0$ be a shift taking values in the open cube $\Cc_{d-1}$.
\begin{enumerate}[\rm (i)] 

\item For even space dimensions $d=2k$, let $\widetilde \sigma_i$ be the irreducible representation of the complex Clifford algebra $Cl_{2k-1}$ on $\mathbb C^{2^{k-1}}$ (with the same conventions as for the bulk), and let the shifted self-adjoint Weyl operator $\widetilde D_{x_0} $ on $\mathbb C^{2^{k-1}} \otimes \ell^2(\mathbb Z^{d-1} \times \mathbb N)$ be defined as
$$
\widetilde D_{x_0} \;=\; \sum_{i=1}^{d-1}\widetilde \sigma_i \otimes (X_i+\tilde x_{0,i})
\;.
$$
Then
$$
\widetilde \Ff_{\rm odd} 
\;=\; 
\Big ({\rm id}\otimes \widetilde \pi_\omega, \mathbb C^{2^{k-1}} \otimes \ell^2(\mathbb Z^2 \times \mathbb N), \sgn (\widetilde D_{\tilde x_0})\Big)_{(\tilde x_0,\omega) \in \Cc_{d-1} \times \Omega}
$$
defines a field of odd Fredholm modules over $\Ee_d$.

\item For odd space dimensions $d=2k+1$, let $\widetilde \gamma_i$ be the irreducible representation of the complex Clifford algebra $Cl_{2k}$ on $\mathbb C^{2^{k}}$ (with the same conventions as for the bulk), and let the shifted Dirac operator $ \widetilde D_{\tilde x_0}$ on $\mathbb C^{2^k} \otimes \ell^2(\mathbb Z^2 \times \mathbb N)$ be defined by
$$
\widetilde D_{x_0} \;=\; \sum_{i=1}^{d-1}\widetilde \gamma_i \otimes (X_i+\tilde x_{0,1})
\;.
$$
Then
$$
\widetilde \Ff_{\rm even} 
\;=\; 
\Big ({\rm id}\otimes \widetilde \pi_\omega, \mathbb C^{2^k} \otimes \ell^2(\mathbb Z^2 \times \mathbb N), \sgn (\widetilde D_{\tilde x_0}), \widetilde\gamma_0 \otimes \one \Big )_{(\tilde x_0,\omega) \in \Cc_{d-1} \times \Omega}
\;,
$$
defines a field of even Fredholm modules over $\Ee_d$.
\end{enumerate}
\end{proposition}

Next let us spell out the summability properties of these Fredholm modules.

\begin{lemma}
\label{BoundarySummability} 
The families of Fredholm modules $\widetilde \Ff_{\rm even}$ and $\widetilde \Ff_{\rm odd}$ are $n$-summable over the smooth sub-algebra $\mathscr E_d$ for any $n \geq d$.
\end{lemma}

\noindent {\bf Proof.}  First recall that elements in $\mathscr E_d$ are of the form 
$$
\tilde a 
\;=\;
\sum_{n,m \in \mathbb N}\; \sum_{x \in \mathbb Z^{d-1}} \tilde a_{n,m} (\omega,x) \hat u^y (\hat u_d)^n \hat e (\hat u_d^\ast)^m
$$ 
with continuous functions $\tilde a_{n,m}(\omega,x)$ in $\omega$ of rapid decay w.r.t. $n$, $m$ and $x$ (see Proposition~\ref{RapidDecay}). It is convenient to form the matrix $\tilde a(\omega,y)$ out of the Fourier coefficients $\tilde a_{n,m}(\omega,y)$. We will  still view $\langle x|$ and $|x\rangle$ for $x\in \ZM^{d-1}$ as partial isometries, so that
$\langle x| [\widetilde F_{x_0}, \widetilde \pi_\omega(\tilde a)]|y\rangle $ is a matrix from $M_Q(\CM) \otimes \Kk$ with $Q$ the dimension of the space of irreducible representations of the Clifford algebras. One can again write out the commutators explicitly: 
\begin{align*}
\langle x| [\widetilde F_{x_0}, \widetilde \pi_\omega(\tilde a)]|y\rangle 
\; = \;
 e^{\I \phi} \Big ( \sgn \langle \nu, x + \tilde x_0 \rangle - \sgn \langle \nu,y + \tilde x_0 \rangle \Big ) \otimes \tilde a(\tau_{x,n} \omega, x-y) \;,
\end{align*}
where $e^{\I \phi}$ is the Peierls factor $e^{\I \langle x-y|\BB_+ | y \rangle }$. The structure of this commutator is very similar to the one appearing in Lemma~\ref{BulkSummability}, hence the estimate
\begin{equation}
\label{S2}
\sum_{x' \in \ZM^{d-1}} \|\langle x|  [\widetilde F_{\tilde x_0}, \widetilde \pi_\omega(\tilde a)]^k| x' \rangle \|_{(1)} 
 \;\leq \;
 {\rm const} \cdot |x+x_0|^{-k}
 \end{equation}
can be proved by applying identical steps as in the proof of \eqref{S1}. Indeed, 
let us again denote the l.h.s. by $Y$. Then, 
\begin{align*}
Y \;\leq\; 
\sum_{x_1,\ldots,x_{k+1} \in \mathbb Z^{d-1}}\, \delta_{x_1, 0}\;  \Big \|\prod_{i=1}^k \big ( & \sgn \langle \nu,x_i + x + x_0 \rangle - \sgn \langle \nu,x_{i+1} + x + x_0 \rangle \big ) \\
& \otimes a(\tau_{x_i} \omega, x_i-x_{i+1})\Big \|_{(1)}
\;.
\end{align*}
By employing the asymptotics from Lemma~\ref{BulkSummability} and H\"older's inequality for the Schatten norms, we arrive at 
\begin{equation}\label{KPowerBounday}
Y
\; \leq \;
 \frac{\mbox{\rm const}}{ |x + x_0|^k}  
\,\sum_{y_1,\ldots,y_k \in \mathbb Z^{d-1}}  
\prod_{i=1}^k \;(1+|y_i|)^k\; \| \tilde a(\omega_i, y_i)\|_{(k)}
\end{equation} 
where $\omega_i$ are translates of $\omega$ and $y_1 = x_i - x_{i+1}$ for $i = 1,\ldots k$. This can be further processed as 
$$
Y
\; \leq \;
 \frac{\mbox{\rm const}}{ |x + x_0|^k}\; \Big (\sum_{y \in \mathbb Z^{d-1}} 
(1+|y|)^k\; \sup_{\omega \in \Omega} \| \tilde a(\omega, y)\|_{(k)} \Big )^k \;, 
$$
and the sum inside the pharanteses  is finite due to the rapid decay of the Fourier coefficients $\tilde a_{n,m}(\omega, y)$, see \eqref{Decay2}. This concludes the proof of \eqref{S2}. Now, take $\tilde a$ self-adjoint and consider the decomposition
$$
\big |[\widetilde F_{\tilde x_0}, \widetilde \pi_\omega(\tilde a)] \big |^{d}
\;=\; 
\big (\I [\widetilde F_{\tilde x_0}, \widetilde \pi_\omega(\tilde a)]\big )^{d} \widetilde U
\;,
$$
with $\widetilde U$ a unitary operator. Then
$$
{\rm Tr}\left (\big |[\widetilde F_{\tilde x_0}, \widetilde \pi_\omega(\tilde a)]\big |^d \right ) 
\; \leq \; 
\sum_{x \in \ZM^{d-1}} \sum_{x' \in \ZM^{d-1}} \big \| \langle x | \big (\I [\widetilde F_{x_0}, \widetilde \pi_\omega(\tilde a)]\big )^d |x' \rangle \langle x' |\widetilde U|x \rangle \big \|_{(1)} 
$$
and, by using $|\langle x' |U|x \rangle |\leq 1$ together with the generic properties of the Schatten norms and \eqref{S2},
$$
{\rm Tr}\left (\big |[\widetilde F_{\tilde x_0}, \widetilde \pi_\omega(\tilde a)]\big |^{d} \right ) \;\leq \;
\mbox{\rm const}\cdot \sum_{x \in \mathbb Z^{d-1}} |x + x_0|^{-d} 
\;<\; \infty
\;,
$$
and $d$-summability follows.\hfill $\Box$

\vspace{0.2cm}

Below, we summarize the properties of the boundary topological invariants resulting from the pairing of the Fredholm modules and the $K$ groups of $\Ee_d$.

\begin{theorem}\label{BoundaryChernCharacters} 
The $(d-1)$-th Connes-Chern characters
$$
\widetilde  \zeta_{d-1}( \tilde a_0, \tilde a_1, \ldots, \tilde a_{d}) 
\;=\;  
{\rm \Gamma}_{d-1} \, \mathrm{Tr}'\Big (\widetilde \pi_\omega(\tilde a_0) [\widetilde F_{\tilde x_0},\widetilde \pi_\omega (\tilde a_1)], \ldots, [\widetilde F_{\tilde x_0},\widetilde \pi_\omega (\tilde a_{d-1})] \Big ) 
$$
are well-defined over $\mathscr E_d$ for both $\widetilde \Ff_{\rm even}$ and $\widetilde \Ff_{\rm odd}$. They pair integrally with the $K$-groups of $\Ee_d$ as described in Proposition~\ref{ChernPairing}. In particular:

\begin{enumerate}[\rm (i)]

\item If $d$ is even and $\hat h = (h,\tilde h) \in M_N(\mathbb C) \otimes \widehat \Aa_d$ is a short-range half-space Hamiltonian such that BGH holds, then
$$
({\rm tr} \otimes \widetilde \zeta_{d-1} ) \big (\tilde u_\Delta^\ast - \bm 1_N,\ldots,\tilde u_\Delta - \bm 1_N \big ) 
\;=\; 
\Ind \big ( \widetilde E_{x_0} \widetilde U_\omega \widetilde E_{x_0} \big )
\;,
$$
where $\tilde u_\Delta$ is the unitary operator defined in Eq.~\eqref{TildeU} and 
$$
\widetilde U_\omega 
\;=\; 
\widetilde \pi_\omega(\tilde u_\Delta) 
\;=\; 
e^{2\pi \I \FFunc (\widehat H_\omega)}
$$
is a physical representation at disorder configuration $\omega$. Recall that $\FFunc $ is a smooth function with $\FFunc = 1$  below and $\FFunc =0$  above the bulk insulating gap.

\item If $d$ is odd and $\hat h = (h,\tilde h) \in M_{2N}(\mathbb C) \otimes \widehat \Aa_d$ is a short-range Hamiltonian such that BGH and CH apply, then
$$
({\rm tr} \otimes \widetilde \zeta_{d-1}) \big (\tilde p_\Delta - s(\tilde p_\Delta),\ldots,\tilde p_\Delta - s(\tilde p_\Delta) \big ) 
\;=\; 
\Ind\big ( \widetilde P_\omega \widetilde G_{x_0} \widetilde P_\omega \big )
$$
where $\tilde p_\Delta$ is the projection defined in Eq.~\eqref{BoundaryProjection1} and
$$
\widetilde P_\omega 
\;=\; 
\widetilde \pi_\omega(p_\Delta)
\;=\; 
\tfrac{1}{2}\,\hat J \,\big (e^{\I \pi \GFunc(\widehat H_\omega)} +\one_{2N} \big )\; +\; {\rm diag}(0_N,\one_N)
$$
is a physical representation at disorder configuration $\omega$. Here $\GFunc$ is a smooth odd functions with $\GFunc=\pm 1$ above/below the bulk insulating map.
\end{enumerate}
In both even and odd cases, the Fredholm indices are independent of $\omega$ and $\tilde x_0$.
\end{theorem}

The above result has the following important physical consequence.

\begin{corollary} Assume that the boundary invariants defined in Theorem~\ref{BoundaryChernCharacters} are not zero. Then the Fermi level must be located in the essential spectrum $\widehat h$.
\end{corollary}

\noindent {\bf Proof.} Indeed, if the Fermi level is located in a gap of the essential spectrum of $\hat h$, then we can choose $\FFunc$  and $\GFunc$ such that the support of their derivative is contained entirely inside this gap. In this case, the Fredholm operators are compact, hence have zero index. \hfill $\Box$ 

\vspace{0.2cm}

Since for solid state systems from the unitary class the Fermi level can be located anywhere inside the bulk gap, it follows that, if the invariant defined in Theorem~\ref{BoundaryChernCharacters} is not zero, then the essential spectrum of $\hat h$ must fill the bulk gap $\Delta$ entirely. For solid state system from the chiral unitary class, the Fermi level is constrained to be at the origin, hence the above result assures that, whenever the invariant defined in Theorem~\ref{BoundaryChernCharacters} is not zero, $\hat h$ has essential spectrum at the origin. Note, however, that, as of yet, we cannot make any assertion about the nature of the spectrum at the Fermi level. 

\section{Equality between Connes-Chern and Chern cocycles}
\label{Sec-Equality}

The Connes-Chern characters from Theorem~\ref{BulkChernCharacters} and the Chern cocycles defined in Theorems~\ref{Th-BulkInv} and \ref{Th-BoundaryInvariants} in Chapter~\ref{Chap-TopologicalInvariants} look very different from one another. The Chern cocycle is local in nature in the sense that it involves only derivations and no convolutions. In contradistinction, the Connes-Chern cocycle is non-local as it involves a convolution with the non-local kernel of $F_{x_0}=\sgn (D_{x_0})$. 

\vspace{.6cm}

\begin{theorem}
\label{EqualityConnesChern} \
\begin{enumerate}[\rm (i)] 
\item For odd or even space dimensions, $\xi_d = (-1)^d \; \zeta_d$ over $\mathscr A_d$.

\vspace{0.2cm}

\item For odd or even space dimensions, $\widetilde \xi_{d-1} = (-1)^{d-1} \; \widetilde \zeta_{d-1}$ over $\mathscr E_d$.
\end{enumerate}
\end{theorem}

\noindent {\bf Proof.} There are four cases to be covered and inherently there will be repetitive arguments. Common to all is the observation that the fields of covariant representations over the {\it smooth sub-algebras} are continuous as functions of $\omega$ and $F$ operators are norm-continuous of $x_0$. As a consequence \cite{Con0}, the Connes-Chern cocycles at different $\omega$ and $x_0$ belong to the same cohomology class. Hence the average over $\omega$ and $x_0$ is still a representative for the same cohomology class. Another useful fact is that the cocycles are continuous over $\mathscr A_d$ or $\mathscr E_d$, hence it is enough to prove the equalities for polynomials which form dense subsets in these algebras. Henceforth, all the arguments of the cocycles below are assumed polynomial. Another common feature of the proofs are the use of two geometric identities (one in each case) which are presented in Lemma~\ref{KeyIdentity} below.

\vspace{.2cm}

(i) {\it The $d$ even case for the bulk invariant.} From the definition,
\begin{equation}
\label{Fedosov1}
\zeta_d
\;=\;
\tfrac{1}{2} {\rm \Gamma}_d \int_{{\mathcal C}^d} d x_0 \int_\Omega  \mathbb P (d\omega)  
\sum\limits_{x \in {\mathbb Z}^d} \tr \Big(\gamma_0 \langle x |F_{x_0} \prod_{i=0}^d [F_{x_0},\pi_\omega( a_i)]  | x \rangle \Big )
\;,
\end{equation}
and we use the magnetic translations to shift all the $\mathbb C^{2^k}$ fibers to the origin,
$$
\zeta_d
\;=\;
\tfrac{1}{2} {\rm \Gamma}_d \int_{{\mathcal C}^d} d x_0 \int_\Omega  \mathbb P (d\omega)  
\sum\limits_{x \in {\mathbb Z}^d} \tr \Big(\gamma_0 \langle 0 |F_{x_0+x} \prod_{i=0}^d [F_{x_0 + x},\pi_{\tau_x \omega}( a_i)]  | 0 \rangle \Big )
\;.
$$
The Fredholm module is $(d+1)$-summable, hence the sums and the integrals above are absolute convergence and we can interchange them. Then one can use the invariance of $\mathbb P(d\omega)$ to replace $\tau_x \omega$ by $\omega$. After that, let us combine the summation over $x$ with the integral over $x_0$ to write:  
\begin{align*}
\zeta_d
\;=\;
\tfrac{1}{2}\, {\rm \Gamma}_d \int_{\mathbb{R}^d} d x \int_\Omega \mathbb P (d\omega) \; 
\tr \Big (\gamma_0
\langle 0 | F_x \prod_{i=0}^d [F_x,\pi_{\omega}(a_i)] |0\rangle \Big)
\;.
\end{align*}
Writing out the commutator $[F_x,\pi_{\omega}(a_0)]$ and using the cyclic property of the trace as well as $F_x[F_x,\pi_\omega(a)]=-[F_x,\pi_\omega (a)]F_x$, $F_x^2=\one $ and $\gamma_0 F_x= - F_x \gamma_0$, we arrive at
$$
\zeta_d
\;=\;
{\rm \Gamma}_d \int_{\mathbb{R}^d} d x \int_\Omega \mathbb P (d\omega)  \,
\tr \Big( \gamma_0 \langle 0| \pi_\omega (a_0) \prod_{i=1}^d [F_x,\pi_\omega (a_i) ] |0 \rangle  \Big)
\;.
$$
Next we insert partitions of unity using the projections $\PI_x=\one_{2^k} \otimes |x\rangle \langle x|$:
$$
\zeta_d 
\;=\;
{\rm \Gamma}_d \int_{\mathbb{R}^d} d x \int_\Omega \mathbb P (d\omega) 
\sum\limits_{x_i \in {\mathbb Z}^d} \tr
\Big(\gamma_0 \langle 0 | \pi_\omega(a_0) \prod_{i=1}^d \PI_{x_i}[F_x,\pi_\omega (a_i)] |0 \rangle\Big)
\;.
$$
Since $a_i$ are polynomials, the sums over $x_i$ contain finite number of terms and they can be interchanged with the integrals, to continue
\begin{align*}
\zeta_d
& \;=\; {\rm \Gamma}_d \sum\limits_{x_1,\ldots,x_d \in {\mathbb Z}^d} \ \int_{\mathbb{R}^d} d x \
\tr \Big( \gamma_0 \prod_{i=1}^d \big (\sgn \langle \gamma,x_i + x\rangle - \sgn \langle \gamma,x_{i+1}+x\rangle \big )  \Big)
\; 
\\
& \;\;\;\;\;\;\;\;\;
\cdot \int_\Omega \mathbb P (d\omega) \, \langle 0 | \pi_\omega( a_0)\prod_{i=1}^d \PI'_{x_i}  \pi_\omega(a_i)| 0 \rangle
\;,
\end{align*}
where $\PI'_x = |x\rangle \langle x |$ and $\pi_\omega$ is just the representation on $\ell^2(\mathbb Z^d)$. Also, in the first line, it is understood that $x_{d+1}$ is fixed at the origin. We now use the first identity from Lemma~\ref{KeyIdentity} to deduce
\begin{align*}
\zeta_d
 \;=\; \I^{d}\, \Lambda_d  \sum\limits_{x_1,\ldots,x_d \in {\mathbb Z}^d} \ \sum_{\rho \in \Ss_d} (-1)^\rho \prod_{i=1}^d x_{i,\rho_i} \int\limits_\Omega \mathbb P (d\omega) \, \langle 0 | \pi_\omega( a_0)\prod_{i=1}^d \PI'_{x_i}  \pi_\omega(a_i)| 0 \rangle
 \;,
\end{align*}
where we already combined the normalization constants. Next we combine the sum over $x_i$ with the projections $\PI'_{x_i}$ to form the position operators
$$
\zeta_d
\;=\;  
\I^{d}\, \Lambda_d  \sum_{\rho\in \Ss_d} (-1)^\rho  
\int\limits_\Omega \mathbb P (d\omega) \langle 0 |  \pi_\omega(a_0) \prod_{i=1}^d X_{\rho_i}\pi_\omega (a_i) |0\rangle
\;.
$$
Due to the anti-symmetrizing factor $(-1)^\rho$, one can actually form commutators:
\begin{align*}
\zeta_d & \;= \; \Lambda_d  \sum_{\rho\in \Ss_d} (-1)^\rho\; 
\int\limits_\Omega \mathbb P (d\omega) \langle 0 | \pi_\omega(a_0) \prod_{i=1}^d \I [X_{\rho_i},\pi_\omega(a_i)]  |0 \rangle \\
& \;=\; \Lambda_d  \sum_{\rho\in \Ss_d} (-1)^\rho\; 
\int\limits_\Omega \mathbb P (d\omega) \langle 0 | \pi_\omega \Big (a_0 \prod_{i=1}^d \partial_{\rho_i}a_i\Big)  |0 \rangle
\;.
\end{align*}
Here one can recognize the trace $\Tt$ over $\Aa_d$ and the statement follows.

\vspace{.2cm}

(i) {\it The $d$ odd case for the bulk invariants.} We only highlight the main points. From definition and after a few elementary steps
\begin{equation}\label{Fedosov2}
\zeta_d
\;=\;
\tfrac{1}{2}\, {\rm \Gamma}_d \int_{{\mathcal C}^d} d x_0 \int_\Omega  \mathbb P (d\omega)  
\sum\limits_{x \in {\mathbb Z}^d} \tr \Big(\langle 0 | F_{x_0+x} \prod_{i=0}^d [F_{x_0 + x},\pi_{\tau_x \omega}( a_i)] | 0 \rangle \Big)
\;,
\end{equation}
As before, this can be processed to  
$$
\zeta_d 
\;=\;
{\rm \Gamma}_d \int_{\mathbb{R}^d} d x \int_\Omega \mathbb P (d\omega)  \,
\tr \Big( \langle 0 | \pi_\omega (a_0) \prod_{i=1}^d [F_x,\pi_\omega (a_i) ] | 0 \rangle \Big)
\;,
$$
and by inserting the partitions of unity
\begin{align*}
\zeta_d
& \;=\; {\rm \Gamma}_d \sum\limits_{x_1,\ldots,x_d \in {\mathbb Z}^d} \ \int_{\mathbb{R}^d} d x \
\tr \Big( \prod_{i=1}^d \big (\sgn \langle \sigma,x_i + x\rangle - \sgn\langle \sigma,x_{i+1}+x\rangle \big )  \Big)
\; 
\\
& \;\;\;\;\;\;\;\;\;
\cdot \int\limits_\Omega \mathbb P (d\omega) \, \langle 0 | \pi_\omega( a_0)\prod_{i=1}^d \PI'_{x_i} \pi_\omega(a_i) | 0 \rangle
\;.
\end{align*}
We now use the second identity from Lemma~\ref{KeyIdentity},
\begin{align*}
\zeta_d 
\;=\; 
-\;\I^d \Lambda_d \sum\limits_{x_i \in {\mathbb Z}^d} \sum_{\rho \in \Ss_d} (-1)^\rho \prod_{i=1}^d x_{i,\rho_i}  \int\limits_\Omega \mathbb P (d\omega) \, \langle 0 | \pi_\omega( a_0)\prod_{i=1}^d \PI'_{x_i} \pi_\omega(a_i) | 0 \rangle
\end{align*}
and combine the sum over $x_i$ with the projections $\PI '_i$ to form the position operators, then the commutators,
$$
\zeta_d 
\;=\; 
-\, \Lambda_d \,
 \sum_{\rho\in \Ss_d} (-1)^\rho\; \int\limits_\Omega \mathbb P (d\omega)
\langle 0 | \pi_\omega(a_0) \prod_{i=1}^d \I [\pi_\omega(a_i),X_{\rho_i}]  |0 \rangle
\;.
$$
The statement now follows.

\vspace{.2cm}

(ii) {\it The $d$ odd case for the boundary invariants.} From the definition of the even Connes-Chern cocycle for the boundary algebra:
\begin{align}
\widetilde \zeta_{d-1}\;=\; &
\tfrac{1}{2}{\rm \Gamma}_{d-1} \int_{\Cc^{d-1}} d \tilde x_0 \int_\Omega  \mathbb P (d\omega)  
\sum\limits_{\tilde x \in {\mathbb Z}^{d-1}} \sum\limits_{n \in \mathbb N}
\nonumber
\\
& \;\;
\tr \Big(\widetilde \gamma_0 \langle 0,n |\widetilde F_{\tilde x_0+\tilde x} \prod_{i=0}^{d-1} [\widetilde F_{\tilde x_0 + \tilde x},\widetilde \pi_{\tau_{\tilde x} \omega}( \tilde a_i)]  | 0,n \rangle \Big )
\;,
\label{Fedosov3}
\end{align}
where the magnetic translations parallel to the boundary were used to shift the fibers. The Fredholm modules are summable, hence the sums and the integrals above are absolute convergence and we can interchange them. In this case, one can use the invariance of $\mathbb P(d\omega)$ to replace $\tau_{\tilde x} \omega$ by $\omega$. After that, we can combine the summation over $\tilde x$ with the integral over $\tilde x_0$ to write:  
\begin{align*}
\widetilde \zeta_{d-1}
\;=\;
\tfrac{1}{2} \,{\rm \Gamma}_{d-1}  \int\limits_{\mathbb{R}^{d-1}} d \tilde x \sum_{n \in \mathbb N} \int_\Omega \mathbb P (d\omega) \; 
\tr \Big (\widetilde  \gamma_0
\langle 0,n | \widetilde F_{\tilde x} \prod_{i=0}^{d-1} [\widetilde F_{\tilde x},\widetilde \pi_{\omega}(\tilde a_i)] |0,n\rangle \Big)
\;.
\end{align*}
Repeating the manipulations performed for the odd bulk case,
$$
\widetilde \zeta_{d-1}
\;=\;
{\rm \Gamma}_{d-1} \int\limits_{\mathbb{R}^{d-1}} d \tilde x \ \sum_{n \in \mathbb N} \ \int_\Omega \mathbb P (d\omega)  \,
\tr \Big(\widetilde  \gamma_0 \langle 0,n| \widetilde \pi_\omega (\tilde a_0) \prod_{i=1}^{d-1} [\widetilde F_{\tilde x},\widetilde \pi_\omega (\tilde a_i) ] |0,n \rangle  \Big)
\;.
$$
Next we insert partitions of unity using the projections $\PI_{\tilde x,n}=\one_{2^k} \otimes |\tilde x,n\rangle \langle \tilde x,n|$:
\begin{align*}
\widetilde \zeta_{d-1} \;=\; & \
{\rm \Gamma}_{d-1} \sum_{n \in \mathbb N} \ \int\limits_{\mathbb{R}^{d-1}} d \tilde x \ \sum_{n \in \mathbb N} \ \int_\Omega \mathbb P (d\omega)  \sum\limits_{\tilde x_i,n_i \in {\mathbb Z}^{d-1}\times \mathbb N}\cdot \\
& \;\;\cdot
\tr
\Big(\widetilde \gamma_0 \langle 0,n | \widetilde \pi_\omega(\tilde a_0) \prod_{i=1}^{d-1} \PI_{\tilde x_i,n_i}[\widetilde F_{\tilde x},\widetilde \pi_\omega (\tilde a_i)] |0,n \rangle\Big)
\;.
\end{align*}
Since $\tilde a_i$'s are polynomials, the sums over $(\tilde x_i,n_i)$ contain a finite number of terms and we can interchange them and the integrals, to continue
\begin{align*}
\widetilde \zeta_{d-1}
 \;=\; 
&{\rm \Gamma}_{d-1}  \sum\limits_{\tilde x_i,n_i \in {\mathbb Z}^{d-1} \times \mathbb N}  
\ \int\limits_{\mathbb{R}^{d-1}} d \tilde x \
\tr \Big( \widetilde \gamma_0 \prod_{i=1}^{d-1} \big (\sgn \langle \widetilde \gamma,\tilde x_i + \tilde x\rangle - \sgn\langle \widetilde \gamma,\tilde x_{i+1}+\tilde x\rangle \big )  \Big)
\cdot 
\\
& 
\;\;\;\;\;\;\;\cdot
\sum_{n \in \mathbb N} \ \int_\Omega \mathbb P (d\omega) \, \langle 0,n | \widetilde \pi_\omega(\tilde a_0)\prod_{i=1}^{d-1} \PI'_{\tilde x_i,n_i}  \widetilde \pi_\omega(\tilde a_i)| 0,n \rangle,
\end{align*}
where $\PI'_{\tilde x,n} = |\tilde x,n \rangle \langle \tilde x,n |$ and $\widetilde \pi_\omega$ is now just the representation on $\ell^2(\mathbb Z^{d-1}\times \mathbb N)$. Also, in the first line, it is understood that $(\tilde x_d,n_d)$ is fixed at $(0,n)$. We now use the first identity from Lemma~\ref{KeyIdentity} with $2k=d-1$
\begin{align*}
\widetilde \zeta_{d-1}
& \;=\; \I^{d-1}{\rm \Lambda}_{d-1}  \sum\limits_{\tilde x_i,n_i \in {\mathbb Z}^{d-1} \times \mathbb N} 
\;\;\sum_{\rho \in \Ss_{d-1}} \;(-1)^\rho \prod_{i=1}^{d-1} \tilde x_{i,\rho_i}
\cdot
\\
& \;\;\;\;\;\;\;\;\;
\cdot \sum_{n \in \mathbb N} \ \int\limits_\Omega \mathbb P (d\omega) \, \langle 0,n | \widetilde \pi_\omega(\tilde a_0)\prod_{i=1}^{d-1} \PI'_{\tilde x_i,n_i}  \widetilde \pi_\omega(\tilde a_i)| 0,n \rangle
\;,
\end{align*}
and combine the sum over $(\tilde x_i,n_i)$ with the projections $\PI'_{\tilde x_i,n_i}$ to form the position operators, then the commutators,
\begin{align*}
\widetilde \zeta_{d-1} & 
\;=\;  {\rm \Lambda}_{d-1}  \sum_{\rho\in \Ss_{d-1}} (-1)^\rho\; 
\sum_{n \in \mathbb N} \ \int\limits_\Omega \mathbb P (d\omega) \;\langle 0,n | \widetilde \pi_\omega(\tilde a_0) \prod_{i=1}^{d-1} \I [X_{\rho_i},\widetilde \pi_\omega(\tilde a_i)]  |0,n \rangle \\
& 
\;=\; {\rm \Lambda}_{d-1}  \sum_{\rho\in \Ss_{d-1}} (-1)^\rho\; 
\sum_{n \in \mathbb N} \ \int\limits_\Omega \mathbb P (d\omega)\; \langle 0,n | \widetilde \pi_\omega \Big (\tilde a_0 \prod_{i=1}^{d-1} \tilde \partial_{\rho_i} \tilde a_i \Big )  |0,n \rangle.
\end{align*}
On the r.h.s. appears the trace $\widetilde \Tt$ over $\Ee_d$ and this concludes the proof.

\vspace{.2cm}

(ii) {\it The $d$ even case for the boundary invariants.} This is a combination of the arguments above and is left to the reader.
\hfill $\Box$

\begin{corollary}[Index Theorems under BGH]\label{IndexBGH}
\begin{enumerate}[\rm (i)]

\item Let $d$ be even and $\hat h=(h,\tilde h) \in M_N(\mathbb C) \otimes \widehat \Aa_d$ be a short-range half-space Hamiltonian for which BGH applies. Then, for all $\omega \in \Omega$,
\begin{equation}\label{IndFormula1}
\boxed{
\;{\rm Ch}_d(p_F) \;=\; \Ind\Big ( P_\omega G_{x_0} P_\omega \Big ) \;,\;
}
\end{equation}
where $p_F$ is the Fermi projection and $P_\omega = \pi_{\omega}(p_F) = \chi(H_\omega \leq \mu)$  is the physical representation at disorder configuration $\omega$. Furthermore, for all $\omega \in \Omega$, 
\begin{equation}\label{IndFormula2}
\boxed{
\widetilde {\rm Ch}_{d-1}(\tilde u_\Delta) 
\;=\; 
-\,\Ind\Big ( \widetilde E_{x_0} \widetilde U_\omega \widetilde E_{x_0}\Big ) 
\;,
}
\end{equation}
where $\tilde u_\Delta$ is the unitary operator defined in Eq.~\eqref{TildeU} and
$$
\widetilde U_\omega 
\;=\; 
\widetilde \pi_\omega(u_\Delta) 
\;=\; 
e^{2\pi \I \FFunc (\widehat H_\omega)}
$$
is a physical representation at disorder configuration $\omega$. Here $\FFunc $ is a smooth function and $\FFunc = 1$ and $\FFunc =0$ below and above the bulk insulating gap respectively. Lastly, for all $\omega \in \Omega$, we have the equality
\begin{equation}\label{MainEq1}
\boxed{
{\rm Ch}_d(p_F) \; =\;\widetilde {\rm Ch}_{d-1}(\tilde u_\Delta).
}
\end{equation}

\item Let $d$ be odd and $\hat h = (h,\tilde h) \in M_{2N}(\mathbb C) \otimes \widehat \Aa_d$ be a short-range half-space Hamiltonian for which BGH and CH apply. Then, for all $\omega \in \Omega$,
Then
\begin{equation}\label{IndFormula3}
\boxed{
\;
{\rm Ch}_d(u_F)\; =\; -\, \Ind\Big (E_{x_0} U_\omega E_{x_0} \Big )
\;,\;}
\end{equation}
where $u_F$ is  the Fermi unitary operator and $U_\omega = \pi_\omega (u_F) $ its physical representation at disorder configuration $\omega$. Furthermore, for all $\omega \in \Omega$,
\begin{equation}\label{IndFormula4}
\boxed{
\widetilde {\rm Ch}_{d-1}(\tilde p_\Delta) 
\;=\; 
\Ind\Big (\widetilde P_\omega \widetilde G_{\tilde x_0} \widetilde P_\omega \Big )
\;,
}
\end{equation}
where $\tilde p_\Delta$ is the projection defined in Eq.~\eqref{BoundaryProjection1} and
$$
\widetilde P_\omega 
\;=\; 
\widetilde \pi_\omega(p_\Delta)
\;=\; 
\tfrac{1}{2} \widehat J \; ( e^{\I \pi \GFunc(\widehat H_\omega)} +\bm 1_{2N} ) + {\rm diag}(\one_N,0_N)
$$
is a physical representation at disorder configuration $\omega$. Recall that $\GFunc$ is smooth odd function and $\GFunc=\pm 1$ above/below the bulk insulating map. Lastly, for all $\omega \in \Omega$, we have the equality
\begin{equation}\label{MainEq2}
\boxed{
{\rm Ch}_d(u_F) \; =\; -\, \widetilde {\rm Ch}_{d-1}(\tilde p_\Delta).
}
\end{equation}
\end{enumerate}
For all cases, the Fredholm indices are independent of $\omega$ or $x_0$ and the strong invariants remain quantized and invariant under continuous deformations of $h$ as long as BGH holds. 
\end{corollary}

\section{Key geometric identities}
\label{Sec-KeyId}

The following identities are direct generalizations \cite{PLB,PS} of an identity which played a pivotal role in the non-commutative geometry of IQHE (see \cite{ASS} and Th.~10 in \cite{BES}). In $d=2$, the identity is due to Connes (see \cite{Con0} pp. 81), who originally used it to compute the Chern characters of the convolution algebras $C_c^\infty(\mathbb R^2)$ and $C_c^\infty(SL(2,\mathbb R))$. A more elementary proof was found by Verdi\'ere, and this proof was reproduced in \cite{AG} and served as inspiration for \cite{PLB,PS}.

\begin{lemma}[Key geometric identities] 
\label{KeyIdentity}

\begin{enumerate}[\rm (i)]

\item Let $x_1, \ldots, x_{2k+1}\in \mathbb{R}^{2k}$ with $x_{2k+1}$ fixed at the origin and $\gamma_1, \ldots, \gamma_{2k}$ be the irreducible representation on $\mathbb C^{2^k}$ of the complex Clifford algebra $Cl_{2k}$ as given in the CCR. Then:
\begin{align}
\int_{\mathbb{R}^{2k} }d x \
\tr & \Big(\gamma_0 \prod_{i=1}^{2k} \big (\sgn \langle \gamma ,x_i +  x \rangle -  \sgn \langle \gamma, x_{i+1}+x \rangle  \big ) \Big) \nonumber \\
& = \; \frac{(2\I \pi)^k}{k!} \sum_{\rho \in \Ss_{2k}} (-1)^\rho \prod_{i=1}^{2k} x_{i,\rho_i}
\;. 
\label{KeyIdentity1}
\end{align}

\item Let $x_1, \ldots, x_{2k+2}\in \mathbb{R}^{2k+1}$ with $x_{2k+2}$ fixed at the origin and $\sigma_1,\ldots,\sigma_{2k+1}$ be the irreducible representation on $\mathbb C^{2^k}$ of the complex Clifford algebra $Cl_{2k+1}$ as given in the CCR. Then:
\begin{align}
\int_{\mathbb{R}^{2k+1} }d x \
\tr & \Big(\prod_{i=1}^{2k+1} \big (\sgn \langle \sigma ,x_i +  x \rangle -  \sgn \langle \sigma, x_{i+1}+x \rangle  \big ) \Big) \nonumber\\
& =\; -\, \frac{2^{2k+1}(\I \pi)^k}{(2k+1)!!} \sum_{\rho \in \Ss_{2k+1}} (-1)^\rho \prod_{i=1}^{2k+1} x_{i,\rho_i}
\;. 
\label{KeyIdentity2}
\end{align}
\end{enumerate}

\end{lemma}

\noindent {\bf Proof.} (i) Let us begin by listing a few identities for the $\gamma$ matrices which follow from the CCR:
\begin{enumerate}[\rm (a)]
\item $\gamma_0 \gamma_1 \cdots \gamma_{2k} = \I^k \one_{2^k}$;
\item $\tr(\gamma_0\gamma_{\rho_1} \cdots \gamma_{\rho_q})=0$ if $q$ even and $q<2k$;
\item $\tr (\gamma_0 \gamma_{\rho_1} \cdots \gamma_{\rho_{2k}} )=0$ unless $\rho$ is a permutation of $1,2,\ldots,2k$;
\item $\tr (\gamma_0\gamma_{\rho_1} \cdots \gamma_{\rho_{2k}})= (2\I)^k (-1)^\rho$ if $\rho$ is such a permutation.
\end{enumerate}
All these identities follow from the defining relations of the Clifford algebra and our conventions on its representations. As another preparation let us establish the following basic geometric identity
\begin{equation}
\label{Id1}
\tr \Big ( \gamma_0\prod_{i=1}^{2k} \langle \gamma ,y_i \rangle \Big )
\;=\;
(2\I)^k  (2k)!  \ \mathrm{Vol}[0,y_1,\ldots,y_{2k}]
\;,
\end{equation}
for any set of points $y_1,\ldots,y_{2k}$ from $\mathbb{R}^{2k}$. Above,   $[y_0,y_1,\ldots,y_{2k}]$ denotes the simplex in $\mathbb{R}^{2k}$ corresponding to this set of points and $\mathrm{Vol}[\ldots]$ the oriented volume of the simplex. Indeed, expending the r.h.s. of \eqref{Id1} and taking into account (c), one has
$$
\tr \Big (\gamma_0 \prod_{i=1}^{2k} \langle \gamma ,y_i \rangle \Big )
\;=\;
\sum_{\rho \in \Ss_{2k}} y_{1,\rho_1} \cdots y_{2k,\rho_{2k}} \; \tr 
\left (\gamma_0 \gamma_{\rho_1} \cdots \gamma_{\rho_{2k}} \right )
\;,
$$
and from (d)
$$
\tr \Big(\gamma_0\prod_{i=1}^{2k} \langle \gamma ,y_i \rangle \Big )
\;=\; 
(2 \I)^k \;\mathrm{Det}[y_1,\ldots,y_{2k}]
\;,
$$
where inside the determinant is the $2k \times 2k$-matrix with columns $y_1, \ldots, y_{2k}$. On the other hand, the volume of a simplex in $\mathbb{R}^{2k}$ can be computed with the formula
$$
\mathrm{Vol}[y_0,y_1,\ldots,y_{2k}]
\; =\;
 \frac{1}{(2k)!}\;\mathrm{Det}\left [
\begin{array}{cccc}
y_0 & y_1   & \ldots & y_{2k} \\
1  &  1 &  \ldots & 1 
\end{array}
\right ]
\;,
$$
hence Eq.~\eqref{Id1} follows by setting $y_0 = 0$ above.

\vspace{.1cm}

For the computation of the l.h.s. of \eqref{KeyIdentity1} let us set
$$
I(x)
\;=\;
\tr \left(\gamma_0\prod_{i=1}^{2k} \big (\sgn \langle \gamma, x_i -  x \rangle  - \sgn \langle \gamma, x_{i+1}-x \rangle \big ) \right)
\;.
$$
We will now use the identity $\sgn \langle \gamma, y \rangle = \langle \gamma, \overline y \rangle$ where $\overline y$ denotes the unit vector  $y/|y|$ for $y\not=0$. Replacing this and writing all terms one finds that
$$
I(x)
\; = \;\sum_{j=1}^{2k+1} (-1)^{j+1} \; \tr \left( \gamma_0 \big \langle \gamma , \overline{x_1- x} \big \rangle 
\cdots \dashuline{ \big \langle \gamma,\overline{x_j- x} \big \rangle}  \cdots \big \langle \gamma, \overline{x_{2k+1}- x} \big \rangle \right) 
\;,
$$
where the underline designates a factor which is omitted. This holds except for the cases when $x=x_i$ which, however, have no contribution to the integral over $x$ to be taken below. Now with \eqref{Id1}
$$
I(x)
\; =\;
(2\I)^k\, (2k)!\;\sum_{j=1}^{2k+1} (-1)^{j+1} \;\mathrm{Vol}\big [0, \overline{ x_1- x}, \ldots ,\dashuline{ \overline{ x_j- x} },  \ldots ,\overline{ x_{2k+1}- x} \big ]
\;.
$$
The vertices can be re-ordered and it is convenient to translate the whole simplex by $x$. Taking into account the sign changes due to the re-ordering,
$$
I(x)
\; =\;
(2\I)^k\, (2k)!\;\sum_{j=1}^{2k+1} \mathrm{Vol}[x+\overline{ x_1- x}, \ldots ,x,  \ldots ,x+\overline{ x_{2k+1}- x}]
\;,
$$
where the vertex $x$ is located in the $j$-th entry. We introduce the notations
$$
\mathfrak{S}_j(x)
\;=\;
\big [x+\overline{x_1- x}, \ldots, x,  \ldots ,x+\overline{x_{2k+1}- x} \big ]
$$
and
$$
\mathfrak{S}
\;=\;
[x_1, x_2,\ldots, x_{2k+1}]
\;,
$$
and recall that $x_{2k+1}=0$. The orientations of these simplexes are the same because each $\mathfrak{S}_j({\bm x})$ can be continuously deformed into $\mathfrak{S}$ without reducing the volume to zero. Now note that, for arbitrarily selected $j$, all vertices of $\mathfrak{S}_j(x)$, except the vertex $x$, are located on the unit sphere centered at $x$. As such, the facets of $\mathfrak{S}_j(x)$ stemming from $x$ define a solid angle sector of the unit ball centered at $x$. This sector will be denoted by $\mathfrak{B}_j(x)$ and its orientation is taken to be the same as that of $\mathfrak{S}_j(x)$. The entire unit ball will be denoted by $\mathfrak{B}$ and its orientation will be taken to be the same as that of $\mathfrak{S}$. One key fact is that
$$
\mathrm{Vol}( \mathfrak{S}_j(x)) - \mathrm{Vol}( \mathfrak{B}_j(x))\; \sim \;|x|^{-(2k+1)}
\;, 
\qquad 
\mbox{\rm as}  \;\; |x| \rightarrow \infty
\;.
$$
This suggests to break the integral into two terms:
\begin{align*}
\int\limits_{\mathbb{R}^{2k}} d x \;I(x)
&\; =\; (2\I)^k\, (2k)!\; \sum_{j=1}^{2k+1} \,\int\limits_{\mathbb{R}^{2k}} d x \; [\mathrm{Vol}(\mathfrak{S}_j(x)) - \mathrm{Vol}(\mathfrak{B}_j(x)) ]   
\\
&  \qquad +\;(2\I)^k\, (2k)!\; \int \limits_{\mathbb{R}^{2k}} d x\; 
\sum_{j=1}^{2k+1} \mathrm{Vol}(\mathfrak{B}_j(x))
\;.
\end{align*}
At this point let us note that
$$
\int \limits_{\mathbb{R}^{2k}} d x \; [\mathrm{Vol} (\mathfrak{S}_j(x)) - \mathrm{Vol} (\mathfrak{B}_j(x)) ]
\;=\;0
\;,
$$
which is a consequence of the odd-symmetry of the integrand relative to the inversion of $x$ relative to the center of the facet $x_1, \ldots,\uline{x_j}, \ldots, x_{2k+1}$ of $\mathfrak{S}$. Furthermore:
$$
\sum_{j=1}^{2k+1} \mathrm{Vol} ( \mathfrak{B}_j(x))
\;=\;
\left\{
\begin{array}{cc}
\mathrm{Vol}(\mathfrak{B})\;,
& \mbox{if} \ x \ \mbox{inside} \ \mathfrak{S}\;,  \\
0 \;,
& \;\;\mbox{if} \ x \ \mbox{outside} \ \mathfrak{S}\;,
\end{array}
\right . 
$$
because the solid angles corresponding to the facets of the simplex $\mathfrak{S}$, as seen from $x$, add up to the full solid angle if $x$ is inside the simplex, and they add up to zero if $x$ is outside the simplex. Hence
\begin{align*}
\int\limits_{\mathbb{R}^{2k}} d x \; I(x)
\; =\; 
(2\I)^k\, (2k)!\; \mathrm{Vol}(\mathfrak{B}) |\mathrm{Vol}(\mathfrak{S})|
\;.
\end{align*}
Now the orientations of $\mathfrak{B}$ and $\mathfrak{S}$ are the same so that
\begin{align*}
\mathrm{Vol}(\mathfrak{B}) \,|\mathrm{Vol}(\mathfrak{S})|
& \;=\;
|\mathrm{Vol}(\mathfrak{B})|\,\mathrm{Vol}(\mathfrak{S})
 \;=\;
\frac{\pi^k}{k!}\;\frac{1}{(2k)!}\;{\rm Det}(x_1,\ldots,x_{2k})
\;,
\end{align*}
and the identity follows.

\vspace{.2cm}

(ii) Again, identities for the representation of the Clifford algebra are used: 
\begin{enumerate}[\rm (a)]
\item $\sigma_1 \sigma_2 \cdots \sigma_{2k+1} = \I^k \, \one_{2^k}$;
\item $\tr(\sigma_{\rho_1} \cdots \sigma_{\rho_q})=0$ if $q$ odd and $q<2k+1$;
\item $\tr (\sigma_{\rho_1} \cdots \sigma_{\rho_{2k+1}} )=0$ unless $\rho$ is a permutation of $1,2,\ldots,2k+1$;
\item $\tr (\sigma_{\rho_1} \cdots \sigma_{\rho_{2k+1}})= (2\I)^k (-1)^\rho$ if $\rho$ is such a permutation.
\end{enumerate}
Based on them, the following basic geometric identity can be established as before:
$$
\tr \Big ( \prod_{i=1}^{2k+1} \langle \sigma ,y_i \rangle \Big )
\;=\;
(2\I)^k  (2k+1)!  \ \mathrm{Vol}[0,y_1,\ldots,y_{2k+1}]
\;.
$$
Let $I(x)$ now denote the integrand on the l.h.s. of \eqref{KeyIdentity2}.  Expanding shows
$$
I(x)
\; = \;\sum_{j=1}^{2k+2} (-1)^j \; \tr \left( \big \langle \sigma , \overline{x_1- x} \big \rangle 
\cdots \dashuline{ \big \langle \sigma,\overline{x_j- x} \big \rangle}  \cdots \big \langle \sigma, \overline{x_{2k+2}- x} \big \rangle \right)
\;,
$$
hence
$$
I(x)
\; =\;
(2\I)^k\, (2k+1)!\;\sum_{j=1}^{2k+2} (-1)^j \;\mathrm{Vol}[0, \overline{ x_1- x}, \ldots ,\dashuline{ \overline{ x_j- x} },  \ldots ,\overline{ x_{2k+2}- x}]
\;.
$$
We reorder the vertices and translate the whole simplex, to write
$$
I(x)
\; =\;
-\,(2\I)^k\, (2k+1)!\;\sum_{j=1}^{2k+2} \mathrm{Vol}[x+\overline{ x_1- x}, \ldots ,x,  \ldots ,x+\overline{ x_{2k+2}- x}]
\;,
$$
where the vertex $x$ is located at the $j$-th position. From here on, the proof continues identically to the even case. 
\hfill $\Box$

\section{Stability of strong bulk invariants under strong disorder}
\label{SecIndexFormulas}

The prior results lead to index theorems for projections and unitaries in the bulk and boundary C$^\ast$-algebras. In this section, these index theorems are extended to much larger classes of projections and unitaries, not necessarily lying in the C$^*$-algebras, but only in the non-commutative Sobolev spaces introduced in Section~\ref{Sec-Sobolev}. This will enable us to replace the assumption of a bulk gap with that of a mobility bulk gap. Let us now formulate the index theorems in a general form.

\begin{theorem}\label{IndexThBulk} Let $F_{x_0}$, $E_{x_0}$  and $G_{x_0}$ be the Weyl (Dirac) phase,  Hardy projection and the upper right corner of $F_{x_0}$ in odd and even dimension, respectively.
 
\begin{enumerate}[\rm (i)]

\item Let the space dimension $d$ be even and consider a projection $e\in M_N(\mathbb C) \otimes L^\infty(\Aa_d,\Tt)$ which is also in the Sobolev space $M_N(\mathbb C) \otimes \Ww'_{s,k}({\Aa_d},\PM)$, with $s=k=d+1$. Then, $\PM$-almost surely, $\pi_\omega(e) \, G_{x_0} \, \pi_\omega(e)$  is a Fredholm operator on the range of $ \pi_\omega(e)$ and its $\PM$-almost sure index is independent of $x_0$ and connected to the  strong even Chern number of $e$ by the following index formula:
\begin{equation}
\label{IndexThBulkEvenChern}
{\rm Ch}_d(e)\;=\;
{\rm \Lambda}_d \sum_{\rho \in \Ss_d} (-1)^\rho \Tt \Big (e \prod_{j=1}^d \partial_{\rho_j} e \Big)
 \; =\; 
 \Ind \Big( \pi_\omega(e) \, G_{x_0} \, \pi_\omega(e) \Big )
 \;.
\end{equation}
Furthermore, the strong even Chern number remains constant during any homotopy  of projections
$$
t \;\;\mapsto\;\; e_t\in M_N(\mathbb C) \otimes \Big ( L^\infty(\Aa_d,\Tt)\cap \mathcal W'_{s,k}(\Aa_d,\mathbb P) \Big )
$$ 
which is continuous w.r.t. the norm $\| \,.\, \|'_{s,k}$ (and not necessarily w.r.t. $\|\,.\, \|_{\infty}$).

 \item Let the space dimension $d$ be odd and consider a unitary element $u \in M_N(\mathbb C) \otimes L^\infty(\Aa_d,\Tt)$ which also belongs to the Sobolev space $M_N(\mathbb C) \otimes \Ww'_{s,k}({\Aa_d},\PM)$, with $s=k=d+1$. Then, $\PM$-almost surely,   $E_{x_0} \, \pi_\omega(u) \, E_{x_0}$ is a Fredholm operator on the range of $E_{x_0}$ and its $\PM$-almost sure index is independent of $x_0$ and connected to the strong odd Chern number of $u$ by the following index formula:
\begin{equation}\label{IndexThBulkOddChern}
{\rm Ch}_d(u)
\;=\;
{\rm \Lambda}_d \sum_{\rho \in \Ss_d} (-1)^\rho \Tt \Big ( (u^\ast -\bm 1_N) \prod_{j=1}^d \partial_{\rho_j} u^{\ast_{j-1}} \Big) 
\;=\; 
- \, \Ind \Big( E_{x_0} \, \pi_\omega(u) \, E_{x_0} \Big )
\;.
\end{equation}
Furthermore, the strong odd Chern number  remains constant under any homotopy of unitaries 
$$
t \;\;\mapsto\;\; u_t\in M_N(\mathbb C) \otimes \Big ( L^\infty(\Aa_d,\Tt)\cap \mathcal W'_{s,k}(\Aa_d,\PM) \Big )
$$ 
which is continuous w.r.t. the norm $\| \,.\, \|'_{s,k}$ (and not necessarily w.r.t. $\|\,.\,\|_{\infty}$).
\end{enumerate}
\end{theorem}

\noindent {\bf Proof.} Let us begin by trivially extending the potential Fredholm operators to the whole Hilbert space. Hence, let $K$ denote either $\pi_\omega(e) \, G_{x_0} \, \pi_\omega(e) + (\one -\pi_\omega(e))$ or $E_{\bm x_0} \pi_\omega (u) E_{\bm x_0} + (\one - E_{\bm x_0})$. Recall the Calderon-Fedosov principle \cite{Cal,Fed}, which states that $K$ is Fredholm provided there is a positive integer $n$ such that $(\one-KK^\ast)^n$ and $(\one-K^\ast K)^n$ are trace class. As already mentioned in Section~\ref{sec-KFredPair}, the Calderon-Fedosov principle for $K$ reduces precisely to the summability condition \eqref{eq-Schattenbound}, regardless of the parity of $d$. Recall the convention by which $\langle x|\cdot | y \rangle$ is viewed as a matrix, which in this context belongs to $M_N(\CM) \otimes M_{Q}(\CM)$, where $Q$ is the dimension of the representation space of the Clifford algebras. We will show that
\begin{equation}
\label{AvBound}
\int_\Omega \mathbb P(d\omega) \;  \mathrm{Tr}\left ( \big |\I [F_{x_0},\pi_\omega (a)] \big |^{d+1} \right)
\; \leq \;
{\rm const} \cdot \ (\|a\|'_{s,k})^{d+1}
\;,
\end{equation}
for any $a\in M_N(\mathbb C) \otimes \mathcal W'_{s,k}(\Aa_d,\PM)$. This ensures that the Calderon-Fedosov principle holds $\mathbb P$-almost surely for $K$. As in the proof of Lemma~\ref{BulkSummability}, it is enough to consider only self-adjoint elements. Furthermore, decomposing $a$ in a basis of $M_N(\mathbb C)$ and using Minkovski inequality for the Schatten norms, one can see that it is enough to take $a$ from $\mathcal W'_{s,k}(\Aa_d,\PM)$ rather than from $M_N(\mathbb C) \otimes \mathcal W'_{s,k}(\Aa_d,\PM)$. For  $a = a^\ast\in \mathcal W'_{s,k}(\Aa_d,\PM)$, we will prove the estimate
\begin{equation}
\label{AvBoundP}
\int_\Omega \mathbb P(d\omega) \,  \sum_{x' \in \mathbb Z^d} \|\langle x \big |[F_{ x_0},\pi_\omega (a)]^{d+1} | x' \rangle \|_{(1)}
\; \leq \;
\frac{\rm const}{| x+x_0|^{d+1}} \ (\|a\|'_{s,k})^{d+1}
\;,
\end{equation}
which as  in Lemma~\ref{BulkSummability} implies \eqref{AvBound}. Let us denote the l.h.s. of \eqref{AvBoundP} by $W$ and start the calculation from Eq.~\eqref{KPower},
\begin{align*}
W 
\;\leq \;
 \frac{\rm const}{ |x + x_0|^{d+1}}  \int_\Omega \mathbb P(d\omega) \sum_{y_1,\ldots,y_{d+1} \in \mathbb Z^d}  
 \prod_{i=1}^{d+1} \;(1+|y_i|)^{d+1}\; |a(\omega_i, y_i)|
 \;. 
\end{align*}
Since all the terms are positive, the sums and the integral can be interchanged and H\"older's inequality gives
\begin{align*}
W 
\;& \leq \;
 \frac{\rm const}{ |x + x_0|^{d+1}} \sum_{y_1,\ldots,y_{d+1} \in \mathbb Z^d} \;\prod_{i=1}^{d+1} (1+|y_i|)^{d+1}\; \left [  \int_\Omega \mathbb P(d\omega)  |a(\omega_i, y_i)|^{d+1} \right ]^\frac{1}{d+1} \\
 &\; \leq \; \frac{\rm const}{ | x + x_0|^{d+1}} \;\Big  (\sum_{y \in \mathbb Z^d} (1+| y|)^{d+1}\; \left [  \int_\Omega \mathbb P(d\omega)  |a(\omega, y)|^{d+1} \right ]^\frac{1}{d+1} \Big )^{d+1},
\end{align*}
and \eqref{AvBoundP} follows. 

\vspace{.1cm}

Next, let us prove the $\mathbb P$-almost sure constancy of the Fredholm indices in $\omega$.  Since $\mathbb P$ is ergodic w.r.t. the lattice shifts, it is sufficient to check constancy along every orbit. Hence, let us compare the indices of $E_{x_0}\pi_\omega(u) E_{x_0}\oplus (\bm 1 -E_{x_0})$ and $E_{x_0} \pi_{\tau_b\omega}(u) E_{x_0}\oplus(\bm 1 -E_{x_0})$ for arbitrary $b \in{\mathbb Z}^d$. Since the index is invariant to conjugations with unitaries, we only need to check equality of the indices of $E_{ x_0}\pi_\omega(u) E_{x_0}+(\bm 1 -E_{ x_0})$ and $E_{b+x_0}\pi_\omega(u) E_{b+x_0}+(\bm 1 -E_{b+x_0})$. But 
\begin{align*}
& \indent E_{b+x_0}\pi_\omega(u) E_{b+x_0} - E_{x_0}\pi_\omega(u) E_{x_0}  
\\
& \;\;\;\;\;\;\;=\;
\tfrac{1}{2}(F_{b+x_0}-F_{x_0}) \pi_\omega(u) E_{b+x_0} +\tfrac{1}{2}E_{x_0}\pi_\omega(u) (F_{b+x_0}-F_{x_0} )
\;
\end{align*}
and the operator difference
$$
F_{b+x_0}-F_{x_0}
\;=\;
\one_N \otimes \big (\sgn \langle \sigma, b+x_0+X\rangle - \sgn \langle \sigma,x_0+X \rangle \big )
$$
is compact due to the asymptotic estimate \eqref{Asy}. For the same reason, $E_{x_0}-E_{b+x_0}$ is also compact. The compact stability of the index now allows to conclude. The invariance of the index in $x_0$ follows by a similar argument. Exactly same arguments apply to the operator $\pi_\omega(e) \, G_{x_0} \, \pi_\omega(e) \oplus (\one -\pi_\omega(e))$.

\vspace{.1cm}

At this point, we have established that $K$ is $\PM$-almost surely in the Fredholm class and with a $\PM$-almost sure index given by the Calderon-Fedosov formula 
$$
{\rm Ind}\big (K\big ) 
\;=\;
{\rm Tr} \Big ( (\one-KK^\ast)^{d+1} \Big) - {\rm Tr} \Big ( (\one-K^\ast K)^{d+1} \Big )
\;.
$$
As well-known \cite{Con}, this formula reduces precisely to the pairing between the Fredholm modules and $K$-groups, spelled out in Proposition~\ref{ChernPairing}. Furthermore, since the index is $\mathbb P$-almost surely constant in $\omega$ and independent of $x_0$, one is allowed to take the average of the r.h.s.. But this leads precisely to \eqref{Fedosov1} and \eqref{Fedosov2}. Hence, from here on, the calculation can proceed as in the proof of Theorem~\ref{EqualityConnesChern}.

\vspace{.1cm}

We now turn to the invariance of the strong Chern numbers under the deformations $e_t$ and $u_t$. The $\mathbb P$-almost sure Fredholm index of $K$ and the strong Chern numbers are linked by Eqs.~\eqref{IndexThBulkEvenChern} and \eqref{IndexThBulkOddChern}, and both sides of these equations must be used to establish the claim. First, Proposition~\ref{SobolevBulk} and the fact that the Chern cocycles are continuous over the Sobolev spaces of first kind assures that the strong Chern numbers are a continuous functions of $t$. Now assume that that they change from one integer value to another as $t$ is varied. Due to continuity with $t$, then there must be at least one value of $t$ for which the strong Chern numbers are not integers. Now since the Fredholm index is {\it de facto} an integer number, we have to conclude that actually there is not a single $\omega$ in the whole $\Omega$ for which Eqs.~\eqref{IndexThBulkEvenChern} and \eqref{IndexThBulkOddChern} hold at this $t$. But this will contradict the $\mathbb P$-almost sure character of these equalities. Hence, the starting assumption must be false and the conclusion is that the strong Chern numbers stay pinned to a single integer value at all $t$'s.\hfill $\Box$


\vspace{.2cm}

Let us now spell out the implications topological solid state systems. 

\begin{corollary}[Index formulas for the bulk Chern numbers under MBGH] \
\begin{enumerate}[\rm (i)]
\item Let $d$ be even and $h \in M_N(\mathbb C) \otimes \Aa_d$ be a short-range Hamiltonian for which MBGH applies. Then, $\PM$-almost surely, 
$$
\boxed{
\;{\rm Ch}_d(p_F) \;=\; \Ind\Big ( P_\omega G_{x_0} P_\omega \Big ) \;,\;
}
$$
where $p_F$ is the Fermi projection and $P_\omega = \pi_{\omega}(p_F) = \chi(H_\omega \leq \mu)$  is the physical representation at disorder configuration $\omega$.

\item Let $d$ be odd and $h \in M_{2N}(\mathbb C) \otimes \Aa_d$ be a short-range Hamiltonian for which either the MBGH and CH apply. Then
$$
\boxed{
\;
{\rm Ch}_d(u_F)\; =\; -\, \Ind\Big (E_{x_0} U_\omega E_{x_0} \Big )
\;,\;}
$$
where $u_F$ is  the Fermi unitary operator and $U_\omega = \pi_\omega (u_F) $ its physical representation at disorder configuration $\omega$.

\end{enumerate}
For both odd and even cases, the Fredholm indices are $\PM$-almost surely independent of $\omega$ and also of $x_0$, and the strong invariants remain quantized and invariant under continuous deformations of $h$ as long as the MBGH holds. 
\end{corollary}

An important physical consequence of the above statements is the fact that the strong invariants can change their quantized value only if MBGH fails, {\it i.e.} if the Fermi level crosses a region of delocalized spectrum. In other words, the topological phases from the unitary and chiral unitary classes, labeled by the strong Chern numbers, are separated by phase boundaries where the Anderson localization length diverges. These phase boundaries can be detected experimentally via transport measurements of the direct conductance. This will be further elaborated in Chapter~\ref{Chap-Conclusions}

\section{Delocalization of the boundary states}

In this section we again assume a bulk spectral gap and then provide a proof of the delocalized character of the boundary states for non-trivial complex topological insulators. The first task is to push the index theorems for the boundary algebras over to Sobolev spaces, just as was done for the bulk in Section~\ref{SecIndexFormulas}. The arguments there used $\ZM^d$-ergodicity of the probability $\PM$ on the space of disorder configurations. For half-space models, however, the representations $\widetilde{\pi}_\omega$ are no longer covariant w.r.t. the shifts along the $d$th direction, and the probability measure $\PM$ is in general not ergodic w.r.t. just the $\ZM^{d-1}$ shifts parallel to the boundary. For this reason, we restrict to boundary disorder only, which allows us to carry over the arguments from the previous section. Let us begin by describing the probability space of boundary disorder configurations.

\begin{proposition}
\label{Prop-ErgodicL} 
Set $I_L = \{-L,\ldots,L\} \subset \ZM$ for $L \in \NM$ and let $(\Omega_L, \PM_L)$ be the probability space 
\begin{equation}
\label{eq-OmegaL} 
\Omega_L\;=\;\Big(\prod_{y\in \mathcal R}\Omega^y_0\Big)^{\ZM^{d-1}\times I_L} 
\;, 
\qquad 
\PM_L(d\omega_L) 
\;=\; 
\prod_{y \in \Rr}\; \prod_{x\in \mathbb Z^{d-1}\times I_L} \,\PM^y_0(d \omega_x^y)
\;,
\end{equation}
where $(\Omega_0^y,\PM_0^y)$ are the probability spaces from Definition~\ref{DisorderConfigSpace}. Let $\tau$ be the action of $\ZM^{d-1}$ on $\Omega_L$, given by the shifts of the first $d-1$ coordinates of $x$ in $\omega_x^y$. Then $(\Omega_L,\tau,\ZM^{d-1},\PM_L)$ is an ergodic dynamical system.
\end{proposition}

\noindent {\bf Proof.} The statement is evident once one realizes that this dynamical system is identical with the original dynamical system, set for the space dimension $d-1$ and with $\Rr$ replaced by $\Rr \times I_L$.\hfill $\Box$

\vspace{0.2cm}

Now, let $p_L : \Omega_L \rightarrow \Omega$ be the map which assigns to $\omega=p_L(\omega_L)$ the same $\omega_x^y$ components if $x \in \ZM^{d-1} \times I_L$ and sets $\omega_x^y=0$ otherwise. It is continuous and pushes forward the probability measure $\PM_L$, and in fact the entire ergodic system defined in Proposition~\ref{Prop-ErgodicL}, on $\Omega$. We will use the notation $(\Omega,\tau,\ZM^{d-1},\PM_L)$ for this ergodic system. Note that the C$^\ast$-algebras of observables remain unaltered and only the probability by which the disorder configurations occur has been changed, for the boundary algebra only. 
We will use the notation $\widetilde \Tt_L$ for the trace on $\Ee_d$ defined with the probability measure $\PM_L$. The weak von Neumann closure of $\Ee_d$ and the Sobolev spaces constructed in Section~\ref{Sec-Sobolev} can be automatically adapted to the present settings. The statements about the boundary algebra in Section~\ref{Sec-Sobolev} are valid for any probability measure over $\Omega$ which is ergodic and invariant w.r.t. the action of $\mathbb Z^{d-1}$, hence they carry over to the present settings without any modifications. 

\begin{lemma}
\label{IndexThBoundary} 
Let $\widetilde{F}_{x_0}$, $\widetilde{E}_{x_0}$ and $\widetilde{G}_{x_0}$ be the boundary Dirac (Weyl) phase,  Hardy projection and the upper right corner of $\widetilde{F}_{x_0}$ in odd and even dimension, respectively.
 
\begin{enumerate}[\rm (i)]

\item Let the space dimension $d$ be even and let $\tilde u \in M_N(\CM)\otimes L^\infty(\Ee_d,\widetilde \Tt_L)$ be a unitary element such that $\tilde u-\bm 1$ belongs to the Sobolev space $M_N(\mathbb C) \otimes \Ww'_{d,d}(\Ee_d,\PM_L)$. Then, $\PM_L$-almost surely, the operator  $\widetilde E_{\tilde x_0}\widetilde \pi_\omega (\tilde u) \, \widetilde E_{\tilde x_0}$  belongs to the Fredholm class and the odd strong Chern number of $\tilde u$ admits the following index formula:
\begin{equation}\label{IndexThBoundaryOddChern}
{\rm \Lambda}_{d-1} \sum_{\rho \in \Ss_{d-1}} (-1)^\rho \widetilde \Tt_L \Big ( (\tilde u^\ast -\bm 1_N) \prod_{j=1}^d \widetilde \partial_{\rho_j} \tilde u^{\ast_{j-1}} \Big) 
\;=\; 
-\, \Ind \Big( \widetilde E_{\tilde x_0} \, \widetilde \pi_\omega(\tilde u) \, \widetilde E_{\tilde x_0} \Big )
\;.
\end{equation}
Furthermore, the strong odd Chern number $\widetilde{\rm Ch}_d(\tilde u_t)$ remains quantized and invariant for any unitary deformation $t \mapsto \tilde u_t\in M_N(\mathbb C) \otimes L^\infty(\Ee_d,\widetilde \Tt_L)$ such that $\tilde u_t - \one$ belongs to $M_N(\mathbb C) \otimes \mathcal W'_{d,d}(\Ee_d,\PM_L) $ and varies continuously w.r.t. the norm $\| \,.\, \|'_{d,d}$ (and not necessarily w.r.t. $\|\,.\, \|_{\infty}$ norm).

\item Let the space dimension $d$ be odd and let $\tilde p\in M_N(\CM)\otimes L^\infty(\Ee_d,\widetilde \Tt_L)$ be a projection such that $\tilde p - s(\tilde p)$ belongs to the Sobolev space $M_N(\mathbb C) \otimes \Ww'_{d,d}(\Ee_d,\PM_L)$. Then, $\PM_L$-almost surely, the operator  $\widetilde \pi_\omega (\tilde p) \, \widetilde G_{\tilde x_0} \, \widetilde \pi_\omega (\tilde p)$  belongs to the Fredholm class and the even Chern number of $\tilde p$ admits the following index formula:
\begin{equation}\label{IndexThBoundaryEvenChern}
{\rm \Lambda}_{d-1} \sum_{\rho \in \Ss_{d-1}} (-1)^\rho \widetilde \Tt_L \Big (\tilde p \prod_{j=1}^{d-1} \widetilde \partial_{\rho_j} \tilde p \Big) 
\;=\; 
\Ind\Big( \widetilde \pi_\omega (\tilde p) \, \widetilde G_{\tilde x_0} \, \widetilde \pi_\omega (\tilde p) \Big )
\;.
\end{equation}
Furthermore, the strong even Chern number $\widetilde{\rm Ch}_d(\tilde p_t)$ remains quantized and invariant for any projection deformation $
t \mapsto \tilde p_t\in M_N(\mathbb C) \otimes L^\infty(\Ee_d,\widetilde \Tt_L)$ such that $\tilde p_t - s(\tilde p_t)$ belongs to $M_N(\mathbb C) \otimes \mathcal W'_{d,d}(\Ee_d,\PM_L)$ and varies continuously w.r.t. the norm $\| \,.\, \|'_{d,d}$ (and not necessarily w.r.t. $\|\,.\, \|_{\infty}$ norm).
\end{enumerate}
\end{lemma}

\noindent {\bf Proof.} The arguments are very similar to those employed for Theorerm~\ref{IndexThBulk}, hence let us only mention the key points. First, starting from \eqref{KPowerBounday}, one can establish that, for any $\tilde a\in M_N(\mathbb C) \otimes L^\infty(\Ee_d,\widetilde \Tt_L)$ with  $\tilde a-s(\tilde a)\in M_N(\mathbb C) \otimes \mathcal W'_{d,d}(\Ee_d,\PM_L)$,
$$
\int_\Omega \PM_L (d\omega) \,  \sum_{x' \in \mathbb Z^{d-1}} \big \|\langle x \big |[\widetilde F_{ \tilde x_0},\widetilde \pi_\omega (\tilde a)]^d | x' \rangle \big \|_{(1)}
\; \leq \;
\frac{\rm const}{| x+\tilde x_0|^d} \ (\|\tilde a-s(\tilde a)\|'_{d,d})^{d}
\;,
$$
which in turn gives
\begin{equation}
\label{AvBoundB}
\int_\Omega \PM_L (d\omega) \;  \mathrm{Tr}\left ( \big |\I [\widetilde F_{\tilde x_0},\widetilde \pi_\omega (\tilde a)] \big |^d \right)
\; \leq \;
{\rm const} \cdot (\|\tilde a - s(\tilde a)\|'_{d,d})^{d}
\;.
\end{equation}
This ensures that the Calderon-Fedosov principle holds $\mathbb P_L$-almost surely for the operators mentioned in the statement. Above, note that the scalar part of $\tilde a$ already drops out when taking the commutator. The $\mathbb P_L$-almost sure constancy of the indices w.r.t. $\omega$ and $\tilde x_0$ follows in exactly the same way as in Theorerm~\ref{IndexThBulk}, after $\Ee_d$ is mapped into $\Aa_{d-1} \otimes \Kk$. Then, by applying Calderon-Fedosov formula and taking the average over $\omega$ and $\tilde x_0$ w.r.t. the measure $\PM_L$, one arrives at Eq.~\eqref{Fedosov3} and its equivalent for even $d$, hence from there on the calculations can  proceed identically. Finally, the continuity of the strong Chern numbers w.r.t. the variable $t$ can be readily established using Proposition~\ref{SobolevBoundary} and the fact that the Chern cocycles are continuous over the Sobolev spaces of first kind. The invariance of the strong Chern numbers under the deformations $\tilde u_t$ and $\tilde e_t$ can be establish using the same argument as in Theorerm~\ref{IndexThBulk}.\hfill $\Box$

\vspace{.2cm}

The following result now shows that, when exposed to disorder, the boundary states of a non-trivial topological insulator behave completely different from the boundary states of a normal insulator. Indeed,  Jaksic and Molchanov \cite{JM} proved by an adaption of the techniques from \cite{AM} that the boundary states of a normal insulator are localized, in the sense that Aizenmann-Molchanov bound holds for the those energies, as soon as a random boundary potential is added. On the other hand, one has the following.

\begin{theorem}[Delocalizion of the boundary states] \label{th-DelocSpec}\
Let $\hat h=(h,\tilde h) \in M_N(\mathbb C) \otimes \widehat \Aa_d$ be a short-range half-space Hamiltonian for which BGH applies.

\begin{enumerate}[\rm (i)]

\item If the space dimension $d$ is even and the bulk invariant $\Ch_d(p_F)$ defined in \eqref{EvenBulkChernNumbers1} is not zero, then the Aizenmannn-Molchanov bound (uniform as $\epsilon \rightarrow 0$)
\begin{equation}\label{AM25}
\int_\Omega \PM_L(d\omega) \; \big | \big (E + \I \epsilon - \hat h \big )^{-1}(\omega,x) \big |^s 
\;\leq \;
\gamma_s \,e^{-\beta_s |x|} \;, \quad s<1,
\end{equation}
with $E$ anywhere in the spectral gap of $h$, cannot hold for any arbitrarily large but finite $L$. In other words, adding disorder in an arbitrarily thick surface layer will not result in the Anderson localization of any part of the boundary spectrum.

\item If the space dimension $d$ is odd, the CH holds and the bulk invariant $\Ch_d(u_F)$ defined in \eqref{OddBulkChernNumbers1} is not zero, then the Aizenmannn-Molchanov bound \eqref{AM25} at $E=0$  cannot hold for any arbitrarily large but finite $L$. In other words, the boundary spectrum at $E=0$, hence at the Fermi energy, cannot be Anderson localized by the addition of disorder in an arbitrarily thick surface layers.

\end{enumerate}

\end{theorem}

\noindent {\bf Proof.} Key to the argument are the equality between the bulk and the boundary invariants and the index formulas for the boundary invariants, all summarized in Corollary~\ref{IndexBGH}. Since the boundary invariants are defined using the functional calculus with smooth functions, the operators appearing inside the Fredholm indices in both \eqref{IndFormula2} and \eqref{IndFormula4} are norm-continuous of $\omega$. As such, the average over $\omega$ performed in the proof of Theorem~\ref{EqualityConnesChern} can be done w.r.t. any probability measure over $\Omega$, in particular, with $\PM_L$. If so, then the Chern number of the boundary unitary operator $\tilde u_\Delta$ or boundary projection $\tilde p_\Delta$ remains unchanged if we replace the trace $\widetilde \Tt$ by $\widetilde \Tt_L$. But once this switch is made, one has an index theorem which remains valid over the Sobolev space $M_N(\mathbb C) \otimes \mathcal W'_{d,d}(\Ee_d,\PM_L) $, by virtue of Lemma~\ref{IndexThBoundary}. Now, it is known (see  \cite{AG,PLB}) that, whenever the Aizenmannn-Molchanov bound \eqref{AM25} applies, the functional calculus with piece-wise smooth functions generates elements belonging to this Sobolev space, provided the discontinuities occur in an interval of energies where \eqref{AM25} applies. As such, we can deform the functions $\FFunc $ and $\GFunc$ used in the definition of the boundary invariants into $\FFunc (t)=\chi(t\leq \mu)$ and $\GFunc(t)= \sgn(t-\mu)$, continuously within the Sobolev space. Lemma~\ref{IndexThBoundary} then assures that the values of the invariants do not change during these deformations. But with the $\FFunc $ and $\GFunc$ assuming these particular forms, one has $\tilde u_\Delta = \one_N$ and $\tilde p_\Delta = {\rm diag}(\one_N,0_N)$ and the boundary invariants are necessarily zero, hence also the bulk invariants. This contradiction shows that the Aizenmannn-Molchanov bound cannot apply at the Fermi energy $\mu$. The statements now follow because the Fermi energy can be anywhere in the bulk spectral gap for topological insulators from class A, while it is always pinned at $\mu = 0$ for topological insulators from class AIII. At the technical level, the constraint $\mu =0$ must be strictly imposed for $\tilde p_\Delta$ to remain a projection inside the smooth algebra or the Sobolev space.  
\hfill $\Box$

\chapter{Invariants as measurable quantities}
\label{Chap-Conclusions}

\abstract{This chapter presents various applications to solid state physics of the mathematical results obtained in the earlier chapters. The topological invariants are connected to linear and nonlinear transport coefficients and the expected physical effects are discussed in depth for class A and class AIII of topological insulators, in several space dimensions. Then we follow with an in depth analysis of orbital polarization and magneto-electric effects, and virtual topological insulators are taken up as a more recent development. As a further novel implication, it is shown that the surface states of approximately chiral systems may exhibit a quantum Hall effect with a Hall conductance imposed by the bulk invariants.
}

\section{Transport coefficients of homogeneous solid state systems}
\label{sec-transport}

The topological invariants are closely related to the transport coefficients. These are briefly reviewed in this section within the operator algebra formalism developed so far. Let us consider a bulk homogeneous solid state system defined by the Hamiltonian $h \in M_N(\CM) \otimes \Aa_d$. Following mainly \cite{BES,SBB1} (see also \cite{Pro7} for a computational perspective), let us assume an effective time evolution $e^{t\Ll}$ on $M_N(\CM) \otimes \Aa_d$ in the presence of a macroscopic electric field $\mathscr E$ and dissipation, generated by the densely defined derivation
$$
\Ll (a) 
\;=\; 
\I[a,h] \;+\; \langle \mathscr E,  \partial a \rangle \;+\; \Gamma(a)
\;,
$$
where $\Gamma$ is the so called collision (super-) operator having adequate dissipation properties \cite{SBB1}. Recall that $\langle\; ,\; \rangle$ denotes the Euclidean scalar product. The temporal evolution of a density matrix is $\rho_t=e^{t\Ll^*}\!\rho_0$ for a  given an initial density matrix $\rho_0$. Now one is interested in computing (or measuring) the time average charge current density
\begin{equation}
\label{CurrentDensity}
\mathfrak J 
\;=\; 
\lim_{T \rightarrow \infty} \tfrac{1}{T} \int_0^T dt \; \Tt \big (  \mathfrak j \, \rho_t \big )
\;,
\end{equation}
where $\mathfrak j = \partial h=\{\partial_j h\}_{j=1,\ldots,d}$ is the observable representing the charge current. 

\begin{proposition}[\cite{BES,SBB1,SBB2}] 
Assume the initial state of the system to be that of thermal equilibrium, namely the initial density matrix is the Fermi-Dirac function 
$$
\rho_0 
\;=\; 
f_{\beta,\mu}(h) 
\;=\; 
\frac{1}{1+ \exp \big (\beta(h - \mu) \big)}
\;.
$$ 
Then:
\begin{enumerate}[\rm (i)]

\item The current density is given by $\mathfrak J_i = \sum_{j=1}^d \sigma_{i,j} \mathscr E_j + o(\mathscr E^2)$, $i=1,\ldots,d$, with the linear conductivity tensor $\sigma$ given by the Kubo formula
$$
\sigma_{i,j} 
\;=\; 
\Tt \Big ( (\partial_i h)\, \Ll^{-1} \big (\partial_j f_{\beta,\mu}(h) \big) \Big )
\;.
$$

\item If BGH or MBGH holds, the off-diagonal components of the linear conductivity tensor converge in the limit $\beta \rightarrow \infty$ and $\Gamma \rightarrow 0$ to
\begin{equation}\label{OffCond}
\boxed{
\;
\sigma_{i,j} 
\;=\; \big \langle [\xi_{\{i,j\}}], [p_F]_0 \big \rangle 
\;=\; 
{\rm Ch}_{\{i,j\}}(p_F)
\;, 
}
\end{equation}
for $1 \leq i \neq j \leq d$,  while the diagonal components vanish in this limit.
\end{enumerate}
\end{proposition}

 The above statement provides a direct link between the 2-cocycles and the linear conductivity tensor. By taking derivatives with respect to the magnetic field of Eq.~\eqref{OffCond} and using the generalized Streda formulas from Corollary~\ref{coro-Streda}, we will be able to establish direct links between higher cocycles and non-linear transport coefficients. This will be quite relevant for the analysis in dimensions higher than $d=2$. 

\vspace{.2cm}
 
We now turn our attention to the charge transport parallel the boundary of a solid state system defined by $\hat h = (h,\tilde h) \in M_N(\CM) \otimes \widehat \Aa_d$. The observable representing the charge current parallel to the boundary is given by $\hat{\mathfrak j}=  \widehat \partial \hat h$, which indeed provides the expected expression when represented on the Hilbert space,
$$
\widehat \pi_\omega ( \hat{\mathfrak j} ) 
\;=\; 
\I [\widehat H_\omega, \widehat X]
\;,
$$
with $\widehat X = (X_1,\ldots,X_{d-1})$. Now, assume that BGH applies and let $\FFunc :\RM\to[0,1]$ be as in Proposition~\ref{ExpMap}, that is, its derivative $\FFunc '$ is positive and supported in the bulk gap and $\int dE\,\FFunc '(E)=1$, and 
$$
[\tilde u_\Delta]_1\;=\;\Exp[p_F]_0\;=\;[\exp(2\pi\I \,\FFunc (\hat{h}))]_1
\;.
$$ 
The function $\FFunc '(\hat h)$ can be regarded as a density matrix, and since $\FFunc '$ is smooth and with support inside the bulk gap, this function is an element from the boundary algebra and in fact from the smooth sub-algebra $\mathscr E_d$. Then
\begin{equation}
\widetilde{\mathfrak J} 
\;=\; 
\widetilde \Tt \big ( \FFunc '(\hat h) \, \widehat \partial \hat h \big )
\end{equation}
is the well-defined charge current density, flowing along the boundary when the quantum states are populated with a statistical weight given by $\FFunc '(E)$. We will refer to $\widetilde{\mathfrak J}$ as the boundary current.

\begin{proposition}[\cite{SKR,KRS}]
\label{1Cocycle} 
The following identity holds for $j =1, \ldots, d-1$: 
\begin{equation}
\label{BoundaryCurrent0}
\I \, \widetilde \Tt \Big ( \big (\exp(-2\pi \I \FFunc (\hat h)) - \one \big ) \, \widehat \partial_j \exp( 2\pi \I \FFunc (\hat h)) \Big ) 
\;=\;
-\, 2 \pi \; \widetilde \Tt \big ( \FFunc '(\hat h) \, \widehat \partial_j \hat h \big )
\;.
\end{equation}
Written differently, 
\begin{equation}
\label{BoundaryCurrent}
\boxed{
\;
\widetilde{\rm Ch}_{\{j\}}(\tilde u_\Delta) \;=\; -\, 2\pi \; \widetilde{\mathfrak J}_j 
\;.\;
}
\end{equation}
\end{proposition}

\noindent {\bf Sketch of Proof.} Let $\Wind$ denote the quantity on the l.h.s. of \eqref{BoundaryCurrent0}. Expanding the exponential under the derivation as a series and using the Leibniz rule
$$
\Wind 
\;=\; 
\I \,  \sum_{m=1}^\infty
\frac{(2\pi\I)^m}{m!}
\;\sum_{l=0}^{m-1}\;
\widehat{\Tt}
\left(({\tilde{u}_\Delta}^*-\one)\,\FFunc ({\hat{h}})^l\,
\widehat{\partial}_j \FFunc ({\hat{h}})\,\FFunc ({\hat{h}})^{m-l-1}
\right)
\mbox{ , }
$$
\noindent where the trace and the infinite sum can be exchanged
because $\tilde u_\Delta-\one$ belongs to the smooth sub-algebra $\mathscr E_d$. Due to cyclicity and the fact that 
$[\tilde u_\Delta,\FFunc ({\hat{h}})]=0$, each summand is equal to 
$\widetilde{\Tt}((\tilde u_\Delta^*-1)\,\FFunc ({\hat{h}})^{m-1}\,
{\widehat{\partial}}_j\FFunc ({\hat{h}}))$. Exchanging the sum and the trace again and
summing up the exponential, 
$$
\Wind = \I \,  \widetilde{\Tt}((\tilde u_\Delta^*-1)\,\widehat{\partial}_j {\tilde u_\Delta})
\;=\;
2\pi \;\widetilde{\Tt}
\left((\one-\tilde u_\Delta)\,{\widehat{\partial}}_j \FFunc ({\hat{h}})\right) \; . 
$$
Now let us use the homomorphism property of the pairing and repeat the same argument for $\tilde u_\Delta^k=\exp(2\pi
\I \,k\,\FFunc ({\hat{h}}))$ with  $k\neq 0$,
$$
\Wind \;=\; \frac{\I}{k}\;
\widetilde{\Tt}\bigl( (\tilde{u}_\Delta^k-\one)^*\widehat{\partial}_j \hat{u}^k\bigr)
\;=\; 2\pi\;
\widetilde{\Tt}
\left((\one-\tilde{u}_\Delta^k)\,{\widehat{\partial}}_j\FFunc ({\hat{h}})\right)
\;.
$$
Writing  $\FFunc (E)=\int_{-\infty}^\infty dt\,{\tFFunc }(t)\,e^{-E(1+\I t)}$ as a Laplace transform with an adequate function ${\tFFunc }$, the last expression can be further processed using Duhamel's formula
$$
\;\Wind\;=\;2\pi\;
\int_{-\infty}^\infty dt\,{\tFFunc }(t)\,(1+\I t)\;
\int^1_0dq\;
\widetilde{\Tt}
\left((\hat{u}^k-\one)\,
e^{-(1-q)(1+\I t){\hat{h}}}
(\widehat{\partial}_j \hat h)
e^{-q(1+\I t){\hat{h}}}
\right)
\;.
$$
Using the cyclic property of the trace and 
$\FFunc '(E)=-\int dt\,(1+\I t) \,{\tFFunc }(t)\,e^{-E(1+\I t)}$, one therefore finds for $k\neq 0$,
$$
\Wind  
\; = \;
2\pi\;
\widetilde{\Tt}
\left((\hat{u}^k-\one)\,\FFunc '({\hat{h}})\,
{\widehat{\partial}}_j\hat{h}
\right)
\;.
$$
For $k=0$, the r.h.s. vanishes, a fact which will be used below.

\vspace{.2cm}

To conclude, let us choose a differentiable function $\phi:[0,1]\to \RM$ 
vanishing at the boundary points $0$ and $1$. Its Fourier coefficients will be denoted by 
$a_k=\int^1_0 dx \,e^{-2\pi\I k x}\phi(x)$. Then 
$\sum_k a_k e^{2\pi\I k x}=\phi(x)$ and thus $\sum_k a_k=0$.
Hence
\begin{eqnarray*}
a_0\;\Wind & = &
-\;\sum_{k\neq 0} a_k\;\Wind
\\
& = &
- 2\pi\;
\sum_{k} a_k\;
\widetilde{\Tt}
\left((\one-\hat{u}^k)\,
\FFunc '({\hat{h}})\,{\widehat{\partial}}_j\hat{h}
\right)
\\
& = &
- 2\pi\;\widetilde{\Tt}
\bigl(\phi(\FFunc (\hat{h}))\,\FFunc '({\hat{h}})\,{\widehat{\partial}}_j\hat{h}\bigr)
\mbox{ . }
\end{eqnarray*}
\noindent Finally, we let $\phi$ converge to the indicator function of
$[0,1]$. Then $a_0\to 1$, 
while on the other hand $\phi(\FFunc (\hat{h}))\FFunc '({\hat{h}})\to 
\FFunc '({\hat{h}})$ (the Gibbs phenomenon is
damped). This concludes the proof.
\hfill $\Box$

\vspace{.2cm}

The above statement establishes a direct link between the boundary 1-cocycles and the charge current density flowing along the boundary. By taking derivatives with respect to the magnetic field of Eq.~\eqref{OffCond} and using the generalized Streda formulas from Corollary~\ref{coro-Streda}, we will be able to establish direct links between higher cocycles and measurable physical quantities. This will again be quite relevant for the analysis in dimensions higher than $d=2$.  Furthermore, let us point out that the calculation of the above proof combined with a homotopy argument can be used to deal with quantized currents at interfaces of two materials with different topological invariants \cite{KSV}.

\section{Topological insulators from class A in $d=2$, $3$ and $4$}

In dimension $d=2$, the topological phases from the unitary class include the classical integer quantum Hall phases and there are many excellent accounts on the physics and mathematics of the integer quantum Hall effect in dimension $d=2$, and we refrain from giving an incomplete list here. The papers of Bellissard \cite{Bel,Bel2} present the bulk theory for tight-binding models and build up the algebraic formalism used in this work. A detailed account of this and an extension to the regime of a MBGH can be found in \cite{BES}. The bulk-boundary principle was first demonstrated by Hatsugai in \cite{Hat1} for the rational magnetic flux case, then \cite{SKR,KRS} used the Pimsner-Voiculescu sequence to extend this result to a more general context (see (iii) of Corollary~\ref{coro-QHE2} below). In particular, \cite{SKR} also contains a detailed description of the physical interpretation and importance of this result as well as many citations to the physics literature. Later on, other rigorous proofs of bulk-boundary correspondence for tight-binding quantum Hall systems were found \cite{EG,EGS} and the techniques were extended to models in continuous physical space \cite{KS0,KS,CG}. An application of the machinery developed in \cite{SKR,KRS} to Chern insulators can be found in \cite{Pro5}. 

\vspace{.2cm}

Below we summarize the main statements available for the topological phases from class A in dimension $d=2$. They follow directly from \cite{BES,SKR,KRS} and they were also known in the physics literature \cite{Halp1}, but here we view them as direct corollaries of the theory developed in the previous chapters. Of course, the input from the previous section is absolutely necessary.

\begin{corollary} 
\label{coro-QHE2} 
Let $\hat h = (h,\tilde h) \in M_N(\CM) \otimes \widehat \Aa_d$ with $d=2$.

\begin{enumerate}[\rm (i)]

\item If BGH holds, then the integrated density of states can take only the discrete values
$$
\Tt(p_F) \;=\; {\rm Ch}_{\emptyset}(p_F) \;\in\; \ZM \;+\; \frac{B_{1,2}}{2\pi} \;  \ZM\;.
$$

\item If MBGH holds, then the off-diagonal element of the bulk conductivity tensor is quantized by the strong bulk invariant 
$$
\sigma_{1,2}\; =\; {\rm Ch}_2 (p_F)\;\in\; \ZM
\;.
$$
Furthermore, as long as MBGH holds, $\sigma_{1,2}$ remains quantized and invariant to the deformations of $h$ defined by Definition~\ref{ContDeformation}. 

\item If BGH holds, then the boundary current is quantized by the bulk and boundary invariants
$$
2\pi \; \widetilde{\mathfrak J}_1
\;=\;
-\,\widetilde{\rm Ch}_1(\tilde u_\Delta)
\;=\;
-\, {\rm Ch}_2(p_F)
\;=\;
\sigma_{1,2} \;\in\; \ZM
\;.
$$
Furthermore, if ${\rm Ch}_2 (p_F) \neq 0$, the entire boundary spectrum is delocalized.
\end{enumerate}
\end{corollary}

Let us point out that (ii) assures us that the topological phases corresponding to the different values of ${\rm Ch}_2 (p_F)$ are separated by a localization-delocalization phase transitions, which can be sharply identified experimentally via transport measurements, as demonstrated in \cite{CYT}.

\vspace{.2cm}

In dimension $d=3$ there are only weak topological phases. Among them are the quantum Hall phases in 3-dimensions. The available results for the latter \cite{Halp2,MK,KH,KHW,KAK,KA} are restricted to the cases where the entries in the $\BB$ matrix (divided by $2\pi$) are rational numbers. The following statements, which are again direct corollaries of the theory of the previous chapters, generalize them to arbitrary $\BB$ and also include the disorder.

\begin{corollary} 
\label{coro-QHE3} 
Let $\hat h = (h,\tilde h) \in M_N(\CM) \otimes \widehat \Aa_d$ with $d=3$ and assume that BGH holds. Then:

\begin{enumerate}[\rm (i)]

\item The integrated density of states can take only the discrete values
$$
\Tt(p_F) \;=\; 
{\rm Ch}_{\emptyset}(p_F)\; \in\; \ZM\; +\; \sum_{1\leq i < j \leq 3 } \frac{B_{i,j}}{2\pi} \;  \ZM \; .
$$

\item The off-diagonal elements of the bulk conductivity tensor are quantized
$$
\sigma_{i,j} 
\;=\; 
{\rm Ch}_{\{i,j\}} (p_F)\;\in\; \ZM\;, 
\qquad 1 \leq i < j \leq 3
\;.
$$
Furthermore, as long as BGH holds, $\sigma_{i,j}$'s remains quantized and invariant to the deformations of $h$ defined by Definition~\ref{ContDeformation}.

\item The boundary currents are quantized too
$$
2\pi \; \widetilde{\mathfrak J}_j
\;=\;
-\, \widetilde{\rm Ch}_j(\tilde u_\Delta)
\;=\;
{\rm Ch}_{\{j,3\}}(p_F)
\;=\;
-\, \sigma_{j,3}\; \in\; \ZM
\;, \qquad j=1,2\;.
$$
\end{enumerate}
\end{corollary}

Since the weak Chern numbers do not accept an index formula, we cannot replace BGH with MBGH at point (i). In other words, with the methods developed here we cannot conclude that weak topological phases defined by  the quantized values of $\sigma_{i,j}$'s  are separated by phase boundaries where the localization length diverges, as it happens in $d=2$. Also, at point (ii), we cannot say anything about the localized/delocalized character of the boundary spectrum, though we can say that is never gapped if any of $\sigma_{\{j,3\}}$ happens to be non-zero. Note that \cite{BF} predicted a certain delocalization of the boundary states, hence it will be important to further investigate the weak topological insulators. 

\vspace{.2cm}

Although purely fictitious, the quantum Hall effect in dimension $d=4$ was conceptually very important in condensed matter theory \cite{ZH,QHZ}. Below we summarize our predictions for the hypothetical topological insulators from class A in $d=4$.

\begin{corollary} 
\label{coro-QHE4} 
Let $\hat h = (h,\tilde h) \in M_N(\CM) \otimes \widehat \Aa_d$ with $d=4$.

\begin{enumerate}[\rm (i)]

\item If BGH holds, the integrated density of states can take only the discrete values
$$
\Tt(p_F) 
\;=\; 
{\rm Ch}_{\emptyset}(p_F) \in \ZM \;+\; 
\sum_{\{i, j\} } \frac{B_{i,j}}{2\pi} \;  \ZM 
\;+\; 
 \frac{{\rm Pf}(\BB)}{(2\pi)^2} \;  \ZM \; ,
$$
where all indices are assumed as being ordered.
 
\item If BGH holds, the off-diagonal elements of the bulk conductivity tensor take only the discrete values
$$
\sigma_{i,j} 
\;=\; 
{\rm Ch}_{\{i,j\}} (p_F) \;\in\; \ZM \;+\; \frac{B_{k,l}}{2\pi} \; \ZM
\;,
$$
where $k<l$ and such that $\{i,j\} \cap \{k,l\} = \emptyset$. Furthermore, as long as BGH holds, $\sigma_{i,j}$'s remains quantized and invariant to the deformations of $h$ defined by Definition~\ref{ContDeformation}.

\item If MBGH holds, the derivatives of the Hall conductivities w.r.t. to the magnetic field are quantized by the strong invariant
$$
2 \pi \; \partial_{B_{i,j}} \sigma_{k,l} 
\;=\; (-1)^\rho\, {\rm Ch}_4(p_F) 
\;\in\; 
\ZM
\;, 
\qquad \{i,j\} \cap \{k,l\}\,=\,\emptyset\;,
$$
where $\rho$ is the permutation which brings $\{i,j,k,l\}$ into $\{1,2,3,4\}$.
Furthermore, as long as MBGH holds, $\partial_{B_{i,j}} \sigma_{k,l}$'s remain quantized and invariant to the deformations of $h$ defined by Definition~\ref{ContDeformation}.

\item If BGH holds, then the boundary currents can take only the discrete values
\begin{equation}
2\pi \; \widetilde{\mathfrak J}_j
\;=\;
- \, \widetilde{\rm Ch}_j(\tilde u_\Delta)
\;=\;
-\,{\rm Ch}_{\{j,4\}}(p_F) \in \ZM \;+\; \frac{B_{k,l}}{2\pi} \;\ZM
\;, \qquad j=1,2,3,
\end{equation}
where $\{k,l\}$ are the unique set of indices such that $\{k,l\} \cap \{j,4\} = \emptyset$.

\item If BGH holds, then the derivatives of the boundary currents w.r.t. the magnetic field are quantized
\begin{equation}
(2\pi)^2 \;\partial_{B_{i,j}}\widetilde{\mathfrak J}_k
\;=\;
-\, (-1)^\rho \,\widetilde{\rm Ch}_3(\tilde u_\Delta) 
\;=\;
-\,(-1)^\rho \, {\rm Ch}_4(p_F) 
\;\in\; \ZM, 
\end{equation}
where $i\neq j \neq k$ and $\rho$ is the permutation which brings $\{i,j,k\}$ into $\{1,2,3\}$. Furthermore, if the above invariants are not zero, then the entire boundary spectrum is necessarily delocalized.

\end{enumerate}
\end{corollary}

Note that $\partial_{B_{i,j}} \sigma_{k,l}$ represents the second-order response function $\partial^2 {\mathfrak J}_k/\partial E_l \partial B_{i,j}$, hence point (iii) predicts the quantization of this physically measurable quantity, in agreement with the original finding in \cite{ZH}.

\section{Topological insulators from class AIII in $d=1$, $2$ and $3$}
\label{sec-ChiralPol}

The experimentally measurable bulk properties relevant to the class of chiral symmetric solid state systems are the chiral (orbital) polarization $P_{\mbox{\rm\tiny C}}$ and the variations of $P_{\mbox{\rm\tiny C}}$ w.r.t. the magnetic field. For a chiral Hamiltonian $H=\{H_\omega\}_{\omega\in\Omega}$ of a solid state system with sub-lattice symmetry, the chiral polarization is defined as the difference between the electric dipole moments per unit cell of the two sub-lattices, which can be written as: 
\begin{equation}
\label{eq-ChiralPol}
P_{\mbox{\rm\tiny C}}
\;=\;
\int_\Omega \PM(d\omega)\;\tr\,\langle 0|P_\omega\,J\,X\,P_\omega|0\rangle
\;, \quad P_\omega=\chi(H_\omega\leq 0).
\end{equation}
Using $X|0\rangle=0$, one can rewrite $P_{\mbox{\rm\tiny C}}$  with the non-commutative analysis tools as
$$
P_{\mbox{\rm\tiny C}}
\;=\;
\I\;\Tt (p_FJ\partial p_F)
\;.
$$
Let up point out that, without the chirality operator $J$, the r.h.s. would vanish identically. Hence, it is impossible to define the total dipole polarization in this manner. The real reason for this is that definition \eqref{eq-ChiralPol} will be ill behaved without $J$. Now, the following result show that $P_{\mbox{\rm\tiny C}}$ is actually of topological nature, namely given by a pairing of a $K_1$-group element with a $1$-cocycle.

\begin{proposition}
\label{prop-PC}
Let $h\in M_{2N}(\CM) \otimes \Aa_d$ and assume CH and BGH hold. Then
$$
P_{\mbox{\rm\tiny C},j}
\;=\;
-\,\frac{1}{2}\;\big \langle [\xi_{\{j\}}],[u_F]_1\big \rangle
\;=\;
-\,\frac{1}{2}\;{\rm Ch}_{\{j\}}(u_F)
\;,
\qquad
j=1,\ldots,d
\;.
$$
\end{proposition}

\noindent {\bf Proof.} Recall from \eqref{eq-PF_UF} that
$$
p_F
\;=\; 
\frac{1}{2}
\begin{pmatrix}
\one & -u_F^* \\
-u_F & \one
\end{pmatrix}
\;,
\qquad
J\;=\;
\begin{pmatrix}
\one & 0 \\
0 & -\one
\end{pmatrix}
\;.
$$
Thus 
$$
P_{\mbox{\rm\tiny C}}
\;=\;
\frac{\I}{4}\,
\Tt
\left(
\begin{pmatrix}
\one & u_F^* \\
-u_F & -\one
\end{pmatrix}
\begin{pmatrix}
0 & -\partial u_F^* \\
-\partial u_F & 0
\end{pmatrix}
\right)
\;=\;
\frac{\I}{4}\,
\Tt
(-u_F^*\partial u_F+u_F\partial u_F^*)
\;.
$$
Now by Proposition~\ref{Indenties1}(iv), $u_F(\partial u_F^*)=-(\partial u_F) u_F^*$, so that by cyclicity
$$
P_{\mbox{\rm\tiny C}}
\;=\;
-\,\frac{\I}{2}\,
\Tt
(u_F^*\partial u_F)
\;,
$$
which is the precisely the claim.
\hfill $\Box$

\vspace{.2cm}

We now have all the tools to characterize the physics of the chiral symmetric solid state systems. The following statements were discussed extensively in Chapter~\ref{Chap-Illustration}, but we state them for completeness. In the published literature, one can find them in \cite{SF} and \cite{MSHP}.

\vspace{-.1cm}

\begin{corollary} 
\label{coro-1D} 
Let $\hat h = (h,\tilde h) \in M_{2N}(\CM) \otimes \widehat \Aa_d$ with $d=1$. Assume that CH holds and recall that, for $d=1$, the spectrum of $\hat h$ inside $\Delta$ is discrete whenever a bulk spectral gap exists.

\begin{enumerate}[\rm (i)]

\item If the MBGH holds, then the chiral polarization is quantized by the strong bulk invariant 
$$
P_{\mbox{\rm\tiny C}}\; =\; -\,\frac{1}{2}\;{\rm Ch}_1 (u_F)
\;\in\; \frac{1}{2}\,
\ZM
\;.
$$
Furthermore, as long as the MBGH holds, $P_{\mbox{\rm\tiny C}}$ remains quantized and invariant to the deformations of $h$ defined by Definition~\ref{ContDeformation}. 

\item If the BGH holds and $P_{\mbox{\rm\tiny C}} \neq 0$, then by Corollary~\ref{coro-GappedEf} there will necessarily be edge states exactly at $E=0$, which are the zero modes discussed in Section~\ref{Sec-ChiralUnitaryClass}. Furthermore 
$$
N_+ \;-\;N_- \;= \; \widetilde{\rm Ch}_{\emptyset} (\tilde p_\Delta)\;=\; -\,{\rm Ch}_1 (u_F)\;=\;  \,2\,P_{\mbox{\rm\tiny C}}
\;,
$$
where $N_\pm$ is the number of zero modes of $\pm$ chirality. 
\end{enumerate}
\end{corollary}

Let us stress that, as for the IQHE, topological phases corresponding to different values of $P_{\mbox{\rm\tiny C}}$ are separated by a localization-delocalization phase transition which can be determined experimentally via transport measurements. Next, in dimension $d=2$, there are only weak chiral systems. Nevertheless, there are some interesting predictions for these systems.

\vspace{-.1cm}

\begin{corollary} 
\label{coro-2D} 
Let $\hat h = (h,\tilde h) \in M_{2N}(\CM) \otimes \widehat \Aa_d$ with $d=2$ and assume that BGH and CH hold. Then everything said in Corollary~\ref{coro-QHE2} holds and, additionally:

\begin{enumerate}[\rm (i)]

\item The components of the chiral polarization are quantized as
$$
P_{\mbox{\rm\tiny C}, j} 
\;=\; 
-\,\tfrac{1}{2}\;{\rm Ch}_{\{j\}} (u_F)\;\in \;\tfrac{1}{2}\;\ZM\;, 
\qquad j=1,2
\;.
$$
Furthermore, as long as BGH and CH hold, the components $P_{\mbox{\rm\tiny C}, j}$ remain quantized and invariant under the deformations of $h$ defined by Definition~\ref{ContDeformation}.

\item The bulk-boundary principle gives
$$
\widetilde \Tt ( \tilde p_\Delta)
\;=\;
\widetilde{\rm Ch}_{\emptyset} (\tilde p_\Delta)
\;=\; -\, {\rm Ch}_{\{2\}} (u_F)
\;=\;  
\,2\,P_{\mbox{\rm\tiny C},2}
\;.
$$
As a consequence, if $P_{\mbox{\rm\tiny C}, 2} \neq 0$, $\hat h$ will have essential spectrum at $E=0$.

 \end{enumerate}
\end{corollary}

\noindent {\bf Proof.} We only need to show point (ii). If the spectrum at the origin is discrete, then we can choose an interval $[-\delta,\delta]$ as in Proposition~\ref{prop-SurfaceStateGap}, and $[-\delta,\delta]$ contains only discrete spectrum of $\hat h$. With the notations from Proposition~\ref{prop-SurfaceStateGap}, the bulk-boundary principle gives
$$
\widetilde\Tt \big (\tilde p_+(\delta) \big) \;-\; \widetilde \Tt \big( \tilde p_-(\delta)\big) \; = \; 2 \,P_{\mbox{\rm\tiny C}, 2}
\;.
$$
Hence for $\tilde p(\delta)=\tilde p_+(\delta)+\tilde p_-(\delta)$ 
$$
\widetilde\Tt \big (\tilde p(\delta) \big) 
\;\geq \;
\Big |\widetilde\Tt \big (\tilde p_+(\delta) \big)\, -\, 
\widetilde \Tt \big(\tilde p_-(\delta)\big) \Big| 
\; = \; \big | 2\, P_{\mbox{\rm\tiny C}, 2}\big |
\;.
$$
But for a spectral projector $\tilde p(\delta)$ onto discrete spectrum one has $\widetilde\Tt \big (\tilde p(\delta) \big)=0$, and this is a contradiction.
\hfill $\Box$

\vspace{0.2cm}

The bulk invariants appearing in (i) of Corollary~\ref{coro-2D} are weak odd Chern numbers, hence we cannot replace the BGH by the MBGH. Consequently, with the methods developed so far, we cannot conclude that weak topological phases defined by  the quantized values of $P_{\mbox{\rm\tiny C}, j}$'s  are separated by phase boundaries where the localization length diverges, as it happens in $d=1$. Also, in item (ii), we cannot say anything about the localized or delocalized character of the boundary spectrum appearing at $E=0$. 

\begin{corollary} 
\label{coro-3D} 
Let $\hat h = (h,\tilde h) \in M_{2N}(\CM) \otimes \widehat \Aa_d$ with $d=3$ and assume that the CH holds. Then: 

\begin{enumerate}[\rm (i)]

\item If the  BGH holds, the components of the chiral polarization take discrete values
$$
P_{\mbox{\rm\tiny C}, i} 
\;=\;
-\,\tfrac{1}{2} \big \langle [\xi_{\{i\}}], [u_F]_1 \big \rangle 
\;=\;
-\,\tfrac{1}{2} {\rm Ch}_{\{i\}}(u_F)
\;\in\; \tfrac{1}{2}\,\ZM \;+\; \frac{B_{j,k}}{4\pi} \,  \ZM \;, 
\qquad i\neq j \neq k\neq i
\;.
$$
Furthermore, as long as BGH holds, the components $P_{\mbox{\rm\tiny C}, i}$ remain quantized and invariant to the deformations of $h$ defined by Definition~\ref{ContDeformation}. 

\item If the MBGH holds, then the chiral magneto-electric response coefficients are quantized by a strong invariant 
$$
\partial_{B_{i,j}}P_{\mbox{\rm\tiny C},k}
\;=\;
\frac{1}{4\pi}\;\big \langle [\xi_{\{i,j,k\}}],[u_F]_1\big \rangle 
\;=\; 
\frac{\eta}{4\pi} \; {\rm Ch}_3(u_F)
\; \in\; 
\frac{1}{4\pi}\, \ZM
\;,
$$
with $\eta$ the sign of the permutation which brings ${i,j,k}$ to the natural order. Furthermore, as long as the MBGH holds, $\partial_{B_{i,j}}\,P_{\mbox{\rm\tiny C},k}$ remains quantized and invariant to the deformations of $h$ defined by Definition~\ref{ContDeformation}.

\item If the BGH holds, then the bulk-boundary principle gives
$$
\widetilde \Tt(\tilde p_\Delta) 
\;=\; 
\widetilde{\rm Ch}_{\emptyset} (\tilde p_\Delta)
\;=\; 
-\,{\rm Ch}_{\{3\}} (u_F)
\;=\; 2\,P_{\mbox{\rm\tiny C},3}
\;\in\; 
\ZM 
\;+\; 
\frac{B_{1,2}}{2\pi} \,  \ZM \;,
$$
and
$$
\widetilde{\rm Ch}_2(\tilde p_\Delta) 
\;=\;
-\, {\rm Ch}_3(u_F) 
\;=\; 
4\pi\,  \partial_{B_{1,2}}P_{\mbox{\rm\tiny C},3} 
\;\in\; \ZM \; .
$$
As a consequence, if $P_{\mbox{\rm\tiny C}, 3} \neq 0$, then $\hat h$ will necessarily display essential spectrum at $E=0$. If instead of or additionally to  $P_{\mbox{\rm\tiny C}, 3} \neq 0$ we have $\partial_{B_{1,2}}\,P_{\mbox{\rm\tiny C},3} \neq 0$, then the boundary spectrum at $E=0$ is necessarily delocalized.

\item Assume the existence of an interval $[-\delta,\delta] \subset \Delta$ such that the ends $\pm \delta$ lie in a region of Anderson localized surface spectrum.  Let $\tilde p(\delta)=\chi(-\delta\leq \hat{h}\leq\delta)$  be the associated spectral projection and decompose it as in Proposition~\ref{prop-SurfaceStateGap} into chiral sectors $\tilde p(\delta)=\tilde p_+(\delta)+\tilde p_-(\delta)$ with $J\tilde p_\pm(\delta)=\pm\tilde p_\pm(\delta)$. Then
$$
\widetilde{\rm Ch}_2\big (\tilde p_+(\delta) \big ) 
\;-\; 
\widetilde{\rm Ch}_2 \big (\tilde p_-(\delta) \big )
\;=\;
-\,{\rm Ch}_3(u_F) 
\;=\; 
\,4\pi\,  \partial_{B_{1,2}}P_{\mbox{\rm\tiny C},3} \;\in\; \ZM \;.
$$
Among other things, this implies that, if the bulk invariant is odd, then necessarily 
$$
\ZM \;\ni\; \widetilde{\rm Ch}_2\big (\tilde p(\delta)\big )\; \neq \;0\;,
$$
so that the surface will display the IQHE with the Hall conductance jumping at least by one unit in the interval $[-\delta,\delta]$.
\end{enumerate} 
\end{corollary}

\noindent {\bf Proof.} Item (ii) follows from Proposition~\ref{prop-deri1coc} and (iv) by choosing the lift as in Proposition~\ref{prop-SurfaceStateGap}. \hfill $\Box$

\vspace{0.2cm}

Let us stress that (ii) assures that the topological phases corresponding to the different values of $\partial_{B_{i,j}}P_{\mbox{\rm\tiny C},k}$ are separated by a localization-delocalization phase transitions which is again visible in transport experiments. This has been confirmed numerically in \cite{SFP}. The statement  (iii) on the delocalized character of the surface states at $E=0$ is in full agreement with the conclusions drawn in Ref.~\cite{EG2}. As already pointed out there, no such statement can be formulated about the states at other energies. For the IQHE predicted in (iv), the methods developed so far give no further information about the values of $\widetilde{\rm Ch}_2\big (\tilde p(\delta)\big )$. Hence we have no general prediction about the value of the Hall conductance of the surface states, though we will make a conjecture on these values in the next section. Nevertheless, let us note that the spectrum away from the origin is expected to be localized (see the discussion in \cite{EG2}) and that (iv) can occur in the absence of a magnetic field. In the latter situation, item (iv) hence predicts an {\it anomalous} quantum Hall effect. Lastly, let us mention that the IQHE at the surface may be absent altogether for an even bulk invariant, as for example would happen if ${\rm Ch}_3(u_F)=2$ and $\widetilde{\rm Ch}_2\big (\tilde p_\pm(\delta)\big ) = \mp 1$. However, there are other interesting particular scenarios which are worth discussing and this is done in the next seciton.

\section{Surface IQHE for exact and approximately chiral systems}
\label{sec-SurfacQHE}

Let us start by formulating a conjecture on the values of $\widetilde{\rm Ch}_2\big (\tilde p(\delta)\big )$ which is compatible with the bulk-boundary principle. For this, we introduce the concept of stable configuration which is best explained for $d=1$. In this case, the bulk-boundary principle states that $N_+-N_- = -\,{\rm Ch}_1(u_F)$, from where one can conclude that the number of edge zero modes $N= N_+ + N_-$ is necessarily larger than or equal to the absolute value of the bulk invariant, but one cannot say what exactly this number is, just from the bulk topology. However, under small perturbations or disorder, pairs of zero modes of opposite chirality can and usually will exit the zero-mode subspace, and this phenomenon will repeat itself until one of the chiral sectors is completely depleted of zero modes. The process cannot continue and the system reached what we call the stable configuration. In $d=3$ and in the absence of disorder, something similar will happen because pairs of zero-energy Dirac points of opposite chirality in the boundary spectrum can annihilate each other or leave the zero-energy level, and a stable configuration can be reached only when one chiral sector is completely depleted of zero-energy Dirac points. For a general chiral system in dimension $d$, we define a stable configuration to be reached if there is a $\delta$ such that one of $\widetilde{\rm Ch}_{d-1}\big (\tilde p_\pm(\delta)\big )$ is zero. We are now ready to formulate our conjectures. The notations from Corollary~\ref{coro-3D} will be used throughout.

\vspace{.2cm}

\noindent {\bf Conjecture [(Anomalous) Surface IQHE]}
{\it Let $\hat h = (h,\tilde h) \in M_{2N}(\CM)\times \widehat \Aa_d$ in dimension $d=3$ be such that BGH and CH apply, and assume ${\rm Ch}_3(u_F) \neq 0$. Then Corollary~\ref{coro-3D} assures us that the boundary spectrum is delocalized at $E=0$. The first conjecture is that, in presence of disorder, the boundary spectrum is everywhere localized except at $E=0$ for $\BB=0$, and for $\BB\not=0$ furthermore at a discrete set of Landau bands symmetrically located around $E=0$. The second conjecture is that, in presence of disorder, the system is always in a stable configuration for all values of the magnetic field.  In these conditions, the Hall conductance of the surface will display a plateau-plateau transition exactly at $E=0$, with a jump equal precisely to $\big | {\rm Ch}_3(u_F) \big |$. For $\BB=0$  this is hence an anomalous surface IQHE with Hall conductance dictated by the bulk invariant.   
}

\vspace{.2cm}

This conjecture can be probed numerically. For vanishing magnetic fields, our initial efforts in this direction unfortunately could not shed any light on these important issues. During these attempts, it became clear that resolving the localized/delocalized character of the surface states will be a large scale computational endeavor. We hope that this will be of interest to the experts in the field. We also hope that the possibility of observing the anomalous IQHE at the surface of a non-magnetic material will renew the experimental and theoretical efforts on identifying a topological solid state system from the AIII class in $d=3$.

\vspace{.2cm}

If an external magnetic field perpendicular to the surface is present, then the situation is more traceable because gaps in the surface spectrum open at weak disorder. Indeed, as it usually happens for two-dimensional electron systems, Landau bands are forming. If the bulk invariant is now odd, then based on item (iv) of Corollary~\ref{coro-3D} we know that a Landau band will be pinned at the origin and that the Hall conductance of the surface will jump by at least one unit as the Fermi level crosses this band. In this situation, we have verified the conjecture numerically for all topological phases of the model presented in Section~\ref{CUExactModel} in $d=3$, under relatively small magnetic fields. Note that there is one phase with even bulk invariant which hence also had a non-vanishing surface Hall conductance. 

\vspace{.2cm}

Let us further elaborate on the importance of the parity of the bulk invariant in the case of a non-vanishing magnetic field, hence supplementing statement (iv) of Corollary~\ref{coro-3D}. Suppose that there is a Landau band at $E \neq 0$. Then, due to the chiral symmetry, there will be another Landau band at $-E$ and the Chern numbers of the two bands are equal. Under small perturbations, these paired Landau bands can, in principle, migrate towards $E=0$ and then join the central Landau band, but note that such process will change the Chern number of the central band by an even number. If $\widetilde{\rm Ch}_2\big (\tilde p(\delta)\big )$ was odd in the first place, then the Chern number of the central Landau band cannot be canceled by the processes just described and it indeed remains odd.

\vspace{.2cm}

The physics described in the above conjecture might remind one of the observations made on graphene at relatively small magnetic fields \cite{ZTS,NGM} where the Hall conductance jumps by four units as the Fermi level crosses the Landau band pinned at the origin. However, this feature of graphene is not stable and at larger magnetic fields where the central Landau band splits into four Landau sub-bands and only jumps by one unit occur for the Hall conductance \cite{ZZK,YDW}.

\vspace{.2cm}

We now turn our attention to the solid state systems with approximate chiral symmetry in dimension $d=3$, that is, the ACH defined in Section~\ref{Sec-MainH} is supposed to hold. By Proposition~\ref{prop-gapACA}, such a system is homotopically connected to a solid state system exhibiting an exact chiral symmetry and thus displaying the physics discussed above on its surface. Since the IQHE is robust against homotopies, we can automatically conclude that this interesting physics will also be observed under weak breaking of the chiral symmetry. More precisely:

\begin{proposition}[(Anomalous) Surface IQHE under ACH] 
Let $\hat h = (h,\tilde h) \in M_{2N}(\CM)\times \widehat \Aa_d$ in dimension $d=3$ be such that BGH and CH apply, and assume that the above Conjecture applies. Let $t\in[0,1]\mapsto \hat h(t)$ be a continuous deformation of $\hat h$ (as defined in Definition~\ref{ContDeformation}) which breaks the chiral symmetry. Further assume that the interval $[-\delta,\delta]$ can be chosen such that its ends resides in a region of localized boundary spectrum for all $t\in[0,1]$ (which is always possible for small deformations). Then the spectral projections $\tilde p(\delta,t)=\chi(-\delta\leq \hat{h}(t)\leq \delta)$ lead to a constant value $\widetilde{\rm Ch}_2\big (\tilde p(\delta,t)\big )$ during the deformations. As such, the system with weakly broken chiral symmetry will continue to display the surface IQHE, which is anomalous if the magnetic field vanishes. However, the divergence of the localization length is not necessarily at $E=0$ any more.
\end{proposition}

\noindent {\bf Proof.} From Proposition~\ref{HFSmoothFunc2}, it follows that $\tilde p(\delta,t)$ varies continuously in the non-commutative Sobolev space $M_N(\CM)\otimes {\mathcal W}_{d-1,1}(\Ee_d,\widetilde \Tt)$. Then the statement follows from Theorem~\ref{IndexThBoundary}.
\hfill $\Box$

\vspace{0.2cm}

When the chiral symmetry is broken, the Hall conductance of the surface should continue to display a net jump of $|{\rm Ch}_3(u_F)|$ over the interval $[-\delta,\delta]$. This net jump, however, will very likely not happen suddenly at a single energy, but instead will be a sum of elementary jumps by one unit. As we already pointed out several times, the chiral symmetry is expected to hold only approximately in real solid state systems, hence the established stability of the physical effects also against weak symmetry breaking should facilitate the experimental observability of the surface IQHE in adequate materials.

\section{Virtual topological insulators}
\label{Sec-VirtualTI}

The topological systems in $d=4$ or higher dimensions are not entirely fictitious since additional dimensions can occur in a parameter space. A special place among such systems is held by the virtual topological insulators, introduced and characterized in \cite{KLR}. Their defining characteristic is a strong topological invariant which is defined in $d + d'$ space dimensions, where $d$ counts the physical dimensions and $d'$ the virtual ones, with an invariant that is yet computable and experimentally measurable inside the $d$ physical dimensions.

\vspace{.2cm}

Let us briefly describe the virtual topological insulators from class A in $3+1$ dimensions, introduced in \cite{Pro6}. For sake of simplicity, the disorder will be neglected. Then the virtual systems are generated by the algebra $\Aa_4=C^\ast(u_1,\ldots,u_4)$ from Definition~\ref{def-rotalg} via the following faithful representation on $\ell^2(\ZM^3)$ invoking only three magnetic translations $U_1,U_2,U_3$:
$$
\pi_\theta(u_j) \;=\;U_j\;, \quad \mbox{\rm for }j = 1,2,3\;,
\qquad
\pi_\theta (u_4)\;=\;e^{\I ( \langle \BB_4, X \rangle + \theta)}
\;, 
$$
where $\BB_4 = (B_{1,4},B_{2,4},B_{3,4})$ now plays the role of frequencies of the perturbation and $\theta\in\RM$ the phase of the representation. As a non-trivial example, let us take
$$
h \;=\; 
\tfrac{1}{2\I}\,\sum_{j=1}^4 \gamma_j \otimes (u_j - u_j^\ast) \,+\, \gamma_0 \otimes \Big(m + \tfrac{1}{2} \sum_{j=1}^4(u_j + u_j^\ast) \Big) \; \in\;M_4(\CM) \otimes \Aa_4\;,
$$
which generates the model already analyzed in Section~\ref{Sec-UExactModel}. There it was also shown to posses a strong topological invariant ${\rm Ch}_4(p_F) \neq 0$. Here focus is on the representations $H_\theta=\pi_\theta(h)$ on $\CM^4 \otimes \ell^2 (\ZM^3)$ rather than $\CM^4 \otimes \ell^2(\ZM^4)$:
\begin{align*}
H_\theta \;=\; & 
\tfrac{1}{2\I}\;
\sum_{j=1}^3 \gamma_j \otimes (U_j - U_j^\ast) 
\;+\; 
\gamma_0 \otimes \Big(m + \tfrac{1}{2} \sum_{j=1}^3(u_j + u_j^\ast \Big) 
\\
& \; +\; \gamma_4 \otimes \sin( \langle \BB_4,X \rangle + \theta) 
\;+\; \gamma_0 \otimes \cos ( \langle \BB_4,X \rangle + \theta)
\;,
\end{align*}
which describes a periodic crystal subjected to a magnetic field and an additional incommensurate periodic potential, namely we require $ \frac{1}{2\pi} B_{j,4}$ to be irrational. As $H_\theta$ acts on a Hilbert space over the three-dimensional lattice and and it depends on an additional parameter $\theta \in \SM^1$ we refer to it as a model in $3+1$ dimensions. Let us now show that the topological invariant can be computed at fixed $\theta$. First of all,
\begin{align*}
\Tt (a) 
\;=\; \int_{\SM^1}  \frac{d \theta}{2\pi} \; \Tr\,\langle 0 | \pi_\theta(a) | 0 \rangle 
&\;=\; \lim_{|V| \rightarrow \infty}  \frac{1}{|V|}\;
\sum_{x \in V} \Tr\,\langle 0 | \pi_{\theta+\langle \BB_4, x \rangle}(a) | 0 \rangle 
\\
&\;=\; \lim_{|V| \rightarrow \infty} \frac{1}{|V|}\; \sum_{x \in V} \Tr\,\langle x | \pi_\theta(a) | x \rangle
\;,
\end{align*}
where Birkhoff's theorem was used combined with $U_j \pi_\theta (a) U_j^\ast = \pi_{\theta +  B_{j,4}}(a)$ and the irrationality of $B_{j,4}$. Hence the topological invariant can be indeed computed at fixed $\theta$:
$$
{\rm Ch}_4(p_F) 
\;=\; \Lambda_4 \,\sum_{\rho \in \Ss_4} (-1)^\rho \,\Tt\Big ( P_\theta \prod_{j=1}^4 \partial_j P_\theta \Big )
\;,
$$
where $P_\theta = \chi(H_\theta \leq \mu)$, $\partial_4 P_\theta = \partial_\theta P_\theta$ and $\partial_j P_\theta = \I [P_\theta,X_j]$ for $j=1,2,3$. This bulk topological invariant was related in \cite{Pro6} to the magneto-electric response function, discussed in the following sections. Another interesting link can be established via the generalized Streda formulas. For example,
$$
{\rm Ch}_4(p_F) \;=\; 2\pi \; \partial_{B_{3,4}} \sigma_{1,2}
\;,
$$
which implies the quantization of the variation of the bulk Hall conductance in the $(1,2)$ plane ({\it i.e.} the non-linear Hall conductivity) w.r.t. the modulation of the incommensurate potential (or of the original lattice) in the third direction. This is a piezo-magneto-electric effect and the prediction could be tested with cold atom physics. Furthermore, assume now a boundary, say at $x_1=0$. Then we can consider the topological invariant $\widetilde{\rm Ch}_3(\tilde u_\Delta)$ and by applying the statement (v) of Proposition~\ref{coro-QHE4} we obtain
$$
(2 \pi)^2 \, \partial_{B_{3,4}} \widetilde{\mathfrak J}_2 
\;=\; 
(2\pi)^2 \, \partial_{B_{2,4}} \widetilde{\mathfrak J}_3 
\;=\; 
-\, \widetilde{\rm Ch}_3(\tilde u_\Delta) 
\;=\; -\, {\rm Ch}_4(p_F) \;\in\; \ZM
\;.
$$
This implies the existence of boundary currents in the second (third) direction whose variation w.r.t. the modulation of the incommensurate potential in the third (second) direction is quantized in  units of $\frac{1}{(2\pi)^2}$.

\section{Quantized electric polarization}
\label{sec-polar}

The electric polarization has two contributions, one from the displacements of the nuclei and one from the electrons. Here we will be dealing only with the latter contribution, which is often called the orbital polarization $P=(P_1,\ldots,P_d)$. It has been realized in the 1990's that $P$ itself is not a gauge-invariant and measurable quantity, but that the variation $\Delta P$ of the orbital polarization during adiabatic deformations of crystals  is gauge-invariant and measurable which is directly related to the flow of charges induced by such deformations (see \cite{RV,Res} for a historical account). If the deformation is periodic in time, it turns out that the orbital polarization is of topological nature and is actually the same quantity considered in charge pumps \cite{Tho}. This well known effect can now be placed in a broader context and several predictions can be made using the tools developed so far.

\vspace{.2cm}

Let be given a closed differentiable path $t\in[0,T]\mapsto h(t)\in\Aa_d$, $h(T)=h(0)$, of Hamiltonians satisfying the BGH at a fixed Fermi lelve $\mu$, and set $p_A(t)=\chi(h(t)\leq \mu)$ to be the instantaneous Fermi projection. Then it is shown in \cite{ST} that, up to arbitrarily small corrections in the adiabatic limit, the change in the electric polarization during one adiabatic cycle is
\begin{equation}
\label{eq-KV}
\Delta P_j\;=\;
\I \int^T_0 dt\;
\Tt \Big(
p_A(t)\,\big [\partial_t p_A(t), \partial_j p_A(t) \big ]\Big)
\;.
\end{equation}
This is the disordered version of the King-Smith-Vanderbilt formula for the orbital polarization \cite{KV}. Note that Eq.~\eqref{eq-KV} is invariant to the scaling of the time, hence $t$ can be seen as taking values on the unit circle $\SM^1\cong[0,2\pi)$. The r.h.s. is, up to a constant, the pairing of the projection $p_A=\big \{p_A(t)\big \}_{t\in \SM^1}$ with a $2$-cocyle over the algebra $C(\SM^1,\Aa_d)$, which is isomorphic to $\Aa_{d+1}$ if the periodic time dependence is interpreted as an extra space direction. To avoid confusion, we choose the time to be in the $0$-th direction. Then, from \eqref{eq-KV},
\begin{equation}
\label{eq-KV2}
\Delta P_j\;=\; 2\pi\,
\big\langle [\xi_{\{0,j\}}],[p_A]_0\big\rangle
 \;=\;
2\pi\;\Ch_{\{0,j\}}(p_A)
\;.
\end{equation}
Based on \eqref{eq-KV2}, Theorem~\ref{theo-PairingRange} and Corollary~\ref{coro-Ranges} gives the following prediction.

\begin{corollary}\label{Coro-DeltaP} The change in the components of the bulk electric polarization, after and adiabatic periodic cycle, depends only on the class $[p_A]_0 \in K_0(\Aa_{d+1})$ of the Fermi projection, and is equal to:
$$
\Delta P_j \;=\; 
 \sum_{ \{0,j\} \subseteq J \subseteq \{0,\ldots,d\} }
\beta_J\,(2\pi)^{ 1-\frac{|J|}{2}}  
\; {\rm Pf}\big (\BB_{J \setminus \{0,j\}} \big)  \;,
$$
with $|J|$ even and $\beta_J$ the integer numbers appearing in the decomposition of $[p_A]_0$ into the generators of the $K_0(\Aa_{d+1})$ group,
$$
[p_A]_0 \;=\; \sum_{J \subset \{0,\ldots,d\} } \beta_J [e_J]_0 \; ,
$$
as elaborated in Section~\ref{Sec-GeneratorsKGroups}. Above, it is assumed that ${\rm Pf}(\BB_\emptyset) =1$.
\end{corollary}

According to the above statement, $\Delta P_j$ can take only discrete values but these values are not necessarily integer. For example, for $d=1$ and $d=2$ the set $J$ can only be $\{0,j\}$, hence $\Delta P_j = \beta_{\{0,j\}}$ is always an integer, while for $d=3$ we have in general
$$
\Delta P_j\; =\; \beta_{\{0,j\}}\; + \;\beta_{\{1,2,3\}\setminus \{j\}} B_{\{1,2,3\}\setminus \{j\}}\;, \qquad j=1,2,3\;.
$$
Note, however, that the variation of the magneto-electric response coefficient 
$$
\partial_{B_{\{1,2,3\}\setminus \{j\}}} \Delta P_j \; =\; \beta_{\{1,2,3\}\setminus \{j\}}
\;, \qquad j=1,2,3\;,
$$
is an integer, a fact which will be addressed in more detail in Section~\ref{sec-MagEl}. Let us mention that, for $d=1$, the above quantization already appeared in the work of Thouless \cite{Tho}, while, for $d=2$, a non-trivial example manifesting this quantization is constructed in \cite{DL}, where an adequate loop of next-nearest hopping Hamiltonians on the hexagonal lattice is constructed. It will definitely be very interesting to test the prediction of Corollary~\ref{Coro-DeltaP} in dimension $d=3$. 

\vspace{.2cm}

Next let us show how the $K$-theoretic result of Section~\ref{sec-adiabatic} can be applied to obtain a further formula for the polarization.
Invoking \eqref{eq-pairsuspodd} in Theorem~\ref{theo-suspension} on the duality of pairings under the suspension map combined with Proposition~\ref{prop-adiabatic}, on obtains:
\begin{align*}
\big \langle {[\xi}_{\{0,j\}}],[p_A]_0 \big \rangle
&
\;=\;
\big \langle {[\xi}_{\{0,j\}}],[p_A]_0-[p_F]_0 \big \rangle
\\
& 
\;=\;
\big \langle [\xi_{\{j\}}],[p_Fv_{A,2\pi}p_F+\one_N-p_F]_1 \big \rangle
\;,
\end{align*}
where $v_{A,2\pi}$ is the Poincar\'e map of the adiabatic time evolution over one cycle, see Section~\ref{sec-adiabatic}. Now the r.h.s. can be written out more explicitly (using the identity $v_{A,2\pi}p_F=p_Fv_{A,2\pi}p_F$):
$$
\Delta P_j
\;=\;
2\pi\I\,\Tt\left(p_Fv_{A,2\pi}^*p_F\,\partial_j\big(p_Fv_{A,2\pi}p_F\big)\right)
\;.
$$
This is the stroboscopic interpretation of the polarization, expressing it in terms of the winding number of the adiabatic evolution over one cycle restricted to the range of the Fermi projection. Yet another formula for the polarization will be given in the next section.

\vspace{.2cm}

Next let us come to periodic loops of chiral systems. The following shows that their polarization vanishes.

\begin{proposition}
\label{prop-PChiral} 
Suppose that $t\in\SM^1\cong [0,2\pi)\mapsto h(t)\in\Aa_d$ is a loop of Hamiltonian satisfying the CH. Then $\Delta P$ given by \eqref{eq-KV} vanishes.
\end{proposition}

\noindent {\bf Proof.} Inserting $J^2=\one$ and using $Jp_AJ=\one-p_A$ on the r.h.s. of \eqref{eq-KV} shows $- \, \Delta P_j=2\pi\;\Ch_{\{0,j\}}(\one-p_A)$. But the homomorphism property of the pairing implies $\Ch_{\{0,j\}}(p_A)+\Ch_{\{0,j\}}(\one-p_A)=\Ch_{\{0,j\}}(\one)=0$ so that $\Delta P_j=-\Delta P_j=0$.
\hfill $\Box$

\vspace{.2cm}

Nevertheless, it is possible to associate a topological quantity to a loop of chiral systems, namely the chiral time polarization defined by
$$
P_{\mbox{\rm\tiny CT}}
\;=\;
\I\,\int^{2\pi}_0 dt\,\Tt(p_F(t)\,J\,\partial_tp_F(t))
\;.
$$
The chiral polarization $P_{\mbox{\rm\tiny C}}$ defined for a given chiral Hamiltonian (and not a loop of them) in Section~\ref{sec-ChiralPol} is quite similar. Following the calculation in the proof of Proposition~\ref{prop-PC} shows
$$
P_{\mbox{\rm\tiny CT}}
\;=\;
\tfrac{1}{2\I}\,\int^{2\pi}_0dt\,\Tt(u_F(t)^*\partial_t u_F(t))
\;=\;
-\,\tfrac{1}{2}\;
\big \langle [\xi_{\emptyset}^s],[u_F(t)_{t\in[0,2\pi)}]_1\big \rangle 
\;.
$$
The r.h.s. is, up to a factor, the winding number of the time-varying Fermi unitary operator, hence it is a stable topological number.  Using the Streda formula, one deduces for $1\leq i,j\leq d$
\begin{equation}
\label{eq-PCTStreda}
\partial_{B_{i,j}}\,
P_{\mbox{\rm\tiny CT}}
\;=\;
-\, \tfrac{1}{4\pi}\;
\big \langle [\xi_{\{i,j\}}^s],[u_F(t)_{t\in[0,2\pi)}]_1\big \rangle 
\;.
\end{equation}
In $d=2$, the r.h.s. is integer valued by the odd index theorem. For $d=3$ it is an integer valued weak invariant under the BGH.

\section{Boundary phenomena for periodically driven systems}
\label{sec-PerDriv}

In this section investigates the implications of the bulk-boundary correspondence for the periodically driven systems used for the definition of the orbital polarization in Section~\ref{sec-polar}. Thus let us consider a time-periodic family of half-space Hamiltonians
$$
t\in\SM^1\cong[0,2\pi)
\;\mapsto\;
\widehat{h}(t)\; =\; 
\big (h(t),\tilde h(t) \big ) 
\;\in\; 
\widehat{\Aa}_d
\;.
$$ 
This family is a lift of $t\in\SM^1\mapsto {h}(t)$ in the exact sequence of time period systems 
\begin{equation}
\label{eq-SeqTimePer}
\begin{diagram}
0 &\rTo & C(\SM^1,\Ee_{d})   &\rTo{i}  & C(\SM^1,\widehat{\mathcal A}_{d})  &\rTo{ \mathrm{ev}}  & C(\SM^1,\mathcal A_{d}) &\rTo &0
\end{diagram}
\;,
\end{equation}
which is just a reformulation of \eqref{ExactSequence}. In fact, if we see the time as another space direction, then \eqref{eq-SeqTimePer} is exactly \eqref{ExactSequence}. Now the bulk-boundary correspondence \eqref{EQ1} implies
$$
\Delta P_d 
\;=\; 
2\pi \, {\rm Ch}_{\{0,d\}}(p_A) 
\;=\;
2\pi \, \widetilde{\rm Ch}_{\{0\}}(\tilde u_\Delta)
\;,
$$
where the $0$-th component is still time and $[\tilde u_\Delta]_1=\Exp[p_A]_0$. Our goal here is to give a physical interpretation of the 1-cocycle appearing on the r.h.s.. According to Proposition~\ref{1Cocycle}
\begin{equation}
\label{OneCocycle}
\widetilde{\rm Ch}_{\{0\}}(\tilde u_\Delta) 
\;=\;
-\, 2\pi\,\int^{2\pi}_0\!\! dt\;\widetilde \Tt
\Bigl(\FFunc ' \big ( \hat{h}(t) \big )\,\partial_t \hat{h}(t)
\Bigr) 
\;.
\end{equation}
Following an argument from \cite{DS} (see Proposition~4 there), in the case $d=1$, the r.h.s. of \eqref{OneCocycle} is just $2\pi$ times the classical spectral flow \cite{Phi} of boundary eigenvalues of the path $t\in\SM^1\mapsto \hat{h}(t)$ through the bulk gap at $\mu$, 
$$
\Delta P_1\;=\; -\,2\pi\;\mbox{Sf}\big(t\in\SM^1\mapsto \hat{h}(t)\;\mbox{\rm by }\mu\big)
\,.
$$
The spectral flow counts the number of eigenvalues crossing the Fermi level from below minus the number of eigenvalues crossing from above during the adiabatic cycle. As one can immediately see, this is precisely the amount of charge pumped from the valence to the conduction states. For $d> 1$, the spectral flow in the above bulk-boundary correspondence has to be understood in a generalized  sense of Breuer-Fredholm operators (see \cite{BCP}), but its physical interpretation remains the same, as the charge per the unit area pumped during the adiabatic cycle. We will use the symbol Sf also for the spectral flow in this generalized sense.

\vspace{.2cm}

Let us briefly comment on the bulk-boundary correspondence for the chiral time polarization $P_{\mbox{\rm\tiny CT}}$ for paths of chiral Hamiltonians. As $P_{\mbox{\rm\tiny CT}}$ itself is given by the pairing with a $0$-cocycle, there is no bulk-boundary correspondence for it. On the other hand, for its derivatives w.r.t. a magnetic field perpendicular to the surface one has due to \eqref{eq-PCTStreda}:
\begin{align*}
\partial_{B_{i,d}}\,
P_{\mbox{\rm\tiny CT}}
& \;=\;
-\, \tfrac{1}{4\pi}\;
\big \langle [\xi_{\{i\}}^s],[\tilde{p}_\Delta(t)_{t\in[0,2\pi)}]_0\big \rangle 
\\
& \;=\;
-\, \tfrac{1}{4\pi}\;
\Big(
\big \langle [\xi_{\{i\}}^s],[\tilde{p}_+(\delta,t)_{t\in[0,2\pi)}]_0\big \rangle 
-\big \langle [\xi_{\{i\}}^s],[\tilde{p}_-(\delta,t)_{t\in[0,2\pi)}]_0\big \rangle 
\Big)
\;,
\end{align*}
where in the second identity it was supposed that $\pm\delta$ lie in gaps of the surface spectrum ({\it e.g.} opened by the magnetic field).

\section{The magneto-electric response in $d=3$}
\label{sec-MagEl}

The magneto-electic effect in an insulating material consists in the change of its electric polarization under a variation of the external magnetic field or, alternatively, the change of the magnetization under a variation of an electro-static potential. As in the previous section, we will be dealing only with the electron contributions to the effect. Now, let us consider a periodically driven system in dimension $d=3$ for which the orbital polarization is given by \eqref{eq-KV}. Then the change in the magneto-electric response coefficients per cycle is
$$
\Delta \, \alpha_{i,j,k}
\;=\;
\partial_{B_{i,j}} \,\Delta P_k
\;, 
\qquad \{i,j,k\}=\{1,2,3\}
\;.
$$
By using the connection given in \eqref{eq-KV2} and applying the generalized Streda formula from Theorem~\ref{theo-StredaGen}, we obtain
$$
\Delta \, \alpha_{i,j,k}
\;=\;
(-1)^\rho \,
\big\langle[\xi_{\{0,1,2,3\}}],[p_F]_0\big\rangle
\;=\;
(-1)^\rho \,
\Ch_4(p_F) \in \ZM
\;,
$$
where $\rho$ is the permutations which sends $\{i,j,0,k\}$ into $\{0,1,2,3\}$. The r.h.s. is the strong even pairing over the algebra $\Aa_{3+1}$ and hence integer-valued. A formula of this type already appeared in \cite{PLB}, but there an average over the space direction $k$ was taken and used. The above statement shows that all $3$ terms are in fact equal to the same invariant. In dimension $d=4$, which will be relevant for the virtual topological insulator discussed above, a similar statement holds, but the even pairing is only a weak invariant in this case.

\vspace{.2cm}

For a crystal with surface in $d=3$, we can use the bulk-boundary principle of \eqref{EQ1} in the following way
$$
\Delta \alpha_{1,2,3} 
\;=\; 
\partial_{B_{1,2}}\Delta P_3 
\;=\; 
\partial_{B_{1,2}}\,\widetilde{\rm Ch}_{\{0\}}(\tilde u_\Delta) 
\;= \;
-\, 2\pi\;\partial_{B_{1,2}}\mbox{Sf}\big(t\in\SM^1\mapsto \hat{h}(t)\;\mbox{\rm by }\mu\big)
\;.
$$
Hence, the spectral flow is not quantized but its variation with respect to the component of the magnetic field perpendicular to the surface is quantized:
$$
-\, 2\pi\;\partial_{B_{1,2}}\;\mbox{Sf}\big(t\in\SM^1\mapsto \hat{h}(t)\;\mbox{\rm by }\mu\big) 
\;=\; 
{\rm Ch}_4(p_F)
\;.
$$
This relations tells that, if ${\rm Ch}_4(p_F)\neq 0$, there is a spectral flow no matter where we place the Fermi level in the bulk gap. This implies that essential spectrum moves across the bulk gap as the time evolves, connecting the upper and lower parts of the bulk spectrum.

\newpage


\end{document}